\documentclass[%
	paper=A4,				
	twoside=true,				
	openright,					
	parskip=full,				
	chapterprefix=true,			
	12pt,						
	headings=normal,			
	bibliography=totoc,			
	listof=totoc,				
	titlepage=on,				
	captions=tableabove,		
	draft=false,				
	spanish,
]{scrreprt}%

\usepackage[utf8]{inputenc}		
\usepackage[T1]{fontenc}
\usepackage[spanish]{babel} 	
\addto\captionsspanish{} 
\usepackage[					
	figuresep=colon,%
	sansserif=false,%
	hangfigurecaption=false,%
	hangsection=true,%
	hangsubsection=true,%
	colorize=full,%
	colortheme=dani,%
	bibsys=bibtex,%
	bibfile=ms.bib,%
	bibstyle=numeric,%
]{dani}
\usepackage{xcolor}

\definecolor{colordani}{RGB}{0,88,94}
\definecolor{colordanibyn}{RGB}{105,105,105}

\newcommand{\esp}[1]{S^3\!/#1}

\usepackage{amsmath,amssymb}
\usepackage{enumitem,cleveref}
\usepackage{slashed}

\usepackage{setspace}

\hypersetup{					
	pdftitle={Teorías Cuánticas de Campos en Espacios Esféricos},	
	pdfsubject={Tesis Doctoral},
	pdfauthor={Daniela D'Ascanio},	
	plainpages=false,			
	colorlinks=false,			
	pdfborder={0 0 0},			
	breaklinks=true,			%
	bookmarksnumbered=true,		%
	bookmarksopen=true			%
}

\begin{document}

\renewcaptionname{spanish}{\figurename}{Fig.}
\renewcaptionname{spanish}{\tablename}{Tab.}

\pagenumbering{Roman}			
\pagestyle{empty}				

\begin{titlepage}
	\pdfbookmark[0]{Título}{Título}
	\tgherosfont
	\centering
	
\begin{minipage}{\textwidth}
\noindent\makebox[\textwidth][c]{%
	\colorbox{colordani}{%
		\parbox{\textwidth}{%
			\begin{minipage}{\textwidth}\centering{
				\includegraphics[width=4.5cm]{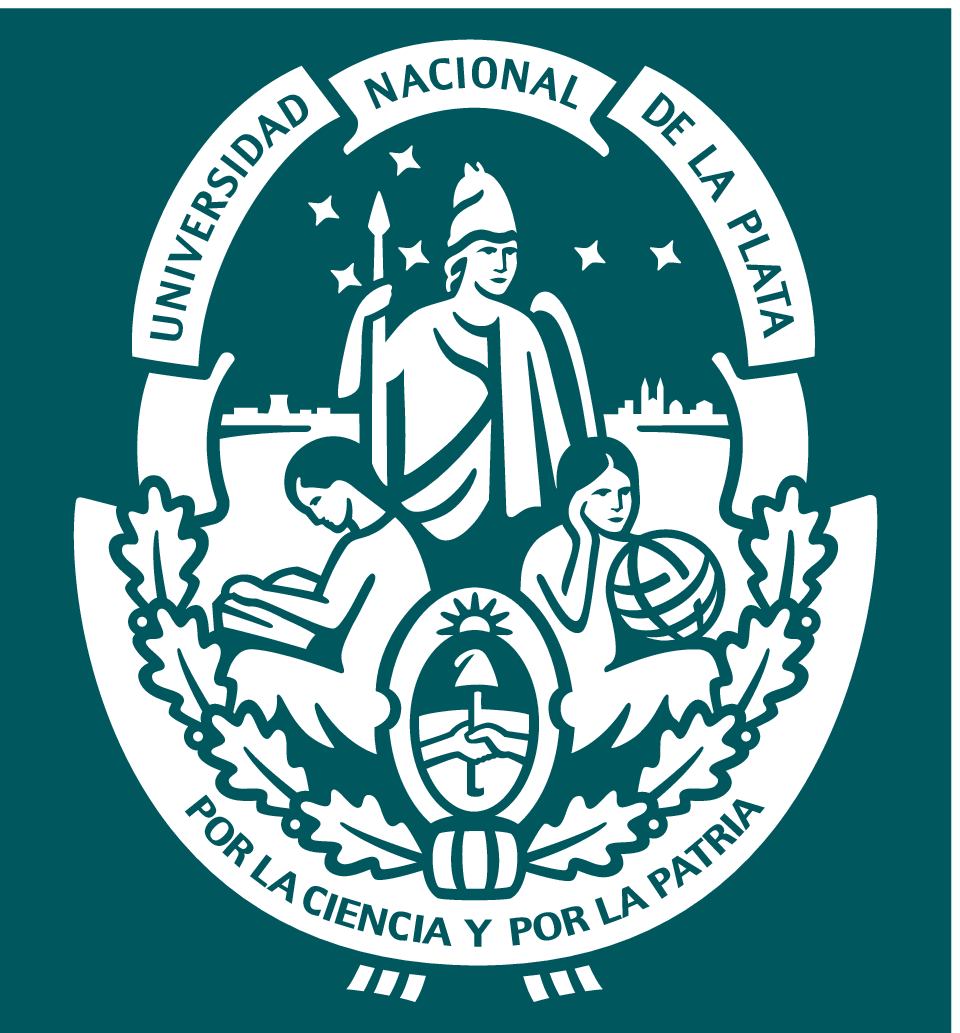}
			}
			\end{minipage}%
		}
	}
}
	\end{minipage}
    \\[18mm]

	\vfill
		{\large Thesis submitted to obtain the PhD degree in Physics}\\[10mm]
		{\LARGE \color{ctcolortitle}\textbf{Quantum Field Theory \\on spherical space forms} \\[4mm]}
        {\large (in Spanish)}\\[12mm]
        
	{\Large by} \\ [4mm]
	{\Large Daniela D'Ascanio} \\ [4mm]
	
	\vfill
	{\large Departamento de Física}\\[4mm]
	{\large Facultad de Ciencias Exactas}\\[4mm]
	{\large Universidad Nacional de La Plata}\\[25mm]
	{\large 2018}

\end{titlepage}

\hfill
\vfill

\pagebreak

\null
\vfill
\begin{flushright}
	\begin{minipage}{.86\textwidth}%
		\begin{flushleft}
			\emph{Para todos nós descerá a noite e chegará a dilig\^encia. Gozo a brisa que me d\~ao e a alma que me deram para gozá-la, e n\~ao interrogo mais nem procuro. Se o que deixar escrito no livro dos viajantes puder, relido um dia por outros, entret\^e-los também na passagem, será bem. Se n\~ao o lerem, nem se entretiverem, será bem também.}
		\end{flushleft}
		\begin{flushright}
			--- \textrm{Bernardo Soares (Fernando Pessoa)} \\
			\emph{Livro do desassossego}
		\end{flushright}
	\end{minipage}\\
\end{flushright}

\pagebreak

\restoregeometry		
\cleardoublepage

\pagestyle{plain}				
\pdfbookmark[0]{Abstract}{Abstract}
\chapter*{Abstract}
\label{sec:abstract_en}
\vspace*{-10mm}

One of the fundamental questions in Quantum Field Theory regards the determination of a measure of the degrees of freedom of theories that is consistent with the Renormalization Group flow. The answer seems to be encoded in the $C$-theorems, which involve quantities that decrease with the Renormalization Group flow to the IR and are stationary at the fixed points, thus ordering the space of theories.

In an originally different problem, inspired by the use of spherical space forms in cosmological models, we study the thermodynamic properties of a free theory at finite temperature defined on such spaces. We start by analyzing the case of a conformal scalar theory: from the zeta regularization of the effective action we compute the entropy, in whose high-temperature expansion we find a term ---often disregarded--- which does not depend on the temperature nor the radius of the covering sphere, which can be obtained also as the determinant of the zero-temperature theory on the spatial manifold, and which we relate to its topological properties. We consider in the case of a massive theory the same expansion, whose temperature-independent term depends on the mass, and we find that dependence resemblant to the behavior of a $C$-quantity. We then analyze the behavior of the same quantity for the free scalar and Dirac theories on real projective spaces in arbitrary dimension.		
\cleardoublepage

\setcounter{tocdepth}{2}		
\tableofcontents				
\cleardoublepage
%
%
\pdfbookmark[0]{Agradecimientos}{Agradecimientos}
\chapter*{Agradecimientos}
\label{sec:acknowledgement}
\vspace*{-10mm}

\begin{minipage}{.865\textwidth}%
	\begin{flushright}
		\begin{minipage}{.65\textwidth}%
			\begin{flushleft}
				\textit{Quiero escribir, pero me siento puma;\\
					quiero laurearme, pero me encebollo.\\
					No hay [voz] hablada que no llegue a bruma\\
					no hay dios ni hijo de dios, sin desarrollo.}
			\end{flushleft}
			\begin{flushright}
				--- \textrm{César Vallejo,}
				\emph{Intensidad y Altura}
			\end{flushright}
		\end{minipage}%
	\end{flushright}
\end{minipage}%

\bigskip

Con estas palabras cumplo a la vez con mi obligación hacia las instituciones y personas que hicieron posible el desarrollo de mi investigación doctoral y con el deseo de incluir explícitamente a las que me marcaron el camino que en algún sentido termina con esta tesis.
La combinación de todas ellas refleja las oportunidades que tuve;
el resultado es enteramente mi responsabilidad.

Agradezco profundamente a la educación pública y gratuita argentina, sin la cual posiblemente no estaría escribiendo estas líneas: a la Escuela número 19 ``Santiago del Estero'' de Villa Nueva, a la Escuela número 7 ``General Enrique Mosconi'' y a la Escuela de Educación Media número 1 ``Raúl Scalabrini Ortiz'' de Berisso y al Departamento de Física y la Facultad de Ciencias Exactas de la Universidad Nacional de La Plata.

Al Consejo Nacional de Investigaciones Científicas y Técnicas y al Departamento de Física de la UNLP por haber financiado las actividades de investigación que dieron lugar a esta tesis. A todas las instituciones que financiaron mis viajes: la UNLP, el ICTP-SAIFR, la Universidade Federal do ABC, el CONACYT mexicano, la Asociación Universitaria Iberoamericana de Postgrado y la Universidad de Zaragoza. 
No quiero dejar de mencionar que fue gracias al empuje que el gobierno nacional dio a las actividades científicas en el período 2003-2015 que mi generación creció en la Física a la luz de las oportunidades para apostar por una carrera en investigación. Agradezco entonces, también, haber podido asistir a ese ``cambio de época''.

\vfill\pagebreak

\ 

\vspace{1cm}

Los resultados contenidos en esta tesis no podrían haber sido obtenidos por mí en soledad. Agradezco a Mariel Santangelo y Gabriela Beneventano, por haber compartido conmigo uno de sus temas de investigación, por la inmensa cantidad de tiempo que pasaron trabajando en él, por el rigor y la responsabilidad con que lo hicieron, y por su tolerancia y guía durante estos años. A Mariel, además, por estar siempre predispuesta a mis preguntas y comentarios acerca de cualquier aspecto de la vida académica, por incentivar mi incursión en otros temas, y por su cuidadosa lectura y correcciones a este manuscrito. Agradezco también a Manuel Asorey e Inés Cavero-Peláez por el trabajo invertido en este proyecto y (junto con Lita) por la calidez con que me recibieron las veces que los visité. A Manolo le agradezco además por todo lo que aprendí trabajando con él, por las ideas con que impulsó este trabajo, por el entusiasmo que me contagió cada vez que el partido parecía perdido y por su paciencia para buscar la explicación que me convenciera, aunque ambos supiéramos que él iba a tener razón.

A mis colaboradores en proyectos que no están presentados en esta tesis: Naser Ahmadiniaz, Olindo Corradini, Sendic Estrada-Jiménez, Peter Gilkey, Pablo Pisani y Dmitri Vassilevich. Por las horas y los temas de trabajo compartidos.

A Horacio Falomir, que sin ser parte directa de este proyecto siempre estuvo relacionado de alguna manera con mi doctorado, por su generosa,  paciente y permanente presencia.

Agradezco a Pablo, a Gabriela y a Adrián Lugo por compartir conmigo su oficina los primeros tiempos de mi doctorado, y a Mariel y Horacio por hacerme un lugar con ellos en los últimos.

Agradezco a Horacio Falomir y Fernando Lombardo por haber evaluado este trabajo en una etapa intermedia. A Fernando Lombardo, Gerardo Rossini y Paula Villar por haber conformado el jurado que evaluó esta tesis.

\vfill\pagebreak

\begin{minipage}{.865\textwidth}%
	\begin{flushright}
		\begin{minipage}{.8\textwidth}%
			\begin{flushleft}
				\emph{Though fast youth's glorious fable flies,\\
					View not the world with worldling's eyes;\\
					Nor turn with weather of the time.\\
					Foreclose the coming of surprise:\\
					Stand where Posterity shall stand;\\
					Stand where the Ancients stood before,\\
					And, dipping in lone founts thy hand,\\
					Drink of the never-varying lore:\\
					Wise once, and wise thence evermore.}
			\end{flushleft}
			\begin{flushright}
				--- \textrm{Herman Melville,}
				\emph{Lone Founts}
			\end{flushright}
		\end{minipage}%
	\end{flushright}
\end{minipage}%

\bigskip

A mi familia: a la Gra y el Caño, porque por ellos tuve una biblioteca antes que un televisor, y porque ya completaron los doce trabajos pero siguen buscando casa en San Clemente para todos. 
A la Gra, por dar sin pedir nada a cambio y preferir palpar a pisar, tomar a pedir, bailar a desfilar y disfrutar a medir.
Al Caño, por predicar el valor de vivir siempre de la misma forma. 
A la tía Pelu, por el ejemplo de cómo vivir sin nunca rendirse.
A la Ara, Martín y la Delfi, porque saben mejor que nadie que nada se consigue sin esfuerzo. 
Al Erni, por no sentirse más de lo que es ni menos de lo que debe ser. 

A mi familia más amplia: a los D'Ascanio, Larrañaga, Machuca, Álvarez, Nagavonsky, Milloc, Segovia, González-Pisani, por acompañarme siempre sin condiciones, y sin dejar de pensar de mí lo mejor. Muy en especial a la Elsi y el Ricky, que junto con mis viejos son mis primeros y grandes maestros, a quienes siempre intentaré imitar la forma de enfrentar la vida.

A Pablo, aunque para agradecerle con justicia tendría que escribir otra tesis. Porque sabe los nombres extraños de las yerbas y las flores, y me hace ver del mundo cosas que sola no podría descubrir. Porque en el trabajo como en la vida sabe capturar la esencia de los problemas más complicados. Porque con paciencia inmensurable aguantó los vaivenes de mi doctorado ---incluyendo el largo proceso de escritura de esta tesis--- sin perder nunca la confianza, enseñándome a tomar con calma tanto los momentos bajos como los altos. Porque sin él no sólo nada de esto habría sido posible, sino que tampoco tendría sentido.

A Néstor Kirchner y Cristina Fernández, por haberme dado la oportunidad de vivir y estudiar en un país del que siento orgullo.

\cleardoublepage

\pagestyle{maincontentstyle} 	
\pagenumbering{arabic}			
\setcounter{page}{1}			

\chapter{Introducción}
\label{sec:intro}

En esta tesis estudiamos propiedades de teorías cuánticas de campos definidas sobre ciertas variedades esféricas con topología no trivial. Presentaremos resultados relacionados con la Termodinámica de campos escalares sobre dichos espacios y su relación con los teoremas $C$ en Teoría Cuántica de Campos.
 
No obstante los grandes avances en Teoría Cuántica de Campos sobre espacios curvos \cite{Birrell:1982ix}, las teorías sobre espacios con topología no trivial no han sido estudiadas en forma sistemática. La historia de la topología en la Física Cuántica se remonta por lo menos al monopolo magnético de Dirac \cite{dirac1931quantised} y, pasando por el efecto Aharonov-Bohm \cite{aharonov1959significance}, es motivo de una gran revolución con la Teoría Cuántica de Campos Topológica \cite{witten1989quantum} y su realización en sistemas de Materia Condensada, que da lugar a las fases topológicas de la materia \cite{haldane1988model,thouless1982quantized}. Por otra parte, en los intentos de formular una teoría que permita comprender los primeros momentos del universo ---en la que los fenómenos cuánticos tendrían un rol preponderante--- no hay indicios que impliquen la necesidad de que la topología del espacio sea trivial \cite{starkman1998topology}. Los efectos cuánticos debidos a dicha topología se manifiestan a bajas energías ---razón por la cual la información acerca de la topología del universo se busca en los momentos multipolares más bajos del espectro angular de variaciones térmicas del fondo cósmico de radiación \cite{Levin:2001fg}---.

Inspirados en el estudio de modelos cosmológicos sobre espacios topológicamente no triviales \cite{Aurich:2013fwa,Luminet:2003dx}, consideramos la teoría cuántica de un campo escalar sobre algunos espacios con esa característica, centrándonos en particular en la obtención de las propiedades termodinámicas de los campos a temperatura finita. En el análisis de la dependencia de dichas propiedades con la masa encontramos un comportamiento semejante al de las llamadas cantidades $C$. 

\subsection*{Teoremas C en Teoría Cuántica de Campos}

De nuestra experiencia cotidiana podemos deducir que la complejidad de un sistema físico aumenta cuando lo miramos en mayor detalle ---lo que en términos más técnicos podríamos enunciar diciendo que el número de grados de libertad del sistema aumenta con la energía de los procesos involucrados en la observación---, lo cual es cierto en principio si la Física que describe al sistema es la misma para cualquier escala de energías. En Teoría Cuántica de Campos, donde las cantidades pueden depender de la escala de manera no intuitiva, esta idea es abordada en el contexto de los llamados \emph{teoremas $C$}, que involucran cantidades que de alguna manera miden el número de grados de libertad de la teoría y que decrecen monótonamente cuando la teoría se mueve entre altas y bajas energías.

En el espacio de teorías, las curvas que conectan distintas teorías se recorren variando la escala de energía en lo que se conoce como flujo del grupo de renormalización; éste se realiza en términos de la dependencia de las constantes de acoplamiento de las teorías con la escala. Típicamente, al mover la escala entre altas (en el ultravioleta) y bajas energías (en el infrarrojo) la teoría se mueve entre dos teorías conformes, las cuales son puntos fijos del flujo. Este cambio con la escala puede ser interpretado como un cambio en el número de grados de libertad de la teoría; en ese contexto, una cantidad $C$ da una medida del comportamiento de las teorías cuánticas de campos a altas y bajas energías (ó, lo que es análogo, a cortas y largas distancias).

El problema de la medida del número de grados de libertad en el espacio de teorías está entendido en dos dimensiones \cite{Zamolodchikov:1986gt}, donde la cantidad $C$ coincide en los puntos fijos con la carga central de la correspondiente teoría conforme, que además es proporcional al coeficiente $c$ de la anomalía de traza de la teoría cuántica. Esta relación inspiró la búsqueda de cantidades con propiedades similares en cuatro dimensiones \cite{Cardy:1988cwa}, donde los coeficientes de la anomalía de traza son dos; en este número de dimensiones se demostró \cite{Komargodski:2011vj} que el coeficiente $a$ (el coeficiente del término de Euler) es el que se comporta como una cantidad $C$. En el resto de las dimensiones pares puede hacerse una propuesta similar. En dimensiones impares, en cambio, la simetría conforme no es anómala \cite{Deser:1993yx}, de modo que hay que considerar el problema desde un enfoque diferente. Además de estudios holográficos \cite{Myers:2010tj}, la propuesta más firme en tres dimensiones \cite{Jafferis:2011zi} proviene de la observación de que la carga central en dos dimensiones puede relacionarse también con la función de partición euclídea $Z$ de la teoría en la esfera; la energía libre $F=-\log Z$ es en ese caso el valor de la cantidad $C$ en el punto fijo ultravioleta, mientras que la función que interpola en las teorías intermedias del flujo se construye a partir de la entropía de entrelazamiento del estado fundamental de la teoría entre un círculo y su complemento en el espacio \cite{Casini:2012ei,Liu:2012eea}. Quedan aún aspectos por resolver, que tienen que ver con el comportamiento en el punto fijo infrarrojo y la estabilidad en el ultravioleta \cite{Beneventano:2017eyu}.

\subsection*{Estructura de la tesis}

En la descripción de las propiedades de teorías cuánticas de campos a temperatura finita, la obtención de magnitudes físicas está relacionada con el cálculo de determinantes funcionales de operadores diferenciales. En el capítulo \ref{sec:mate} hacemos un resumen de la formulación perturbativa de la Teoría Cuántica de Campos, mostrando cómo aparecen los determinantes funcionales en el cálculo de correcciones cuánticas a una teoría clásica de campos, y repasamos la relación entre las correcciones al orden de un loop y la Mecánica Estadística del sistema. Mencionamos además la necesidad de una regularización en la definición de dichos determinantes funcionales y describimos la que usaremos a lo largo de la tesis. 

Los espacios esféricos tridimensionales, sobre los que definiremos las teorías, son introducidos en el capítulo \ref{sec:espacios}; comenzamos con una presentación del área de la Cosmología que estudia la topología del universo, que esperamos sirva de motivación para la consideración de teorías sobre espacios con topología no trivial. Describimos luego en detalle una forma de construir dichos espacios a partir de las simetrías de la esfera tridimensional, con la que mostramos cómo obtener el espectro del laplaciano sobre ellos.

Usando dichos espectros, en el capítulo \ref{sec:termo} calculamos las propiedades termodinámicas de una teoría escalar libre sin masa acoplada conformemente a la métrica en cada uno de los espacios esféricos. Comenzamos analizando el caso de la esfera, cuyo conocimiento simplificará el posterior tratamiento del resto de los espacios. En todos los casos, usamos propiedades de dualidad de las series infinitas que aparecen al calcular el determinante funcional de interés para obtener dos expresiones diferentes de la acción efectiva, ambas válidas en todo el rango de temperaturas, una de las cuales nos permite tomar fácilmente el límite de altas temperaturas, y la otra el de bajas temperaturas. Analizamos la estructura de las cantidades termodinámicas en ambos límites, y observamos la aparición en el límite de altas temperaturas de un término que es independiente de las escalas de la teoría ---la temperatura y el volumen del espacio---. En la entropía de los campos sobre los espacios esféricos con topología no trivial, la comparación de este término con el correspondiente en la esfera da lugar a la definición de una cantidad que llamamos \textit{entropía de holonomía} \cite{Asorey:2012vp}.

En el capítulo \ref{sec:masivo} generalizamos estos cálculos al caso de un campo masivo, analizando la dependencia de las cantidades termodinámicas de la teoría con la masa del campo. En el capítulo \ref{sec:holonomia} sugerimos la posibilidad de que la entropía de holonomía cumpla las condiciones para representar una medida del número de grados de libertad de las teorías en el flujo de masa del campo escalar conforme sobre los distintos espacios esféricos \cite{Asorey:2014gsa}. Dicho capítulo comienza con un repaso histórico de los teoremas $C$ en Teoría Cuántica de Campos y continúa con el análisis de algunas cantidades candidatas a funciones $C$ sobre variedades esféricas tridimensionales, con énfasis en el comportamiento de las mismas en los puntos fijos ultravioleta e infrarrojo \cite{Beneventano:2017eyu}, para pasar finalmente al estudio detallado de la entropía de holonomía. Luego de analizar el comportamiento de la entropía de holonomía con la masa, testeamos la mencionada posibilidad en el caso del flujo de masa de teorías escalares en dimensiones más altas y en el caso de un campo de Dirac sobre uno de los espacios esféricos en dimensión arbitraria. 

Finalmente, en el capítulo \ref{sec:conclusiones} presentamos un resumen de los resultados obtenidos, junto con las conclusiones del trabajo y las posibles direcciones de acción.
 
%
\chapter{Elementos de la Teoría Cuántica de Campos}
\label{sec:mate}

\begin{minipage}{.885\textwidth}%
	\begin{flushright}
		\begin{minipage}{.7\textwidth}%
			\begin{flushleft}
				\emph{Mas no conseguiréis amurallar la ciudad prometida antes de que un hambre cruel os obligue, por la ofensa que nos habéis causado,\\ a devorar a dentelladas vuestras propias mesas.}
			\end{flushleft}
			\begin{flushright}
				--- \textrm{(Celeno a Eneas)} \\
				Virgilio, \emph{La Eneida}
			\end{flushright}
		\end{minipage}%
	\end{flushright}
\end{minipage}%
\bigskip

Como ya hemos mencionado, esta tesis se centra en el estudio de las propiedades termodinámicas ---y cantidades derivadas de ellas--- de teorías cuánticas de campos sobre espacios esféricos con topología no trivial. Estas cantidades serán calculadas a partir de las propiedades espectrales de operadores asociados a la teoría. Es objetivo de este capítulo hacer un resumen de algunos de los conceptos involucrados en los cálculos que describiremos, de modo de hacer autocontenida la lectura de la tesis, a la vez que introducimos algo de notación.

\section{La acción efectiva}
\label{sec:mate:seff}

Dada una teoría de campos definida por la acción clásica $S[\Phi]$ ---donde $\Phi$ denota de manera colectiva al conjunto de campos que componen la teoría---, la acción efectiva es otra funcional de $\Phi$ que contiene las correcciones cuánticas a la teoría de modo que el principio de mínima acción (efectiva) conduce a las ecuaciones de movimiento para los valores de expectación de vacío de los campos.

A partir de aquí consideramos ---por economía de notación--- la teoría de un campo escalar real $\phi$ definido sobre una variedad de sección espacial $\mathcal{M}$, cuya acción clásica euclídea podemos escribir como 
\begin{align}\label{eq:sclas}
S[\phi] = \frac12\int_{\mathbb{R}\times\mathcal{M}} d^dx
\sqrt{g} \left\{ g^{\mu\nu} \partial_\mu \phi\,\partial_\nu \phi +m^2\phi^2
+ \xi R\,\phi^2
\right\}\,,
\end{align}
donde hemos supuesto que el campo tiene una masa $m$ y está acoplado a la curvatura escalar $R$ de la variedad, siendo $\xi$ el coeficiente de dicho  acoplamiento. El acoplamiento se denomina mínimo si $\xi=0$. En el caso de un acoplamiento conforme ---en el que la teoría a masa cero posee la simetría de Weyl y el valor de vacío de la traza del tensor de energía impulso se anula--- a la métrica de la variedad $d$-dimensional, tendremos $\xi=(d-2)/4(d-1)$.

En presencia de una fuente externa (clásica) $J(x)$ la función de partición $Z[J]$ y la funcional generatriz $W[J]$ de la teoría euclídea pueden obtenerse a partir de la integral funcional
\begin{align}\label{eq:particion}
Z[J]\equiv e^{-\frac1\hbar W[J]}:=\int \mathcal{D}\phi\ e^{-\frac1\hbar S[\phi]+\frac1\hbar \left(J,\phi\right)}\,,
\end{align}
donde hemos denotado $(\cdot,\cdot)$ al producto escalar de funciones en la variedad $\mathbb{R}\times\mathcal{M}$ con métrica euclídea, y donde la normalización se ha elegido de modo que las derivadas funcionales de $Z[J]$ den lugar a las funciones de correlación de la teoría según la expresión
\begin{align}\nonumber
\langle \phi(x_1)\phi(x_2)\ldots\phi(x_N)\rangle = \frac{\hbar^N}{Z[0]}\left. \frac{\delta}{\delta J(x_N)}\ldots\frac{\delta}{\delta J(x_2)}\frac{\delta}{\delta J(x_1)} Z[J]\right\vert_{J=0}\,.
\end{align}
Es en este sentido que decimos que $Z[J]$ es la funcional generatriz de las funciones de correlación de la teoría, siendo $W[J]$ la funcional generatriz de los correladores que en el formalismo de Wick-Feynman corresponden a diagramas conexos. 

La acción efectiva se obtiene como la transformada de Legendre (funcional) de $W[J]$ en términos del campo medio $\phi_J(x):=-\delta W[J]/\delta J(x)$:
\begin{align}\label{eq:seff}
\Gamma[\phi_J]:= W[J]+\left(J,\phi_J\right)\,,
\end{align}
de donde vemos que
\begin{align}\label{eq:seffder}
\frac{\delta \Gamma[\phi_J]}{\delta\phi_J(x)}=J(x)\,,
\end{align}
de modo que en ausencia de fuentes externas la acción efectiva es extremizada por el campo medio $\phi_J$. Este campo medio no es otra cosa que el valor de expectación de vacío del campo cuántico $\phi$ en presencia de la corriente externa $J$. La acción efectiva resulta además la funcional generatriz de las funciones de correlación correspondientes a diagramas de Feynman irreducibles de una partícula. Estos diagramas son suficientes para construir todos los diagramas de la teoría a partir de los de orden árbol, de modo que la acción efectiva resulta adecuada para describir todo el comportamiento cuántico de la teoría.

\section{Correcciones a un loop a la teoría clásica}

Si bien en algunos casos es posible calcular la acción efectiva en forma exacta, la forma usual de encarar la mayoría de los problemas es la utilización de una aproximación semiclásica ---el desarrollo en potencias de $\hbar$---. En lo que concierne a los resultados contenidos en esta tesis, como veremos más adelante, la primera corrección a la acción clásica en este desarrollo permite obtener la termodinámica de los campos.

Para obtener dicho desarrollo hacemos en \eqref{eq:particion} el cambio de variables $\phi(x)\rightarrow \phi_{J}(x)+\phi(x)$, con el que obtenemos
\begin{align}
Z[J] &= e^{-\frac1\hbar S[\phi_{J}]+\frac1\hbar\left(J,\phi_{J}\right)} \int \mathcal{D}\phi\ e^{-\frac1\hbar\left(\delta S-J,\phi\right)-\frac{1}{2\hbar}\left(\phi,\left(\delta^2S,\phi\right)\right)}\,,
\end{align}
donde hemos denotado
\begin{align}
\delta S(x):= \left.\frac{\delta S[\phi]}{\delta\phi(x)}\right\vert_{\phi=\phi_{J}}\,\qquad \delta^2S(x,y):= \left.\frac{\delta^2 S[\phi]}{\delta\phi(x)\delta\phi(y)}\right\vert_{\phi=\phi_{J}}\,.
\end{align}

Luego del cambio de escala $\phi(x)\rightarrow\sqrt{\hbar}\,\phi(x)$ obtenemos para la acción efectiva
\begin{align}
\Gamma[\phi_J] = S[\phi_J] -\hbar\log\int\mathcal{D}\phi\, e^{-\frac12 \left(\phi,\left(\delta^2S,\phi\right)\right)} +\mathcal{O}(\hbar^2)\,,
\end{align}
expresión que, como la corriente externa es una función arbitraria, y como ésta determina de manera unívoca el campo medio, está definida para configuraciones arbitrarias del campo. Notemos además que en el límite $\hbar\rightarrow 0$ la integral funcional en \eqref{eq:particion} está dominada por la configuración clásica $\phi_{\mathrm{clás}}(x)$, que satisface
\begin{align}
\left.\frac{\delta S[\phi]}{\delta\phi(x)}\right\vert_{\phi=\phi_{\mathrm{clás}}} = J(x)\,,
\end{align}
y que es, en esta aproximación, igual al campo medio $\phi_J(x)$.

La integral funcional en la primera corrección a la acción clásica puede entenderse como el determinante funcional del \emph{operador de fluctuaciones cuánticas} $A$ cuyo núcleo es $\delta^2S$: \emph{grosso modo}, suponiendo que las autofunciones $\varphi_n$ de $A$ forman un conjunto completo en el espacio de funciones sobre la variedad,\footnote{ \,Esto está asegurado en particular en los casos en que la métrica del espacio sea definida positiva. Notemos que en este caso el operador $A$ es hermítico, puesto que su núcleo es simétrico.} podemos desarrollar el campo como 
\begin{align}
\phi(x) =\sum_n c_n \varphi_n(x)\,, \qquad \mathrm{con} \quad c_n=(\varphi_n,\phi)\,,
\end{align}
con lo que
\begin{align}
\left(\phi,A\phi\right) = \sum_{n,m} c_n c_m (\varphi_n,A\varphi_m) = \sum_n \lambda_n c_n^2\,,
\end{align}
donde $\lambda_n$ es el autovalor de $A$ correspondiente a la autofunción $\varphi_n$. Por otra parte, la medida en la integral funcional puede calcularse ---a menos de alguna normalización--- como el producto \cite{Hawking:1976ja}
\begin{align}
\mathcal{D}\phi=\prod_n dc_n\,,
\end{align}
de modo que podemos escribir
\begin{align}
\int\mathcal{D}\phi e^{-\frac12 \left(\phi,A\phi\right)} &\sim \int \left(\prod_n dc_n\right) e^{-\frac12 \sum_n\lambda_nc_n^2} = \prod_n\int dc_ne^{-\frac12 \lambda_nc_n^2}\\\nonumber
& \sim \prod_n\lambda_n^{-1/2}\,.
\end{align}
Identificando finalmente el producto de los autovalores del operador con su determinante funcional, resulta
\begin{align}\label{eq:logdet}
\Gamma^{(1\mathrm{-loop})}[\phi] = \frac12 \log\mathrm{Det}\,A\,.
\end{align}
En adelante denotaremos a esta corrección simplemente $\Gamma$. Además, consideraremos unidades en las que $\hbar=1$. Notemos que esta expresión para la corrección a un loop a la acción efectiva es válida para campos bosónicos; en el caso de un campo fermiónico $\psi$, las propiedades de la integral funcional sobre variables de Grassmann conducen en cambio a la corrección
\begin{align}\label{eq:logdetferm}
\Gamma^{(1\mathrm{-loop})}[\psi] = - \log\mathrm{Det}\,A\,.
\end{align}

Como las teorías cuánticas de campos son en general locales, los operadores de fluctuaciones cuánticas son operadores diferenciales, como el de Laplace-Beltrami en el primer término entre llaves en \eqref{eq:sclas}. Estos operadores son no acotados, lo que significa en particular que sus autovalores crecen sin límite, y el determinante resulta entonces una cantidad infinita. Es por eso que es necesario adoptar alguna forma de regularización de esa cantidad de modo de obtener a partir de ella una cantidad finita que podamos asociar con el determinante. Más adelante en el capítulo veremos una forma de regularizar el determinante funcional de un operador a partir de sus autovalores, que es la que usaremos a lo largo de la tesis. Antes de eso, supongamos que podemos calcular de alguna forma la corrección a un loop a la acción efectiva de una teoría, y veamos cómo podemos obtener a partir de ella las propiedades termodinámicas de los campos a temperatura finita.

\section{Campos cuánticos a temperatura finita}

En primer lugar, tenemos que introducir el concepto de temperatura en teoría de campos. Para ello, partiendo de la teoría euclídea, confinamos el tiempo euclídeo $\tau$ a un intervalo $[0,\beta]$ e imponemos ciertas condiciones de contorno sobre el campo en los extremos del intervalo: periódicas para reproducir la mecánica estadística de un campo bosónico y antiperiódicas cuando queramos describir la teoría de un campo fermiónico a temperatura finita \cite{bernard}. En el caso de un campo bosónico $\phi$ la teoría queda entonces definida sobre el espacio $S^1\times\mathcal{M}$, donde la circunferencia $S^1$ tiene radio $\beta/2\pi$. La función de partición $Z[\phi]$ resultante es la función de partición gran canónica de la mecánica estadística del campo a temperatura $1/\beta$.

\vfill\pagebreak

En el caso de la teoría escalar \eqref{eq:sclas} ---que es el que ocupará la mayor parte de la tesis---, luego de la introducción de una temperatura $1/\beta$ el operador de fluctuaciones cuánticas tiene la forma
\begin{align}
A = -\partial_{\beta}^2 - \triangle + \xi R +m^2\,,
\end{align}
donde
\begin{align}\label{eq:lapl}
\triangle=\frac1{\sqrt{g}}\, \partial_i \left(g^{ij}{\sqrt{g}}\,\partial_j\right)\,
\end{align}
es el operador de Laplace-Beltrami sobre la variedad $\mathcal{M}$ de métrica $g_{ij}$, al que en adelante llamaremos simplemente laplaciano. 
 
Separando la variable $\tau$ en la ecuación de autovalores y teniendo en cuenta la condición de periodicidad para dicha variable vemos que las contribuciones correspondientes ---que podemos llamar ``térmicas''--- están dadas por las frecuencias de Matsubara $\omega _l =2\pi l/\beta$ con $l\in\mathbb{Z}$, mientras que las contribuciones espaciales son, en el caso en que la curvatura escalar sea constante, los autovalores del laplaciano sobre $\mathcal{M}$ desplazados una cantidad $\xi R+m^2$.

Teniendo en cuenta que la acción efectiva $\Gamma$ está dada por (menos) el logaritmo de la función de partición de la teoría, las propiedades termodinámicas del campo pueden obtenerse a partir de ella como en el formalismo canónico de la Mecánica Estadística. En particular, nos interesan la energía $E=\partial_{\beta}\Gamma$, la energía libre $F=\Gamma/\beta$ y la entropía $S=\beta (E-F)$.

\section{Cálculo del determinante funcional. Funciones espectrales}

Como ya dijimos, para los operadores de fluctuaciones cuánticas que nos interesan el determinante funcional entendido como el producto de los autovalores no está bien definido. Con el objetivo de obtener para él una cantidad finita, en esta tesis utilizaremos la llamada \emph{regularización zeta}\footnote{ \,Una nota sobre la bibliografía: aunque hay antecedentes del uso de funciones zeta en la regularización de cantidades físicas \cite{Brown:1969na,ray1971r,Salam:1974xe}, es generalmente aceptado que la introducción de la regularización zeta como tal fue realizada en \cite{Dowker:1975tf}.}, en la que el logaritmo del determinante es entendido como la derivada
\begin{align}\label{eq:logdetzeta}
\log\mathrm{Det}\,A :=-\left.\frac{d}{ds}\zeta_A(s)\right\vert_{s=0}\,,
\end{align}
donde la \emph{función zeta} del operador $A$ se construye como 
\begin{align}\label{eq:zeta}
\zeta_A(s):=\mathrm{Tr}\,A^{-s}\,
\end{align}
en la región del plano complejo $s$ donde resulte convergente, o su extensión analítica en el resto del plano. En el caso en que el espectro del operador sea discreto, $\{\lambda_n\}_{n\in\mathbb{N}}$, la función zeta es la suma
\begin{align}
\zeta_A(s)=\sum_{n\in\mathbb{N}}\lambda_n^{-s}\,.
\end{align}

En rigor \cite{seeley,seeleyzeta}, para un operador diferencial elíptico $A$ de orden $n$ sobre una variedad $d$-dimensional, que sea autoadjunto y definido positivo, la función zeta como la definimos es holomorfa para $\Re(s)>d/n$ y admite una extensión analítica al resto del plano complejo que es meromorfa ---con un número finito de polos, que no pueden estar en puntos que no sean $s=s_k=(d-k)/n$, con $k=0,1,\ldots$ y $k\neq d$ \cite{gilkey1995invariance}---. En particular, la función zeta \eqref{eq:zeta} es en todos los casos analítica en $s=0$, por lo que el determinante en \eqref{eq:logdetzeta} es siempre finito.

Esta definición permite obtener el determinante de un operador, una vez conocidos sus autovalores, por medio del cálculo directo de la función zeta, lo que puede involucrar, entre otros métodos, integrales de contorno en el plano complejo: representaciones integrales como la \emph{del tiempo propio de Schwinger}
\begin{align}\label{eq:mellin}
\lambda^{-s} = \frac{1}{\Gamma(s)}\int_0^\infty dt\, t^{s-1} e^{-t}\,;
\end{align}
o fórmulas de inversión de series como la \emph{de Poisson},
\begin{align}\label{eq:poissongeneral}
\sum_n f(n) = \sum_m\tilde{f}(m)\,,
\end{align}
donde hemos denotado $\tilde{f}$ a la transformada de Fourier de $f$. Entre éstas últimas, encontraremos especial utilidad en el caso $f(x)=\exp[-(x+c)^2t]$, con $c,t\in\mathbb{C}$ y $\Re(t)>0$, conocido a veces como \emph{fórmula de inversión para la función Theta de Jacobi}: 
\begin{align}\label{eq:poisson}
\sum_{k\in\mathbb{Z}}e^{-(k+c)^2t}=\sqrt{\frac{\pi}{t}}\sum_{k\in\mathbb{Z}}e^{-\pi^2 k^2\!/t}e^{2\pi i kc}\,.
\end{align}

En los casos en que un cálculo directo no sea posible, es útil considerar la representación \eqref{eq:mellin} para escribir la función zeta como la transformada de Mellin inversa
\begin{align}
\zeta_A(s) = \frac{1}{\Gamma(s)} \int_0^\infty dt\,t^{s-1}\mathrm{Tr}\,e^{-tA}\,,
\end{align}
de otra función espectral: la \emph{traza del heat-kernel},
\begin{align}
K_A(t):=\mathrm{Tr}\,e^{-tA}=\sum_{n\in\mathbb{N}} e^{-\lambda_n t}\,.
\end{align}
En términos de esta cantidad, el determinante funcional de un operador puede calcularse como
\begin{align}\label{eq:logdethk}
\log\mathrm{Det}\,A:=-\int_0^\infty \frac{dt}{t} K_A(t)\,.
\end{align}
Esta representación permite obtener resultados a partir del desarrollo asintótico del \emph{heat-kernel} \cite{seeley1966singular,seeley1969resolvent}, que se obtienen como función de coeficientes que son integrales sobre la variedad y su borde de invariantes geométricos locales \cite{dewitt1963dynamical,gilkey1975spectral} y pueden calcularse en una gran cantidad de casos \cite{Vassilevich:2003xt}.

Volviendo a la regularización zeta, notamos que es usual la introducción de un parámetro $\mu$ con dimensiones de masa con el fin de adimensionalizar el argumento de la potencia compleja: por ejemplo, para un operador como el de la acción escalar \eqref{eq:sclas}, los autovalores tienen unidades de masa al cuadrado, por lo que escribimos la función zeta como
\begin{align}\label{eq:zetamu}
\zeta_A(s) := \sum_n \left(\lambda_n/\mu^2\right)^{-s}\,.
\end{align}
De este modo, cuando la función zeta del operador no se anule en $s=0$ el determinante tendrá una dependencia en la escala $\mu$, que deberá ser ajustada mediante el procedimiento físico de renormalización de alguno de los parámetros de la acción original\footnote{ \,El método de regularización mediante el \emph{heat-kernel} no está exento de este problema: el integrando en \eqref{eq:logdethk} diverge cuando $t\rightarrow0^+$, de modo que en rigor la integral tiene que ser efectuada a partir de un valor positivo de $t$, digamos un \emph{cutoff} ultravioleta $1/\Lambda^2$.}.

Hasta aquí hemos considerado operadores definidos positivos, cuyos autovalores son todos números reales positivos, de modo que la potencia compleja en la función zeta tiene un valor bien definido. En el caso en que hubiera además algún autovalor nulo, es posible extender esa definición mediante la exclusión de dicho autovalor de la suma. Si consideramos operadores con un número finito o infinito de autovalores negativos ---como el de Dirac--- tendremos una ambigüedad en la potencia compleja $\lambda_n^{-s}$, que corresponde a la elección del corte del logaritmo en $\exp(-s\log\lambda_n)$. En la función zeta
\begin{align}
\zeta (s) = \sum_{\lambda_n>0} \lambda_n^{-s} + (-1)^{-s} \sum_{\lambda_n<0} (-\lambda_n)^{-s} 
\end{align}
dicha ambigüedad se traduce en la elección de la fase del factor $(-1)^{-s}$; esta elección tendrá que hacerse de acuerdo con alguna prescripción, que en las aplicaciones en Teoría Cuántica de Campos corresponde a alguna propiedad deseable de la teoría. 

Resulta también de utilidad la \emph{función eta} \cite{aps},
\begin{align}
\eta (s) = \sum_{\lambda_n>0} \lambda_n^{-s} - \sum_{\lambda_n<0} (-\lambda_n)^{-s}\,,
\end{align}
que caracteriza la asimetría espectral del operador, siendo nula en los casos en que el espectro es simétrico con respecto al origen de la recta real. Su valor en $s=0$ se conoce como \emph{invariante eta} y forma parte de la fase del determinante de un operador de Dirac $D$: en efecto, fijando la fase de la potencia compleja como $(-1)^{-s}=\exp(-i\pi s)$ puede verse que
\begin{align}\label{eq:zetadirac}
\zeta_D'(0) = -i\frac{\pi}{2}\left[\zeta_{D^2}(0)+\eta_D(0)\right] + \frac12\zeta_{D^2}'(0)\,.
\end{align}

Para finalizar esta discusión acerca de los determinantes funcionales, notemos que muchas veces éstos no satisfacen las propiedades de un determinante de matrices usual. En particular, el determinante funcional del producto de dos operadores no necesariamente coincide con el producto de los determinantes funcionales de los factores, y es útil el cálculo de la \emph{anomalía multiplicativa} \cite{kontsevich1995geometry}
\begin{align}
a(A,B):= \log\frac{\mathrm{Det}\,AB}{\mathrm{Det}\,A\,\mathrm{Det}\,B}\,,
\end{align}
para la que existe una expresión en términos del \emph{residuo de Wodzicki} de un producto de potencias de los operadores $A$ y $B$ \cite{wodzicki}. Un caso particular de esta anomalía es la aparición de la fase en el determinante de un operador de Dirac, como puede leerse de la ecuación \eqref{eq:zetadirac}.

\section{Algunas funciones zeta}

La función zeta más célebre es sin duda la \emph{de Riemann},
\begin{align}
\zeta_R(s) := \sum_{n=1}^\infty n^{-s}\,,
\end{align}
que en el lenguaje de la sección anterior puede considerarse, por ejemplo, como la función zeta del operador $A=-i\partial_x$ sobre el espacio de funciones de una variable real con condiciones de contorno periódicas en el intervalo $[0,2\pi]$. Como antes, la definición se entiende como la serie para $\Re(s)>1$ y su extensión analítica en otro caso. Puede verse que $\zeta_R(s)$ tiene un único polo simple en $s=1$. Además, se anula en todos los enteros negativos pares; de acuerdo con la conjetura de Riemann, los demás ceros de $\zeta_R(s)$ estarían situados en la recta $\Re(s)=1/2$.

Como ejemplo del tipo de cálculos al que nos enfrentaremos cuando usemos la regularización zeta, notemos que haciendo uso de la representación \eqref{eq:mellin} podemos escribir
\begin{align}
\zeta_R(s) = \sum_{n=1}^\infty \frac{1}{\Gamma(s)} \int_0^\infty dt\,t^{s-1}\,e^{-nt}\,,
\end{align}
que, una vez resuelta explícitamente la serie geométrica de exponenciales, resulta
\begin{align}\label{eq:zetariemannint}
\zeta_R(s) = \frac{1}{\Gamma(s)} \int_0^\infty dt\,\frac{t^{s-1}}{e^{t}-1}\,.
\end{align}
Con esta expresión es fácil calcular por ejemplo el valor de la zeta de Riemann alrededor del polo $s=1$: suponiendo en primer lugar que $\Re(s)<1$, utilizamos el desarrollo \cite[página 1040]{Gradshteyn}
\begin{align}
\frac{t}{e^t-1} = \sum_{n=0}^\infty \frac{B_n}{n!} t^n\,,
\end{align}
donde $B_n$ son los números de Bernoulli, para reescribir \eqref{eq:zetariemannint} como
\begin{align}\label{eq:zetariemannbern}
\zeta_R(s) = \frac{1}{\Gamma(s)} \int_1^\infty dt\,\frac{t^{s-1}}{e^{t}-1} + \frac{1}{\Gamma(s)} \sum_{n=0}^\infty \frac{B_n}{n!}\frac{1}{s+n-1}\,.
\end{align}
En esta expresión, el primer término es una función entera de $s$, mientras que el segundo término tiene un único polo en $s=1$, correspondiente al término $n=0$ en la suma.\footnote{ \,Los ceros del denominador en los enteros negativos no dan lugar a polos para este término puesto que la función $1/\Gamma(s)$ tiene también ceros en esos valores.} De este modo tenemos la extensión analítica buscada. Usando el hecho de que $B_0=1$, podemos obtener el comportamiento en las inmediaciones del polo:
\begin{align}\label{eq:riemannzpolo}
\zeta_R(s) = \frac{1}{s-1} + \mathcal{O}(1)\,.
\end{align} 

Entre las generalizaciones de la función zeta de Riemann está la \emph{función zeta de Hurwitz}
\begin{align}
\zeta_H(s,q) := \sum_{n=0}^\infty (n+q)^{-s}\,, \qquad q\neq 0, -1, -2, \ldots\,,
\end{align}
que, entendida otra vez como la extensión analítica de la serie del lado derecho cuando ésta no converja, es una función meromorfa en el plano complejo $s$ con un único polo simple en $s=1$.

Otros tipos de funciones zeta que aparecen a menudo en problemas físicos son las de Epstein y Barnes. Para más detalles, el lector puede consultar el libro \cite{kirsten2001spectral}.

%

\chapter{Los espacios esféricos tridimensionales}
\label{sec:espacios}

Los modelos del universo más estudiados tienen secciones espaciales homogéneas e isótropas, en acuerdo con el principio cosmológico. La curvatura espacial en estos casos es constante, y dependiendo de su valor las secciones espaciales son planas, esféricas o hiperbólicas. Observemos aquí que la elección de un valor para la curvatura de una variedad no determina su geometría global, pudiendo ésta corresponder a diferentes topologías.
En efecto, las variedades tridimensionales de curvatura constante pueden ser pensadas como cocientes $\mathcal{M} = \tilde{\mathcal{M}}/H$ de su espacio de cubrimiento universal con algún subgrupo $H$ de su grupo de isometrías. El espacio de cubrimiento $\tilde{\mathcal{M}}$ puede ser, dependiendo de la curvatura, el espacio euclídeo $E^3$, el espacio hiperbólico $H^3$ o la esfera $S^3$. El subgrupo $H$ debe ser discreto y actuar sobre la variedad sin dejar puntos fijos, puesto que de otra forma el espacio cociente no sería una variedad.

En este capítulo nos centraremos en el estudio de teorías cuánticas de campos escalares a temperatura finita sobre variedades homogéneas cuyo grupo de cubrimiento universal es la esfera $S^3$, a las que, por no buscar un nombre más complicado y dado que no habrá confusión, llamaremos \emph{espacios esféricos}.\footnote{ \,En inglés nos referimos a estos espacios como \emph{spherical spaceforms}, término que es más específico que el que hemos elegido.} Luego de establecer una motivación física para el estudio de teorías sobre estas variedades, haremos una clasificación exhaustiva de los espacios posibles, para luego obtener los espectros del operador laplaciano sobre cada uno de ellos. En los capítulos siguientes, dicho espectro será utilizado para calcular determinantes funcionales de operadores de fluctuaciones cuánticas de ciertas teorías de campos sobre estos espacios.

\vfill\pagebreak

\section{Motivación: topología del universo}
\label{sec:espacios:motivacion}

El problema de la topología del universo ha sido estudiado desde los albores de la cosmología relativista. En efecto, luego de la primera solución cosmológica de las ecuaciones de Einstein ---un modelo estático cuyas secciones espaciales son esferas tridimensionales---, de Sitter mostró que un modelo similar con espacios proyectivos $\mathbb{R}\mathrm{P}^3$ como secciones espaciales también resuelve las ecuaciones \cite{deSitter1917}. $S^3$ y $\mathbb{R}\mathrm{P}^3$ tienen la misma métrica pero sus propiedades globales son diferentes; en particular, la esfera es simplemente conexa, mientras que el espacio proyectivo no lo es. No obstante, de Sitter pensaba que el espacio proyectivo era el modelo más adecuado, basándose en el hecho de que en éste dos líneas paralelas se cruzan una única vez \cite{de1917curvature}, mientras que en la esfera lo hacen dos veces ---en puntos antipodales---.\footnote{ \,Como todas las ideas de la época, esta propuesta despertó controversia, como puede apreciarse de la lectura de la correspondencia de Einstein en el período \cite{einsteincorr}, en particular de sus conversaciones con de Sitter durante el año 1917 y con Weyl durante ese año y el siguiente.}

Un siglo después, la pregunta acerca de la topología del universo conforma un área de estudio, principalmente desde el punto de vista de la cosmología observacional. El lector interesado en conocer en más detalle el estudio de este problema querrá comenzar consultando las referencias \cite{LachiezeRey:1995kj,Levin:2001fg}. La pregunta principal tiene que ver con la posibilidad de inferir información acerca de la topología del universo usando los datos cosmológicos de los que disponemos actualmente (en particular, los que provienen de las observaciones de la radiación del fondo cósmico), de la misma forma como fueran explorados algunos parámetros geométricos locales \cite{Kamionkowski:1993aw}.

En la teoría de Einstein, la geometría del universo no está determinada de antemano, sino que depende del contenido de materia, radiación y energía en sus diferentes formas. En particular, para un modelo del universo que sea homogéneo e isótropo a gran escala, las secciones espaciales (que tienen curvatura constante) dependerán de la densidad de materia de éste: hay un valor de la densidad, llamado densidad crítica, para el cual la materia es exactamente la necesaria para frenar la expansión del universo pero no es suficiente como para hacerlo recolapsar, situación que se corresponde con un universo localmente plano; si la densidad es menor que ese valor el universo no dejará de expandirse, lo que corresponde a una geometría hiperbólica, y si la densidad es mayor que la densidad crítica el universo se expandirá a un ritmo cada vez menor, hasta llegar en algún momento a frenar su expansión y comenzar a contraerse, correspondiendo esta situación a una geometría esférica. 

El universo observable es aproximadamente homogéneo e isótropo a gran escala; las soluciones a las ecuaciones de Einstein con estas propiedades son los modelos de Friedmann-Lema\^{\i}tre-Robertson-Walker, que están descriptos genéricamente por una métrica de la forma
\begin{align}\label{eq:flrwmetric}
ds^2 = -dt^2 + a^2\!\left(\eta\right) \left[\frac{dr^2}{1-\kappa r^2}+r^2\left(d\theta^2+\sen^2\!\theta\, d\phi^2\right)\right]\,,
\end{align}
donde $t$ es la variable temporal y $r = \sen\psi$, con $(\psi,\theta,\phi)$ las coordenadas (hiper-)esféricas para las variables espaciales; la función $a(t)$ es el factor de escala de la expansión del universo, y el parámetro $\kappa$ representa la curvatura del espacio: $\kappa=0$ corresponde a un espacio plano, $\kappa=1$ a uno con curvatura positiva (elíptico), y $\kappa=-1$ a uno con curvatura negativa (hiperbólico).

Las ecuaciones de Einstein refieren a propiedades locales del espacio-tiempo ---relacionando en particular la métrica y sus derivadas en un dado punto con el contenido de materia del espacio en ese punto---, pero no determinan sus propiedades globales, como su topología. Dicho de otra forma, a una dada geometría del espacio en la teoría de Einstein pueden corresponder diferentes topologías, sin que se altere la dinámica del espacio-tiempo ni su curvatura. Esto implica por ejemplo que cualquiera sea la curvatura del espacio, es siempre posible modelarlo con una variedad de volumen finito.

Una de las propiedades observadas en la radiación del fondo cósmico por el satélite COBE \cite{Smoot:1992td} es la presencia de inhomogeneidades a gran escala en su temperatura. Éstas pueden ser explicadas por la presencia de pequeñas fluctuaciones cuánticas en el universo en sus primeros momentos, que por causa de la inflación se volvieron inhomogeneidades en la densidad del plasma primordial: luego de 380000 años de expansión del universo con su consiguiente enfriamiento, la temperatura fue tal que éste se volvió transparente a la luz, y los primeros fotones son los que forman el fondo cósmico, que refleja esas inhomogeneidades en la anisotropía a gran escala de su distribución de fluctuaciones de temperatura. Los observatorios WMAP \cite{Hinshaw:2012aka} y Planck \cite{Ade:2015xua} determinaron el mapa de anisotropía con más precisión y detectaron, entre muchísimas otras observaciones, que el modo cuadrupolar en la descomposición en armónicos esféricos de las fluctuaciones de temperatura tiene una amplitud menor que la esperada por los modelos $\Lambda$CDM planos, que son los que con más precisión describen la mayor parte de las características observadas del universo. 
Esta anomalía ha sido tenida en cuenta en distintas hipótesis (véase, por ejemplo, \cite{Contaldi:2003zv,Iqbal:2015tta}); en particular, su existencia podría ser explicada aceptando un universo cuyas secciones espaciales tengan una topología no trivial. Aunque aún no hay evidencia concluyente en favor de esta hipótesis, diferentes espacios con topología no trivial han sido objeto de estudio en este contexto \cite{Aurich:2013fwa,Luminet:2003dx,Riazuelo:2003ud}.

Ciertamente no es objeto de esta tesis explorar en detalle el problema de la topología del universo. En todo caso, hemos sugerido una motivación para la consideración de espacios con topología no trivial en el marco del estudio de teorías físicas. En lo que sigue del capítulo nos ocuparemos de los espacios que nos interesan, dando además de una descripción de sus propiedades geométricas y topológicas un resumen del cálculo del espectro del operador laplaciano definido sobre ellos.

\section{Los espacios esféricos tridimensionales}
\label{sec:espacios:espacios}

Los espacios esféricos en tres dimensiones (esto es, todas las posibles variedades tridimensionales con curvatura constante positiva) fueron clasificados en los años 30 del siglo pasado \cite{birman1980seifert}. Se trata de variedades compactas, como lo es su espacio de cubrimiento $S^3$. Nos ocuparemos aquí de los espacios esféricos homogéneos: comenzaremos haciendo un repaso de la geometría de la esfera y sus simetrías, y a partir de estas últimas clasificaremos los espacios posibles. El lector interesado en detalles matemáticos adicionales puede consultar por ejemplo la referencia \cite{wolf2011spaces}.

\subsection{La esfera: su geometría}

Como es bien sabido, la esfera tridimensional de radio $a$ es el conjunto de puntos en el espacio $\mathbb{R}^4$ distantes del origen en la cantidad positiva $a$: el punto $(x^i, i=0,1,2,3)\in\mathbb{R}^4$ pertenece a la esfera $S^3$ si y sólo si la suma de los cuadrados de sus coordenadas $x^i$ es el cuadrado del radio de la esfera. En coordenadas esféricas $(\psi,\theta,\phi)$, la métrica de la esfera tiene la forma de la parte espacial de la métrica \eqref{eq:flrwmetric} para $\kappa=1$. 

\vfill\pagebreak

Otra forma de la métrica puede obtenerse a partir de las coordenadas de Hopf ---también llamadas coordenadas polares dobles--- $(\chi,\theta,\phi)$, relacionadas con las coordenadas cartesianas de $\mathbb{R}^4$ por medio de las expresiones
\begin{align}\nonumber
x^0 & = a\cos\!\chi\cos\theta\\ \label{eq:hopfcoord}
x^1 & = a\sen\!\chi\cos\phi\\ \nonumber
x^2 & = a\sen\!\chi\sen\phi\\ \nonumber
x^3 & = a\cos\!\chi\sen\theta\,,
\end{align}
en las que los puntos de la esfera corresponden a los valores $0\leq\chi\leq\pi/2$; y $0\leq\theta,\phi<2\pi$. En efecto, para $\chi\in(0,\pi/2)$ fijo, las coordenadas $(\theta,\phi)$ describen el producto cartesiano de dos circunferencias con radios respectivos $a\cos\!\chi$ y $a\sen\!\chi$ en $\mathbb{R}^4$;  haciendo una proyección estereográfica puede verse que al mover la coordenada $\chi$ estos toros cubren la esfera $S^3$.

La métrica de la esfera se escribe en estas coordenadas como
\begin{align}
\frac{ds^2}{a^2} = d\chi^2 + \cos^2\!\chi\, d\theta^2 + \sen^2\!\chi\, d\phi^2\,,
\end{align}
esto es, $g^{ij} = g^{i}\delta^{ij}$, con
\begin{align}
g^{1} = \frac{1}{a^2} \, ; \,\, g^{2} = \frac{1}{a^2\cos^2\!\chi}\, ; \,\, g^{3} = \frac{1}{a^2\sen^2\!\chi}
\end{align}
de donde puede verse que $\sqrt{g} = a^3\cos\!\chi\sen\!\chi$. El laplaciano sobre la esfera resulta entonces
\begin{align}\label{eq:hopflaplacian}
\triangle_{S^3} = \frac{1}{a^2} \left[ \frac{1}{\cos\!\chi\sen\!\chi} \,\partial_{\chi} \!\left( \cos\!\chi\sen\!\chi\,\partial_{\chi} \right) +
\frac{1}{\cos^2\!\chi} \partial^2_{\theta} + \frac{1}{\sen^2\!\chi} \partial^2_{\phi}\right].
\end{align}

\medskip

Hasta aquí hemos considerado a la esfera desde un punto de vista extrínseco ---esto es, a partir de su definición como subconjunto de puntos de $\mathbb{R}^4$---. También podemos pensarla como definida intrínsecamente en tres dimensiones: considerando la operación inversa de la proyección estereográfica tridimensional vemos que es posible pensar a la esfera como la compactificación por un punto del espacio $\mathbb{R}^3$. En esta descripción es fácil ver que la esfera es simplemente conexa, puesto que en tres dimensiones cualquier contorno cerrado puede ser deformado de manera de evitar que encierre un dado punto ---en este caso el infinito---.

Una tercera forma de ver a la esfera $S^3$ es como el espacio homogéneo $SO(4)/SO(3)$. En efecto, cuando la pensamos como subconjunto de $\mathbb{R}^4$ vemos que el grupo $SO(4)$ actúa sobre ella de forma transitiva, siendo $SO(3)$ su grupo de isotropía. Esto es equivalente al espacio homogéneo $SU(2)\times SU(2)/SU(2)$, donde el denominador actúa sobre el producto de manera diagonal. En este esquema, se puede usar técnicas de análisis armónico en espacios homogéneos para obtener el espectro del laplaciano sobre la esfera \cite{Camporesi:1995fb,David:2009xg}.

\subsection{La esfera: sus isometrías}
\label{sec:espacios:espacios:esfera}

El grupo de isometrías de la esfera $S^3$ es $SO(4)$. Esto es evidente si vemos a la esfera inmersa en $\mathbb{R}^4$: las transformaciones que la dejan invariante son las rotaciones en cuatro dimensiones. También puede verse del hecho de que la esfera es la variedad del grupo $SU(2)$ y, como tal, su grupo de isometrías está determinado por la operación del grupo: los elementos pueden actuar sobre los puntos de la esfera a derecha o a izquierda, y por eso las isometrías pertenecen al producto $SU(2)\times SU(2)$. Ahora bien; la acción doble con un dado elemento coincide con la acción doble con el elemento correspondiente al punto de la esfera antipodalmente opuesto, y puede verse que ésta es la única redundancia, lo que quiere decir que el grupo de isometrías de la esfera es $SU(2)\times SU(2)/\{\mathrm{Id}_4,-\mathrm{Id}_4\}$ ---con $\mathrm{Id}_4$ el elemento identidad y $-\mathrm{Id}_4$ el elemento que lleva cada punto a su antipodal---, que no es otro que $SO(4)$. A la inversa, todo elemento de $SO(4)$ tiene una única descomposición como producto de una traslación de Clifford ---esto es, una isometría que traslada todos los puntos de la esfera en la misma distancia--- a izquierda y una a derecha, módulo la multiplicación de ambos factores por $-\mathrm{Id}_4$. El lector interesado en obtener una descripción aún más simple de este esquema puede consultar la formulación en términos de cuaterniones unitarios en \cite{Thurston:1997}. 

Para construir los espacios esféricos necesitamos los subgrupos discretos e invariantes de $SO(4)$ que no dejen puntos de la esfera fijos. Requeriremos además que correspondan a la acción sólo a derecha o sólo a izquierda sobre puntos de la esfera, ya que la acción doble da lugar a un espacio no homogéneo \cite{Gausmann:2001aa}. En particular, esto quiere decir que podemos clasificar todos los subgrupos de interés analizando los subgrupos discretos de $SU(2)$.

Comenzaremos estudiando los subgrupos discretos de $SO(3)$. El grupo $SO(3)$ es el grupo de las simetrías de rotación alrededor de un origen fijo en $\mathbb{R}^3$; la clasificación de sus subgrupos finitos constituye un problema matemático clásico (véase, por ejemplo, \cite{Coxeter:1973}) y es conocido que los posibles subgrupos son: 
\begin{itemize}
	\item[$\ast$] los grupos cíclicos $Z_n$ de orden $n$, $n=2,3,\ldots$, que comprenden las rotaciones en un ángulo múltiplo de $2\pi/n$ alrededor de un único eje;
	
	\item[$\ast$] los grupos diédricos $D_p$ de orden $2p$, $p=2,3,\ldots$, que corresponden, para $p\geq3$, a las simetrías del polígono regular de $p$ lados en alguno de los planos coordenados de $\mathbb{R}^3$ y están generados por las rotaciones en un ángulo múltiplo de $2\pi/p$ alrededor del eje perpendicular a aquel plano y las rotaciones de ángulo $\pi$ alrededor de alguno de los $p$ ejes de simetría del polígono. Para $p=2$, tenemos $D_2\approx Z_2\times Z_2$;
	
	\item[$\ast$] el grupo tetraédrico $T$, que corresponde a las rotaciones que dejan invariante un tetraedro regular. No debe confundirse con el grupo tetraédrico completo, que consiste en todas las simetrías del tetraedro regular, incluyendo aquellas que cambian su orientación. El orden del grupo $T$ es 12, siendo éste el número de rotaciones que no alteran la figura del tetraedro: la identidad, las dos rotaciones en ángulos que sean múltiplos de $2\pi/3$ alrededor de cada uno de sus cuatro ejes de simetría que pasan por un vértice y las rotaciones en ángulo $\pi$ alrededor de cada uno de los tres ejes de simetría que no pasan por un vértice. Estos elementos pueden a su vez ser identificados con las correspondientes permutaciones de los vértices del tetraedro: fijando uno de los vértices, la figura permanece invariante y con la misma orientación ante cualquiera de las tres permutaciones cíclicas de los tres vértices restantes. Dado que una permutación de los vértices que contenga un número impar de transposiciones simples corresponde a una reflexión del tetraedro ---que no puede ser realizada como una rotación y no es, por lo tanto, un elemento de $T$---, vemos que el grupo es isomorfo al grupo alternante $A_4$ de las permutaciones pares de cuatro elementos;
	
	\item[$\ast$] el grupo octaédrico $O$, formado por las rotaciones que dejan invariante un octaedro regular (o, lo que es lo mismo, las que dejan invariante un cubo): la identidad, las dos rotaciones en ángulos múltiplos de $2\pi/3$ alrededor de cada uno de los cuatro ejes de simetría que pasan por dos de sus caras, las tres rotaciones en ángulos múltiplos de $\pi/2$ alrededor de cada uno de los tres ejes de simetría que pasan por dos de sus vértices y las rotaciones en ángulo $\pi$ alrededor de cada uno de los seis ejes de simetría que pasan por el centro de dos de sus aristas. El orden de este grupo es 24, al igual que el número de permutaciones de sus vértices que lo dejan invariante sin alterar su orientación. El grupo es isomorfo al grupo $S_4$ de permutaciones de cuatro elementos, que pueden realizarse como permutaciones de las diagonales del cubo, y
	
	\item[$\ast$] el grupo icosaédrico $I$, formado por las rotaciones que dejan invariante un icosaedro regular (o, equivalentemente, las que dejan invariante un dodecaedro regular). El orden de este grupo es 60, correspondiendo, además de la identidad, a las rotaciones en ángulos múltiplos de $2\pi/5$ alrededor de cada uno de los seis ejes de simetría que pasan por dos de sus vértices, las rotaciones en ángulos múltiplos de $2\pi/3$ alrededor de cada uno de los diez ejes de simetría que pasan por dos de sus caras y las rotaciones de ángulo $\pi$ alrededor de cada uno de los 15 ejes de simetría que pasan por el centro de dos de sus aristas. Es posible demostrar que el grupo es isomorfo al grupo alternante $A_5$ de permutaciones pares de cinco elementos.
\end{itemize} 

\begin{figure}[h!]
	\centering
	\includegraphics[width=12cm]{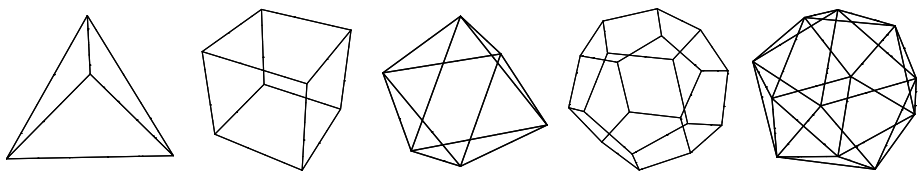}
	\caption{Los cinco sólidos platónicos: de izquierda a derecha, el tetraedro, el hexaedro (o cubo), el octaedro, el dodecaedro y el icosaedro.}
	\label{fig:poly}
\end{figure}

\begin{figure}[h]
	\centering
	\includegraphics[height=4cm]{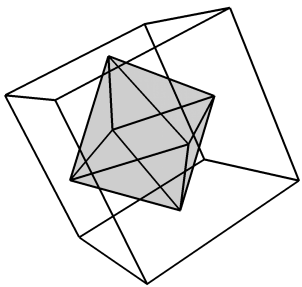}
	\caption{Dos sólidos platónicos duales entre sí: el cubo y el octaedro.}
	\label{fig:polydual}
\end{figure}

Los últimos tres grupos, a los que llamaremos genéricamente \emph{poliédricos}, corresponden a las simetrías de los sólidos platónicos. Notemos que mientras éstos últimos son cinco, los grupos de simetría en cuestión son tres: esto refleja el hecho de que dos poliedros tienen el mismo grupo de simetría si son duales entre sí ---esto es, si los vértices de uno corresponden a las caras del otro--- y, mientras que el tetraedro es autodual, el hexaedro es dual al octaedro y el dodecaedro es dual al icosaedro. En la figura \ref{fig:poly} mostramos los cinco sólidos platónicos, y en la figura \ref{fig:polydual} puede verse un ejemplo de la mencionada dualidad.

Para obtener los correspondientes subgrupos de $SU(2)$ hacemos uso del doble cubrimiento $\phi:SU(2)\rightarrow SO(3)$. La imagen inversa por este homomorfismo \cite{Castro:2011xb} de cada uno de los subgrupos de $SO(3)$ enumerados anteriormente es uno de los siguientes subgrupos finitos de $SU(2)$:
\begin{itemize}
	\item[$\ast$] los grupos cíclicos $Z_n$ de orden $n$, para $n=2,3,\ldots$;
	\item[$\ast$] los grupos diédricos binarios $D_{p}^*$ de orden $4p$, para $p=2,3,\ldots$;
	\item[$\ast$] los grupos poliédricos binarios: el tetraédrico $T^*$ de orden 24, el octaédrico $O^*$ de orden 48 y el icosaédrico $I^*$ de orden 120.
\end{itemize}
Los grupos diédricos y poliédricos binarios y los cíclicos de orden par corresponden a la imagen inversa por $\phi$ de los subgrupos de $SO(3)$ que contienen un subgrupo $Z_2$, aplicación que es dos a uno en estos casos. En el caso de los grupos cíclicos de orden impar, que corresponden a los subgrupos de $SO(3)$ que no tienen un subgrupo $Z_2$, la aplicación es uno a uno \cite{Gausmann:2001aa}.

\subsection{La esfera: sus cocientes}

Como ya se ha dicho, los espacios esféricos se obtienen a partir del cociente entre la esfera tridimensional y algún subgrupo finito de sus isometrías que no deje puntos fijos en su acción sobre la esfera. Dependiendo de la forma en la que este subgrupo actúe sobre la esfera, los cocientes se dividen en variedades de acción simple (en las que los elementos del subgrupo actúan como traslaciones de Clifford sólo a derecha o sólo a izquierda), doble (en las cuales algunos elementos actúan a derecha y otros a izquierda) o ligada (en las que la acción doble sólo admite determinadas combinaciones de elementos). Nos ocuparemos solamente de las variedades de acción simple, que son homogéneas. Los correspondientes subgrupos de $SO(4)$ pueden realizarse como el producto $H\times\{\mathrm{Id}_2\}$, con $H$ alguno de los subgrupos discretos de $SU(2)$ que enumeramos en la sección anterior, e $\mathrm{Id}_2$ el elemento identidad de $SU(2)$. En lo que sigue llamaremos a este producto simplemente $H$ y el lector deberá quedar advertido a partir de este momento de que cuando digamos $H\in SO(4)$ estaremos queriendo decir $H\times\{\mathrm{Id}_2\}\in SO(4)$.

Dado su espacio de cubrimiento universal, la topología de una variedad queda determinada por su \textit{dominio fundamental} ---el subconjunto del espacio de cubrimiento que contiene un punto y sólo uno de cada órbita, que en tres dimensiones es un poliedro convexo--- y su \textit{grupo de isotropía}, formado por el conjunto de transformaciones geométricas que identifican caras del dominio fundamental, de modo que las $n=|H|$ imágenes del dominio fundamental bajo la acción de los $n$ elementos del grupo de isotropía se unen unas con otras para formar el propio espacio de cubrimiento.

Si el grupo $H$ no es trivial, entonces la variedad $\esp{H}$ no es topológicamente equivalente a la esfera $S^3$. En efecto, podemos ver fácilmente que $\esp{H}$ no es simplemente conexa: como el grupo es no trivial, hay al menos un elemento $h\in H$ que no deja invariante ningún $x\in S^3$, de forma que $x$ y $h x$ son puntos diferentes; ahora bien, como bajo la acción de $H$ ambos puntos están en el mismo coset, un camino en $S^3$ que vaya de $x$ a $h x$ corresponde en $\esp{H}$ a un camino cerrado, que no puede ser contraído a un punto porque las órbitas de $H$ no son continuas. Esto quiere decir que la variedad $\esp{H}$ no es simplemente conexa. Más aún, esto implica también que el primer grupo de homotopía de $\esp{H}$ no es otro que $H$.

En la Tabla \ref{tabla:espacios} hacemos una lista de los posibles espacios esféricos en función de su grupo de holonomía $H\in SU(2)$.

\begin{table}[h!]
	\noindent
\begin{tabular}{c|lcl}
	$H$ 	& $\esp{H}$ 						& $n$ 	& Dominio fundamental\\
	\hline
	$Z_n$ 		& Espacios lente $L(n,1)$ 			& $n$ 	& Sólido con forma de lente\\
	$D_p^{*}$ 	& Espacios prisma 					& $4p$ 	& Prisma cuya base es \\
	 			& 									& 		& un polígono de $2p$ caras\\
	$T^{*}$ 	& Espacio octaédrico 				& 24 	& Octaedro regular\\
	$O^{*}$ 	& Espacio cúbico 					& 48 	& Cubo truncado\\
	$I^{*}$ 	& Espacio dodecaédrico de Poincaré 	& 120 	& Dodecaedro regular
\end{tabular}\\
	\caption{Espacios esféricos.}
\label{tabla:espacios}
\end{table}

\section{Espectro del laplaciano sobre espacios esféricos tridimensionales}
\label{sec:espacios:espectro}

\begin{minipage}{.885\textwidth}%
	\begin{flushright}
		\begin{minipage}{.6\textwidth}%
			\begin{flushleft}
				\emph{As a rule, it is more elegant to use as much group theory [...] as possible.}
			\end{flushleft}
			\begin{flushright}
				--- \textrm{J. S. Dowker, hep-th/0404093} 
			\end{flushright}
		\end{minipage}%
	\end{flushright}
\end{minipage}%
	
Obtendremos ahora el espectro del laplaciano sobre espacios esféricos de dos maneras diferentes. Digamos en primer lugar que es posible ``escuchar'' las variedades esféricas tridimensionales: se ha demostrado que si dos de estas variedades tienen el mismo espectro entonces son isométricas \cite{ikeda1980}. Por otra parte, en este punto será ya claro que el cociente $\esp{H}$ puede verse como un conjunto de puntos, cada uno identificado con $|H|$ puntos de $S^3$. Las funciones sobre el espacio homogéneo $\esp{H}$ se obtienen eligiendo entre todas las funciones sobre $S^3$ aquellas que son invariantes bajo la acción de $H$. Ahora bien; como $S^3$ y $\esp{H}$ son variedades localmente isométricas, el operador laplaciano es el mismo para ambas. Las autofunciones de este operador sobre $\esp{H}$ son entonces aquellas autofunciones de $S^3$ que son invariantes bajo la acción de $H$. El proceso de proyección de las autofunciones sobre el espacio de cubrimiento al espacio cociente se conoce en algunas áreas como \emph{symmetry adaptation} \cite{klein1970symmetry}; desde el punto de vista de la teoría cuántica de campos, esta proyección puede ser llevada a cabo sobre las funciones espectrales en la forma de un método de imágenes, como en \cite{Dowker:1972np}. Los autovalores y degeneraciones del laplaciano sobre espacios esféricos fueron calculados por primera vez en \cite{ikeda1995}.

Para obtener las autofunciones en forma explícita se puede recurrir al procedimiento usual de seleccionar un sistema coordenado conveniente, separar variables y resolver el sistema de ecuaciones diferenciales ordinarias resultante. Luego puede analizarse cuáles de estas autofunciones son invariantes bajo la acción de los elementos de $H$, y así obtener el espectro del laplaciano sobre $\esp{H}$. En la  sección \ref{sec:espacios:espectro:esfera} mostramos, por completitud, el cálculo explícito de las autofunciones sobre la esfera usando las coordenadas de Hopf, y en la sección \ref{sec:espacios:espectro:deg} hacemos un análisis \emph{naive} de la invarianza de éstas frente a la acción de los espacios lente, con el que obtenemos las degeneraciones sobre dichos espacios.

Sin embargo, tratándose de la esfera, y como para calcular acciones efectivas sólo nos interesa conocer los autovalores y degeneraciones del operador de fluctuaciones cuánticas de la teoría, hay una forma más simple de proceder. El operador laplaciano sobre la esfera es proporcional al invariante de Casimir del grupo $SO(4)$, por lo que sus autovalores $\lambda_k$ sobre funciones de $L^2(S^3)$ pueden obtenerse de las representaciones irreducibles de ese grupo, resultando $a^2\lambda_k=k^2-1$, con $k=1,2,\ldots$ y $a$ el radio de la esfera; las degeneraciones vienen dadas por la dimensión de la representación correspondiente,  $d_k=k^2$. La proyección a $\esp{H}$ deja, como hemos dicho, un subconjunto del total de las autofunciones, lo que se traduce en una disminución de las degeneraciones con respecto a las de la esfera.
Para obtener las nuevas degeneraciones, notemos en primer lugar que ante la acción de los elementos de $H$ las $k^2$ autofunciones con autovalor $\lambda_k$ se transforman linealmente entre sí, dando lugar a una representación de $H$ de dimensión $k^2$, que llamaremos $D_k$. El número de autofunciones con autovalor $\lambda_k$ invariantes frente a la acción de $H$ coincide con el número de veces que la representación trivial aparece en la descomposición de $D_k$ en representaciones irreducibles. Llamando $\chi_{\alpha}(h)$ al carácter del elemento $h$ en la representación irreducible $D^{(\alpha)}$ (con $\alpha = 1, 2, \ldots, s$, y $s$ el número de representaciones irreducibles de $H$), el correspondiente carácter en la representación $D_k$ es la combinación entera 
\begin{align}
\chi_k(h) = \sum_{\alpha=1}^sa_{\alpha}\chi^{(\alpha)}(h)\,.
\end{align}
Usando las relaciones de ortogonalidad de caracteres de grupos finitos vemos que la multiplicidad de la representación $D^{(\alpha)}$ en la descomposición de $D_k$ en representaciones irreducibles puede escribirse como \cite{horaciogrupos}
\begin{align}
a_{\alpha} =\frac{1}{|H|} \sum_{h\in H}  \,\chi_k(h)\,\chi^{(\alpha)^*}(h)\,,
\end{align}
de modo que la multiplicidad de la representación trivial en dicha descomposición resulta
\begin{align}
a_1 = \frac{1}{|H|}\sum_{h\in H}\chi_k(h)\,.
\end{align}
Para calcular el carácter de la representación $D_k$, notemos que como $H\in SO(4)$ la acción de $h\in H$ sobre el subespacio característico correspondiente al autovalor $\lambda_k$ corresponde a la representación irreducible de $SO(4)$ de dimensión $k^2$. Ahora bien; el carácter del elemento de $SO(4)$ producto de los elementos $g(\theta_L,\hat{n}_L)$ y $g(\theta_R,\hat{n}_R)$ de $SU(2)$ en esta representación es 
\begin{align}
\chi_k(\theta_L,\theta_R) = \frac{\sen(k\,\theta_L/2)}{\sen(\theta_L/2)} \frac{\sen(k\,\theta_R/2)}{\sen(\theta_R/2)}\,,
\end{align}
y un elemento de $H\times\{\mathrm{Id}_2\}$ tiene $\theta_R=0$, de modo que 
\begin{align}
\chi_k(h) = k \frac{\sen(k\,\theta_h/2)}{\sen(\theta_h/2)}\,,
\end{align}
con $\theta_h$ el ángulo de rotación en $SU(2)$ del elemento $h$. Tenemos entonces finalmente para la degeneración de $\lambda_k$ sobre $\esp{H}$
\begin{align}\label{eq:deggrupos}
d_k^{(H)} = \frac{k}{|H|}\sum_{h\in H} \frac{\sen(k\,\theta_h/2)}{\sen(\theta_h/2)}\,.
\end{align}

Esta expresión para las degeneraciones en espacios esféricos fue obtenida en \cite{Dowker:2004nh} mediante un método similar. En \cite{Nash:1992sf} se obtuvo, con argumentos similares, una expresión equivalente, sólo en el caso de los espacios lente.

En la sección \ref{sec:espacios:espectro:prisma} veremos cómo puede usarse esta expresión para calcular las degeneraciones en el caso en que $H$ contenga un subgrupo $Z_2$ ---o, lo que es lo mismo, que contenga al elemento $-\mathrm{Id}\in SU(2)$---, y calcularemos explícitamente las degeneraciones sobre los espacios prisma y poliédricos.

\subsection[Autofunciones sobre la esfera]{Cálculo explícito de las autofunciones del laplaciano sobre la esfera}
\label{sec:espacios:espectro:esfera}

A continuación mostramos el cálculo explícito de las autofunciones del laplaciano sobre la esfera tridimensional de radio $a$. Para eso utilizamos las coordenadas de Hopf \eqref{eq:hopfcoord} y separamos variables para escribir una función $f$ sobre la esfera en la forma
\begin{align}
f(\chi,\theta,\phi) =
X(\chi) \Theta(\theta) \Phi(\phi)\,,
\end{align}
de modo que la ecuación en derivadas parciales $-\triangle f = \lambda f$ se transforma en el sistema de ecuaciones ordinarias
\begin{align} \nonumber
-\frac{1}{\Phi}\frac{d^2\Phi}{d\phi^2} & = m_1^2 \\ \label{eq:sistoro}
-\frac{1}{\Theta}\frac{d^2\Theta}{d\theta^2} & = m_2^2 \\ \nonumber
-\frac{1}{\cos\!\chi\sen\!\chi}\frac{d}{d\chi}\left(\cos\!\chi\sen\!\chi\frac{dX}{d\chi}\right) +
\left(\frac{m_1^2}{\sen^2\!\chi} + \frac{m_2^2}{\cos^2\!\chi}\right)X & = a^2\lambda X\,,
\end{align}
para algunas constantes $m_1$ y $m_2$ a determinar.
Las primeras dos ecuaciones tienen como soluciones a los armónicos circulares
\begin{align}
\Phi(\phi) = e^{im_1\phi}\\
\Theta(\theta) = e^{im_2\theta}\,,
\end{align}
y la condición de periodicidad sobre las coordenadas $\theta$ y $\phi$ exige que $m_1$ y $m_2$ sean enteros. Para obtener la solución de la tercera ecuación proponemos una función de la forma
\begin{align}
X(\chi) = \cos^{\alpha}\!\chi \sen^{\beta}\!\chi \tilde{X}(\chi)\,,
\end{align}
y tomamos $\alpha = \beta = -\tfrac{1}{2}$ para anular el término que contiene a la derivada primera $\tilde{X}'$, con lo que obtenemos
\begin{align}
\tilde{X}'' - \left[ \frac{m_1^2-\tfrac{1}{4}}{\sen^2\!\chi} + \frac{m_2^2-\tfrac{1}{4}}{\cos^2\!\chi} - (a^2\lambda+1)
\right] \tilde{X} = 0\,.
\end{align}
Las soluciones de esta ecuación son \cite{abramowitz}
\begin{align}
\tilde{X}_{n, m_1, m_2}(\chi) = (\sen\chi)^{|m_1|+\frac{1}{2}} (\cos\chi)^{|m_2|+\frac{1}{2}} P_n^{(|m_1|,|m_2|)}(\cos2\chi)\,,
\end{align}
donde $P_n^{(p,q)}(x)$, $n=0,1,\ldots$, son los polinomios de Jacobi, ortogonales con respecto a la función de peso
$(1-x)^p(1+x)^q$, y donde $n$ está relacionado con $\lambda$, $m_1$ y $m_2$ por medio de la ecuación
\begin{align}\label{lambdam}
a^2\lambda = (2n+|m_1|+|m_2|+1)^2-1\,.
\end{align}
Llamando $k$ al entero entre paréntesis que toma valores $k=1,2,\ldots$, y juntando las tres soluciones podemos escribir las autofunciones, que resultan
\begin{align}\label{autofesfera}
f_{k,m_1,m_2}(\chi,\theta,\phi) = e^{im_1\phi}e^{im_2\theta}\sen^{|m_1|}\!\chi\cos^{|m_2|}\!\chi\, P_{(k-|m_1|-|m_2|-1)/2}^{(|m_1|,|m_2|)}(\cos 2\chi)\,.
\end{align}
Los autovalores correspondientes vienen dados por la relación \eqref{lambdam}, y resultan $a^2\lambda_k = k^2-1$.

Para obtener la degeneración de cada autovalor notamos que a $n$ fijo la degeneración del $k$-ésimo autovalor es el número de combinaciones diferentes de enteros
$m_1$ y $m_2$ que dan el mismo valor de $k$. Como $n$ toma valores entre 0 y $\left\lfloor\frac{k-1}{2}\right\rfloor$, con $\lfloor x\rfloor$ la parte entera del número $x$ ---el mayor entero menor o igual a $x$---, tenemos las siguientes posibilidades:
\begin{itemize}
	\item[$\ast$]{Si $k$ es impar:} a $n$ fijo, $|m_2|$ puede tomar valores entre $0$ y $k-1-2n$, lo cual fija $|m_1|$. Cada
	vez que uno de los $|m_i|$ vale cero hay que contar dos modos (correspondientes a los dos valores opuestos
	posibles para el otro), y cuando ambos son no nulos se cuentan cuatro modos. La única excepción es
	$|m_1|=|m_2|=0$ (esto es, $n=\frac{k-1}{2})$, que corresponde a un único modo. La degeneración es entonces
	\begin{align}
	 d_k = 1+\sum_{n=0}^{\frac{k-1}{2}-1} 4(k-1-2n) = k^2
	\end{align}
	\item[$\ast$]{Si $k$ es par:} el análisis es similar al caso $k$ impar, con la diferencia de que ahora
	$n=\left\lfloor\frac{k-1}{2}\right\rfloor = \frac{k-2}{2}$ está asociado a cuatro modos diferentes, puesto que en este caso
	ambos $|m_i|$ son no nulos. De esta forma, resulta también en este caso
	\begin{align}
	 d_k = \sum_{n=0}^{\frac{k-2}{2}} 4(k-1-2n) = k^2
	\end{align}
\end{itemize}

Esto es, la degeneración del autovalor $\lambda_k$ es $d_k=k^2$. Esto era evidente desde el punto de vista de la teoría de grupos: el laplaciano sobre la esfera es proporcional al Casimir cuadrático de $SU(2)$, que toma valores $2j(2j+1)/4$, con $j\in\mathbb{Z}/2$; el cálculo prolijo muestra que los autovalores son entonces $(l+1)^2-1$, con $l=2j\in\mathbb{Z}$. La degeneración correspondiente al $l$-ésimo autovalor es la dimensión de la representación irreducible correspondiente, $(2j+1)^2 = (l+1)^2$.

\subsection[Degeneraciones sobre espacios lente]{Cálculo directo de las degeneraciones del laplaciano sobre espacios lente}
\label{sec:espacios:espectro:deg}
 
Definiremos ahora la acción de los grupos cíclicos sobre los puntos de la esfera en las coordenadas de Hopf para determinar las degeneraciones del laplaciano sobre los espacios lente homogéneos, para lo cual contaremos cuántas de las autofunciones \eqref{autofesfera} obtenidas en la sección anterior son invariantes frente a la acción de un dado grupo cíclico $Z_p$.

En estas coordenadas la acción homogénea del elemento $m\in Z_p$, con $m=0,1,\ldots,p-1$, sobre el punto de la esfera $g(\chi,\theta,\phi)$ corresponde a los cambios
\begin{align}
\theta\rightarrow\theta+\frac{2\pi m}{p},\quad \phi\rightarrow\phi+\frac{2\pi m}{p}
\end{align}
de modo que la acción de $Z_p$ deja invariante la autofunción $f_{k,m_1,m_2}$ siempre que
$e^{2\pi i(m_1+m_2)\frac{m}{p}} = 1$, o bien $m_1 + m_2 = lp$, con $l\in \mathbb{Z}$. La degeneración del autovalor $\lambda_k$ en $\esp{Z_p}$ viene dada entonces por el número de maneras diferentes de elegir el par de enteros
$(m_1,m_2)$ de modo de satisfacer simultáneamente las ecuaciones
\begin{align}\label{eq:relacdeg}
k &= 2n + |m_1| + |m_2| + 1\,, \\ \nonumber
lp &= m_1 + m_2\,,
\end{align}
con $l$ un número entero.

Estudiaremos a continuación los casos de $p$ par e impar por separado.
 
\subsubsection{Caso $p$ impar}
 
Para el caso en el que $p$ es impar, digamos $p=2q+1$ con $q\in\mathbb{N}_0$, buscamos para cada valor de $k$ el número de maneras diferentes de combinar $m_1$ y $m_2$ para satisfacer \eqref{eq:relacdeg}, con $l\in\mathbb{Z}$ y $n\in\mathbb{N}_0$. 

\vfill\pagebreak
 
Consideramos en primer lugar el caso en el que $k$ es impar, $k = 2b+1, b\in \mathbb{N}_0$; las relaciones \eqref{eq:relacdeg} se escriben entonces
\begin{align}\label{eq:relacdegimp}
2b &= 2n + |m_1| + |m_2|\,, \\ \nonumber
(2q+1)l &= m_1+m_2\,.
\end{align}
Debido a la presencia de valores absolutos en la primera ecuación, será útil separar el análisis en casos de acuerdo a los signos de $m_1$ y $m_2$:
\begin{enumerate}[label=\Roman*.,ref={caso~\Roman*}]
	\item $m_1\geq 0$ y $m_2\geq 0$: las relaciones anteriores implican que $2b = 2n+2ql+l$, y entonces $l$ resulta par. Además, para un dado valor de $b$, $l$ puede tomar valores entre $0$ y $\left\lfloor \frac{2b}{2q+1} \right\rfloor$.
 	Ahora bien: para un dado $l$ tenemos $(2q+1)l+1$ posibilidades para el par $(m_1,m_2)$, y entonces la contribución de este caso a la degeneración del $(2b+1)$-ésimo autovalor del laplaciano es
 	\begin{align}\label{eq:deglenteimp1}
 	d_{2b+1}^{(\mathrm{I})} = \sum_{\substack{l=0\\l\mathrm{\,\,par}}}^{l_{\mathrm{máx}}} \left[(2q+1)l+1\right]\,,
 	\end{align}
 	con $l_{\mathrm{max}}$ el mayor entero positivo par menor o igual que $\left\lfloor\frac{2b}{2q+1}\right\rfloor$.
 	Escribiendo $2b = \left\lfloor \frac{2b}{2q+1} \right\rfloor (2q+1) + r$, con $r\in \mathbb{Z}$, $0\leq r< 2q+1$, tenemos, para $r$ par, $l_{\mathrm{máx}} = \left\lfloor \frac{2b}{2q+1} \right\rfloor$, y para $r$ impar, $l_{\mathrm{máx}} = \left\lfloor \frac{2b}{2q+1} \right\rfloor-1$, por lo que el número de modos, que puede calcularse directamente mediante \eqref{eq:deglenteimp1}, depende de la paridad de $r$. \label{item:deglenteimp1}
 	\item $m_1\leq 0$ y $m_2\leq 0$: imitando el análisis realizado en el \ref{item:deglenteimp1} obtenemos el mismo número de modos:
 	\begin{align}
 	d_{2b+1}^{(\mathrm{II})} = d_{2b+1}^{(\mathrm{I})}\,.
 	\end{align}
 	\label{item:deglenteimp2}
 	\item $m_1\geq 0$ y $m_2\leq 0$, pero no simultáneamente nulos: de la combinación de las relaciones \eqref{eq:relacdegimp} podemos ver que $l$ tiene que ser par, y que para $n$ fijo $(0\leq n\leq k-1)$ toma valores entre $-\frac{2j}{2q+1}$ y $\frac{2j}{2q+1}$, donde hemos llamado $j = b-n$ $(j=1,2,\ldots,k)$. La contribución de este caso a la degeneración se calcula entonces como la suma
 	\begin{align}
 	d_{2b+1}^{(\mathrm{III})} = b + 2 \sum_{j=1}^{b}\left\lfloor\frac{j}{2q+1}\right\rfloor\,,
 	\end{align}
 	a la que tenemos que restar el número de modos ya contados en el \ref{item:deglenteimp1} (co\-rres\-pon\-dientes a $m_1\geq0$ y
 	$m_2=0$) y en el \ref{item:deglenteimp2} (correspondientes a $m_1=0$ y $m_2\leq0$), que es
 	\begin{align}
 	d_{2b+1}^{(\mathrm{III},-)} =
 	\begin{cases}
 		\left\lfloor \frac{2b}{2q+1} \right\rfloor + 1,  & \mbox{si }r\mbox{ es par}\,, \\
 		\left\lfloor \frac{2b}{2q+1} \right\rfloor, & \mbox{si }r\mbox{ es impar}\,.
 	\end{cases}
 	\end{align}
 	\label{item:deglenteimp3}
 	\item $m_1\leq 0$ y $m_2\geq 0$, pero no simultáneamente nulos: el número de modos en este caso coincide con el del caso anterior, y si sustraemos como antes los ya contados (que coinciden con los doblemente contados en el \ref{item:deglenteimp1} y en el \ref{item:deglenteimp3}, con excepción de $m_1=m_2=0$, que ahora no contamos), obtenemos
 	\begin{align}
 	d_{2b+1}^{(\mathrm{IV})} =
 	\begin{cases}
 		d_{2b+1}^{(\mathrm{III})} - \left\lfloor \frac{2b}{2q+1} \right\rfloor,  & \mbox{si }r\mbox{ es par}\,, \\
 		d_{2b+1}^{(\mathrm{III})} - \left\lfloor \frac{2b}{2q+1} \right\rfloor +1, & \mbox{si }r\mbox{ es impar}\,.
 	\end{cases}
 	\end{align}
 \end{enumerate}
 
 Sumando todas estas contribuciones y recuperando $2b+1=k$ tenemos finalmente para $k$ impar las degeneraciones
 \begin{align}
 d_k^{k\mathrm{\,\,impar}} =
 \begin{cases}
 	k\left(\left\lfloor\frac{k-1}{2q+1}\right\rfloor+1\right),  & \mbox{si }r\mbox{ es par}\,, \\
 	k\left\lfloor\frac{k-1}{2q+1}\right\rfloor, & \mbox{si }r\mbox{ es impar}\,.
 \end{cases}
 \end{align}

Para el caso en el que $k$ es par se puede imitar paso por paso el cálculo del caso impar y se obtiene la misma forma para las degeneraciones, de manera que finalmente tenemos
\begin{align}\label{eq:deglenteimpar}
 d_k^{(p=2q+1)} =
 \begin{cases}
 	k\left(\frac{k-r}{2q+1} + 1\right),  & \mbox{si }r\mbox{ es impar}\,, \\
 	k\left(\frac{k-r}{2q+1}\right), & \mbox{si }r\mbox{ es par}\,,
 \end{cases}
\end{align}
donde $r$ es el resto en el cociente entre $k$ y $2q+1$.

\vfill\pagebreak
 
\subsubsection{Caso $p$ par}
 
Para el caso en que $p$ es par, $p=2q$ con $q\in\mathbb{N}$, buscamos el número de pares de enteros $(m_1,m_2)$ que satisfagan en simultáneo las relaciones
\begin{align}
 k &= 2n + |m_1| + |m_2| + 1\,,\\ \nonumber
  2lq &= m_1+m_2\,.
\end{align}
De inmediato se observa que la degeneración del autovalor $\lambda_k$ es nula si $k$ es par, puesto que en ese caso no hay forma de satisfacer la primera de las ecuaciones. Introduciendo entonces el entero $b =
\frac{k-1}{2}$, reescribimos las relaciones anteriores como
\begin{align}\label{eq:degenpar}
2b &= 2n + |m_1| + |m_2|\,,\\ \nonumber
2lq &= m_1 + m_2\,.
\end{align}
Separamos otra vez el análisis en casos de acuerdo a los signos de $m_1$ y $m_2$:
\begin{enumerate}[label=\Roman*.,ref={caso~\Roman*}]
	\item $m_1 = m_2 = 0$: en este caso $n=b$ y la segunda relación se satisface trivialmente, por lo que tenemos un único modo.
 	\item $m_1\geq 0$ y $m_2\geq 0$, pero no simultáneamente nulos: en este caso la validez simultánea de las relaciones \eqref{eq:degenpar} requiere que la diferencia $b-n = lq\geq 0$, por lo que $l$ tomará valores entre $1$ ---en el caso en que $n=b-1$--- y
 	$\left\lfloor b/q\right\rfloor$ ---cuando $n=0$---. Para un valor dado de $l$, $m_1$ toma valores entre $0$ hasta $2lq$, lo cual fija $m_2$.
 	El número de modos en este caso es entonces
 	\begin{align}
 	 d_{2b+1}^{(\mathrm{II})} = \sum_{l=1}^{\left\lfloor b/q\right\rfloor} (2lq+1)\,.
 	 \end{align}
 	 \label{item:deglentepar2}

 	\item $m_1\leq 0$ y $m_2\leq 0$, pero no simultáneamente nulos: podemos imitar el análisis realizado en el \ref{item:deglentepar2} (con la diferencia de que $n-b$ es ahora un entero positivo), y obtenemos la misma contribución que en ese caso.
 	\item $m_1\geq 0$ y $m_2\leq 0$, pero no simultáneamente nulos: de las relaciones \eqref{eq:degenpar} podemos ver que para un valor dado de $n$, $l$ puede tomar cualquier valor entero entre $-\frac{b-n}{q}$ y
 	$\frac{b-n}{q}$. Como además $n$ puede tomar cualquier valor entero entre $0$ y $b-1$, tenemos
 	\begin{align}
 	 d_{2b+1}^{(\mathrm{IV})} = b + 2 \sum_{n=0}^{b-1}\left\lfloor\frac{b-n}{q}\right\rfloor\,.
 	\end{align}
 	\label{item:deglentepar4}
 	\item $m_1\leq 0$ y $m_2\geq 0$, pero no simultáneamente nulos: podemos repetir el análisis realizado en el \ref{item:deglentepar4}, obteniendo el mismo número de modos.
 \end{enumerate}
 
La degeneración para $k$ impar viene dada entonces por la suma de estas contribuciones, a la que hay que sustraer el número de modos que fueron contados dos veces: $(m_1,m_2)=(lq,0)$ y viceversa, que son en total $4 \left\lfloor b/q \right\rfloor$. Recuperando $k=2b+1$ obtenemos finalmente
\begin{align}\label{eq:deglentepar}
d_k^{(p=2q)} =
\begin{cases}
k \left( 1 + 2 \left\lfloor \frac{k}{2q} \right\rfloor \right),  & \mbox{si }k\mbox{ es impar}\,, \\
0, & \mbox{si }k\mbox{ es par}\,.
\end{cases}
\end{align}
%
%

\subsection[Degeneraciones sobre espacios prisma y poliédricos]{Cálculo de las degeneraciones del laplaciano sobre espacios prisma y poliédricos}
\label{sec:espacios:espectro:prisma}

En lo que sigue haremos uso de la expresión \eqref{eq:deggrupos} para calcular las degeneraciones de los autovalores del laplaciano sobre los espacios prisma $\esp{D_p^*}$ y los poliédricos $\esp{T^*}$, $\esp{O^*}$ y $\esp{I^*}$. 

Para comenzar, notamos que si el grupo $H^*\subset SU(2)$ contiene al elemento $-\mathrm{Id}$, 
un elemento del grupo de rotaciones $H\subset SO(3)$ con ángulo $\theta$ corresponde a dos elementos del grupo binario $H^*$ con ángulos $\theta$ y $\theta+2\pi$ respectivamente. Ahora bien; el carácter del elemento de ángulo $\theta+2\pi$ en la $k$-ésima representación irreducible de $SU(2)$ es, a menos de un signo, el carácter del elemento de ángulo $\theta$:
\begin{align}
\frac{\sen(k\,(\theta+2\pi)/2)}{\sen((\theta+2\pi)/2)} = \frac{\sen(k\pi+k\,\theta/2)}{\sen(\pi+\theta/2)} = (-1)^{k+1} \frac{\sen(k\,\theta/2)}{\sen(\theta/2)}\,,
\end{align}
por lo que las degeneraciones \eqref{eq:deggrupos} pueden escribirse
\begin{align}\label{eq:deggrupos2}
d_k^{(H^*)} = \frac{k}{|H^*|}[1+(-1)^{k+1}]\sum_{h\in H} \frac{\sen(k\,\theta_h/2)}{\sen(\theta_h/2)}\,.
\end{align}
Vemos inmediatamente que para todos estos espacios la degeneración de los autovalores de índice par es nula. Calcularemos entonces las degeneraciones para los autovalores de índice impar.

Consideramos en primer lugar el espacio $D_p^*$; como ya vimos, los elementos de $D_p$ son $p$ rotaciones en ángulos $\theta_m=2\pi m/p$, $m=0,1,\ldots,p-1$, alrededor de algún eje fijo, y $p$ rotaciones de ángulo $\theta=\pi$ alrededor de $p$ ejes perpendiculares al primero. La expresión \eqref{eq:deggrupos2} se escribe entonces, para $k$ impar,
\begin{align}
d_k^{(p)} = \frac{k}{2p} \Bigg[ \lim_{\theta\rightarrow 0}\frac{\sen(k\,\theta/2)}{\sen(\theta/2)} + \sum_{m=1}^{p-1}\frac{\sen(km\pi/p)}{\sen(m\pi/p)} + p\frac{\sen(k\pi/2)}{\sen(\pi/2)} \Bigg]\,.
\end{align}
La expresión entre corchetes se puede calcular fácilmente; el único término que involucra un cálculo indirecto es el segundo: para $k$ impar la suma en cuestión resulta \cite{bromwich,Dowker:2004nh}
\begin{align}
\sum_{m=1}^{p-1}\frac{\sen(km\pi/p)}{\sen(m\pi/p)} = p - k +2p\left\lfloor\frac{k}{2p}\right\rfloor\,,
\end{align}
de donde
\begin{align}\label{eq:degprisma}
d_k^{(p)} =
\begin{cases}
k \left( \left\lfloor \frac{k}{2p} \right\rfloor +\frac{1+(-1)^{\left\lfloor (k-1)/2\right\rfloor}}{2} \right),  & \mbox{si }k\mbox{ es impar}\,, \\
0, & \mbox{si }k\mbox{ es par}\,.
\end{cases}
\end{align}
Las degeneraciones no nulas pueden escribirse también como
\begin{align}\label{eq:degprismaimp}
d_{2n+1}^{(p)} =
\begin{cases}
(2n+1)\left\lfloor \frac{n}{p} \right\rfloor,  & \mbox{si }n\mbox{ es impar}\,, \\
(2n+1) \left( \left\lfloor \frac{n}{p} \right\rfloor +1 \right), & \mbox{si }n\mbox{ es par}\,.
\end{cases}
\end{align}

En el caso de los espacios poliédricos, teniendo en cuenta los elementos de los subgrupos de rotación correspondientes ---que hemos listado en la sección \ref{sec:espacios:espacios:esfera}---, un cálculo similar al que acabamos de mostrar da como resultado para el espacio tetraédrico 
\begin{align}\label{eq:degtetra}
d_{2n+1}^{(T^*)} = (2n+1)\left\{ 2\left\lfloor\frac{n}{3}\right\rfloor + \left\lfloor\frac{n}{2}\right\rfloor - n + 1 \right\}\,,
\end{align}
para el octaédrico 
\begin{align}\label{eq:degocta}
d_{2n+1}^{(O^*)} = (2n+1)\left\{\left\lfloor\frac{n}{4}\right\rfloor + \left\lfloor\frac{n}{3}\right\rfloor + \left\lfloor\frac{n}{2}\right\rfloor - n + 1 \right\}\,,
\end{align}
y para el icosaédrico 
\begin{align}\label{eq:degicosa}
d_{2n+1}^{(I^*)} = (2n+1)\left\{ \left\lfloor\frac{n}{5}\right\rfloor + \left\lfloor\frac{n}{3}\right\rfloor + \left\lfloor\frac{n}{2}\right\rfloor - n + 1 \right\}\,.
\end{align}
%
%

\chapter{Termodinámica del campo escalar conforme sobre espacios esféricos}
\label{sec:termo}

Usando los resultados del capítulo anterior estudiaremos ahora una teoría escalar libre sin masa sobre cada uno de los espacios esféricos: con el espectro del operador asociado a la teoría obtendremos las acciones efectivas y con ellas las entropías y energías de la teoría sobre los diferentes espacios. Comenzaremos analizando el caso de la esfera, cuyo conocimiento detallado simplificará el posterior tratamiento del resto de los espacios. En todos los casos, usaremos propiedades de dualidad de las series infinitas que aparecen al calcular el determinante del operador de fluctuaciones cuánticas de la teoría para obtener dos expresiones diferentes para la acción efectiva, ambas válidas en todo el rango de temperaturas, una de las cuales nos permite tomar fácilmente el límite de altas temperaturas, y la otra el de bajas temperaturas. Con ellas obtendremos las propiedades termodinámicas de la teoría en ambos límites. En el caso de la esfera, calcularemos además la acción efectiva por otro método, con el objetivo de aclarar algunos aspectos confusos en la literatura.

Los cálculos que presentaremos en este capítulo fueron publicados en \cite{Asorey:2012vp}. Cálculos similares fueron realizados para teorías sin temperatura sobre espacios lente en \cite{Dowker:2004nh,Dowker:1978vy,Dowker:1988pa}. El caso de temperatura finita fue estudiado en \cite{DeFrancia:2000xm} sobre ciertas modificaciones de los espacios lente.

La acción clásica para un campo escalar sin masa con acoplamiento conforme en cuatro dimensiones a la métrica de un espacio-tiempo cuyas secciones espaciales están dadas por el espacio esférico $\mathcal{M}=\esp{H}$ se escribe 
\begin{align}\label{eq:accescalarclas}
S[\phi] = \frac12\int_{\mathbb{R}\times\mathcal{M}} d^4x
	\sqrt{g} \left( g^{\mu\nu} \partial_\mu \phi\,\partial_\nu \phi
	+ \frac{1}{a^2}\,\phi^2
	\right)\,,
\end{align}
donde $a$ es el radio de la esfera cubrimiento de $\mathcal{M}$ y donde hemos usado el hecho de que la curvatura escalar de $\mathcal{M}$ es la misma que la de la esfera, $R=6/a^2$. Como ya hemos visto, las propiedades termodinámicas de la teoría pueden obtenerse de las correcciones a un loop a la acción efectiva luego del confinamiento del tiempo euclídeo al intervalo finito $[0,\beta]$, imponiendo al campo condiciones de contorno periódicas en esta variable. Como describimos en el capítulo \ref{sec:mate}, la acción efectiva puede ser calculada entonces como el determinante del operador de fluctuaciones cuánticas de la teoría, que en este caso se escribe
\begin{align}\label{eq:laplconf}
A=-\frac1{\sqrt{g}}\, \partial_\mu \left(g^{\mu\nu}{\sqrt{g}}\,\partial_\nu\right) + \frac{1}{a^2}\,,
\end{align}
actuando sobre el espacio $S^1\times\mathcal{M}$; dicho determinante puede ser calculado por medio de la regularización de la función zeta del operador.

Usando los resultados del capítulo \ref{sec:espacios} vemos que los autovalores del operador \eqref{eq:laplconf} sobre el espacio $S^1\times\mathcal{M}$ vienen dados por $\lambda_{l,k}=(2\pi l/\beta)^2+(k/a)^2$, con $\lambda_{0,k}=(k/a)^2$ el $k$-ésimo autovalor del laplaciano conforme sobre el espacio esférico $\mathcal{M}$. La parte térmica no introduce ninguna degeneración adicional, de modo que las degeneraciones en este caso son las mismas que sobre el correspondiente espacio esférico, que hemos calculado en aquel capítulo y que denotamos $d^{(H)}_k$. La presencia del acoplamiento conforme a la curvatura escalar se traduce en una traslación de los autovalores de la parte espacial con respecto a los del laplaciano \eqref{eq:lapl}, que, como veremos, resulta fundamental para evitar la aparición de modos cero y la consiguiente presencia de ambigüedades en la derivación de las cantidades termodinámicas.

La acción efectiva a temperatura finita puede calcularse entonces a partir de la función zeta
\begin{align}\label{eq:escalartempzeta}
\zeta_{S^1\times\mathcal{M}}(s)={\mu}^{2s} \sum_{l\in\mathbb{Z}}\sum_{k=1}^{\infty} d^{(H)}_k\left[\left(\frac{2\pi l }{\beta}\right)^2+\left(\frac{k}{a}\right)^2 \right]^{-s}\,, 
\end{align}
donde, por no ahorrar prolijidad, hemos introducido el regulador $\mu$, aunque no será relevante porque la teoría es conforme y entonces la función zeta se anula en $s=0$ ---como es conocido que sucede para teorías en dimensión impar \cite{Deser:1993yx}\footnote{ \,Esto es así porque en dimensión impar no es posible construir un invariante local a partir de la métrica y sus derivadas que tenga las dimensiones adecuadas, pero podría cambiar si el espacio tuviera borde \cite{Blau:1988kv}.} y como podremos verificar al final de los cálculos---.

En lo que queda del capítulo calcularemos las acciones efectivas a temperatura finita para la teoría escalar conforme sobre los distintos espacios esféricos. Por simplicidad, denotaremos a las cantidades correspondientes a la teoría sobre el espacio esférico $S^1\times\mathcal{M}$ a veces con un índice $\mathcal{M}$ y otras veces con un índice $H$, omitiendo en la notación toda referencia a la temperatura.

\section{Acción efectiva sobre la esfera}
\label{sec:termo:esfera}

Para el caso en que la variedad espacial $\mathcal{M}$ es la esfera $S^3$, usando las degeneraciones obtenidas en la sección \ref{sec:espacios:espectro}, la función zeta se escribe
\begin{align}\label{eq:zetas3}
\zeta_{S^3}(s)= \mu^{2s} \sum_{l\in\mathbb{Z}}\sum_{k=1}^{\infty} k^2 \left[\left(\frac{2\pi  l}{\beta}\right)^2+\left(\frac{k}{a}\right)^2 \right]^{-s}\,.
\end{align}
Como ya dijimos, queremos usar la función zeta de la teoría para escribir la acción efectiva de formas que nos permitan obtener los límites de altas y bajas temperaturas de las cantidades termodinámicas. Separaremos entonces el problema en dos partes: para obtener el desarrollo a altas temperaturas de la acción efectiva haremos una inversión en la suma sobre $k$ con la que tendremos una expresión donde sólo un número finito de términos no está suprimido exponencialmente con $a/\beta$, mientras que para obtener su desarrollo a bajas temperaturas tendremos que hacer una inversión de la suma sobre $l$, como resultado de la cual tendremos una expresión donde la mayoría de los términos está suprimida exponencialmente con $\beta/a$.

\subsubsection{Expresión útil para tomar el límite de altas temperaturas}
\label{sec:termo:esfera:high}

Con el fin de obtener un desarrollo a altas temperaturas de la acción efectiva, comenzamos escribiendo la función zeta \eqref{eq:zetas3} como
\begin{align}
\zeta_{S^3}(s)= ({\mu a})^{2s} \sum_{l\in\mathbb{Z}}\sum_{k=1}^{\infty} k^2 \left[\left(\frac{2\pi a l}{\beta}\right)^2+k^2 \right]^{-s}\,.
\label{eq:zetas3high}
\end{align}
Como el sumando es invariante ante el cambio $l\rightarrow-l$, obtendremos una simplificación en los cálculos si separamos de esta expresión el término $l=0$, escribiendo $\zeta_{S^3}(s)=\zeta_{S^3}^{l=0}(s)+\zeta_{S^3}^{l\neq 0}(s)$. Vemos inmediatamente que la contribución del término $l=0$ se escribe
\begin{align}
\zeta_{S^3}^{l=0}(s)= ({\mu a})^{2s}\sum_{k=1}^{\infty}k^{-(2s-2)}= ({\mu a})^{2s}{\zeta}_R (2s-2)\,.
\label{eq:zetas3l0}
\end{align}

Notemos que la contribución de este término a la acción efectiva
coincide con la acción efectiva de la misma teoría a temperatura cero sobre el espacio euclídeo $S^3$ ---una propiedad que vale para todos los espacios ``térmicos'' de la forma $S^1\times \mathcal{M}$ y que encontraremos con frecuencia a lo largo de la tesis---. Dicha contribución es
\begin{align}
\Gamma_{S^3}^{l=0} = \frac{\zeta_R(3)}{4\pi^2}\,.
\end{align}

La contribución de los modos $l\neq0$ puede ser dividida en dos términos,
\begin{align}
\zeta_{S^3}^{l\neq 0}(s) = ({\mu a})^{2s} \Bigg\{ &\sum_{l=1}^\infty\sum_{k\in\mathbb{Z}} \left[k^2+\left(\frac{2l\pi a}{\beta}\right)^{\!\!2}\right]^{-(s-1)}\\
 & -\sum_{l=1}^\infty\left(\frac{2l\pi a}{\beta}\right)^{\!\!2} \sum_{k\in\mathbb{Z}}\left[k^2+\left(\frac{2l\pi a}{\beta}\right)^{\!\!2}\right]^{-s}\Bigg\}\,,\nonumber
\end{align}
donde hemos usado el hecho de que el término $k=0$ es nulo para extender las sumas sobre $k$ al conjunto de todos los enteros.
Llamemos $\zeta_1(s)$ a la primera de las sumas dobles entre llaves en la expresión anterior, y $\zeta_2(s)$ a la segunda. Podemos escribir la potencia de la cantidad entre corchetes en $\zeta_1(s)$ en la representación del tiempo propio de Schwinger,
\begin{align}
\zeta_{1}(s) = \frac{({\mu a})^{2s}}{\Gamma(s-1)} \sum_{l=1}^\infty\sum_{k\in\mathbb{Z}}\int_0^{\infty} dt\,t^{s-1-1}e^{-\left[k^2+\left(2\pi al/\beta\right)^2\right]t}\,,
\end{align}
e invertir la suma sobre $k$ usando la fórmula de Poisson \eqref{eq:poisson} para obtener
\begin{align}
\zeta_{1}(s) = \frac{({\mu a})^{2s}\sqrt{\pi}}{\Gamma(s-1)} \sum_{l=1}^\infty\sum_{k\in\mathbb{Z}}\int_0^{\infty} dt\, t^{s-\frac32-1}e^{-\pi^2k^2\!/t}e^{-(2l\pi a)^2t/\beta^2}\,.
\end{align}
Escribiendo explícitamente el resultado de las integrales \cite{abramowitz} y teniendo en cuenta que $s(s-1)\Gamma(s-1)=\Gamma(s+1)$, obtenemos 
\begin{align}\label{eq:zetas3ln01}
\zeta_{1}(s) =  \,\frac{({\mu a})^{2s}(s-1)\sqrt{\pi}}{\Gamma(s)}\,
\Bigg\{ & \Gamma(s-\tfrac32)\!\left(\frac{2\pi a}{\beta}\right)^{\!\!3-2s}\!\!\zeta_R(2s-3)\\
& + 2\sum_{l=1}^\infty\sum_{k=1}^{\infty} \left(\frac{2al}{\beta k}\right)^{\!\!\frac32-s} \!\!K_{s-\frac32}\!\left(4{\pi}^2lka/\beta\right)\Bigg\}\,,\nonumber
\end{align}
donde $K_{\nu}(x)$ es la función de Bessel modificada de orden $\nu$ y argumento $x$.

\vfill
\pagebreak

Procediendo análogamente puede obtenerse la extensión analítica de $\zeta_{2}(s)$, para la que resulta
\begin{align}\label{eq:zetas3ln02}
\zeta_{2}(s) = - \frac{({\mu a})^{2s}\sqrt{\pi}}{\Gamma(s)}\, \Bigg\{ & \Gamma(s-\tfrac12)\!\left(\frac{2\pi a}{\beta}\right)^{\!\!3-2s}\!\!\zeta_R(2s-3)\\
& + 2\sum_{l=1}^\infty\sum_{k=1}^{\infty}\left(\frac{2\pi al}{\beta}\right)^{\!\!2}\left(\frac{2al}{\beta k}\right)^{\!\!\frac12-s}\!\!K_{s-\frac12}\!\left(4{\pi}^2lka/\beta\right)\Bigg\}\,.\nonumber
\end{align}

Combinando las expresiones \eqref{eq:zetas3l0}, \eqref{eq:zetas3ln01} y \eqref{eq:zetas3ln02} puede verificarse que la zeta se anula cuando su argumento es cero, como tiene que ser tratándose de un problema sobre un espacio tridimensional sin borde, en el que no hay anomalía de traza. Por esa misma razón su derivada se efectúa fácilmente, y obtenemos para la acción efectiva la expresión
\begin{align}
\Gamma_{S^3} =  \frac{\zeta_R (3)}{4{\pi}^2} &-\frac{{\pi}^4}{45} \left(\frac{a}{\beta}\right)^{\!\!3}\nonumber\\
 & +  \frac{1}{4\pi^2}\sum_{k,l=1}^{\infty}\frac{1}{k^3} \left[ 2 + 2\frac{4\pi^2kla}{\beta} + \left( \frac{4\pi^2kla}{\beta} \right)^{\!\!2} \right] e^{-4\pi^2kla/\beta}\,.
\label{eq:seffs3high}
\end{align}

Notemos que esta expresión es válida en todo el rango de temperaturas, y que permite obtener directamente el límite de altas temperaturas: para $\beta\rightarrow0$, la segunda línea tiende exponencialmente a cero y sólo quedan los dos términos de la primera línea. De éstos, el segundo término ---que diverge--- es el de volumen, que es extensivo y tiene la dependencia usual de Stefan-Boltzmann con la temperatura. El primero es la contribución del modo $l=0$, que no depende de la temperatura ni del volumen del espacio, que llamaremos en adelante simplemente \emph{término constante} y que, como veremos, tendrá que ver con una contribución topológica en los demás espacios esféricos. Como se menciona en \cite{Gustavsson:2018sgi}, este mismo término es el que en \cite{Kim:2012qf} ---donde se calculan funciones de partición en la esfera $S^5$--- se obtiene como la contribución perturbativa. Al final de esta sección haremos un comentario sobre la no aparición de este término en algunos artículos previos a nuestro trabajo.

\vfill
\pagebreak

\subsubsection{Expresión útil para tomar el límite de bajas temperaturas}
\label{sec:termo:esfera:low}

Consideramos ahora nuevamente la función zeta \eqref{eq:zetas3} con el objetivo de obtener para ella una expresión que permita tomar el límite $\beta/a\rightarrow\infty$. Para eso la escribimos en primer lugar como
\begin{align}
\zeta_{S^3}(s) = \left(\frac{\mu\beta}{2\pi}\right)^{\!\!2s}\sum_{l\in\mathbb{Z}} \sum_{k=1}^{\infty} k^2 \left[
\left(\frac{\beta k}{2\pi a}\right)^{\!\!2} + l^2 \right]^{-s}
\nonumber\end{align}
o, en términos del tiempo propio de Schwinger,
\begin{align}\label{eq:zetas3low}
\zeta_{S^3}(s) = \frac{1}{\Gamma(s)} \left(\frac{\mu\beta}{2\pi}\right)^{\!\!2s} \sum_{l\in\mathbb{Z}}\sum_{k=1}^{\infty}
k^2 \int_0^{\infty} dt\, t^{s-1} e^{-\left[
	\left(\beta k/2\pi a\right)^2 + l^2 \right]t}\,.
\end{align}
Usando ahora la inversión \eqref{eq:poisson} en la suma sobre $l$ obtenemos
\begin{align}\nonumber
\zeta_{S^3}(s) = \frac{\sqrt{\pi}}{\Gamma(s)} \left(\frac{\mu\beta}{2\pi}\right)^{\!\!2s} \sum_{k=1}^{\infty} k^2 \sum_{l\in\mathbb{Z}} \int_0^{\infty} dt\, t^{s-\frac{3}{2}} e^{-\left(\beta k/2\pi a\right)^2t}
e^{-(\pi l)^2\!/t}\,.\nonumber
\end{align}
El término $l=0$ de la suma resulta
\begin{align}\label{eq:zetas3lowl0}
\zeta_{S^3}^{l=0}(s) = \frac{(\mu a)^{2s}\beta}{2\sqrt{\pi}a} \frac{\Gamma(s-\tfrac{1}{2})}{\Gamma(s)} \zeta_R(2s-3)\,,
\end{align}
mientras que para los términos con $l\neq0$ podemos aprovechar la simetría de reflexión en $l$ y escribir, una vez resueltas explícitamente las integrales,
\begin{align}\label{eq:zetas3lowlno0}
\zeta_{S^3}^{l\neq 0}(s) =
\frac{4\sqrt{\pi}}{\Gamma(s)}\left(\frac{2\pi}{\mu\beta}\right)^{\!\!-2s}\sum_{k=1}^{\infty}k^2 \sum_{l=1}^{\infty}
\left(\frac{2\pi^2al}{\beta k}\right)^{\!\!s-\frac{1}{2}}\!\!K_{\frac{1}{2}-s}\!\left(kl\beta /a\right)\,.
\end{align}
Combinando las ecuaciones \eqref{eq:zetas3lowl0} y \eqref{eq:zetas3lowlno0} vemos que se verifica $\zeta_{S^3}(0)=0$, y podemos obtener la acción efectiva
\begin{align}
\Gamma_{S^3} = \frac{\beta}{2a}\zeta_R(-3) -\sqrt{\frac{2\beta}{\pi a}}\,\sum_{k=1}^{\infty} k^2 \sum_{l=1}^{\infty}
\sqrt{\frac{k}{l}}\, K_{\frac{1}{2}}\!\left(k l\beta/a\right)\,,
\end{align}
o, usando la expresión para la función de Bessel modificada $K_{\frac12}(x)$ en términos de $\exp(-x)$,
\begin{align}\label{eq:seffs3low}
\Gamma_{S^3} = \frac{\beta}{240 a} -\sum_{k=1}^{\infty} k^2 \sum_{l=1}^{\infty}
\frac{e^{-kl\beta/a}}{l} = \frac{\beta}{240a} + \sum_{k=1}^{\infty} k^2 \log(1-e^{-k\beta/a})\,.
\end{align}

\vfill
\pagebreak

Haremos ahora una digresión en el camino hacia la obtención de las cantidades termodinámicas sobre los distintos espacios esféricos para comentar brevemente cómo la aplicación de la llamada fórmula de Abel-Plana a la serie en el desarrollo \eqref{eq:seffs3low} de la acción efectiva a bajas temperaturas da como resultado el desarrollo \eqref{eq:seffs3high} a altas temperaturas. Una sutileza en este cálculo hizo que los autores de \cite{Elizalde:2003cv,Elizalde:2002ak} no obtuvieran el término constante.

\subsubsection{Otra forma de regularizar series (o por qué algunos autores no obtienen el término constante)}
\label{sec:termo:high:esfera:topologico}

En esta tesis usamos principalmente la regularización zeta para dar sentido físico a expresiones en principio divergentes. Como ya vimos, podemos usar representaciones integrales como la del tiempo propio de Schwinger \eqref{eq:mellin} para transformar una serie en una integral de la que podamos aislar el valor en el punto de interés. Sin embargo, existen tantos métodos de regularización analítica como formas de definir una extensión analítica a un punto uno pueda imaginar. Uno de estos métodos consiste en usar el teorema de los residuos para escribir la serie como la integral sobre un contorno en el plano complejo de una función con los polos adecuados. Esto da lugar a la llamada \emph{fórmula de Abel-Plana}.

\begin{figure}[h!]
		\centering
		\includegraphics[height=8cm]{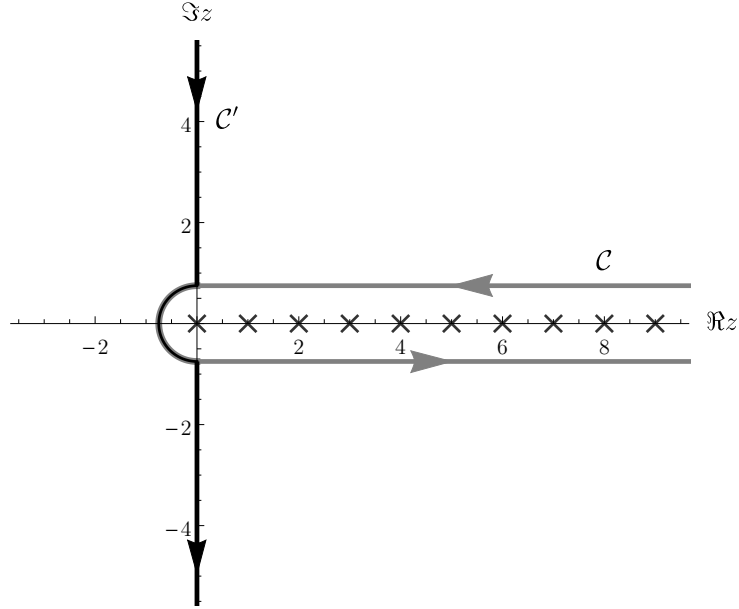}
	\caption{Contornos inicial y deformado en el método de Abel-Plana.} \label{fig:ap1}
\end{figure}

Sea $f(z)$ una función analítica en el semiplano $x\geq0$ del plano complejo $z=x+iy$, tal que $e^{-2\pi y}|f(x+iy)| \rightarrow0$ uniformemente para todo valor de $x$ cuando $|y|\rightarrow\infty$. Consideremos la función $f(z)/(e^{2\pi i z}-1)$, que tiene polos simples en los puntos $z_n=n\in\mathbb{Z}$, con residuos $R_n=f(n)/2\pi i$ respectivamente. Sea $\epsilon\in(0,1)$, y $\mathcal{C}$ el contorno en el plano complejo formado por la semirrecta $y=\epsilon$, la semicircunferencia de radio $\epsilon$ que rodea al origen por la izquierda y la semirrecta $y=-\epsilon$, como se muestra en la figura \ref{fig:ap1}. Del teorema de los residuos resulta
\begin{align}\label{eq:ap1}
\sum_{n=0}^{\infty} f(n) = \oint_{\mathcal{C}} dz \frac{f(z)}{e^{2\pi iz}-1}\,.
\end{align}

Como la función $f(z)$ es analítica en el semiplano derecho del plano complejo y se comporta apropiadamente en el infinito, podemos deformar el contorno $\mathcal{C}$ al contorno $\mathcal{C}'$ de la figura \ref{fig:ap1} ---el eje imaginario, evitando nuevamente el origen con una semicircunferencia de radio $\epsilon$ hacia la izquierda, y cerrado por una semicircunferencia en el infinito---. La integral \eqref{eq:ap1} se escribe entonces
\begin{align}\label{eq:ap2}
\oint_{\mathcal{C'}} dz \frac{f(z)}{e^{2\pi iz}-1} ={} & \,i\int_\epsilon^\infty dy \frac{f(iy)}{1-e^{-2\pi y}} + \frac12\, \mathrm{Res}\!\left(\frac{f(z)}{e^{2\pi iz}-1},0\right) \\ \nonumber
& +i\int_\epsilon^\infty dy \frac{f(-iy)}{1-e^{2\pi y}} 
\,.
\end{align}

Ahora bien; para la primera de las integrales en $y$ tenemos trivialmente
\begin{align}\label{eq:ap3}
\int_\epsilon^\infty dy \frac{f(iy)}{1-e^{-2\pi y}} & = \int_\epsilon^\infty dy\, f(iy)\, \frac{1-e^{-2\pi y}+e^{-2\pi y}}{1-e^{-2\pi y}}\\ \nonumber
& = -i\int_0^\infty dx\, f(x+i\epsilon) + \int_\epsilon^\infty dy \frac{f(iy)}{e^{2\pi y}-1}
\,,
\end{align}
que, con el resto de la expresión \eqref{eq:ap2}, nos permite escribir finalmente, tomando el límite $\epsilon\rightarrow 0^+$, la fórmula de Abel-Plana
\begin{align}\label{eq:ap}
\sum_{n=0}^{\infty} f(n) = \frac12\,f(0) + \int_0^\infty dx\, f(x)\, + i\int_0^\infty dy\, \frac{f(iy)-f(-iy)}{e^{2\pi y}-1}
\,.
\end{align}

\smallskip

\begin{figure}[h!]
	\centering
	\includegraphics[height=8.5cm]{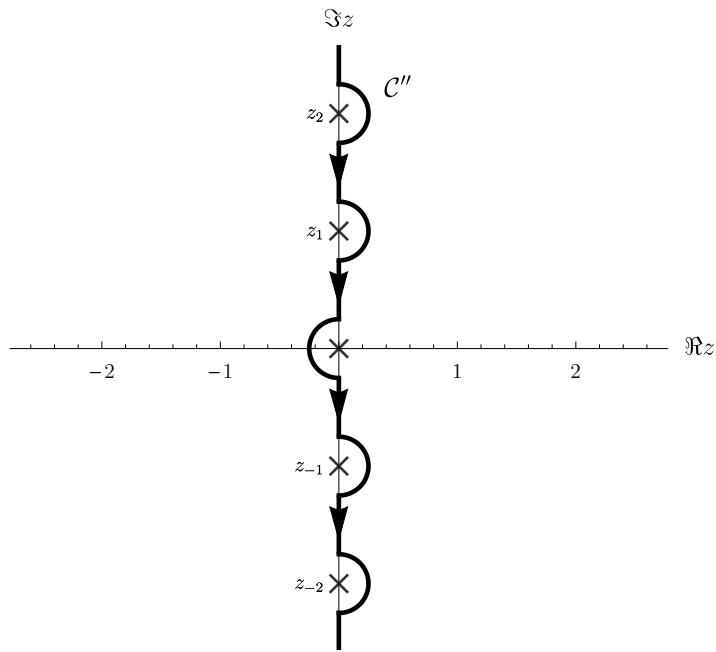}
	\caption{Contorno deformado en el método de Abel-Plana ``modificado''.} \label{fig:ap2}
\end{figure}

Queremos ahora aplicar la fórmula de Abel-Plana a la serie en \eqref{eq:seffs3low}, para lo que tenemos que considerar la función $g(z) = z^2\log(1-e^{-z\beta/a})$. Observamos en primer lugar que esta función no es analítica en todo el semiplano derecho del plano complejo: en efecto, $g(z)$ es multivaluada y, aunque podemos elegir sus cortes de ramificación hacia afuera del semiplano derecho, no podemos evitar los puntos de ramificación en los valores $z_n=\frac{2\pi a n}{\beta}i$, $n\in\mathbb{Z}$, de su argumento. Estos puntos yacen sobre el eje imaginario, de modo que la deformación del contorno $\mathcal{C}$ en el método de Abel-Plana debe efectuarse esquivándolos. Lo haremos usando, en lugar del contorno $\mathcal{C'}$ considerado en el caso analítico, el contorno $\mathcal{C''}$ que se muestra en la figura \ref{fig:ap2}, en el que hemos introducido una semicircunferencia $\mathcal{C}_\epsilon^{(n)}$ de radio $\epsilon\in(0,1)$ hacia la derecha de cada uno de los puntos de ramificación $z_n$, $n\neq0$.

El cambio del contorno deformado podría introducir alguna modificación en la fórmula de Abel-Plana. Para ver si ese es el caso, volvamos por un momento a la derivación anterior y notemos que la expresión \eqref{eq:ap2} se vería modificada por la introducción de las infinitas integrales sobre las semicircunferencias,
\begin{align}
\int_{\mathcal{C}_\epsilon^{(n)}} dz\,\frac{z^2\log(1-e^{-z\beta/a})}{e^{2\pi i z}-1}\,,
\end{align}
con $n\in\mathbb{Z}$. Parametrizando $\mathcal{C}_\epsilon^{(n)}$ como $z=z_n+\epsilon e^{i\varphi}$, con $-\pi/2\leq\varphi\leq\pi/2$, puede verse fácilmente que esta integral se anula para $\epsilon\rightarrow0^+$, de modo que la integral sobre $\mathcal{C''}$ coincide con la integral sobre $\mathcal{C'}$ y la fórmula de Abel-Plana puede aplicarse a la función $g(z)$ en su forma usual \eqref{eq:ap}.

La primera de las integrales en la fórmula de Abel-Plana se resuelve directamente para $g(x)$ y resulta
\begin{align}
\int_0^{\infty} dx\, x^2 \log(1-e^{-x \beta/a})= -\frac{\pi^4 a^3}{45 \beta^3}\,.
\end{align}
Sin embargo, la segunda integral tiene que ser tratada con cuidado debido a la presencia de los puntos de ramificación. El resultado que se obtiene es
\begin{align}
i \int_0^\infty dy\,\frac{g(iy)-g(-iy)}{e^{2\pi y}-1} =& \,\int_0^\infty dy\,\frac{y^2(\pi-\beta x/a)}{e^{2\pi y}-1}\\ 
&+2\pi \sum_{n=0}^\infty n \int_{2\pi na/\beta}^{2\pi(n+1)a/\beta} dy\,\frac{y^2}{e^{2\pi y}-1}\,.\nonumber
\end{align}
La primera de las integrales en el lado derecho puede ser resuelta directamente y se obtiene
\begin{align}
\int_0^\infty dy\,\frac{y^2(\pi-\beta x/a)}{e^{2\pi y}-1} = \frac{\zeta_R(3)}{4\pi^2}-\frac{\beta}{240 a}\,.
\end{align}
Para resolver la integral en el término $n$-ésimo de la serie podemos usar en el integrando el desarrollo en serie
\begin{align}
\frac{1}{e^{{2\pi x}-1}} = e^{-2\pi x} \frac{1}{1-e^{-2\pi x}} = e^{-2\pi x} \sum_{k=0}^{\infty} e^{-2\pi kx}\,,
\end{align}
e intercambiar esta nueva serie con la integral, con lo que esta última se resuelve fácilmente, obteniéndose finalmente para \eqref{eq:seffs3low} la expresión alternativa
\begin{align}\nonumber
\Gamma_{S^3} =  -\frac{\pi^4 a^3}{45 \beta^3} + \frac{\zeta_R(3)}{4\pi^2} + \frac{1}{4\pi^2} \sum_{l,k=1}^{\infty} \frac{1}{k^3} \left[ 2 + 2\frac{4\pi^2lka}{\beta} + \left( \frac{4\pi^2lka}{\beta}\right)^{\!\!2} \right] e^{-4\pi^2lka/\beta} \,,
\end{align}
que no es otra cosa que el desarrollo \eqref{eq:seffs3high} de la acción efectiva a altas temperaturas.

En el artículo \cite{Elizalde:2002ak} se obtuvo, como aquí, el desarrollo de la acción efectiva a altas temperaturas a partir del de bajas temperaturas por medio de la fórmula de Abel-Plana. Sin embargo, los autores de ese artículo no tienen en cuenta el cambio en los argumentos de los logaritmos, por lo que pierden el término constante y también los términos exponencialmente suprimidos con la temperatura, aunque a estos últimos los obtienen en \cite{Elizalde:2003cv}. La ausencia del término constante en los artículos \cite{Brevik:2002gh,Kutasov:2000td} obedece a razones diferentes: allí se regulariza una derivada de la acción efectiva de la teoría y luego se integra el resultado. En el proceso, los autores descartan la constante de integración, que no es otra cosa que el término constante.

El lector que quiera conocer otras aplicaciones de la fórmula de Abel-Plana a la regularización de cantidades divergentes en Teoría Cuántica de Campos encontrará interesante el artículo \cite{Saharian:2007ph}, donde además se estudia una de sus generalizaciones, con énfasis en cálculos en el marco del efecto Casimir. Una generalización de la fórmula a funciones con singularidades no integrables puede encontrarse en \cite{Fialkovski:2007jh}.

\section{Acciones efectivas en el límite de altas temperaturas}
\label{sec:termo:high}

A continuación obtendremos un desarrollo a altas temperaturas de la acción efectiva de teorías escalares sin masa sobre los distintos espacios esféricos. Comenzamos escribiendo la función zeta \eqref{eq:escalartempzeta} como
\begin{align}
\zeta_{\mathcal{M}}(s)= ({\mu a})^{2s} \sum_{l\in\mathbb{Z}}\sum_{k=1}^{\infty} d_k^{(H)} \left[\left(\frac{2\pi a l}{\beta}\right)^2+k^2 \right]^{-s}\,,
\label{eq:zetaaltas}
\end{align}
donde, como hicimos en el caso de la esfera, hemos extraído un factor $a$ para obtener términos exponencialmente suprimidos con $\beta$. En los apartados siguientes consideraremos por separado los espacios lente impares y pares, los espacios prisma y los poliédricos.

\subsection{Espacios lente impares}
\label{sec:termo:high:oddlens}

La degeneración del $k$-ésimo autovalor del laplaciano sobre el espacio lente impar $\mathcal{M}=\esp{Z_{2q+1}}$ viene dada por la expresión \eqref{eq:deglenteimpar}, que reescribimos como 
\begin{align}\label{eq:deglenteimparrees}
	d_k^{(Z_{2q+1})} =
	\begin{cases}
		\frac{k^2}{2q+1}-\frac{rk}{2q+1} +k,  & \mbox{si }r\mbox{ es impar}\,, \\
		\frac{k^2}{2q+1}-\frac{rk}{2q+1}, & \mbox{si }r\mbox{ es par}\,,
	\end{cases}
\end{align}
donde $r$ es el resto de dividir $k$ por $2q+1$.
El primer término en ambas líneas no es otra cosa que la degeneración  del autovalor $k$-ésimo del laplaciano sobre la esfera $S^3$, dividida por el orden $2q+1$ del grupo cíclico. En consecuencia, insertando esta expresión en la función zeta \eqref{eq:zetaaltas} podemos escribir
\begin{align}
	\zeta_{\esp{{\mathbb Z}_{2q+1}}}(s) = \frac{\zeta_{S^3}(s)}{2q+1} +\delta\zeta_{2q+1}(s)\,,
\label{eq:zetaimpares}
\end{align}
de donde vemos que el conocimiento de la función zeta sobre la esfera nos permite obtener una parte de las funciones zeta sobre espacios lente impares. A continuación extenderemos la otra parte de la función zeta de forma similar a como hicimos con la esfera. Escribimos primeramente
\begin{align}\nonumber
\delta\zeta_{2q+1}(s) ={} & 
-\frac{({\mu a})^{2s}}{p}\sum_{l\in\mathbb{Z}}\sum_{n=0}^{\infty}\Bigg\{ \sum_{r=1}^q 2r\left(np+2r\right)\!\left[\left(np+2r\right)^2 +\left(\frac{2\pi al}{\beta}\right)^{\!\!2}\right]^{-s} \\ \nonumber
&+ \sum_{r=0}^{q-1} 2(r-q)\left(np+2r+1\right)
\!\left[\left(np+2r+1\right)^2 +\left(\frac{2\pi al}{\beta}\right)^{\!\!2}\right]^{-s}\Bigg\}\,,
\end{align}
donde hemos definido nuevamente $p:=2q+1$, con $k=np+r$ para $0\leq r\leq 2q$, y donde además hemos separado los casos en que dicho resto es par de los casos donde es impar y ---en un abuso de notación--- hemos denotado a ese resto respectivamente $2r$ o $2r+1$. Haciendo un cambio en el índice de la segunda suma entre llaves y reordenando factores $p$ obtenemos
\begin{align}\nonumber
\delta\zeta_{2q+1}(s) ={} & -2{\left(\frac{\mu a}{p}\right)}^{\!\!2s}
\sum_{l\in\mathbb{Z}}\sum_{n=0}^{\infty}  \Bigg\{\sum_{r=1}^q r\left(n+\tfrac{2r}{p}\right)\!\left[\left(n+\tfrac{2r}{p}\right)^{\!2} +\left(\frac{2\pi a l}{p\beta}\right)^{\!\!2}\right]^{-s} \\ \label{eq:zetaimpares2} 
&- \sum_{r=1}^{q} r\!\left(n+1-\tfrac{2r}{p}\right) \!\left[\left(n+1-\tfrac{2r}{p}\right)^{\!2} +\left(\frac{2\pi a l}{p\beta}\right)^{\!\!2}\right]^{-s}\Bigg\}\,.
\end{align}

Como en el caso de la esfera, separamos de la última expresión el término $l=0$, escribiendo $\delta\zeta_p=\delta\zeta_p^{l=0}+\delta\zeta_p^{l\neq0}$. Comenzamos evaluando la primera parte ---la más simple---, que da directamente
\begin{align}
\delta\zeta_{2q+1}^{l=0}(s) = -2\left(\frac{\mu a}{p}\right)^{\!\!2s} \sum_{r=1}^{q}r \Bigg[\zeta_H\!\left(2s-1,\tfrac{2r}{p}\right)- \zeta_H\!\left(2s-1,1-\tfrac{2r}{p}\right)\!\Bigg]\,. 
\end{align}
Usando la representación de la zeta de Hurwitz como suma de series de senos y cosenos que puede encontrarse por ejemplo en \cite[página 1037]{Gradshteyn} y teniendo en cuenta que $\cos{(s-\frac12)\pi}=\pi/\Gamma(s)\Gamma(1-s)$, podemos escribir
\begin{align}\label{eq:deltazetaimparesl0}
\delta\zeta_{2q+1}^{l=0}(s)=-4 \left(\frac{\mu a}{p}\right)^{\!\!2s}\frac{\Gamma(2-2s)({2\pi})^{2s-1}}{\Gamma(s)\Gamma(1-s)}
\sum_{r=1}^{q}r\sum_{n=1}^{\infty}n^{2s-2} \sen\!\left(\frac{4\pi nr}{p}\right)\,,
\end{align}
de donde vemos que esta parte de la función zeta se anula en $s=0$.

En cuanto a la contribución de los términos $l\neq 0$, comenzamos cambiando el índice en la segunda suma entre llaves en \eqref{eq:zetaimpares2} para obtener
\begin{align}\label{eq:zetaimpareslnot0}
\delta\zeta_{2q+1}^{l\neq 0}(s)= -4{\left(\frac{\mu a}{p}\right)}^{\!\!2s}\!\sum_{r=1}^{q} r\sum_{n\in\mathbb{Z}}\sum_{l=1}^{\infty} \left(n+\tfrac{r}{p}\right) \!\Bigg[\left(n+\tfrac{r}{p}\right)^{\!\!2} +\left(\frac{2\pi a l}{p\beta}\right)^{\!\!2}\Bigg]^{-s},
\end{align}
que para $\Re(s)$ suficientemente grande también puede escribirse
\begin{align}\label{eq:zetaimpareslnot01}
\delta\zeta_{2q+1}^{l\neq 0}(s) ={} & -\frac{{(\mu a)}^{2s}{p}^{1-2s}}{(1-s)}\sum_{r=1}^{q} r  \\ \nonumber
& \qquad\qquad\qquad\times \left.\frac{d}{d\alpha} \sum_{n\in\mathbb{Z}}\sum_{l=1}^{\infty} \Bigg[\left(n+\tfrac{2r+2\alpha-2}{p}\right)^{\!\!2}
+\left(\frac{2\pi a l}{p\beta}\right)^{\!\!2}\Bigg]^{-s+1}\right\vert_{\alpha=1}.
\end{align}
Podemos ahora extender analíticamente la doble suma infinita para evaluarla en $s=0$ y calcular la acción efectiva de manera similar a como hicimos en el caso de la esfera. En efecto, escribiendo la potencia entre corchetes en la representación del tiempo propio de Schwinger e invirtiendo la suma sobre $k$, y luego de evaluar la derivada con respecto a $\alpha$, obtenemos
\begin{align}\label{eq:deltazetaimpareslnot02}
\delta\zeta_{2q+1}^{l\neq 0}(s) ={} & -\frac{16{(\mu a)}^{2s}{\pi}^\frac32}{\Gamma(s){p}^{2s}}\sum_{r=1}^{q} r\!\sum_{l,n=1}^{\infty} n \sen\!\left(\frac{4\pi n r}{p}\right)
\\ \nonumber
&\qquad\qquad\qquad\qquad\qquad\qquad\times\left(\frac{p\beta n}{al}\right)^{\!\!s-\frac32}
\!K_{s-\frac32}\!\left(4{\pi}^2aln/p\beta\right)\,,
\end{align}
que también se anula en $s=0$.

A partir de las expresiones \eqref{eq:deltazetaimparesl0} y \eqref{eq:deltazetaimpareslnot02} podemos obtener la contribución de $\delta\zeta_{2q+1}(s)$ a la acción efectiva. Usando además el resultado \eqref{eq:seffs3high} para la esfera tenemos, según \eqref{eq:zetaimpares},
\begin{align}\label{eq:seffimpareshigh}
\Gamma_{\esp{{Z_{p=2q+1}}}} ={} &\frac{\zeta_R(3)}{4{\pi}^2p} +
\frac{1}{\pi}\sum_{r=1}^{q}r \sum_{n=1}^{\infty}\frac{1}{n^2}\sen\!\left(\frac{4\pi nr}{p}\right)
-\frac{{\pi}^4}{45p} \left(\frac{a}{\beta}\right)^{\!\!3}\\ \nonumber 
&+ \frac{1}{4\pi^2p}\sum_{k,l=1}^{\infty}\frac{1}{k^3} \left[ 2 + 2\frac{4\pi^2akl}{\beta} + \left( \frac{4\pi^2akl}{\beta} \right)^{\!\!2} \right] e^{-4\pi^2akl/\beta} \\ \nonumber
&+\frac{2}{\pi p}\sum_{r=1}^{q}r\sum_{n,l=1}^{\infty}\frac{1}{n^2}
\sen\!\left(\frac{4\pi n r}{p}\right)\!\left(1+\frac{4\pi^2 aln}{p\beta}\right)e^{-4\pi^2 anl/{p\beta}}\,.
\end{align}

Podemos ver que en el límite $\beta\rightarrow0$ ---en el que sólo queda la primera línea de la expresión anterior--- los dos primeros términos son independientes de la temperatura y el volumen del espacio y,  recordando que el volumen de $\esp{Z_{2q+1}}$ es el volumen de la esfera $S^3$ dividido por $2q+1$, el tercer término es una contribución extensiva, con una dependencia tipo Stefan-Boltzmann con la temperatura.

\subsection{Espacios lente pares}
\label{sec:termo:high:evenlens}

Para los espacios lente de orden par, la degeneración de los autovalores del laplaciano con índice par se anula. En el caso particular del más grande de esos espacios ---$\esp{Z_2}$, que se conoce usualmente como el \emph{espacio proyectivo} real en tres dimensiones, y se denota también $\mathbb{R}\mathrm{P}^3$---, además, la degeneración de los autovalores con índice impar es la misma que la del autovalor correspondiente del laplaciano sobre la esfera. Esto hace que la acción efectiva sobre este espacio pueda obtenerse a partir de la de la esfera; para verlo, escribamos en primer lugar la función zeta 
\begin{align}\label{eq:zetaz2high}
\zeta_{\esp{Z_2}}(s)= ({\mu a})^{2s} \sum_{l\in\mathbb{Z}}\sum_{n=0}^{\infty} (2n+1)^2 \!\left[\tilde{\omega}_l^2+(2n+1)^2 \right]^{-s}\,,
\end{align}
donde por comodidad hemos introducido las frecuencias de Matsubara adimensionalizadas $\tilde{\omega}_l = 2\pi al/\beta$. Sumando y restando una contribución equivalente de los modos pares y renombrando según el caso $k=2n$ ó $k=2n+1$ obtenemos
\begin{align}\label{eq:zetaz2high1}
\zeta_{\esp{Z_2}}(s)={} ({\mu a})^{2s} \sum_{l\in\mathbb{Z}}\sum_{k=1}^{\infty}\Big\{k^2 \!\left(\tilde{\omega}_l^2+k^2 \right)^{-s}-
(2k)^2 \!\left[\tilde{\omega}_l^2+(2k)^2 \right]^{-s}\Big\}\,.
\end{align}
Haciendo explícita en la notación la dependencia en $\beta$, identificamos al primer término en esta última expresión con $\zeta_{S^3}(s,\beta)$, y observamos que el segundo término puede escribirse como $-2^{2-2s} \zeta_{S^3}(s,2\beta)$. Si tenemos en cuenta además que $\zeta_{S^3}(0)=0$ para cualquier valor de $\beta$, vemos que es posible obtener la acción efectiva sobre este espacio como
\begin{align}\label{eq:seffs3z2}
\Gamma_{\esp{Z_2}}(\beta)=\Gamma_{S^3}(\beta)-4\Gamma_{S^3}(2\beta)\,.
\end{align}

Queremos ahora obtener las acciones efectivas sobre los demás espacios lente pares. En esa dirección, partimos de las degeneraciones \eqref{eq:deglentepar}, y redefinimos el índice de aquellas que no son nulas:
\begin{align}\label{eq:deglentepar1}
d_{2k+1}^{(Z_{2q})}=\left(2\left[\frac{2k+1}{2q}\right]+1\right)\!(2k+1)\,.
\end{align}
Haciendo $k=nq+r$ con $n\in\mathbb{N}_0$ y $r=0,1,...,q-1$, podemos escribir
\begin{align}\label{eq:degeven}
d_{2k+1}^{(Z_{2q})} = \left(2\frac{k-r}{q}+1\right)\!(2k+1) = \frac{(2k+1)^2}{q}+\frac{q-2r-1}{q}(2k+1)\,,
\end{align}
donde en la última igualdad recuperamos, a menos de un factor $q$, la degeneración sobre $\esp{Z_2}$, de modo que la función zeta puede escribirse
\begin{align}\label{eq:zetahigheven}
\zeta_{\esp{Z_{2q}}}(s)=\frac{\zeta_{\esp{Z_2}}(s)}{q}+\delta\zeta_{2q}(s)\,,
\end{align}
con
\begin{align}\label{eq:deltazetahigheven}  
\delta\zeta_{2q}(s) ={} & 2 {\left(\frac{\mu a}{2q}\right)}^{\!\!2s} \sum_{r=0}^{q-1} \left[(q-r-1)-r\right] \\ \nonumber
&\qquad\qquad\qquad\qquad\times\sum_{l\in\mathbb{Z}}\sum_{n=0}^{\infty} \left(n+\tfrac{2r+1}{2q}\right) \!\!\left[\left(n+\tfrac{2r+1}{2q}\right)^{\!2}  + \left(\frac{\pi al}{q\beta}\right)^{\!\!2}\right]^{-s} \,.
\end{align}
Separando ahora el término entre paréntesis en el factor de la primera línea dentro de la suma finita y renombrando en ese término el índice de dicha suma, podemos escribir
\begin{align}\nonumber
\delta\zeta_{2q}(s) ={} & 2{\left(\frac{\mu a}{2q}\right)}^{\!\!2s} \sum_{r=1}^{q-1} r \sum_{l\in\mathbb{Z}}\sum_{n=0}^{\infty} \Bigg\{\!\left(n+1-\tfrac{2r+1}{2q}\right) \!\left[\left(n+1-\tfrac{2r+1}{2q}\right)^{\!2} +\left(\frac{\pi al}{q\beta}\right)^{\!\!2}\right]^{-s}\\ \label{eq:deltazetahigheven1}
&\qquad\qquad\qquad-\left(n+\tfrac{2r+1}{2q}\right) \!\left[\left(n+\tfrac{2r+1}{2q}\right)^{\!2} +\left(\frac{\pi al}{q\beta}\right)^{\!\!2}\right]^{-s} \Bigg\}\,.
\end{align}

Como en los casos anteriores, evaluamos en primer lugar la contribución del término $l=0$. Tenemos
\begin{align}\label{eq:deltazetahighevenl0}
\delta\zeta_{2q}^{l=0}(s) ={} & -2{\left(\frac{\mu a}{2q}\right)}^{\!\!2s} \sum_{r=1}^{q-1}r\, \Big\{ \zeta_H\!\left(2s-1,\tfrac{2r+1}{2q}\right) \\ \nonumber
&\qquad\qquad\qquad\qquad\qquad -\zeta_H\!\left(2s-1,1-\tfrac{2r+1}{2q}\right)\!\Big\}\,.  
\end{align}
\vfill
\pagebreak

Esta expresión puede ser reescrita, luego de renombrar índices y usando el teorema de multiplicación de la función zeta de Hurwitz \cite[página 13]{apostol}, como
\begin{align}\nonumber 
\delta\zeta_{2q}^{l=0}(s) = -2{\left(\frac{\mu a}{2q}\right)}^{\!\!2s}\Bigg[\sum_{r=0}^{q-1}(2r+1)\zeta_H\!\left(2s-1,\tfrac{2r+1}{2q}\right)-q^{2s}\zeta_H\!\left(2s-1,\tfrac12\right)\Bigg] \,, 
\end{align}
que, sumada a la contribución correspondiente de $\esp{Z_2}$, reproduce el resultado reportado en \cite{Dowker:2004nh}.

La contribución de los modos con $l\neq 0$ puede evaluarse de la misma forma que en el caso impar, resultando
\begin{align}\label{eq:deltazetahighevenlnot0}
\delta\zeta_{2q}^{l\neq 0}(s) ={} & - \frac{16\pi^{3/2}}{q\Gamma(s)} {\left(\frac{\mu a}{2q}\right)}^{\!\!2s}
\sum_{r=1}^{q-1} r \sum_{l,n=1}^{\infty} n\sen\!\left(2\pi n \tfrac{2r+1}{2q}\right) \\ \nonumber
& \qquad\qquad\qquad\qquad\quad\qquad\qquad\qquad
\times\left(\frac{q\beta n}{al}\right)^{\!\!s-\frac32}\!K_{s-\frac{3}{2}}\! \left(\frac{2\pi^2aln}{q\beta}\right)\,.
\end{align}

Reuniendo los distintos resultados podemos finalmente calcular la acción efectiva sobre los espacios lente pares, que se escribe
\begin{align}\label{eq:seffhigheven}
\Gamma_{\esp{Z_{p=2q}}} ={} &
-\frac{6\zeta_R(3)}{4\pi^2p} +\frac{1}{\pi}\sum_{r=1}^{q-1}r\sum_{n=1}^{\infty}\frac{1}{n^2}\sen\!\left(2\pi n\tfrac{2r+1}{p}\right) -\frac{\pi^4}{45p} \left(\frac{a}{\beta}\right)^{\!\!3} \\ \nonumber
&+\frac{1}{2\pi^2 p}\sum_{k,l=1}^{\infty}\frac{1}{k^3} \left[ 2 + 2\frac{4\pi^2akl}{\beta} + \left( \frac{4\pi^2akl}{\beta} \right)^{\!\!2} \right] e^{-4\pi^2akl/\beta} \\ \nonumber
&-\frac{2}{\pi^2p}\sum_{k,l=1}^{\infty}\frac{1}{k^3} \left[ 2 + 2\frac{2\pi^2akl}{\beta} + \left( \frac{2\pi^2akl}{\beta} \right)^{\!\!2} \right] e^{-2\pi^2akl/\beta}\\ \nonumber
&+\frac{4}{\pi p} \sum_{r=1}^{q-1}r\sum_{n,l=1}^{\infty}\frac{1}{n^2} \sen\!\left(2\pi n\tfrac{2r+1}{p}\right) \!\left(1+\frac{4{\pi}^2aln}{\beta p}\right)e^{-4\pi^2aln/\beta p}\,.
\end{align}

Vemos que la acción efectiva tiene en este caso la misma estructura que en el caso impar: los dos primeros términos en la primera línea son independientes de la temperatura y la escala espacial del espacio, el tercero es extensivo, y el resto se anula exponencialmente cuando $\beta\rightarrow0$.

\vfill\pagebreak

\subsection{Espacios prisma}
\label{sec:termo:high:prisma}

Las degeneraciones sobre el espacio prisma $\esp{D_p^*}$ se anulan para los autovalores con índice par, mientras que para los autovalores con índice impar, según \eqref{eq:degprisma}, tenemos
\begin{align}\label{eq:degprismaimpar}
d_{2k+1}^{(D_p^*)} =
\begin{cases}
(2k+1) \left\lfloor \frac{k}{p} \right\rfloor ,  & \mbox{si }k\mbox{ es impar}\,, \\
(2k+1) \left( \left\lfloor \frac{k}{p} \right\rfloor + 1 \right), & \mbox{si }k\mbox{ es par}\,,
\end{cases}
\end{align}
que puede reescribirse en términos de las degeneraciones \eqref{eq:deglentepar1} sobre los espacios lente de orden par como
\begin{align}\label{eq:degprismaimpar1}
d_{2k+1}^{(D_p^*)} =
\frac12 d_{2k+1}^{(Z_{2p})}+\frac{(-1)^k}{2}(2k+1)
\end{align}

De este modo, la función zeta resulta
\begin{align}\label{eq:zetahighprisma}
\zeta_{\esp{D_p^*}}(s) = \frac12\zeta_{\esp{Z_{2p}}}(s)+\delta\zeta(s)\,,
\end{align}
donde denotamos $\delta\zeta$ a la contribución correspondiente al segundo término en \eqref{eq:degprismaimpar1}, que no depende explícitamente de $p$: 
\begin{align}\label{eq:deltazetahighprisma}
\delta\zeta(s) = \frac{(\mu a)^{2s}}{2} \sum_{l\in\mathbb{Z}}\sum_{k=0}^\infty (-1)^k (2k+1) \!\left[ (2k+1)^2 + \left( \frac{2\pi al}{\beta} \right)^{\!\!2} \right]^{-s}\,.
\end{align}

Procedemos con esta contribución imitando lo que hicimos con los espacios lente. Para empezar, separamos de la suma sobre $k$ los términos con índice par de los términos con índice impar, de modo de aislar el signo negativo, con lo que tenemos
\begin{align}\label{eq:deltazetahighprisma1}
\delta\zeta(s) = \frac{(\mu a)^{2s}}{2} \sum_{l\in\mathbb{Z}} \Bigg\{& \sum_{k=0}^\infty (4k+1) \!\left[ (4k+1)^2 + \left( \frac{2\pi al}{\beta} \right)^{\!\!2} \right]^{-s}\\ \nonumber 
&\qquad\qquad\qquad- \sum_{k=0}^\infty (4k+3) \!\left[ (4k+3)^2 + \left( \frac{2\pi al}{\beta} \right)^{\!\!2}  \right]^{-s} \Bigg\}\,.
\end{align}
Consideramos primeramente el término $l=0$, que resulta
\begin{align}\label{eq:deltazetahighprismal0}
\delta\zeta^{l=0}(s) = \frac{(\mu a)^{2s}4^{1-2s}}{2}  \left[ \zeta_H\!\left(2s-1,\tfrac14\right) -\zeta_H\!\left(2s-1,\tfrac34\right) \right]\,.
\end{align}
Para obtener el desarrollo a altas temperaturas del resto de los términos, renombramos el índice en la segunda suma entre llaves en \eqref{eq:deltazetahighprisma1}, con lo que tenemos
\begin{align}\label{eq:deltazetahighprismalnot0}
\delta\zeta^{l\neq 0}(s) = (\mu a)^{2s} \sum_{l=1}^\infty \sum_{k\in\mathbb{Z}} (4k+1) \!\left[ (4k+1)^2 + \left( \frac{2\pi al}{\beta} \right)^{\!\!2} \right]^{-s}\,.
\end{align}
La extensión analítica de esta expresión a una región que contenga al valor $s=0$ puede obtenerse escribiendo en primer lugar
\begin{align}\label{eq:deltazetahighprismalnot01}
\delta\zeta^{l\neq 0}(s) = \frac{(\mu a)^{2s}}{2(1-s)}  \sum_{l=1}^\infty \sum_{k\in\mathbb{Z}} \left.\frac{d}{d\alpha} \left[ (4k+\alpha)^2 + \left( \frac{2\pi al}{\beta} \right)^{\!\!2} \right]^{-s+1}\right\vert_{\alpha=1}\,,
\end{align}
donde para valores de $\Re(s)$ suficientemente grandes podemos intercambiar el orden de la derivada y las series. Por otra parte, la doble suma de la potencia entre corchetes puede reescribirse, de manera análoga a la utilizada en los casos anteriores, en la forma
\begin{align}\nonumber
\delta\zeta^{l\neq 0}(s) = -\frac{\sqrt{\pi}(\mu a)^{2s}}{8\Gamma(s)} \left.\frac{d}{d\alpha} \sum_{l=1}^\infty \sum_{k\in\mathbb{Z}} e^{i\frac{\pi}{2}\alpha k}\int_0^\infty dt\, t^{s-\frac32-1} e^{-t(2\pi al/\beta)^2-(\pi k)^2/16t}\,\right\vert_{\alpha=1} \,,
\end{align}
o, resolviendo explícitamente la integral sobre $t$ y evaluando la derivada con respecto a $\alpha$, 
\begin{align}\label{eq:deltazetahighprismalnot03}
\delta\zeta^{l\neq 0}(s) = \frac{\pi^{\frac32}(\mu a)^{2s}}{4\Gamma(s)}  \sum_{\substack{k,l=1\\k\mathrm{\,\,impar}}}^\infty (-1)^{\lfloor\frac{k}{2}\rfloor} k \left(\frac{8al}{k\beta}\right)^{\!\!\frac32-s} \!K_{\frac32-s}(\pi^2kla/\beta)\,.
\end{align}
Luego de sumar todas las contribuciones, la acción efectiva resulta
\begin{align}\label{eq:seffhighprisma}
\Gamma_{\esp{D_p^*}} ={} &
-\frac{3\zeta_R(3)}{16\pi^2p} +\frac{1}{2\pi}\sum_{r=1}^{p-1}r\sum_{n=1}^{\infty}\frac{1}{n^2}\sen\!\left(2\pi n\tfrac{2r+1}{2p}\right) -\frac{G}{\pi} \\ \nonumber
& -\frac{\pi^4}{180p} \left(\frac{a}{\beta}\right)^{\!\!3} -\frac{2}{\pi} \sum_{\substack{k,l=1\\k\mathrm{\,\,impar}}}^\infty (-1)^{\lfloor\frac{k}{2}\rfloor} \frac{1}{k^2} \left(1+\frac{\pi^2 akl}{\beta}\right)e^{-\pi^2kla/\beta}\\ \nonumber
&+\frac{1}{2\pi^2p}\sum_{k,l=1}^{\infty}\frac{1}{k^3} \left[ 2 + 2\frac{4\pi^2akl}{\beta} + \left( \frac{4\pi^2kla}{\beta} \right)^{\!\!2} \right] e^{-4\pi^2akl/\beta} \\ \nonumber
&-\frac{1}{2\pi^2p}\sum_{k,l=1}^{\infty}\frac{1}{k^3} \left[ 2 + 2\frac{2\pi^2akl}{\beta} + \left( \frac{2\pi^2kla}{\beta} \right)^{\!\!2} \right] e^{-2\pi^2akl/\beta}\\ \nonumber
&+\frac{1}{\pi p} \sum_{r=1}^{p-1}r\sum_{n,l=1}^{\infty}\frac{1}{n^2} \sen\!\left(2\pi n\tfrac{2r+1}{2p}\right) \left(1+\frac{2{\pi}^2aln}{\beta p}\right)e^{-2\pi^2lna/\beta p}\,,
\end{align}
donde $G$ es la constante de Catalan, $G=0.915966\ldots$. Esta acción efectiva tiene la misma estructura de términos que en los demás espacios: la línea de arriba contiene las contribuciones independientes de la temperatura y el tamaño del espacio, el primer término de la segunda línea es extensivo y el resto es suprimido exponencialmente a altas temperaturas.

\subsection{Espacios poliédricos}
\label{sec:termo:high:poly}

En el caso de los espacios poliédricos, podemos evitar un nuevo cálculo si notamos que las degeneraciones del laplaciano \eqref{eq:degtetra} a \eqref{eq:degicosa} sobre estos espacios pueden escribirse en términos de las degeneraciones \eqref{eq:deglentepar} sobre espacios lente de orden par. En efecto, en todos los casos mencionados las degeneraciones de autovalores de índice par son nulas, mientras que las de autovalores con índice impar pueden reescribirse fácilmente como
\begin{align}\label{eq:degpolycyclic}
d_{2n+1}^{(T^*)} &= d_{2n+1}^{(Z_6)} + \frac12d_{2n+1}^{(Z_4)} - \frac12d_{2n+1}^{(Z_2)}\,, \\ \nonumber
d_{2n+1}^{(O^*)} &= \frac12d_{2n+1}^{(Z_8)} + \frac12d_{2n+1}^{(Z_6)} + \frac12d_{2n+1}^{(Z_4)} - \frac12d_{2n+1}^{(Z_2)}\,, \\ \nonumber
d_{2n+1}^{(I^*)} &= \frac12d_{2n+1}^{(Z_{10})} + \frac12d_{2n+1}^{(Z_6)} + \frac12d_{2n+1}^{(Z_4)} - \frac12d_{2n+1}^{(Z_2)}\,.
\end{align}

Esto nos permite escribir las funciones zeta sobre estos espacios como las correspondientes combinaciones de funciones zeta sobre espacios lente pares. Este resultado puede encontrarse en la literatura bajo el nombre de \emph{descomposición cíclica} y fue utilizado en \cite{Chang:1992fu} para calcular la energía de vacío de una teoría escalar sobre ciertos cocientes de la esfera bidimensional.

Con esto, las acciones efectivas en este límite pueden obtenerse fácilmente como las respectivas combinaciones de acciones efectivas \eqref{eq:seffhigheven}. Sería engorroso escribir aquí explícitamente todos los casos; digamos solamente que la estructura de términos es similar a la de los espacios lente pares: un término extensivo proporcional al cubo de la temperatura, términos constantes (independientes de la temperatura y el volumen del espacio) y términos que decaen exponencialmente con la temperatura. Que el término extensivo en las acciones efectivas sobre los espacios lente pares corresponde a un término extensivo en los espacios poliédricos puede verse del hecho de que los coeficientes en las combinaciones \eqref{eq:degpolycyclic} son tales que en los tres casos la suma de $1/2q$ sobre los valores de $q$ que aparecen en la correspondiente combinación da como resultado la inversa del orden del grupo correspondiente.

\section{Acciones efectivas en el límite de bajas temperaturas}
\label{sec:termo:low}

Calculamos ahora los desarrollos a bajas temperaturas de la acción efectiva de la teoría escalar conforme sobre los distintos espacios esféricos. Si llamamos $\lambda_k^2$ a los autovalores del laplaciano conforme con $d_k$ las correspondientes degeneraciones, la acción efectiva a temperatura finita para el campo escalar conforme se escribe
\begin{align}\label{eq:sefflowgeneral}
\Gamma=\beta E_0+\sum_{k=1}^\infty d_k \log(1-e^{-\beta \lambda_k})\,,
\end{align}
donde $E_0$ es la energía de vacío (apropiadamente regularizada) a temperatura cero. Para demostrar que esto es así buscamos un desarrollo a bajas temperaturas de la acción efectiva, para lo cual comenzamos escribiendo la función zeta como
\begin{align}
\zeta_{\esp{H}}(s) = \left(\frac{\mu\beta}{2\pi}\right)^{\!\!2s}\sum_{l\in\mathbb{Z}} \sum_{k=1}^{\infty} d_k \left[
\left(\frac{\lambda_k\beta}{2\pi}\right)^{\!\!2} + l^2 \right]^{-s}
\nonumber\end{align}
o, usando la expresión para la potencia entre corchetes en términos del tiempo propio de Schwinger,
\begin{align}\label{eq:zetalowgeneral}
\zeta_{\esp{H}}(s) = \frac{1}{\Gamma(s)} \left(\frac{\mu\beta}{2\pi}\right)^{\!\!2s}\sum_{l\in\mathbb{Z}} \sum_{k=1}^{\infty}
d_k \int_0^{\infty} dt\, t^{s-1} e^{-t\left[l^2 +
	\left(\lambda_k\beta/2\pi\right)^2 \right]}\,.
\nonumber\end{align}
Hacemos ahora una inversión de Poisson de la suma sobre $l$ y separamos en la expresión resultante el término $l=0$. Tenemos, una vez resuelta la integral sobre $t$,
\begin{align}
\zeta_{\esp{H}}^{l=0}(s) = \frac{\beta\mu^{2s}}{2\sqrt{\pi}} \frac{\Gamma(s-\frac{1}{2})}{\Gamma(s)}\sum_{k=1}^{\infty} d_k {\lambda_k}^{1-2s}\,,
\end{align}
y análogamente para los términos con $l\neq0$,
\begin{align}
\zeta_{\esp{H}}^{l\neq 0}(s) =
\frac{2\sqrt{\pi}}{\Gamma(s)}\left(\frac{2\pi}{\mu\beta}\right)^{\!\!-2s}\sum_{k=0}^{\infty}d_k \sum_{l=1}^{\infty} 2
\left(\frac{2\pi^2l}{{\lambda_k}\beta}\right)^{\!\!s-\frac{1}{2}}\!K_{\frac{1}{2}-s}({\lambda_k}\beta l).
\end{align}
Tomando ahora la derivada con respecto a $s$, y teniendo en cuenta la definición de la energía de vacío,
\begin{align}\label{eq:vacuum}
E_0= \frac12\lim_{s\rightarrow0}\sum_{k=1}^\infty d_k \left(\lambda_k^2\right)^{\frac12-s}\,,
\end{align}
tenemos
\begin{align}
\Gamma_{\esp{H}} =
\beta E_0 -2\sqrt{\pi} \sum_{k=1}^{\infty} d_k \sum_{l=1}^{\infty}
\left(\frac{2\pi^2l}{{\lambda_k}\beta}\right)^{\!\!-1/2} \!K_{\frac{1}{2}}({\lambda_k}\beta l)\,.
\end{align}
Usando la expresión para la función de Bessel $K_\frac12(x)$ en términos de $\exp(-x)$, e identificando la serie en $l$ resultante con la serie de Taylor del logaritmo, podemos poner esta última expresión en la forma \eqref{eq:sefflowgeneral}.

Vemos entonces que la única contribución a la acción efectiva que no se anula exponencialmente cuando la temperatura tiende a cero es la correspondiente a la energía de Casimir. A continuación calcularemos esta contribución para los distintos espacios esféricos.

\subsection{Espacios lente}
\label{sec:termo:low:lente}

Comenzamos calculando la energía de vacío sobre espacios lente de orden impar. Usando las degeneraciones del laplaciano sobre el espacio $\esp{Z_{2q+1}}$ en la forma \eqref{eq:deglenteimparrees} vemos que, como pasaba antes para la función zeta a temperatura finita, la energía de vacío puede escribirse como
\begin{align}
E_{0,\esp{Z_{2q+1}}} = \frac{E_{0,S^3}}{2q+1}+\delta E_{0,2q+1}\,,
\end{align}
donde $E_{0,S^3}=1/240a$ es la energía de vacío en la esfera, que es bien conocida \cite{Ford:1975su}, y además puede leerse de la ecuación \eqref{eq:seffs3low}. La parte restante viene dada por
\begin{align}\nonumber
\delta E_{0,2q+1} = \frac12 \lim_{s\rightarrow0} a^{2s-1}\sum_{n=0}^\infty \Bigg\{ &\sum_{r=0}^{q-1} \left(1-\frac{2r+1}{2q+1}\right) \left[n(2q+1)+2r+1\right]^{2-2s}\\
& \quad- \sum_{r=0}^{q} \frac{2r}{2q+1} \left[n(2q+1)+2r\right]^{2-2s} \Bigg\}\,,
\end{align}
donde, como en la sección \ref{sec:termo:high:oddlens}, escribimos $k=n(2q+1)+r$, con $0\leq r\leq 2q$, y renombramos a este resto llamándolo $2r$ ó $2r+1$ según sea par o impar. Siguiendo un camino similar al recorrido en aquella sección y tomando el límite $s\rightarrow0$ podemos obtener la expresión
\begin{align}
\delta E_{0,2q+1} = \frac{2q+1}{a\pi^3} \sum_{r=1}^q r\sum_{n=1}^\infty \frac{1}{n^3} \sen\left(\frac{4\pi nr}{2q+1}\right)\,.
\end{align}
Finalmente, podemos resolver explícitamente las sumas restantes, con lo que obtenemos una expresión para esta contribución que sumada a la correspondiente en la esfera da como resultado
\begin{align}\label{eq:casimirodd}
E_{0,\esp{Z_{2q+1}}} = -\frac{(2q+1)^4+10(2q+1)^2-14}{720 a(2q+1)}\,.
\end{align}

Pasamos ahora al cálculo de la energía de vacío sobre espacios lente pares $\esp{Z_{2q}}$. Consideramos en primer lugar la energía de vacío sobre $\esp{Z_2}$, que puede obtenerse fácilmente de la definición \eqref{eq:vacuum}:
\begin{align}\label{eq:casimirz2}
E_{0,\esp{Z_2}} = \frac{1}{2}\lim_{s\rightarrow0} a^{2s-1} \sum_{k=0}^{\infty} (2k+1)^{3-2s} = -\frac{7}{240a}\,.
\end{align}

Para analizar los demás espacios lente pares comenzamos escribiendo la energía de vacío como
\begin{align}\label{eq:casimirlentepares}
E_{0,\esp{Z_{2q}}}=\frac{E_{0,\esp{Z_2}}}{q}+\delta E_{0,2q}\,,
\end{align}
de manera análoga a la que nos llevó a la expresión  \eqref{eq:zetahigheven} para la función zeta. El segundo término viene dado por 
\begin{align}\nonumber
\delta E_{0,2q} = \frac12 \lim_{s\rightarrow0} a^{2s-1} \sum_{r=0}^{q-1} \left[(q-r-1)-r\right] \sum_{n=0}^\infty  \left(2nq+2r+1\right)^{2-2s}\,,
\end{align}
donde hemos escrito $k=2nq+2r+1$, con $0\leq r\leq q-1$. 
Procediendo como en la sección \ref{sec:termo:high:evenlens}, tomando el límite y resolviendo las sumas resultantes, y sumando la contribución \eqref{eq:casimirz2} con el factor adecuado, obtenemos finalmente
\begin{align}\label{eq:casimireven}
E_{0,\esp{Z_{2q}}} = -\frac{(2q)^4+10(2q)^2-14}{720 a(2q)}\,.
\end{align}

\vfill\pagebreak

Notemos que las expresiones \eqref{eq:casimirodd} y \eqref{eq:casimireven} tienen la misma forma; en efecto, podemos escribir para la energía de vacío sobre el espacio lente $\esp{Z_p}$
\begin{align}\label{eq:casimirlens}
E_{0,\esp{Z_{p}}} = -\frac{p^4+10p^2-14}{720 ap}\,,
\end{align}
que coincide con el resultado calculado en \cite{Dowker:2004nh,Dowker:1978vy} directamente desde la teoría sin temperatura ---en un caso usando las funciones generatrices de las degeneraciones y en el otro a partir del cálculo de la densidad de energía $\langle T_{00}\rangle$ en términos de los generadores de $SU(2)$---.

La acción efectiva sobre el espacio lente $\esp{Z_p}$ tiene entonces la forma
\begin{align}
\Gamma_{\esp{Z_{p}}}=\frac{14 - 10p^2 - p^4}{720 p}\frac{\beta}{a} + \mathcal{O}{(e^{-\beta/a})}\,.
\end{align}

\subsection{Espacios prisma}
\label{sec:termo:low:prisma}

Para obtener la energía de vacío sobre el espacio prisma $\esp{D_p^*}$ usamos la expresión \eqref{eq:degprismaimpar1}, que nos permite escribir
\begin{align}\label{eq:casimirprisma}
E_{0,\esp{D_p^*}} = \frac12E_{0,\esp{Z_{2p}}}+\delta E_0\,,
\end{align}
con
\begin{align}
\delta E_0 = \frac14\lim_{s\rightarrow0} a^{2s-1} \sum_{k=0}^\infty (-1)^k (2k+1)^{2-2s}\,.
\end{align}

Procedemos ahora de manera similar a la utilizada en el apartado anterior y, luego de tomar el límite, obtenemos fácilmente $\delta E_0 = -1/8a$, con lo que tenemos
\begin{align}\label{eq:e0prism}
E_{0,\esp{D_p^*}} = -\frac{8p^4+20p^2+180p-7}{1440 ap}\,,
\end{align}
que coincide con los valores reportados en \cite{Dowker:2004nh} ---lo que a primera vista no es evidente, pero puede deducirse rastreando allí un error de tipeo en la expresión final---.

\vfill\pagebreak

\subsection{Espacios poliédricos}
\label{sec:termo:low:poly}

Para los espacios poliédricos, al igual que en el caso de altas temperaturas, utilizamos las expresiones \eqref{eq:degpolycyclic} para las degeneraciones, con lo que la energía de vacío en cada caso puede escribirse como una combinación de energías de vacío sobre determinados espacios lente pares. Haciendo explícitamente cada combinación resultan los valores
\begin{align}
E_{0,\esp{T^*}} = -\frac{3761}{360\times 24 a}\,,&\qquad E_{0,\esp{O^*}} = -\frac{11321}{360\times 48 a}\,,\\ \nonumber
E_{0,\esp{I^*}} &= -\frac{43553}{360\times 120 a}\,,
\end{align}
que son los mismos que los obtenidos en \cite{Dowker:2004nh,Dowker:1978vy}.

\section{Propiedades termodinámicas de la teoría}
\label{sec:termo:props}

A partir de los resultados de lo que va del capítulo podemos obtener las propiedades termodinámicas de la teoría escalar conforme a temperatura finita sobre los distintos espacios esféricos. A riesgo de ser reiterativos, notemos para comenzar que las acciones efectivas sobre todos los espacios esféricos tienen la misma estructura de términos: a altas temperaturas, hay un término divergente ---que es extensivo y, como veremos, tiene la potencia de $\beta$ correspondiente a la ley de Stefan-Boltzmann---, un término constante y términos que se anulan exponencialmente cuando $\beta/a\rightarrow0$. A bajas temperaturas, en cambio, la estructura de términos viene dada por la expresión \eqref{eq:sefflowgeneral}: además de términos que se anulan exponencialmente en el límite $\beta/a\rightarrow\infty$, aparece una contribución divergente proporcional a la energía de vacío de la teoría.

Puede verse que en todos los casos en el límite $\beta/a\rightarrow\infty$ de bajas temperaturas la energía y la energía libre coinciden, y la entropía se anula exponencialmente. Esto es compatible con la formulación de Planck de la tercera ley de la Termodinámica \cite[sección 8.2]{huang}, y puede explicarse por el hecho de que el estado fundamental es no degenerado \cite{Wreszinski2009}. Los valores de la energía en este límite para los distintos espacios esféricos pueden leerse en la sección anterior y, como se ha dicho allí, coinciden con los obtenidos anteriormente por otros autores \cite{Dowker:2004nh,Dowker:1978vy}.

\begin{figure}[h!]
	\centering
	\includegraphics[height=6cm]{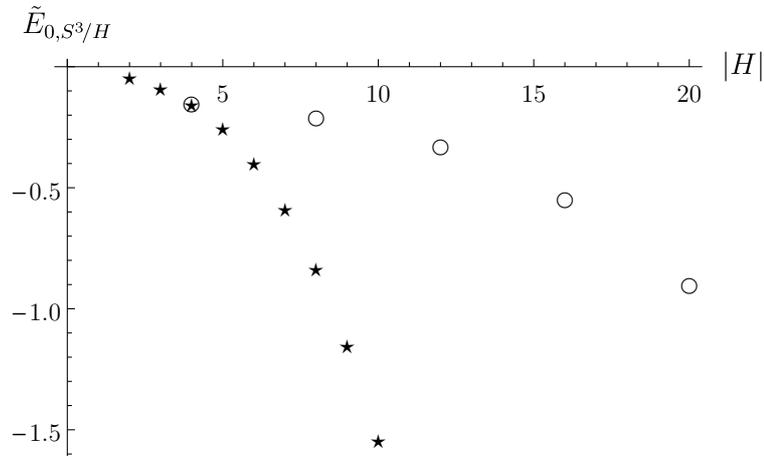}
	\caption{Energía de vacío en función del orden del grupo para los primeros espacios lente (estrellas) y prisma (círculos).}
	\label{fig:casimirquotients}
\end{figure}

A modo de resumen de los resultados en este límite, en la figura \ref{fig:casimirquotients} mostramos los valores para la energía de vacío adimensionalizada $\tilde{E}_0=aE_0$ en función del orden del grupo de isotropía para los espacios lente y prisma de orden más bajo. Los gráficos son compatibles con el comportamiento asintótico cúbico de dicha cantidad para ambas familias de espacios, como era evidente de las expresiones analíticas \eqref{eq:casimirlens} y \eqref{eq:e0prism}. Del gráfico descubrimos además la coincidencia de las energías de vacío sobre el espacio prisma $\esp{D_1^*}$ y el espacio lente $\esp{Z_4}$: esto se entiende al notar que los grupos de isotropía correspondientes son isomorfos; en efecto, evaluando las expresiones \eqref{eq:deglentepar} y \eqref{eq:degprisma} para estos casos particulares podemos ver que las degeneraciones del laplaciano sobre ambos espacios coinciden.

En cuanto al límite $\beta/a\rightarrow0$ de altas temperaturas, la energía resulta
\begin{align}
E_{\esp{H}} \sim \frac{\pi^4 a^3}{15|H|\beta^4}\,,
\end{align}
que es una cantidad extensiva. En particular, esto significa que ---distinto a lo que sucedía en el límite de bajas temperaturas--- cocientes de la esfera con grupos del mismo orden tienen en el límite de altas temperaturas la misma energía.

Usando las cantidades adimensionalizadas $\xi=2\pi a/\beta$ y $\tilde{E}=aE$, podemos ver que para la esfera los límites de bajas y altas temperaturas de la energía satisfacen la relación ${\xi}^{-2}\tilde{E}(\xi)={\xi}^{2}\tilde{E}(1/\xi)$. Esta simetría de inversión en la temperatura se cumple no sólo en los límites sino para valores arbitrarios de $\xi$, como puede verse haciendo las correspondientes derivadas en las expresiones \eqref{eq:seffs3high} y \eqref{eq:seffs3low}. Esta propiedad fue estudiada por primera vez en \cite{Brown:1969na} para la energía libre en el efecto Casimir a temperatura finita. En el caso que nos ocupa, desde un punto de vista matemático la simetría se debe al hecho de que los autovalores del laplaciano conforme en cuatro dimensiones son cuadrados perfectos ---dando a la función zeta una forma similar a la que tendría en un toro--- y fue discutida originalmente en \cite{Cardy1991403} para teorías conformes en dimensión par, y en \cite{Dowker2002405,Kutasov:2000td} para el caso particular de campos escalares sobre la esfera tridimensional. Físicamente, este tipo de simetrías relaciona una cantidad universal ---el límite de altas temperaturas de la energía--- con una cantidad ---la energía de vacío--- que en espacios curvos podría depender \emph{a priori} del esquema de regularización utilizado \cite{Assel:2015nca}. 

En cuanto al resto de los espacios esféricos, vemos que las correspondientes energías no cumplen esta relación. No obstante, puede verse que para los espacios lente la acción efectiva satisface una relación similar donde la cantidad $\xi$ involucra al orden $p$ del grupo de isotropía en el límite $p\rightarrow\infty$ \cite{Shaghoulian:2016gol}.

Volviendo al cálculo de las propiedades termodinámicas de la teoría, analizamos ahora la entropía, que observamos que en el límite de altas temperaturas tiene en todos los casos una estructura de términos similar:
\begin{align}\label{eq:entropyhigh}
S_{\esp{H}} \sim \frac{4\pi^4}{45|H|}\left(\frac{a}{\beta}\right)^3 + S_{0,\esp{H}}\,,
\end{align}
donde el primer término es extensivo y tiene la forma usual de la entropía de Stefan-Boltzmann ---en particular, es positivo y diverge en el límite $\beta\rightarrow0^+$---, y $S_0$ no depende de la temperatura ni del volumen del espacio, y donde el resto de los términos, que no hemos escrito, se anula exponencialmente con la temperatura.

El término constante $S_0$ es negativo en el caso de la esfera y positivo en el resto de los espacios. Como ya señalamos, en la entropía sobre la esfera resultados anteriores de otros autores \cite{Brevik:2002gh,Elizalde:2002ak,Klemm:2001db,Kutasov:2000td} no tenían este término, que sí se obtuvo en \cite{Dowker:2013ia} para espacios lente y en el caso de la esfera en \cite{Myers:2012tz}, donde además se mostró que es necesario para que el cociente entre la entropía y la energía en teorías libres en cuatro dimensiones satisfaga la cota de Bekenstein \cite{Bekenstein:1980jp} independientemente del número de especies presentes, y para que dicho cociente no resulte divergente a bajas temperaturas en el caso de la teoría $\mathcal{N}=4$ SYM.  Por otra parte, puede verse inmediatamente que $S_0$ es una cantidad subaditiva; en efecto, la suma de $S_{0,\esp{H}}$ para $|H|$ copias de
$\esp{H}$ es siempre mayor que $S_{0,S^3}$.

Para evitar el término divergente y aislar de alguna manera la información acerca de la topología de los espacios, podemos tener en cuenta la diferencia entre la entropía sobre el espacio esférico y la de la esfera dividida por el orden del correspondiente grupo de isotropía, a la que, luego de tomar el límite $\beta\rightarrow0$ para eliminar los términos exponenciales, llamamos \emph{entropía topológica}:
\begin{align}
S_{\mathrm{top},\esp{H}} = \lim_{\beta\rightarrow 0} \left[ S_{\esp{H}} - \frac{1}{|H|} S_{S^3} \right]\,.
\end{align}

\begin{figure}[h!]
	\centering
	\includegraphics[height=6cm]{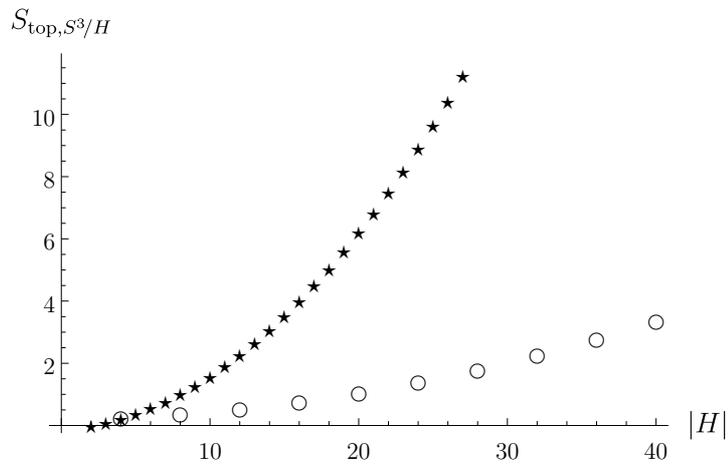}
	\caption{Entropía  topológica en función del orden del grupo para los primeros espacios lente (estrellas) y prisma (círculos).}
	\label{fig:s0}
\end{figure}

En la figura \ref{fig:s0} mostramos los valores de $S_{\mathrm{top}}$ para algunos de los espacios esféricos en función del orden del grupo correspondiente.

Para finalizar el presente capítulo notemos que, a pesar del nombre que le hemos dado, no debe confundirse la entropía topológica con una verdadera entropía, al menos en el sentido usual del término; en particular, este término no tiene un comportamiento creciente con la temperatura (lo que sí sucede para las entropías completas $S_{\esp{H}}$, para las que además se cumple $\partial F/\partial T = -S$, como puede verificarse fácilmente en ambos límites). Si, por ejemplo, para el caso particular de los espacios lente, usáramos una representación integral para las sumas en \eqref{eq:seffhigheven} como en \cite{Dowker:2013ia}, obtendríamos una expresión para la entropía topológica como función del orden $p$ del grupo cíclico correspondiente que puede extenderse a valores reales del parámetro $p$ ---dejando de lado el problema de la interpretación del espacio---, de la que puede verse que la entropía topológica es una función creciente de $p$. En ese sentido ---y, por lo que sabemos, sólo en ése--- es plausible una interpretación de la entropía topológica como una entropía, con la temperatura dada por $p$. Volveremos sobre esto más adelante.

%
\chapter{Campos masivos sobre espacios esféricos}
\label{sec:masivo}

\begin{minipage}{.885\textwidth}%
	\begin{flushright}
		\begin{minipage}{.67\textwidth}%
			\begin{flushleft}
				\emph{Oh, their wild rapture! oh, their eyes like stars and their souls again in Eden, if the next station were unaccountably Baker Street!}
			\end{flushleft}
			\begin{flushright}
				--- \textrm{G. K. Chesterton} \\
				\emph{The man who was Thursday}
			\end{flushright}
		\end{minipage}%
	\end{flushright}
\end{minipage}%

\bigskip

A continuación extenderemos los resultados del capítulo anterior al caso de un campo escalar masivo sobre los distintos espacios esféricos, con el objetivo de analizar la dependencia de las propiedades termodinámicas de la teoría con la masa. Estos cálculos fueron presentados en el artículo \cite{Asorey:2014gsa}.

Consideramos entonces la teoría de un campo escalar $\phi$ de masa $m$ y con un acoplamiento $\xi$ a la curvatura del espacio esférico $\esp{H}$. La acción euclídea para dicha teoría se escribe
\begin{align}
S = \frac12 \int_{\mathbb{R}\times\esp{H}} d^4x \sqrt{g} \,\Big\{ g^{\mu\nu}\partial_{\mu}\phi\, \partial_{\nu}\phi + \frac{R}{6}\phi^2 + \left[ m^2 + \left(\xi - \frac{1}{6} \right) R \,\right] \phi^2 \Big\},
\end{align}
donde $g^{\mu\nu}$ es la métrica del espacio-tiempo $\mathbb{R}\times\esp{H}$ y $R=6/a^2$ la curvatura escalar del espacio $\esp{H}$, siendo $a$ el radio de la esfera $S^3$. En la expresión anterior hemos separado un término en el acoplamiento a la métrica que corresponde a la teoría conforme en cuatro dimensiones, de forma que el acoplamiento restante sumado al cuadrado de la masa del campo puede ser pensado como el cuadrado de la masa de un campo escalar conforme. En lo sucesivo consideraremos el caso conforme $\xi=1/6$, recordando que los resultados que obtengamos pueden ser extendidos a acoplamientos arbitrarios siempre que $ m^2 + \left(\xi - \frac{1}{6} \right) R \geq 0$. La extensión a otros casos ---como el usualmente considerado caso de masa nula y acoplamiento mínimo--- no es directa, y debe estudiarse on otro cuidado. 
	
Al igual que en el caso sin masa, introducimos la temperatura $1/\beta$, con lo que los autovalores del operador de fluctuaciones cuánticas de la teoría pueden obtenerse como suma de los cuadrados de los modos de Matsubara y los modos espaciales, que están desplazados en $m^2$ con respecto a los del capítulo anterior. Las degeneraciones, en tanto, coinciden con las de la parte espacial del operador, que hemos calculado para los distintos espacios en la sección \ref{sec:espacios:espectro}.

La función zeta se escribe entonces
\begin{align}\label{eq:zetamass}
\zeta_{\esp{H}}(s) = \mu^{2s} \sum_{k=1}^{\infty}\sum_{l\in\mathbb{Z}} d_k^{(H)} \left[ \left(\frac{k}{a}\right)^{\!\!2} + m^2 + \left(\frac{2\pi l}{\beta}\right)^{\!\!2} \right]^{-s}\,,
\end{align}
donde, como antes, $\mu$ es una escala de masa que actuará como regulador del procedimiento. En lo que sigue calcularemos las acciones efectivas de la teoría sobre los espacios esféricos. Comenzaremos por el caso más simple de la esfera, analizando por separado la regularización de la función zeta \eqref{eq:zetamass} en los distintos límites de temperatura y masa: por un lado, separaremos el estudio en los casos de bajas y altas temperaturas, y en cada uno de esos casos distinguiremos los regímenes de masas chicas y grandes.\footnote{ \,Aquí, por supuesto, se entiende que todas las escalas se miden con respecto a algún otro parámetro con dimensiones, en este caso el radio $a$ de la esfera que cubre al espacio.} A partir de consideraciones geométricas puede verse \cite{Blau:1988iz} que la función zeta en este caso no se anula en $s=0$, lo que quiere decir en particular que su derivada con respecto a $s$ ---y en consecuencia la acción efectiva--- dependerá del regulador, y tendremos que dar argumentos físicos para una prescripción de renormalización que elimine esa dependencia.

\section{Acción efectiva sobre la esfera en el límite de bajas temperaturas}

Comenzamos analizando el límite $\beta\gg a$ de bajas temperaturas. Para obtener una extensión analítica de la función zeta que sea útil para tomar este límite hacemos, como en el caso no masivo, una inversión de Poisson en la suma sobre los modos térmicos. Para eso reescribimos la función zeta como
\begin{align}
\zeta_{S^3}(s) = \left( \frac{\mu\beta}{2\pi} \right)^{\!\!2s} \sum_{k=1}^{\infty}\sum_{l\in\mathbb{Z}} k^2 \left[ \left(\frac{k\beta}{2\pi a}\right)^{\!\!2} + \left(\frac{\beta m}{2\pi}\right)^{\!\!2} + l^2 \right]^{-s},
\end{align}
de la que obtenemos, usando la representación del tiempo propio de Schwinger y la fórmula de inversión de Poisson,
\begin{align}
\zeta_{S^3}(s) = \left( \frac{\mu\beta}{2\pi} \right)^{\!\!2s} \frac{\pi^{\frac12}}{\Gamma(s)} \sum_{k=1}^{\infty}\sum_{l\in\mathbb{Z}} k^2 \int_0^{\infty}dt\, t^{s-\frac12-1} e^{-\left[ \left(\frac{k\beta}{2\pi a}\right)^2 + \left(\frac{\beta m}{2\pi}\right)^2 \right]t - (\pi l)^2\!/t}.
\end{align}
De la misma forma que en el caso no masivo, luego de resolver explícitamente las integrales en los casos $l=0$ y $l\neq 0$ obtenemos
\begin{align}\label{eq:zetamassll0}
\zeta_{S^3}(s) ={} & \!\left( \frac{\mu\beta}{2\pi} \right)^{\!\!2s} \!\frac{\pi^{\frac12}}{\Gamma(s)} \Bigg\{ \Gamma(s-\tfrac12) \sum_{k=1}^{\infty} k^2 \!\left[ \left( \frac{k\beta}{2\pi a} \right)^{\!\!2} + \left( \frac{\beta m}{2\pi} \right)^{\!\!2} \right]^{\frac12-s} \\ \nonumber
& \!\!\!\!\! + 4\sum_{k,l=1}^{\infty} k^2 (\pi l)^{s-\frac12} \!\left[ \left( \frac{k\beta}{2\pi a} \right)^{\!\!2} + \left( \frac{\beta m}{2\pi} \right)^{\!\!2} \right]^{\frac14-\frac{s}{2}} \!\!\!K_{\frac12-s}\!\left( \sqrt{k^2 + (am)^2} \beta l/a\right)
\!\Bigg\}\,.
\end{align}
	
La contribución de los términos $l\neq 0$ a la acción efectiva puede calcularse fácilmente, dando como resultado
\begin{align}
\Gamma_{S^3}^{l\neq 0}(\beta) = \sum_{k=1}^{\infty} k^2 \log\!\left( 1 - e^{-\frac{\beta}{a}\sqrt{k^2+(am)^2}} \right)\,.
\label{eq:seffmassbajasl}
\end{align}
Como en el caso no masivo, esta parte de la acción efectiva se anula exponencialmente en el límite $\beta/a\rightarrow\infty$.

La contribución del término $l=0$ involucra una suma divergente, que tendrá que ser extendida analíticamente; haremos esto de dos maneras diferentes, cada una de las cuales será útil para tomar uno de los límites de la masa adimensionalizada $am$. En lo que sigue, haciendo un abuso de lenguaje, llamaremos a estos límites ``de masa grande'' y ``de masa pequeña'', aunque ---insistimos--- lo que resulte grande o pequeño según el caso sea $am$.

\subsection{Régimen de masa grande}
	
En primer lugar, analizamos la contribución del término $l=0$ en \eqref{eq:zetamassll0} en una forma que nos permitirá tomar el límite $am\gg 1$ de masas grandes. Para eso, reescribimos dicho término en la forma
\begin{align}\nonumber
\zeta^{l=0}_{S^3}(s) = \frac{(\mu a)^{2s}\beta}{4\pi^{\frac12}a} \frac{\Gamma(s-\tfrac12)}{\Gamma(s)} \sum_{k\in\mathbb{Z}} \left\{ \left[ k^2 + (am)^2 \right]^{\frac32-s} - (am)^2 \left[ k^2 + (am)^2 \right]^{\frac12-s} \right\},
\end{align}
y usamos la representación del tiempo propio de Schwinger de las potencias entre corchetes y la fórmula de Poisson para la suma sobre $k$, con lo que obtenemos, suponiendo $ma\neq0$,
\begin{align}
\zeta^{l=0}_{S^3}(s) ={} & \frac{(\mu a)^{2s}}{\Gamma(s)}\frac{\beta}{4a} \Bigg\{\frac12 (am)^{4-2s} \,\Gamma(s-2)\\ \nonumber & + 4(s-\tfrac32) \sum_{k=1}^{\infty} \left( \frac{\pi k}{am} \right)^{\!\!s-2} \!K_{s-2}(2\pi amk)  \\ \nonumber & \qquad\qquad\qquad\qquad\qquad-  4(am)^2\sum_{k=1}^{\infty} \left(\frac{\pi k}{am}\right)^{\!\!s-1} \!K_{s-1}(2\pi amk) \Bigg\}.
\end{align}
Usando el desarrollo de $\Gamma(s-2)/\Gamma(s)$ para $s$ pequeño podemos calcular fácilmente la derivada con respecto a $s$, de lo que resulta para la acción efectiva
\begin{align}\label{eq:seffmassbajas0}
\Gamma_{S^3}^{l=0}(\beta) =& -\frac{(am)^4}{32}\frac{\beta}{a} \left( \frac{3}{2} + 2\log\frac{\mu}{m} \right)
\\\nonumber
& + \frac{(am)^3}{2\pi} \frac{\beta}{a} \sum_{k=1}^{\infty} \frac{1}{k} K_1(2\pi amk) + \frac{3(am)^2}{4\pi^2} \frac{\beta}{a} \sum_{k=1}^{\infty} \frac{1}{k^2} K_2(2\pi amk).
\end{align}
Vemos que, como esperábamos, la acción efectiva contiene una dependencia en el regulador $\mu$. Se nos presenta entonces el problema de renormalizar esta cantidad; en otras palabras, tenemos que decidir qué términos sustraer a la acción efectiva de modo de obtener una cantidad con sentido físico. En ese contexto, notamos que en el límite de volumen infinito ---$am\rightarrow\infty$, $a/\beta\rightarrow\infty$, en el que el radio de la esfera es el parámetro más grande de la teoría---, donde es esperable obtener la teoría a temperatura cero sobre el espacio plano, los términos en la primera línea de la ecuación anterior no se anulan; tomamos entonces como prescripción de renormalización la correspondiente a la mínima sustracción que elimina estos términos. Dicho de otro modo, agregamos a la acción los contratérminos necesarios para obtener una acción efectiva que se anule en el límite de volumen infinito, donde la prescripción de orden normal de los operadores de la teoría cuántica así lo requiere \cite{kay1979casimir}. Luego de la renormalización obtenemos
\begin{align}\nonumber
\Gamma_{S^3}(\beta) ={} & \frac{3(am)^2}{4\pi^2}\frac{\beta}{a} \sum_{k=1}^{\infty} \frac{1}{k^2} K_2(2\pi amk) + \frac{(am)^3}{2\pi}\frac{\beta}{a} \sum_{k=1}^{\infty} \frac{1}{k} K_1(2\pi amk)  \\ \label{eq:seffmassbajas1}
&  + \sum_{k=1}^{\infty} k^2\log\!\left(1-e^{-\frac{\beta}{a}\sqrt{k^2+(am)^2}}\right).
\end{align}

La entropía calculada a partir de esta acción efectiva contiene sólo la contribución de los modos con $l\neq 0$, por lo que se anula exponencialmente en el límite de bajas temperaturas. La energía de vacío resulta
\begin{align}
E_0 = \frac{3(am)^2}{4\pi^2 a} \sum_{k=1}^{\infty} \frac{1}{k^2} K_2(2\pi amk) + \frac{(am)^3}{2\pi a} \sum_{k=1}^{\infty} \frac{1}{k} K_1(2\pi amk) \,,
\label{eq:casimirmass}
\end{align}
y se anula exponencialmente en el límite $am\rightarrow\infty$, lo cual es también consistente con la prescripción de orden normal para la teoría sin temperatura. Esta misma expresión para la energía de vacío fue obtenida en \cite{Elizalde:2003wd} sumando los modos espaciales, sin pasar por la teoría a temperatura finita.
	
\subsection{Régimen de masa pequeña}

Consideraremos ahora el caso en que la masa adimensionalizada $am$ es pequeña. Comenzamos reescribiendo el término $l=0$ en la función zeta como
\begin{align}
\zeta^{l=0}_{S^3}(s) = (\mu a)^{2s} \frac{\beta}{2\pi^{\frac12} a} \frac{\Gamma(s-\tfrac12)}{\Gamma(s)} \sum_{k=1}^{\infty} k^{3-2s} \left[ 1+ \left( \frac{am}{k} \right)^{\!\!2} \right]^{\frac12 -s},
\end{align}
y, considerando $am<1$, hacemos el desarrollo del binomio entre corchetes para obtener
\begin{align}\nonumber
\zeta^{l=0}_{S^3}(s) = (\mu a)^{2s} \frac{\beta}{2\pi^{\frac12} a} \frac{\Gamma(s-\tfrac12)}{\Gamma(s)} \sum_{n=0}^{\infty} \frac{\Gamma(\tfrac32 -s)(am)^{2n}}{n!\,\Gamma(\tfrac32 -n-s)}  \zeta_R(2s+2n-3).
\end{align}
La derivada con respecto a $s$ de esta cantidad puede obtenerse fácilmente: la única sutileza proviene del término con $n=2$, que derivamos usando el desarrollo \eqref{eq:riemannzpolo} de $\zeta_R(2s+1)$ para $s$ pequeño; la correspondiente contribución a la acción efectiva se escribe	
\begin{align}\nonumber
\Gamma_{S^3}^{l=0} ={} & - \frac{(am)^4}{16}\frac{\beta}{a} \left[ \log(\mu a/2){+1}+\gamma \right]+ \frac{\beta}{2a} \sum_{\substack{n=0\\n\neq 2}}^{\infty} \frac{\Gamma(\tfrac32)(am)^{2n}}{n!\,\Gamma(\tfrac32-n)}\zeta_R(2n-3)\,.
\end{align}
Sustrayendo la misma cantidad que en el régimen de masas grandes obtenemos la acción efectiva renormalizada en este régimen, que resulta
\begin{align}\label{eq:seffmassbajas2}
\Gamma_{S^3}(\beta) ={} & \frac{\beta}{240a} - \frac{(am)^2}{48}\frac{\beta}{a} - \frac{(am)^4}{16}\frac{\beta}{a} \left[ \log\!\left(\frac{m a}{2}\right){+\frac14}+\gamma \right]  \\ \nonumber
&  + \frac{\beta}{2a} \sum_{n=3}^{\infty} \frac{\Gamma(\tfrac32)(am)^{2n}}{n!\,\Gamma(\tfrac32 -n)}\zeta_R(2n-3) + \sum_{k=1}^{\infty} k^2\log\!\left(1-e^{-\frac{\beta}{a}\sqrt{k^2+(am)^2}}\right).
\end{align}

Notemos que en el límite $m\rightarrow0$ esta expresión se reduce al límite de bajas temperaturas \eqref{eq:seffs3low} de la acción efectiva en el caso conforme que obtuvimos en el capítulo previo. Puede verse también que es posible obtener esta misma expresión tomando el límite $am\rightarrow0$ en el desarrollo \eqref{eq:seffmassbajas1} que obtuvimos en la sección anterior para $am\neq0$. De la misma forma, la energía de vacío de la teoría sin temperatura puede ser recuperada de esta misma expresión y está de acuerdo con los resultados ya conocidos \cite{Elizalde:2003wd,Ford:1975su,Myers:2012tz}.  En lo que respecta a la entropía, observamos que no hay contribución independiente de la temperatura, y que los únicos términos que quedan se anulan exponencialmente en el límite $\beta\rightarrow\infty$ en el que la temperatura es el parámetro más grande.

En lo que sigue usaremos la prescripción de renormalización encontrada para obtener la acción efectiva en una expresión útil para tomar el límite de altas temperaturas, donde esperamos obtener para la entropía una contribución independiente de la temperatura como el término constante del capítulo anterior.

\section{Acción efectiva sobre la esfera en el límite de altas temperaturas}

Para analizar el límite de altas temperaturas, $\beta\ll a$, comenzamos reescribiendo la función zeta \eqref{eq:zetamass} como
\begin{align}
\zeta_{S^3}(s) ={} & \frac{\mu^{2s} a^2}{2} \sum_{k,l\in\mathbb{Z}} \left[ \left(\frac{k}{a}\right)^{\!\!2} + m^2 + \left(\frac{2\pi l}{\beta}\right)^{\!\!2} \right]^{-s+1} \\  \nonumber & -\frac{\mu^{2s} a^2}{2} \sum_{k,l\in\mathbb{Z}} \left[m^2 + \left(\frac{2\pi l}{\beta}\right)^{\!\!2} \right]\left[ \left(\frac{k}{a}\right)^{\!\!2} + m^2 + \left(\frac{2\pi l}{\beta}\right)^{\!\!2} \right]^{-s}\,,
\end{align}
donde, haciendo uso del hecho de que las degeneraciones se anulan en todos los casos para $k=0$ ---lo que dicho con otras palabras quiere decir que no hay modos cero del laplaciano conforme en estas variedades---, hemos completado las sumas sobre $k$ a todos los enteros. Con esto, luego de escribir la potencia entre corchetes en cada línea en la representación del tiempo propio de Schwinger podemos utilizar la fórmula de Poisson para la suma sobre $k$, de lo que obtenemos para la función zeta la expresión
\begin{align}
\zeta_{S^3}(s) ={} & \frac{ {\pi}^{\frac12}({\mu a})^{2s}}{2} \sum_{l\in\mathbb{Z}} \Bigg[(a^2\lambda_{l,0})^{-s+\frac32}\frac{\Gamma(s-\frac32)}{2\Gamma(s)}\\ \nonumber
& + \frac{4}{\Gamma(s-1)}\sum_{k=1}^{\infty} (k\pi)^{s-\frac32}(a^2\lambda_{l,0})^{-\frac{s}{2}+\frac34}
K_{s-\frac32}\!\left(2k\pi\sqrt{a^2\lambda_{l,0}}\right)\\ \nonumber
& - \frac{4}{\Gamma(s)}\sum_{k=1}^{\infty} (k\pi)^{s-\frac12}(a^2\lambda_{l,0})^{-\frac{s}{2}+\frac54}
K_{s-\frac12}\!\left(2k\pi\sqrt{a^2\lambda_{l,0}}\right)\Bigg]\,,
\end{align}
en la que, por razones de brevedad, hemos usado para los autovalores del laplaciano en el caso masivo la notación $\lambda_{l,k}=(2\pi l/\beta)^2+(k/a)^2+m^2$.
Debido a su dependencia con $l$, el término en la primera línea (que no es otra cosa que la contribución del modo $k=0$ luego de la inversión de la suma sobre $k$) debe ser analizado de forma diferente en los casos $m\beta< 2\pi$ y $m\beta> 2\pi$, que llamaremos, con las disculpas del caso, ``régimen de masa pequeña'' y ``régimen de masa grande'' respectivamente, queriendo decir en el primer caso que la masa puede ser grande, pero si tiende a infinito tendrá que hacerlo de forma de nunca superar a $2\pi/\beta$.
	
\subsection{Régimen de masa pequeña}
	
En el caso $m\beta<2\pi$, luego de separar el término $l=0$ en la contribución correspondiente a $k=0$ podemos desarrollar el binomio en la suma $l\neq 0$ restante de modo de escribir dicha contribución como
\begin{align}\label{eq:zetamasshigh1}
\zeta^{k=0}_{S^3}(s) ={} & (\mu a)^{2s} \frac{\pi^{\frac12}}{4} \frac{\Gamma(s-\tfrac32)}{\Gamma(s)} \Bigg[ (am)^{3-2s}   
\\ \nonumber
& + 2 \left(\frac{2\pi a}{\beta}\right)^{\!\!3-2s} \sum_{n=0}^{\infty} \frac{\Gamma(\tfrac52 - s)}{n!\,\Gamma(\tfrac52-n-s)} \left(\frac{m\beta}{2\pi}\right)^{\!\!2n} \!\zeta_R(2s+2n-3) \Bigg]\,.
\end{align}
Separando en primer lugar el término $n=2$, del desarrollo \eqref{eq:riemannzpolo} de la zeta de Riemann para $s$ pequeño obtenemos para la contribución a la acción efectiva
\begin{align}\nonumber
\Gamma_{S^3}^{k=0}(\beta) =& -\frac{\pi}{6} (am)^3 -\frac{\pi}{3} \left(\frac{2\pi a}{\beta}\right)^{\!\!3} \sum_{\substack{n=0\\n\neq 2}}^{\infty} \frac{\Gamma(\tfrac52)}{n!\,\Gamma(\tfrac52-n)}\!\left(\frac{m\beta}{2\pi}\right)^{\!\!2n}\!\zeta_R(2n-3) \\ \label{eq:seffmasshigh0}
& -\frac{(am)^4}{32}\frac{\beta}{a} \left[ 2\log(\mu\beta/2\pi) +2\gamma -2\log 2 \right].
\end{align}

Podemos ahora agregar las contribuciones restantes y los contratérminos correspondientes a la prescripción de renormalización elegida: luego de hacerlo la acción efectiva se escribe
\begin{align}\label{eq:seffmasshigh1}
\Gamma_{S^3} =& -\frac{\pi}{6} (am)^3 -\frac{\pi^4}{45}\left(\frac{a}{\beta}\right)^{\!\!3} + \frac{\pi^2}{12} (am)^2 \frac{a}{\beta}\\ \nonumber
&+ \frac{1}{4\pi^2}\sum_{k=1}^{\infty} \frac{1}{k^3} \left[ 1 + 2\pi ma k + \frac12 (2\pi mak)^2 \right] e^{-2\pi ma k}+{\cal O}(m\beta)\,,
\end{align}
donde los términos que no hemos escrito se anulan como potencias positivas de $m\beta$ (la suma sobre $n$ en \eqref{eq:seffmasshigh0}, que corresponde a los modos con $k=0$ y $l\neq0$) o exponencialmente con $a/\beta$ (los términos con $k\neq0$).

En el límite $m\rightarrow0$ en el que la masa se anula más rápido que cualquier otro parámetro con sus dimensiones, esta expresión coincide con el resultado obtenido en el caso conforme en la sección \ref{sec:termo:esfera:low}. Observamos como antes la presencia de términos independientes de la temperatura, que ahora, si $m\neq0$, no son independientes del volumen del espacio sino que dependen del radio $a$ de la esfera a través del producto adimensional $am$. 

Como veremos a continuación, el límite de altas temperaturas es completamente diferente cuando la masa crece más rápido que $2\pi/\beta$.
	
\subsection{Régimen de masa grande}

En el caso $m\beta\geq2\pi$ ---esto es, cuando la temperatura, incluso siendo grande, no es mayor que la masa del campo---, ya no es útil el desarrollo en potencias de $m\beta$, de modo que en el término $k=0$ en lugar de hacer el desarrollo binomial debemos regularizar la suma sobre $l$ de otra forma. Utilizamos para dicha suma la representación del tiempo propio de Schwinger y aplicamos la fórmula de Poisson, con lo que obtenemos
\begin{align}\nonumber
\zeta^{k=0}_{S^3} (s) = \frac{\pi(\mu a)^{2s}}{4\Gamma(s)} \!\left( \frac{2\pi a}{\beta} \right)^{\!\!3-2s}\!\Bigg[ \!\!\left(\frac{m\beta}{2\pi}\right)^{\!\!4-2s}\!\!\Gamma(s-2)  +4\sum_{l=1}^{\infty} \!\left(\frac{2\pi^2l}{m\beta}\right)^{\!\!s-2} \!\!K_{2-s}(m\beta l) \Bigg],
\end{align}
para $m\beta\neq 0$. Nuevamente, usando el desarrollo para $s$ pequeño del cociente $\Gamma(s-2)/\Gamma(s)$ podemos obtener directamente la contribución de este término a la acción efectiva:
\begin{align}
\Gamma_{S^3}^{k=0}(\beta) = -\frac{(am)^4}{32} \frac{\beta}{a} \left[\frac32 + 2\log\left(\frac{\mu}{m}\right)\right] - (am)^2\frac{a}{\beta} \sum_{l=1}^{\infty} \frac{1}{l^2} K_2\left(m\beta l\right)\,.
\end{align}

Una vez más, añadimos el contratérmino correspondiente a la prescripción de renormalización elegida, y teniendo en cuenta también los términos con $k\neq0$ obtenemos
\begin{align}\label{eq:seffmasshigh2}
\Gamma_{S^3}(\beta) =& -(am)^2 \frac{a}{\beta} \sum_{l=1}^{\infty} \frac{1}{l^2} K_2\left(m\beta l\right) +  \mathcal{O} \left(e^{- a/\beta}\right)\,.
\end{align}
Notemos que en el límite de masa infinita, con $m\gg\beta^{-1}\gg a^{-1}$ la acción efectiva ---y, en particular, la parte independiente de la temperatura--- se anula, lo que se conoce como desacoplamiento de los modos masivos de la teoría \cite{Ambjorn:1978hz,Appelquist:1974tg}. Esto implica que en ese caso no se cumple la coincidencia entre los términos independientes de la temperatura en la acción efectiva en el límite de altas temperaturas y la acción efectiva de la misma teoría definida sobre la esfera tridimensional, lo que sí sucedía en el caso $m\beta<2\pi$. La diferencia entre los dos regímenes no puede entenderse si sólo se considera la teoría en tres dimensiones, sino que es necesario partir de la teoría en cuatro dimensiones, donde la temperatura cumple el papel de la escala que permite discriminar entre los mismos.

\section{Acciones efectivas en el límite de altas temperaturas}
\label{sec:masivo:espacios}

Para continuar obtendremos el desarrollo a altas temperaturas de las acciones efectivas de la teoría sobre los distintos espacios esféricos. Como en el caso no masivo, separamos los espacios lente según la paridad del orden del grupo cíclico correspondiente. Consideramos además los espacios prisma y los espacios poliédricos; a estos últimos los reducimos como en aquel caso a una combinación de espacios lente de orden par.

\subsection{Espacios lente de orden impar}

En el caso de los espacios lente de orden impar, llamando $p=2q+1$, podemos escribir la función zeta \eqref{eq:zetamass} como
\begin{align}\nonumber
\zeta_{\esp{Z_{2q+1}}}(s) ={} & \mu^{2s}\sum_{l\in\mathbb{Z}}\sum_{n=0}^{\infty} \left[  \sum_{\substack{r=0\\ r \mathrm{\,par}}}^{p-1} (np+r) n + \sum_{\substack{r=1\\ r \mathrm{\,impar}}}^{p-2} (np+r) \left(n+1\right)  \right]  \lambda_{l,np+r}^{-s} \\[2mm]\label{eq:zetamassodd}
={} & \frac{1}{p}\zeta_{S^3}(s)+\delta\zeta_{2q+1}(s)\,,
\end{align}
donde, al igual que en el caso no masivo, hemos escrito $k=np+r$ con $r=0,1,\dots, p-1$, y donde
\begin{align}\nonumber
\delta\zeta_{2q+1}(s) = -\frac{\mu^{2s}}{p} \sum_{l\in\mathbb{Z}} \sum_{n=0}^{\infty} \Big\{ & \sum_{r=1}^{q} 2r \left(np+2r\right) \lambda_{l,np+2r}^{-s} \\ \nonumber
& + \sum_{r=0}^{q-1} 2(r-q) \left(np+2r+1\right) \lambda_{l,np+2r+1}^{-s}\Big\}\,.
\end{align}
Haciendo el cambio $q-r\rightarrow r$ en la segunda de las sumas sobre $r$ y extrayendo convenientemente algunos factores obtenemos
\begin{align}\label{eq:deltazetaoddmass}
\delta\zeta_{2q+1}(s) =& -2\left(\frac{\mu a}{p}\right)^{\!\!2s} \sum_{l\in\mathbb{Z}} \sum_{n=0}^{\infty} \sum_{r=1}^{q} r  
\\ \nonumber &  \times \Bigg\{ \!\!\left(n+\tfrac{2r}{p}\right) \!\left[\left(n+\tfrac{2r}{p}\right)^{\!2} +\left(\frac{am}{p}\right)^{\!\!2} +\left(\frac{2\pi a l}{p\beta}\right)^2 \right]^{-s} \\ \nonumber	
& - \left(n+1-\tfrac{2r}{p}\right) \!\left[\left(n+1-\tfrac{2r}{p}\right)^{\!2} +\left(\frac{am}{p}\right)^{\!\!2} +\left(\frac{2\pi a l}{p\beta}\right)^{\!\!2} \right]^{-s} \Bigg\}\,.
\end{align}

Antes de obtener una extensión analítica de esta parte de la función zeta que nos permita tomar el límite de altas temperaturas, queremos ver cuál es su contribución al límite $am\rightarrow\infty$ a temperatura cero, para saber si es necesario modificar la prescripción de renormalización que usamos para la esfera y, en todo caso, qué nuevos contratérminos, si los hubiera, habría que tener en cuenta. Para eso comenzamos escribiendo en cada uno de los términos entre llaves la potencia de la cantidad entre corchetes en la representación del tiempo propio de Schwinger, y luego utilizamos la inversión de Poisson de la suma sobre $l$, de la que nos interesa solamente el término $l=0$, puesto que en la acción efectiva los demás términos se anulan en el límite $\beta/a\rightarrow\infty$ para cualquier valor de la masa. Este término puede ser escrito en la forma
\begin{align}\nonumber
\delta\zeta_{2q+1}^{l=0}(s) = & -2\sqrt{\pi}\frac{\Gamma(s-\tfrac12)}{\Gamma(s)}\!\left(\frac{\mu a}{p}\right)^{\!\!2s} \!\sum_{n=0}^{\infty} \sum_{r=1}^{q} r \Bigg\{ \!\!\left(n+\tfrac{2r}{p}\right) \!\left[\left(n+\tfrac{2r}{p}\right)^{\!2} \!+\!\left(\frac{am}{p}\right)^{\!\!2} \right]^{\frac12-s} \\ 
& - \left(n+1-\tfrac{2r}{p}\right) \!\left[\left(n+1-\tfrac{2r}{p}\right)^{\!2} +\left(\frac{am}{p}\right)^{\!\!2} \right]^{\frac12-s} \Bigg\}\,.
\end{align}
Podemos ahora reescribir cada uno de los términos como una derivada de modo de absorber los factores fuera de los corchetes y hacer el cambio de índice $n+1\rightarrow-n$ en la suma del segundo término para obtener la expresión
\begin{align}\nonumber
\delta\zeta_{2q+1}^{l=0}(s) = \sqrt{\pi}\frac{\Gamma(s-\tfrac32)}{\Gamma(s)}\left(\frac{\mu a}{p}\right)^{\!\!2s} \!\sum_{n\in\mathbb{Z}} \sum_{r=1}^{q}  r\left.\frac{d}{d\alpha}  \!\left[\left(n+\tfrac{2r}{p}+\alpha\right)^{\!2} +\left(\frac{am}{p}\right)^{\!\!2} \right]^{\frac32-s}\right\vert_{\alpha=0}
\end{align}
que, si utilizamos la representación del tiempo propio de Schwinger para la potencia entre corchetes y una inversión de Poisson de la suma sobre $n$, se reduce a
\begin{align}\nonumber
\delta\zeta_{2q+1}^{l=0}(s) = \frac{8(am)^2}{p^2\Gamma(s)}\left(\frac{\pi\mu^2}{pm^2}\right)^{\!\!s} \!\sum_{n=1}^{\infty} \frac{1}{n^{1-s}} \sum_{r=1}^{q} r \sen(4\pi nr/p) K_{2-s}(2\pi amn/p)\,.
\end{align}
De aquí vemos que en la derivada con respecto a $s$ todos los términos tienen una función de Bessel $K_2$ de argumento proporcional a $am$, de modo que la acción efectiva se anula exponencialmente cuando $am\rightarrow\infty$, no siendo necesaria entonces la adición de ningún contratérmino adicional a los correspondientes a la prescripción de renormalización utilizada en el caso de la esfera.

Volvemos ahora a la expresión \eqref{eq:deltazetaoddmass} para obtener su desarrollo a altas temperaturas. Otra vez, es posible verificar que los términos con $l\neq 0$ se anulan exponencialmente en el límite $a/\beta\rightarrow\infty$ para cualquier valor de la masa, por lo que sólo calcularemos el término $l=0$, que puede escribirse
\begin{align}\nonumber
\delta\zeta_{2q+1}^{l=0}(s) = 
\frac{1}{s-1} \left(\frac{\mu a}{p}\right)^{\!\!2s} \!\sum_{n\in\mathbb{Z}} \sum_{r=1}^{q} r  \left.\frac{d}{d\alpha} \!\left[\left(n+\tfrac{2r}{p}+\alpha\right)^{\!2} +\left(\frac{am}{p}\right)^{\!\!2} \right]^{-s+1}\right\vert_{\alpha=0}\,.
\end{align}
Para obtener una extensión analítica de esta expresión a partir de la cual se pueda calcular su derivada en $s=0$ usamos, como ya es costumbre, la representación del tiempo propio de Schwinger y la inversión de la suma sobre $n$, lo que da
\begin{align}\nonumber
\delta\zeta_{2q+1}^{l=0}(s) = \frac{-8}{\Gamma(s)} \left(\frac{am}{p}\right)^{\!\!\frac32}
\!\left(\frac{\pi\mu^2a}{pm}\right)^{\!\!s}\!\sum_{r=1}^{q} r  \sum_{n=1}^{\infty} n^{s-\frac12} \sen(4\pi rn/p) K_{\frac32-s}(2\pi amn/p)\,.
\end{align}
Usando ahora la expresión de la función de Bessel $K_{\frac32}(x)$ en términos de la exponencial $e^{-x}$ podemos ver que la contribución de este término a la acción efectiva resulta
\begin{align}\label{eq:seffodd}
\delta\Gamma_{\esp{Z_{p=2q+1}}}^{l=0} = \frac{1}{\pi}\sum_{r=1}^{q}r\sum_{n=1}^{\infty}\frac{1}{n^2}\left(\frac{2\pi amn}{p}+1\right)\sen{\left(4\pi rn/p\right)}e^{-2\pi amn/p}\,.
\end{align}
De este modo, la acción efectiva sobre el espacio lente de orden impar $\esp{Z_{p=2q+1}}$ puede obtenerse, según la ecuación \eqref{eq:zetamassodd} y a menos de términos que se anulan exponencialmente a altas temperaturas, sumando a esta contribución la acción efectiva \eqref{eq:seffmasshigh1} sobre la esfera dividida por el orden $p=2q+1$ del grupo cíclico correspondiente al espacio lente considerado.

\subsection{Espacios lente de orden par}

Estudiaremos ahora las acciones efectivas sobre los espacios lente correspondientes a grupos cíclicos de orden par. Al igual que en el caso no masivo, comenzamos considerando el caso más simple del espacio proyectivo $\esp{Z_2}$, para el que las degeneraciones son las mismas que en la esfera para los autovalores de índice impar, y son nulas para los autovalores de índice par. Como en aquel caso, la función zeta puede escribirse en términos de la función zeta del mismo operador sobre la esfera:
\begin{align}\nonumber
\zeta_{\esp{Z_2}}(s) &=  \mu^{2s} \sum_{l\in\mathbb{Z}} \Bigg\{ \sum_{k=1}^{\infty} k^2 \!\left[\left(\frac{k}{a}\right)^{\!\!2} + m^2 + \left(\frac{2\pi l}{\beta}\right)^{\!\!2}\right]^{-s} \\ \nonumber
& \phantom{\mu^{2s} \sum_{l\in\mathbb{Z}} \Bigg\{ \sum_{a}} - \sum_{k=1}^{\infty} (2k)^2 \!\left[ \left(\frac{2k}{a}\right)^{\!\!2} + m^2 + \left(\frac{2\pi l}{\beta}\right)^{\!\!2} \right]^{-s} \Bigg\} \\
& =  \zeta_{S^3}(s;\beta,a,m) - 2^{2-2s}\zeta_{S^3}(s;2\beta,a,m/2)\,,
\end{align}
de donde resulta para la acción efectiva la combinación
\begin{align}\label{eq:seffmass2}
\Gamma_{\esp{Z_2}}(\beta,a,m) = 
\Gamma_{S^3}(\beta,a,m) - 4\Gamma_{S^3}(2\beta,a,m/2)\,.
\end{align}
Vemos entonces que es posible obtener expresiones para la acción efectiva sobre este espacio que permitan tomar los diferentes límites de temperaturas y masas a partir de las correspondientes expresiones obtenidas anteriormente para la esfera. En particular, la prescripción de renormalización para esta acción efectiva puede obtenerse también de la correspondiente a la esfera.

Nos ocuparemos ahora de los restantes espacios lente correspondientes a grupos cíclicos de orden par. Comenzamos usando la expresión \eqref{eq:degeven} de las degeneraciones para obtener, similarmente a lo que sucedía en el caso no masivo, la expresión
\begin{align}\label{eq:zetamasseven}
\zeta_{\esp{Z_{2q}}}(s) = {}& 
\frac{1}{q}\zeta_{\esp{Z_2}}(s) + \delta\zeta_{2q}(s),
\end{align}
donde ahora
\begin{align}
\delta\zeta_{2q}(s) = {}& \mu^{2s} \sum_{n=0}^{\infty} \sum_{r=0}^{q-1} \left( \frac{q-r-1}{q} - \frac{r}{q} \right) (2nq+2r+1) \\ \nonumber
& \phantom{\mu^{2s} \sum_{n=0}^{\infty} \sum_{r=0}^{q-1} \frac{q-r-1}{q}} \times \sum_{l\in\mathbb{Z}} \left[ \left(  \frac{2nq+2r+1}{a}\right)^{\!\!2} + m^2 + \left(\frac{2\pi l}{\beta} \right)^{\!\!2} \right]^{-s}\,.
\end{align}

Redefiniendo $q-r-1\rightarrow r$ en el primer término del primer factor entre paréntesis y extrayendo de manera conveniente algunos factores, podemos escribir
\begin{align}\label{eq:deltazetaevenmass}
\delta\zeta_{2q}(s) = {}& 2q\left(\frac{\mu a}{2q}\right)^{\!\!2s} \!\sum_{l\in\mathbb{Z}} \sum_{n=0}^{\infty} \Bigg\{ \sum_{r=0}^{q-1} \frac{r}{q} \left(n+1-\tfrac{2r+1}{2q}\right)\\ \nonumber & 
\times\left[ \left(n+1-\tfrac{2r+1}{2q}\right)^{\!2} + \left(\frac{am}{2q}\right)^{\!\!2} + \left(\frac{2\pi a l}{2q\beta}\right)^{\!\!2}\right]^{-s}  \\ \nonumber
& - \sum_{r=0}^{q-1} \frac{r}{q} \left(n+\tfrac{2r+1}{2q}\right) \left[ \left(n+\tfrac{2r+1}{2q}\right)^{\!2} + \left(\frac{am}{2q}\right)^{\!\!2} + \left(\frac{2\pi a l}{2q\beta}\right)^{\!\!2}\right]^{-s} \Bigg\}\,.
\end{align}

Como en el caso de los espacios lente de orden impar, es posible mostrar que esta parte de la función zeta se anula para $am\rightarrow\infty$ en el límite de temperatura cero; al igual que en aquel caso, consideremos la expresión anterior luego de la inversión de la suma sobre $l$ de la cantidad entre llaves escrita en la representación del tiempo propio de Schwinger: los términos con $l\neq0$ se anulan exponencialmente en el límite $\beta/a\rightarrow\infty$ de bajas temperaturas cualquiera sea el valor de la masa, mientras que el término $l=0$ puede ser escrito como
\begin{align}\nonumber
\delta\zeta_{2q}^{l=0}(s) = \sqrt{\pi}\frac{\Gamma(s-\tfrac12)}{\Gamma(s)}\!\left(\frac{\mu a}{2q}\right)^{\!\!2s} \!\sum_{n\in\mathbb{Z}} \sum_{r=1}^{q-1}  r \!\left.\frac{d}{d\alpha}\!\left[\left(n+\tfrac{2r+1}{2q}+\alpha\right)^{\!2} \!+\!\left(\frac{am}{2q}\right)^{\!\!2} \right]^{\frac12-s}\right\vert_{\alpha=0}.
\end{align}
Otra vez, luego de una inversión de Poisson de la suma sobre $n$ una vez escrita la potencia en la representación del tiempo propio de Schwinger, tenemos
\begin{align}\nonumber
\delta\zeta_{2q}^{l=0}(s) = \frac{2\pi am}{\Gamma(s)} \left(\frac{\pi a\mu^2}{2qm}\right)^{\!\!s} \sum_{r=1}^{q-1} r \sum_{n=1}^{\infty} n^s \sen\!\left(2\pi n\tfrac{2r+1}{2q}\right) \!K_{1-s}(\pi amn/q)\,,
\end{align}
de donde vemos, como esperábamos, que la acción efectiva a temperatura cero se anula exponencialmente  en el límite $am\rightarrow\infty$.

Volviendo a la expresión \eqref{eq:deltazetaevenmass}, podemos ver que la contribución a la acción efectiva de los términos con $l\neq0$ es proporcional a $\exp(-a/\beta)$, anulándose en el límite de altas temperaturas. El término $l=0$ puede escribirse
\begin{align}\nonumber
\delta\zeta_{2q}^{l=0}(s) = 
\frac{2}{s-1} \left(\frac{\mu a}{2q}\right)^{\!\!2s} \!\sum_{n\in\mathbb{Z}} \sum_{r=1}^{q-1} r  \!\left.\frac{d}{d\alpha} \!\left[\left(n+\tfrac{2r+1}{2q}+\alpha\right)^{\!2} +\left(\frac{am}{2q}\right)^{\!\!2} \right]^{-s+1}\right\vert_{\alpha=0}\,.
\end{align}
Usando una vez más la representación del tiempo propio de Schwinger para la potencia entre corchetes y la inversión de Poisson de la suma sobre $n$ obtenemos
\begin{align}\nonumber
\delta\zeta_{2q}^{l=0}(s) &= \frac{-8}{\Gamma(s)} \left(\frac{am}{2q}\right)^{\!\!\frac32}
\!\left(\frac{\pi\mu^2a}{2qm}\right)^{\!\!s} \sum_{r=1}^{q-1} r \sum_{n=1}^{\infty} n^{s-\frac12} \sen\!\left(2\pi n\tfrac{2r+1}{2q}\right) \!K_{\frac32-s}(\pi amn/q)\,,
\end{align}
que contribuye a la acción efectiva como
\begin{align}\label{eq:seffeven}
\delta\Gamma_{\esp{Z_{2q}}}^{l=0} = \frac{1}{\pi}\sum_{r=1}^{q-1}r\sum_{n=1}^{\infty}\frac{1}{n^2}\left(\frac{2\pi amn}{2q}+1\right)\sen\!\left(2\pi n\tfrac{2r+1}{2q}\right)e^{-\pi amn/q}\,.
\end{align}
La acción efectiva sobre el espacio lente de orden par $\esp{Z_{2q}}$ puede obtenerse entonces ---excepto términos que se anulan exponencialmente a altas temperaturas--- a partir de la ecuación \eqref{eq:zetamasseven}, sumando la contribución $\delta\Gamma_{\esp{Z_{2q}}}^{l=0}$ y la acción efectiva \eqref{eq:seffmass2} sobre el espacio proyectivo dividida por la mitad $q$ del orden del grupo cíclico correspondiente.

\subsection{Espacios prisma}
	
Para calcular las acciones efectivas sobre espacios prisma en el límite de altas temperaturas, comenzamos usando la expresión \eqref{eq:degprismaimpar1} para las degeneraciones de los autovalores de laplaciano sobre estos espacios, con la que la función zeta puede escribirse como
\begin{align}\label{eq:zetaprismamass}
\zeta_{\esp{D_p^*}}(s) = {}& \frac{1}{2}\zeta_{\esp{Z_{2p}}}(s)+\delta\zeta(s)\,,
\end{align}
donde
\begin{align}\nonumber
\delta\zeta(s) = {}&   \frac{\mu^{2s}}{2}\sum_{l\in\mathbb{Z}} \sum_{k=0}^{\infty} \left[ (4k+1)\lambda_{l,4k+1}^{-s} - (4k+3)\lambda_{l,4k+3}^{-s} \right]\,,
\end{align}
que puede escribirse, redefiniendo el índice $k\rightarrow -k-1$ en la segunda suma, como
\begin{align}\label{eq:deltazetaprisma}
\delta\zeta(s) =  \frac{(\mu a)^{2s}}{4(-s+1)}\sum_{k,l\in\mathbb{Z}} \left.\frac{d}{d\alpha}\left[\left(\frac{2\pi al}{\beta}\right)^{\!\!2} + (am)^2 + (4k+1)^2\right]^{-s+1}\right\vert_{\alpha=1}\,.
\end{align}
Nuevamente, podemos ver que esta parte de la zeta da una contribución a la acción efectiva que se anula para temperatura cero en el límite $am\rightarrow\infty$: para ello comenzamos escribiendo como en los demás casos la potencia entre corchetes en la representación del tiempo propio de Schwinger e invertimos la suma sobre $l$, con lo que vemos inmediatamente que los términos con $l\neq0$ se anulan exponencialmente en el límite de bajas temperaturas, mientras que el término $l=0$ queda escrito en la forma
\begin{align}
\delta\zeta^{l=0}(s) = -\frac{2(am)^2}{\Gamma(s)}\left(\frac{\pi\mu^2a}{4m}\right)^{\!\!s} \sum_{k=1}^{\infty} \frac{1}{k^{1-s}} \sen(\pi k/2) K_{2-s}(\pi amk/2)\,,
\end{align}
de donde se observa que en la acción efectiva la contribución de $\delta\zeta$ se anula para $am\rightarrow\infty$. Esto quiere decir que la renormalización de la acción efectiva sobre el espacio prisma $\esp{D_p^*}$ se efectúa sustrayendo a la acción original la mitad del término que sustrajimos al renormalizar la acción efectiva sobre el espacio lente $\esp{Z_{2p}}$. 

Para obtener un desarrollo de la acción efectiva a altas temperaturas, volvemos a la expresión \eqref{eq:deltazetaprisma} e invertimos la suma sobre $k$, obteniendo una vez más una parte $l\neq 0$ que contribuye a la acción efectiva como términos que se anulan exponencialmente con la temperatura, y una parte $l=0$ que se escribe
\begin{align}\nonumber
\delta\zeta^{l=0}(s) = {}& \frac{\pi^{\frac32}\left(\mu a\right)^{2s}}{16\Gamma(s)}\sum_{k=1}^{\infty} k \sen(\pi k/2) \left(\frac{4am}{\pi k}\right)^{\!\frac32-s} K_{\frac32-s}(\pi amk/2)\,.
\end{align}
La contribución de este último término a la acción efectiva resulta
\begin{align}
\delta\Gamma^{l=0}_{\esp{D_p^*}} = -\frac{1}{\pi}\sum_{k=1}^{\infty} \sen(\pi k/2)\!\left(1+\frac{\pi a m k}{2}\right) e^{-\pi amk/2}\,,
\end{align}
y la acción efectiva completa se obtiene de la expresión \eqref{eq:zetaprismamass} usando los resultados del apartado anterior.

\subsection{Espacios poliédricos}

En los espacios restantes usamos, como en el caso no masivo, las descomposiciones \eqref{eq:degpolycyclic}, con lo que la acción efectiva puede obtenerse en cada caso como la correspondiente combinación de acciones efectivas sobre espacios lente pares. Como las expresiones no aportan información sustancial, para ahorrar espacio no las mostraremos explícitamente, aunque en lo que analizaremos las correspondientes cantidades termodinámicas.

\section{Acciones efectivas en el límite de bajas temperaturas}
\label{sec:masivo:low}

En cuanto a los desarrollos a bajas temperaturas de las distintas acciones efectivas, un cálculo similar al de la sección \ref{sec:termo:low} da como resultado en todos los casos la expresión general
\begin{align}\label{eq:sefflowmass}
\Gamma=\beta E_0^{(H)} + \sum_{k=1}^\infty d_k^{(H)} \log(1-e^{-\beta \sqrt{\lambda_k^2+m^2}})\,,
\end{align}
donde, como en aquella sección, $\lambda_k^2=(k/a)^2$ son los autovalores del laplaciano cambiado de signo sobre la esfera y $d_k^{(H)}$ las correspondientes degeneraciones sobre el espacio esférico $\esp{H}$, y donde $E_0^{(H)}$ es la extensión al caso masivo de la energía de vacío definida en \eqref{eq:vacuum}. 

\vfill\pagebreak

\section{Propiedades termodinámicas}
\label{sec:masivo:termo}

Como hicimos para el campo sin masa, analizaremos ahora las propiedades termodinámicas de la teoría en los límites de bajas y altas temperaturas. Al igual que en aquel caso, daremos expresiones generales de la entropía en ambos límites e ilustraremos los resultados más interesantes por medio de gráficos numéricos.

A bajas temperaturas, la energía y la energía libre difieren en términos que se anulan exponencialmente cuando $\beta/a$ crece, por lo que la entropía se anula exponencialmente y valen las observaciones que hicimos para el caso no masivo en la sección \ref{sec:termo:props}. En efecto, la entropía puede escribirse como
\begin{align}\label{eq:entropylowmass}
S_{\esp{H}} = \sum_{k=1}^{\infty} d_k^{(H)} \left[\frac{\frac{\beta}{a} \sqrt{k^2+(am)^2}}{e^{\frac{\beta}{a}\sqrt{k^2+(am)^2}}-1}-\log\left(1-e^{-\frac{\beta}{a}\sqrt{k^2+(am)^2}}\right)\right]\,,
\end{align}
de donde vemos que tiende a cero en ambos regímenes de masa cuando la temperatura se anula.

En el límite de altas temperaturas, la entropía tiene la forma
\begin{align}\label{eq:entropyhighmass}
S_{\esp{H}} \sim \frac{4\pi^4}{45|H|}\!\left(\frac{a}{\beta}\right)^{\!\!3} - \frac{\pi^2}{6|H|}(am)^2\frac{a}{\beta} + S_{0,\esp{H}}(am)\,,
\end{align}
donde $S_{0,\esp{H}}$ contiene términos que dependen de la masa adimensionalizada $am$, así como términos independientes de la masa y el tamaño del espacio, no dependiendo en forma alguna de la temperatura. Esta parte de la entropía es diferente para los distintos espacios, coincidiendo en el límite $am\rightarrow0$ con el término constante en la entropía que obtuvimos en el caso no masivo del capítulo anterior. Observamos la presencia de términos extensivos de tipo Stefan-Boltzmann, de los cuales el dominante es el mismo que aparecía en el caso no masivo \cite{Elizalde:2003cv,Kutasov:2000td}. El término subdominante tiene el signo opuesto al dominante, lo que en el caso de la esfera está en acuerdo con los resultados reportados en \cite{Myers:2012tz}.

\begin{figure}[h!]
	\centering
	\begin{minipage}{.4\textwidth}
		\centering
		\includegraphics[height=40mm]{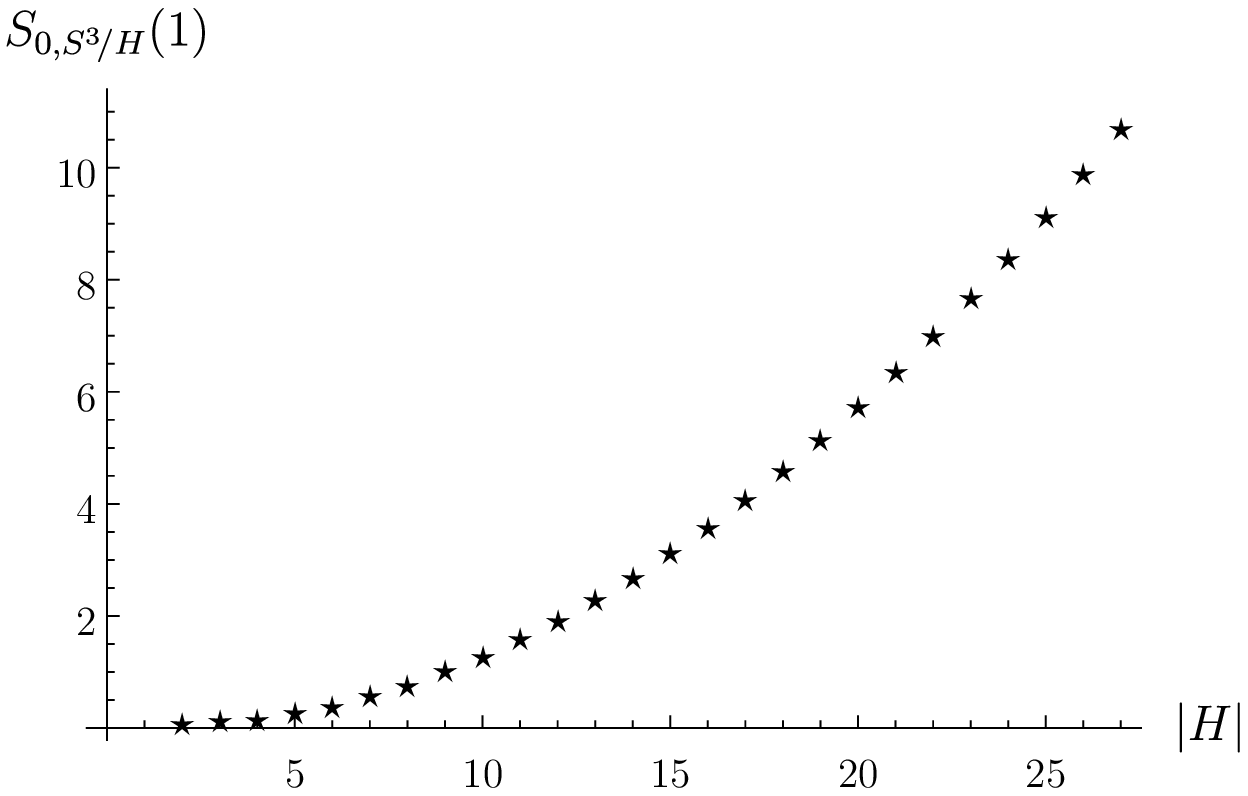}
	\end{minipage}
	\hspace{0.05\textwidth}
	\begin{minipage}{.4\textwidth}
		\centering
		\includegraphics[height=40mm]{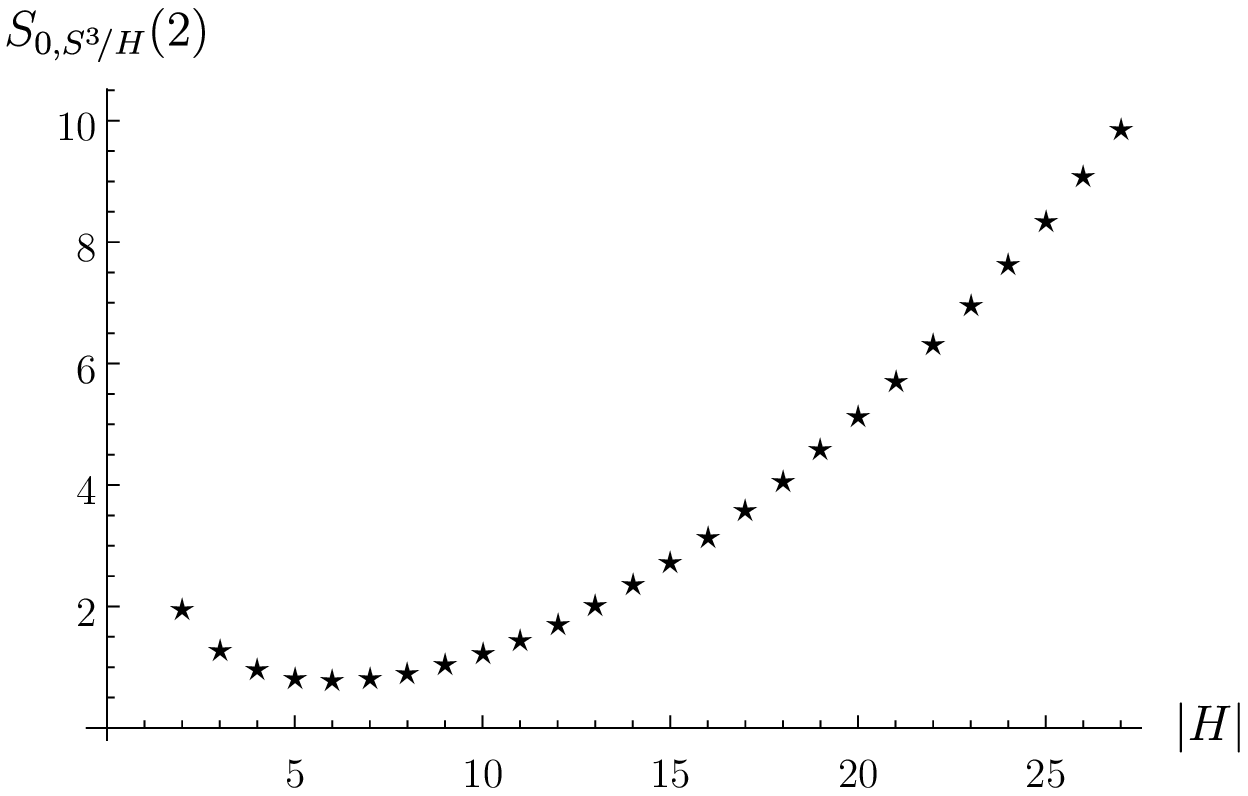}
	\end{minipage}\\
	\begin{minipage}{.4\textwidth}
		\centering
		\includegraphics[height=40mm]{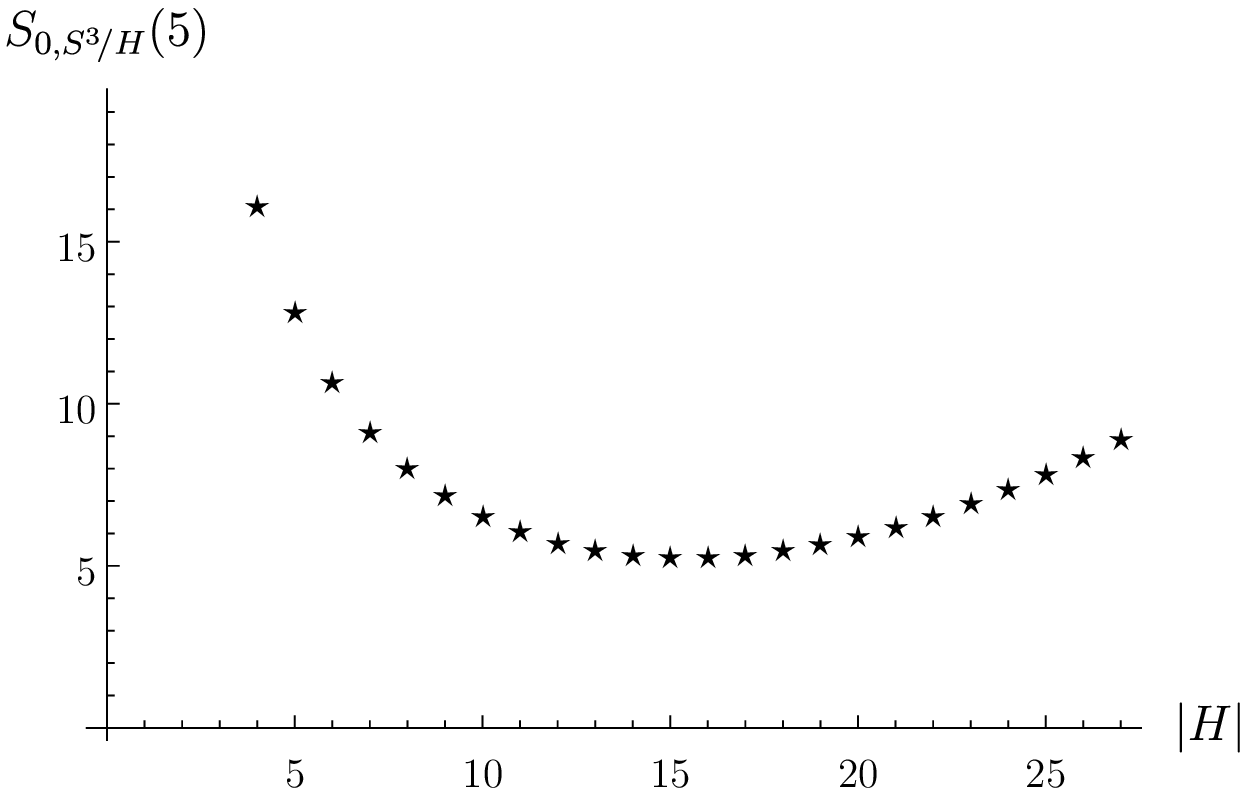}
	\end{minipage}
	\hspace{0.05\textwidth}
	\begin{minipage}{.4\textwidth}
		\centering
		\includegraphics[height=40mm]{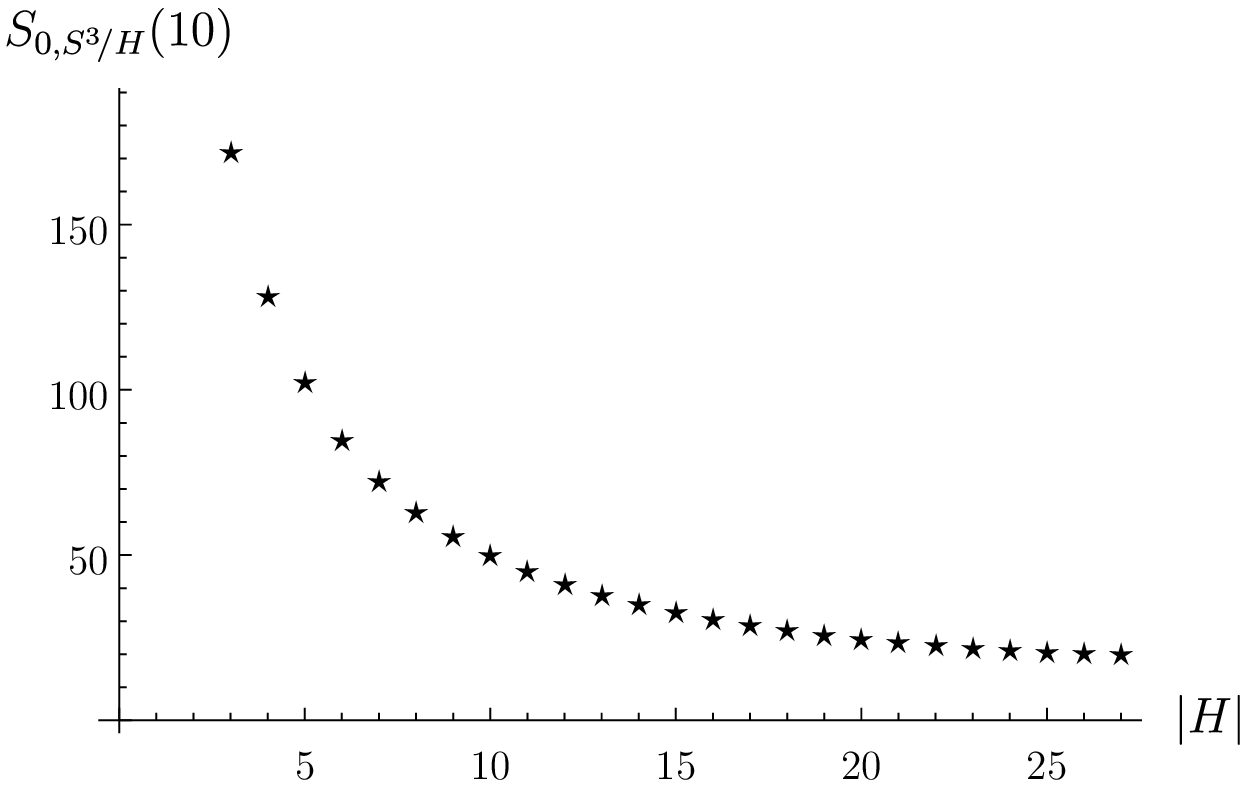}
	\end{minipage}
\caption{\small Parte independiente de la temperatura de la entropía sobre espacios lente como función del orden del grupo para distintos valores de $ma$.}
\label{fig:lentemass}
\end{figure}

El término $S_{0,\esp{H}}(am)$, que por cuestiones de espacio no escribiremos explícitamente para los distintos espacios, depende de la masa adimensionalizada de forma diferente para los distintos espacios; ilustramos esto mostrando en la figura \ref{fig:lentemass} gráficos de su valor para distintos valores de la masa en el caso de los espacios lente de orden más bajo. Allí podemos ver que la dependencia con el orden del espacio tiene una forma muy diferente según el valor de la masa. Una exploración numérica sencilla muestra que rápidamente se alcanza la ``forma asintótica'': para $am=10$ el gráfico tiene la misma forma que para valores mayores de la masa. 

\begin{figure}[h!]
	\centering
	\begin{minipage}{.4\textwidth}
		\centering
		\includegraphics[height=42mm]{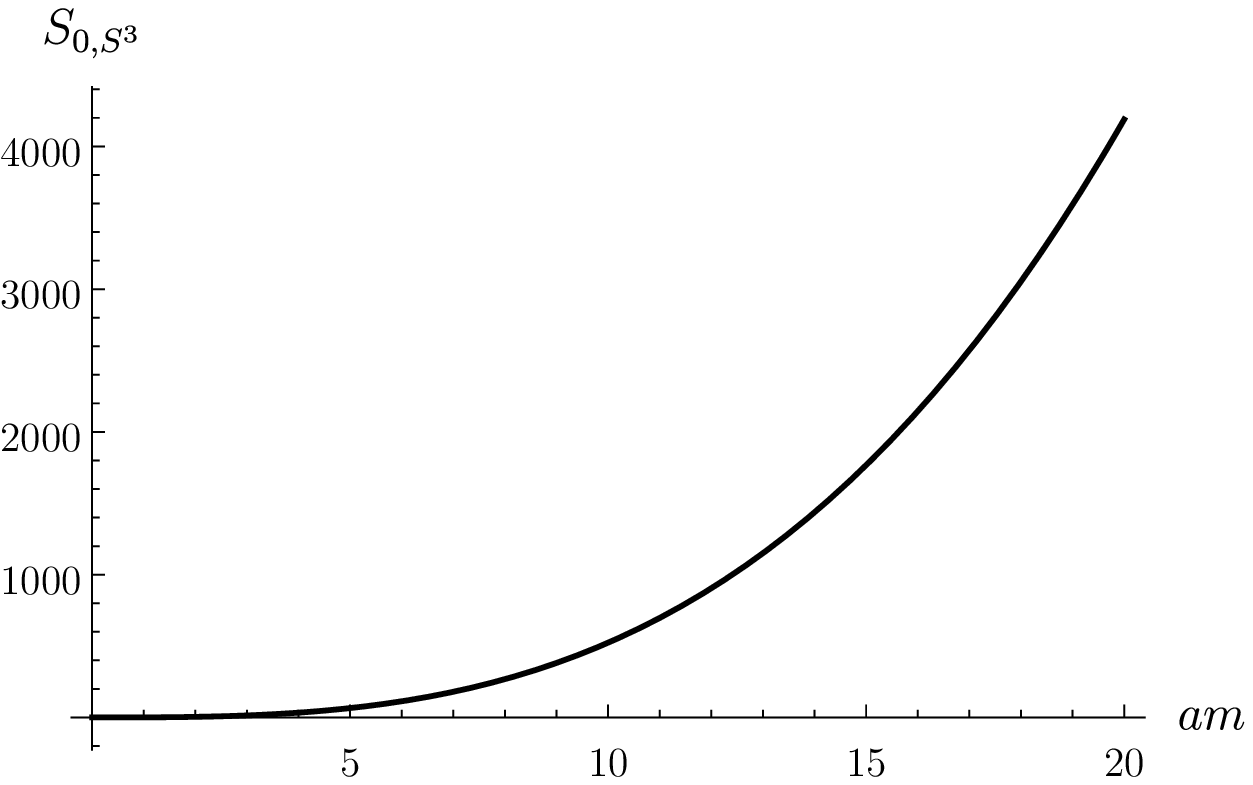}
	\end{minipage}
	\hspace{0.05\textwidth}
	\begin{minipage}{.4\textwidth}
		\centering
		\includegraphics[height=42mm]{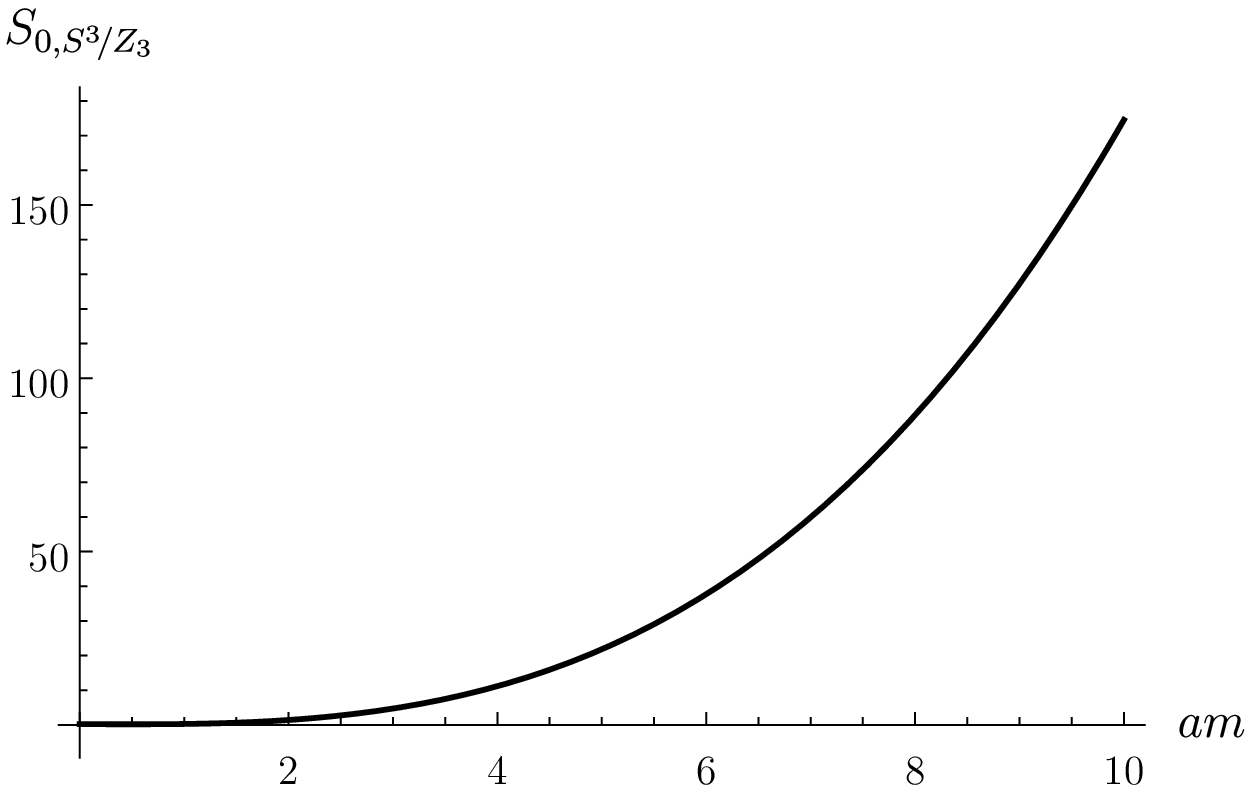}
	\end{minipage}\\
	\begin{minipage}{.4\textwidth}
		\centering
		\includegraphics[height=42mm]{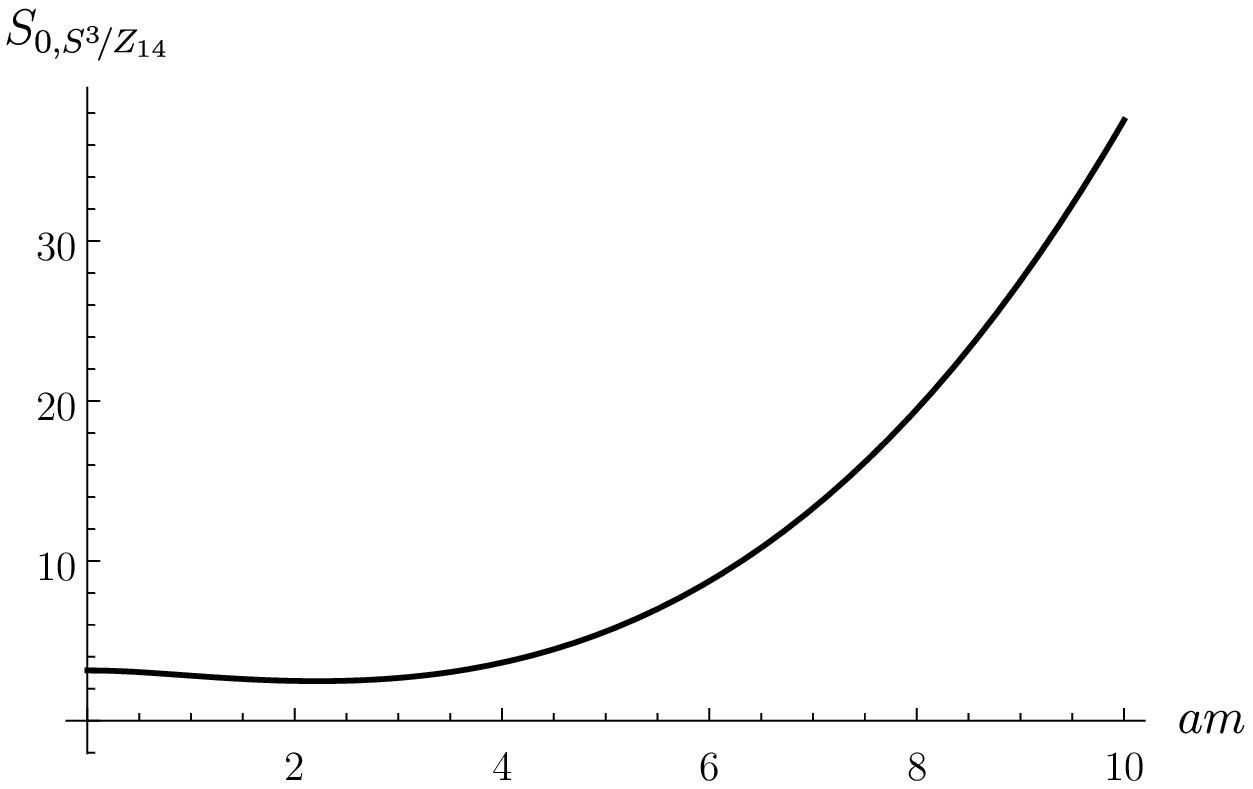}
	\end{minipage}
	\hspace{0.05\textwidth}
	\begin{minipage}{.4\textwidth}
		\centering
		\includegraphics[height=42mm]{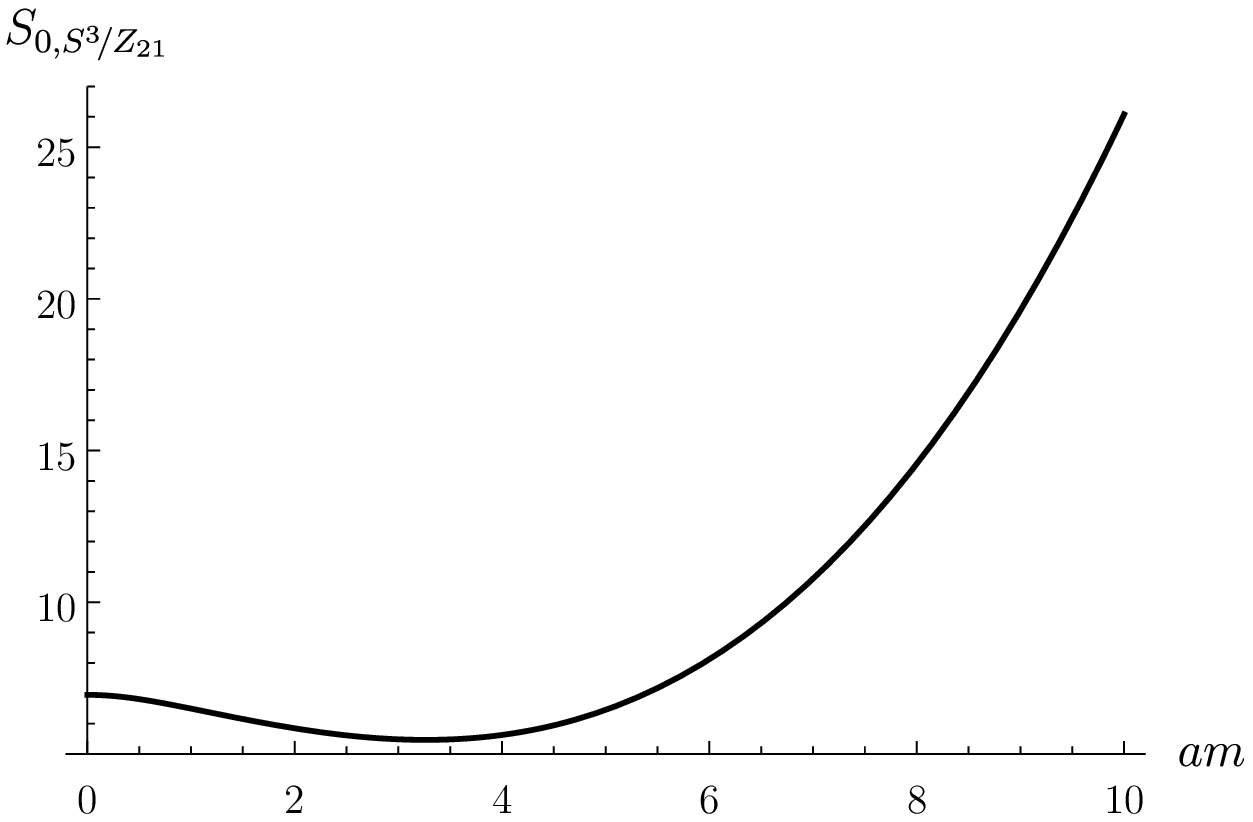}
	\end{minipage}\\
	\begin{minipage}{.4\textwidth}
		\centering
		\includegraphics[height=42mm]{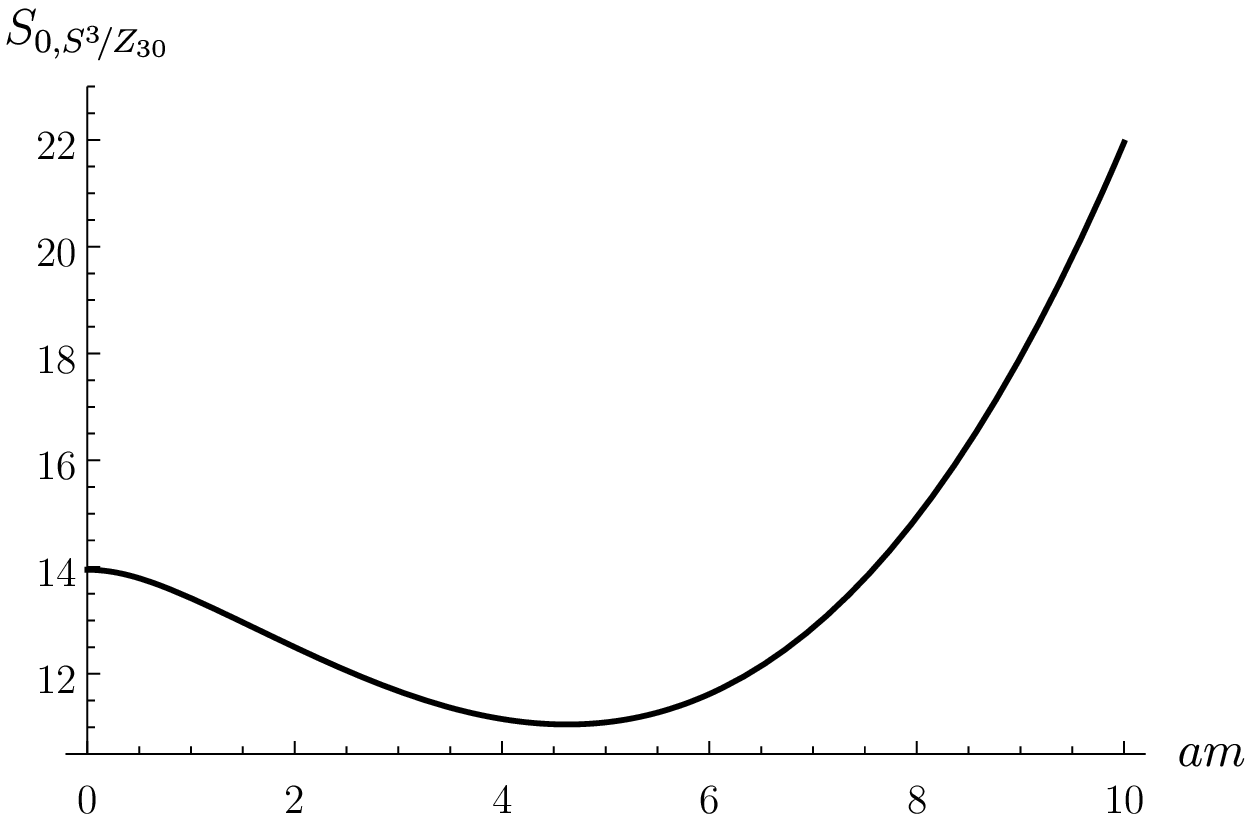}
	\end{minipage}
	\hspace{0.05\textwidth}
	\begin{minipage}{.4\textwidth}
		\centering
		\includegraphics[height=42mm]{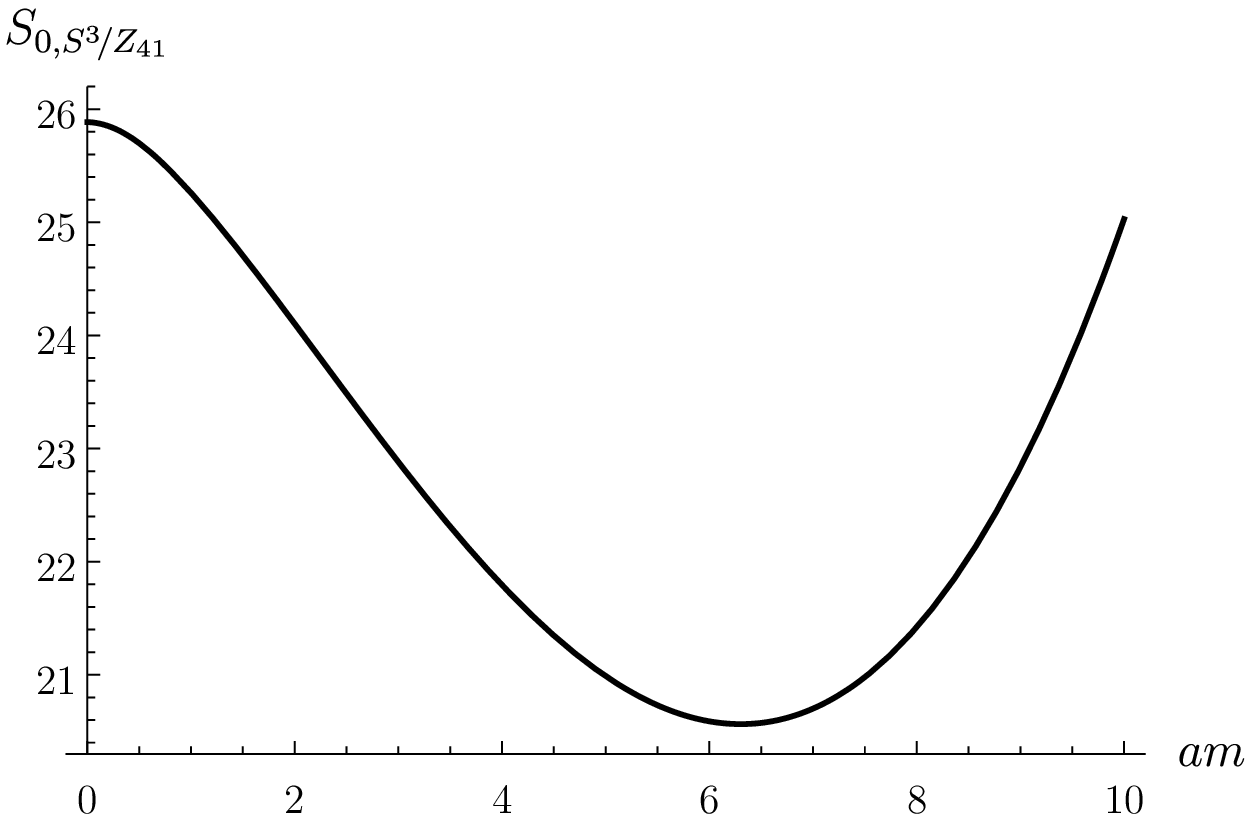}
	\end{minipage}
	\caption{\small Parte constante en el límite de altas temperaturas de la entropía en función de la masa adimensionalizada para distintos espacios lente.}
	\label{fig:entropymass}
\end{figure}

Con el objetivo de estudiar la diferencia en la dependencia con la masa en los distintos espacios, exploramos la forma del término constante en el desarrollo de la entropía a altas temperaturas en función de la masa para algunos espacios lente. En la figura \ref{fig:entropymass} pueden verse tales gráficos, de los que se observa que a medida que el orden del espacio lente crece la forma de la curva se parece menos a la de la esfera. En todos los casos, el término constante en la entropía diverge en el límite $am\rightarrow\infty$. 

En el capítulo siguiente analizaremos en más detalle la dependencia del término $S_{0,\esp{H}}$ con la masa. En particular, veremos que es posible definir a partir de él una cantidad que comparte algunas propiedades con las cantidades involucradas en los teoremas $C$.

%
\chapter{La entropía de holonomía}
\label{sec:holonomia}

\begin{minipage}{.885\textwidth}%
	\begin{flushright}
		\begin{minipage}{.68\textwidth}%
			\begin{flushleft}
				\emph{``Contrariwise,'' continued Tweedledee, ``if it was so, it might be; and if it were so, it would be: but as it isn't, it ain't. That's logic.''}
			\end{flushleft}
			\begin{flushright}
				--- \textrm{L. Carroll},
				\emph{Through the looking-glass,}\\ 
				\emph{and what Alice found there}
			\end{flushright}
		\end{minipage}%
	\end{flushright}
\end{minipage}%

\bigskip

En este capítulo describiremos la propuesta de una cantidad $C$ en tres dimensiones que se construye a partir de la entropía sobre espacios esféricos que calculamos anteriormente. Comenzamos con un resumen del estado del arte en lo referido a los teoremas $C$ en Teoría Cuántica de Campos, que esperamos sirva también como motivación para nuestra contribución.

\section[Teoremas $C$ en Teoría Cuántica de Campos]{Teoremas C en Teoría Cuántica de Campos}
\label{sec:holonomia:c}

El grupo de renormalización describe cómo cambian con la escala las teorías cuánticas de campos. Típicamente, al mover la escala de la teoría entre altas (en el ultravioleta) y bajas energías (en el infrarrojo) la teoría se mueve entre dos teorías de campos invariantes de escala,\footnote{ \,Bajo las suposiciones de unitariedad e invarianza de Poincaré, se espera que dichas teorías sean conformes \cite{Nakayama:2013is}, lo cual está garantizado en dos dimensiones \cite{Polchinski:1987dy}.} siendo ambas puntos fijos del flujo. Este cambio de escalas puede ser pensado como un cambio en el número de grados de libertad de la teoría: en el esquema de Wilson del grupo de renormalización \cite{Wilson:1973jj}, el flujo de altas a bajas energías corresponde a la transformación de la teoría en una teoría efectiva que no contiene los grados de libertad de altas energías. En efecto, desde el punto de vista de la integral funcional \cite{Polchinski:1983gv}, la funcional generatriz puede pensarse como función de una escala de energías ---un \emph{cutoff}---, de modo que la integración funcional se hace sobre las componentes de Fourier de los campos con impulso menor que este valor. Al haber menos configuraciones posibles, el resultado de este proceso puede interpretarse como una teoría efectiva con menos grados de libertad que la original.

De manera quizás más intuitiva, el proceso puede entenderse en términos de una escala de longitudes: dada una teoría con una longitud característica, digamos la distancia entre átomos en una red periódica, podemos hacer una descripción del sistema en la que consideramos el promedio sobre grupos de átomos cercanos, sin incluir la estructura interna de esos grupos; esta descripción corresponderá a una teoría con una longitud característica mayor, y repitiendo este proceso de \textit{coarse-graining} estaremos recorriendo el flujo hacia el infrarrojo del grupo de renormalización. 

En ambos casos, el proceso involucra a teorías intermedias cuyas constantes de acoplamiento dependen de la escala. Si pensamos en un espacio de hamiltonianos posibles ---en el que las coordenadas son las constantes de acoplamiento---, la variación de la escala se corresponde con un flujo en ese espacio. En los extremos de las líneas de flujo tendremos teorías invariantes de escala: los ya mencionados puntos fijos del flujo.

Un problema de larga data en Teoría Cuántica de Campos es el de definir una medida del número de grados de libertad de las teorías que sea consistente con este esquema. En particular, se espera que la cantidad que mida los grados de libertad decrezca con el flujo entre el ultravioleta y el infrarrojo del grupo de renormalización y sea estacionaria\footnote{ \,En el sentido de que no cambie cuando la teoría es perturbada por un operador relevante.} en los puntos fijos del flujo. Si tal cantidad existe, el flujo del grupo de renormalización puede pensarse como un proceso irreversible entre ambos puntos fijos. En el año 1986, Alexander Zamolodchikov \cite{Zamolodchikov:1986gt} obtuvo una función con estas propiedades para cualquier teoría relativista ---esto es, invariante de Poincaré--- y unitaria\footnote{ \,Estrictamente hablando, la prueba de Zamolodchikov utiliza la propiedad de positividad de la teoría, para lo cual es suficiente pero no necesaria su unitariedad.} en dos dimensiones, a partir de las funciones de dos puntos del tensor de energía-impulso $T_{\mu\nu}$. En particular, esa función, a la que llamó $c$, es mayor en el punto fijo ultravioleta que en el infrarrojo: $c_{UV}-c_{IR}>0$. En estos puntos, esta función puede relacionarse con propiedades de la teoría conforme correspondiente: por un lado, coincide con la carga central de dicha teoría, y por otro lado resulta proporcional al valor de la anomalía de traza (esto es, el valor de expectación de la traza del tensor de energía-impulso, que en espacios curvos no se anula, aunque la teoría clásica sea conforme),
\begin{align}
\langle T_\mu^{\,\mu}\rangle = -\frac{cR}{12}\,,
\end{align}
donde $R$ es la curvatura escalar de la variedad sobre la que está definida la teoría.

Este teorema tiene importantes aplicaciones en Teoría Cuántica de Campos, dado que al imponer restricciones bastante generales sobre los flujos del grupo de renormalización ---los flujos sólo pueden conectar teorías conformes si la teoría en el ultravioleta tiene carga central mayor que la teoría en el infrarrojo--- proporciona un mapa del espacio de teorías posibles en dos dimensiones, en el que por ejemplo no puede haber curvas cerradas.
Esto aporta información de la relación entre diferentes teorías conformes, permitiendo ordenar las teorías según clases de universalidad \cite{Cappelli:1990yc}, y puede ser usado fuera de los puntos fijos para obtener información acerca de la dependencia de escala del modelo como en \cite{kastor1989rg}.

Como la prueba del teorema requiere propiedades muy generales de la teoría como la existencia de un tensor de energía-impulso conservado, la anomalía de traza, la invarianza de Poincaré y la unitariedad, de inmediato se buscó una generalización a teorías en más dimensiones. La extensión no necesariamente sería directa, puesto que por ejemplo en el caso bidimensional el grupo conforme tiene dimensión infinita, lo que no ocurre en dimensiones más altas. Dos años después del trabajo fundacional, John Cardy \cite{Cardy:1988cwa} propuso una versión más débil del teorema para teorías en cuatro dimensiones; en ese caso, hay dos invariantes de Weyl locales que contribuyen a la anomalía de traza\footnote{ Un término proporcional a $\square R$ es también posible, pero puede ser eliminado por medio de la introducción de un contratérmino local en la acción \cite{Deser:1993yx}.} ---el tensor de Weyl $W_{\mu\nu\rho\sigma}$ y la densidad de Euler $E_4$\footnote{ En las convenciones de \eqref{eq:anomalia4d} resulta \begin{align}\nonumber
	W^2=\frac{1}{16\pi^2}\left(R^{\mu\nu\rho\sigma}R_{\mu\nu\rho\sigma}-2R^{\mu\nu}R_{\mu\nu}+\frac13 R^2\right)\,,\\ \nonumber E_4=\frac{1}{16\pi^2}\left(R^{\mu\nu\rho\sigma}R_{\mu\nu\rho\sigma}-4R^{\mu\nu}R_{\mu\nu}+ R^2\right)\,.
	\end{align}}---:
\begin{align}\label{eq:anomalia4d}
\langle T_\mu^{\,\mu}\rangle =  c\, W_{\mu\nu\rho\sigma}W^{\mu\nu\rho\sigma} - a E_4 \,.
\end{align}
Observando que en dos dimensiones el coeficiente $c$ puede obtenerse también por medio de la integral en todo el espacio de $\langle T_\mu^{\,\mu}\rangle$, y que un cálculo similar en cuatro dimensiones da el coeficiente $a$ de la anomalía conforme, Cardy estudió el comportamiento de dicho coeficiente bajo el flujo del grupo de renormalización, encontrando que al orden más bajo en teoría de perturbaciones (para algunas trayectorias del grupo de renormalización) se verifica la desigualdad $a_{UV}-a_{IR}>0$. Este ``teorema $a$'' fue testeado en numerosos ejemplos \cite{Anselmi:1997am,Anselmi:1997ys,Osborn:1989td}, y
fue probado décadas más tarde \cite{Komargodski:2011xv}, encontrándose además una función interpolante entre los valores $a_{UV}$ y $a_{IR}$ (véase también \cite{Komargodski:2011vj,Luty:2012ww}).
En cuanto al coeficiente $c$ en \eqref{eq:anomalia4d}, rápidamente se encontraron teorías en las cuales no tiene el comportamiento esperado \cite{Anselmi:1997ys,Cappelli:1990yc}. 

En general, en dimensión $d$ par la anomalía de traza tiene la forma \cite{Deser:1993yx}
\begin{align}\label{eq:anomaliaevend}
\langle T_\mu^{\,\mu}\rangle =  \sum_i c_i I_i - (-1)^{d/2} a E_d \,,
\end{align}
donde $E_d$ es la densidad de Euler y los $I_i$ son invariantes de Weyl. La prueba del teorema $a$ puede extenderse en principio al coeficiente $a$ en dimensión par arbitraria \cite{Elvang:2012yc,Elvang:2012st}.

En dimensión impar la simetría conforme no es anómala, por lo que la traza del tensor de energía-impulso se anula en los puntos fijos del flujo del grupo de renormalización. Esto impide la extensión obvia de la propuesta de Cardy a dimensiones impares. Sin embargo, volviendo a dimensión par, el coeficiente $a$ puede obtenerse también como la parte logarítmica (que es universal, en el sentido de que no depende de la regularización) de la energía libre $F=-\log Z_{S^d}$ de la teoría conforme sobre la esfera de radio $r$ como \cite{Cardy:1988cwa}
\begin{align}
a= \frac{(-1)^{d/2}}{d}\frac{dF}{d\log r}\,.
\end{align}
Además, este término coincide con el término universal en la entropía de entrelazamiento del estado fundamental de la teoría entre la esfera $S^{d-2}$ en $d$ dimensiones y su complemento \cite{Casini:2011kv}, lo que sugiere la posibilidad de considerar la misma cantidad en dimensión impar ---en ese caso, no habiendo término logarítmico, el término equivalente en la energía libre es el constante---. 

\vfill\pagebreak

En \cite{Freedman:1999gp,Girardello:1998pd} se estudiaron flujos del grupo de renormalización en el contexto de la correspondencia AdS/CFT, encontrando para ciertas\footnote{ \,La monotonicidad de las cantidades $C$ en este caso se prueba suponiendo que el sector de materia satisface la \textit{null energy condition} (condición de energía para vectores tipo luz).} teorías duales a la gravedad de Einstein acoplada a campos de materia en dimensión $d$ arbitraria cantidades que satisfacen la versión débil del teorema $C$ ---esto es, que son mayores en el punto fijo ultravioleta que en el infrarrojo--- y que en dichos puntos fijos coinciden con el ya mencionado coeficiente $a$.  No obstante, en este tipo de teorías las cargas centrales son muy particulares: en el caso $d=4$, por ejemplo, en todas ellas el coeficiente $a$ de la anomalía conforme coincide con el $c$. Buscando una formulación que distinguiera el coeficiente $a$ en toda dimensión, en  \cite{Myers:2010tj,Myers:2010xs} se estudiaron teorías con duales holográficos más generales,\footnote{ \,Más específicamente, las teorías en cuestión tienen dual gravitatorio ``cuasi-topológico'' ---en el que la acción de gravedad incluye, además de la parte de Einstein, términos que involucran el cuadrado y el cubo de la curvatura escalar--- \cite{Myers:2010ru,Oliva:2010eb}. Además, los parámetros de la acción gravitatoria se eligen de modo que las correspondientes teorías conformes no involucren operadores no unitarios, y se supone como en los trabajos previos que la acción de materia satisface la \textit{null energy condition}.} encontrando cantidades $a^*_d$ con la propiedad deseada. Para teorías en dimensión par, estas cantidades coinciden con el correspondiente coeficiente $a$, lo que sugirió una extensión de los teoremas $C$ ya conocidos al caso impar: la cantidad relevante en el caso conforme que puede ser calculada en cualquier dimensión resulta ser el término universal en la entropía de entrelazamiento entre dos mitades de la esfera $S^{d-1}$ ---separadas por un ``círculo máximo'' $S^{d-2}$--- en la teoría sobre $\mathbb{R}\times S^{d-1}$ \cite{Nishioka:2009un,Ryu:2006ef}. En este caso, el flujo del grupo de renormalización se implementa usando como escala el radio de la esfera.

Además de la observación de que el coeficiente $a$ de la anomalía conforme puede obtenerse de la función de partición euclídea sobre la esfera, hay otra propiedad que sugiere a la función de partición en esferas como la cantidad a considerar en el caso tridimensional. Antes de que se probara el teorema $a$, dentro de la evidencia a favor del mismo se encontró el llamado principio de maximización de $a$ \cite{Intriligator:2003jj}: para teorías supersimétricas, $a$ está determinado por las correspondientes cargas $U(1)_R$, y se encontró evidencia de que esa simetría maximiza localmente el coeficiente $a$. En \cite{Jafferis:2010un} se propuso un análogo tridimensional al principio de maximización de $a$, que se estudió para teorías superconformes $\mathcal{N}=2$, en el que la cantidad relevante es justamente $F$. Con estas propiedades, un grupo de investigadores propuso  \cite{Jafferis:2011zi} en tres dimensiones el llamado ``teorema $F$'', en el que la cantidad que es más grande en el ultravioleta que en el infrarrojo es el término constante $F$ en la acción efectiva de la correspondiente teoría conforme sobre la esfera. Esta conjetura corresponde, otra vez, a una formulación ``débil'' del teorema $C$, dado que no involucra ninguna propiedad de las teorías intermedias del flujo entre las dos teorías conformes en cuestión.

En \cite{Casini:2004bw}, Horacio Casini y Marina Huerta dieron una demostración alternativa del teorema $c$ en dos dimensiones en el cual la cantidad relevante es justamente la parte universal de la entropía de entrelazamiento en círculos; la prueba es válida para cualquier teoría relativista unitaria y utiliza únicamente las propiedades de la entropía de entrelazamiento. En los puntos fijos del flujo del grupo de renormalización dicha cantidad es proporcional a la carga central de la correspondiente teoría conforme.\footnote{ \,Si bien el valor en los puntos conformes de las cantidades $c$ de Zamolodchikov y Casini-Huerta es el mismo, las funciones interpolantes son diferentes \cite{Casini:2006es}.} Con algunos cuidados y detalles técnicos adicionales, la prueba puede extenderse a tres dimensiones \cite{Casini:2012ei}, lo que se completa con la observación de que este término en la entropía de entrelazamiento de la teoría tridimensional coincide, en el caso de teorías conformes, con el término constante en la acción efectiva de la teoría sobre la esfera \cite{Liu:2012eea}, en la prueba del teorema $F$. Hay que hacer aquí alguna salvedad, puesto que en el flujo de masa de campos libres $F$ diverge en el punto fijo infrarrojo, y la sustracción de la divergencia afecta el cambio de $F$ en el ultravioleta; estudiaremos este problema con algo más de detalle en el apartado siguiente. Desde un punto de vista filosófico, el hecho de que los teoremas $C$ involucren a la entropía de entrelazamiento es en sí mismo interesante, puesto que permite un paralelo con la teoría clásica ---en la que la entropía de Shannon mide la falta de información que se tiene de un sistema dado---, y porque la idea de pérdida de información a lo largo del flujo del grupo de renormalización se asemeja a la operación de traza sobre configuraciones de los campos mediante la que se define la entropía de von Neumann.

Llegados a este punto podríamos pensar que el problema en tres dimensiones está agotado. Sin embargo, quedan cuestiones por entender. En particular, notemos que si bien da una función interpolante entre las teorías conformes en los extremos del flujo, esta prueba del teorema $F$ no involucra ningún cálculo explícito en las teorías de campos intermedias; esto es, no tenemos una cantidad, como la que había en la propuesta de Zamolodchikov en dos dimensiones, de la que podamos conocer el valor a lo largo de todo el flujo. Por otra parte, para el caso de teorías conformes, la teoría en el espacio plano puede mapearse a la esfera para calcular allí su valor de $F$, pero en las teorías intermedias no hay una prescripción definida para el mapeo a la esfera. Los autores de \cite{Klebanov:2011gs} estudiaron la posibilidad de definir la función interpolante entre los valores conformes como la energía libre de la teoría sobre la esfera para flujos de masa de campos escalares y de Dirac libres, encontrando que la función diverge en el límite de masa infinita, por lo que es necesario considerar alguna modificación de la misma. La posibilidad que sugieren es la sustracción manual de los términos divergentes, que son funciones no constantes de la masa. En el artículo \cite{Beneventano:2017eyu} analizamos cómo esa propuesta afecta la estabilidad en los puntos fijos, y qué otras funciones interpolantes pueden considerarse. No expondremos aquí en detalle los cálculos en dicho artículo, pero hacemos a continuación un resumen de los resultados.

\subsection[Una cantidad $C$ para teorías en la esfera tridimensional]{Una cantidad C para teorías en la esfera tridimensional}

Como resumen de los requisitos para una cantidad $C$, que nos permitirá también introducir algo de notación, supongamos, en la representación de Wilson del grupo de renormalización, que hemos integrado funcionalmente  las configuraciones del campo con impulso mayor a $\mu$ y tenemos entonces una teoría efectiva a esa escala, y llamemos $g^i(\mu)$ a las constantes de acoplamiento de dicha teoría. Moviendo la escala $\mu$ realizamos el flujo del grupo de renormalización como un flujo en el espacio de parámetros $g^i$. Si parametrizamos el flujo con la cantidad $t=-$log$\,\mu$, que crece hacia el infrarrojo, vemos que éste está generado por el vector
\begin{equation}
\frac{d g^i}{dt}=-\mu\frac{d g^i(\mu)}{d\mu}=:-\beta^i(g(\mu))\,,\nonumber
\end{equation}
donde $\beta$ depende de $\mu$ sólo a través de las constantes de acoplamiento. En otras palabras, el flujo del grupo de renormalización es un movimiento de un solo parámetro en el espacio de las constantes de acoplamiento renormalizadas de la teoría, en el que las velocidades son las funciones beta:
\begin{align}
\frac{\partial}{\partial t} = -\beta^i(g)\frac{\partial}{\partial g_i}\,.
\end{align}
	
Además, estos flujos tienen puntos fijos, en los cuales la función beta se anula. Allí, las constantes de acoplamiento no dependen de la escala; dicho de otro modo, en esos puntos la teoría es invariante de escala, y bajo ciertas condiciones corresponde a una teoría conforme.
	
Recapitulando las formulaciones de teoremas $C$ en distintas dimensiones, observamos diferentes niveles de profundidad en las propiedades exigibles a las cantidades involucradas \cite{Barnes:2004jj}:
	
\begin{enumerate}
	\item Como condición más débil, podemos pensar en una cantidad que cuente grados de libertad en teorías conformes conectadas por una trayectoria del grupo de renormalización; en ese caso, buscaremos una cantidad $C\geq 0$ con $C_{IR}< C_{UV}$. \label{item:reqc1}
	\item Por otra parte, podemos buscar una función que interpole monótonamente entre sus valores en los puntos conformes, que dé una medida de los grados de libertad de todas las teorías intermedias: si $g(t)$ denota colectivamente a las constantes de acoplamiento de la teoría como funciones de una escala $t$ que controla el flujo hacia el infrarrojo del grupo de renormalización, querremos entonces una función $C$ tal que
	\begin{align}\label{eq:reqcr2}
	\dot{C}(g(t)) = -\beta^{i}(g)\frac{\partial}{\partial {g}_i}C(g(t))\leq 0\,,
	\end{align} 
	con $\dot{C}= 0$ si y sólo si la teoría es conforme. \label{item:reqcr2}
	\item Finalmente, podríamos imponer la condición más restrictiva
	\begin{align}
	\dot{C}(g(t)) = -G_{ij}(g)\beta^{i}(g)\beta^{j}(g)\,,
	\end{align}
	con $G_{ij}$ una métrica ---definida positiva--- en el espacio de constantes de acoplamiento, y $\dot{C}= 0$ si y sólo si la teoría es conforme. \label{item:reqc3}
	\end{enumerate}
	
Siguiendo la idea al final de \cite{Klebanov:2011gs}, en \cite{Beneventano:2017eyu} estudiamos el flujo de masa de campos escalares y de Dirac libres sobre la esfera tridimensional, con el objetivo de analizar el comportamiento bajo dicho flujo de algunos posibles candidatos a cantidades $C$ que conecten los valores de $F$ en las teorías conformes correspondientes a los puntos fijos ultravioleta (el campo sin masa) e infrarrojo (en el límite de masa infinita). 

Comenzamos estudiando la acción efectiva de la correspondiente teoría sobre la esfera. Encontramos, en coincidencia con \cite{Dowker:2014xca,Klebanov:2011gs}, que para el campo de Dirac dicha acción efectiva satisface la primera parte de la condición \ref{item:reqcr2}, pero su derivada con respecto a la constante de acoplamiento no se anula en el punto fijo ultravioleta. Para el campo escalar, en cambio, posee una divergencia en el infrarrojo que al ser renormalizada o bien conduce a una contribución imaginaria en el punto fijo ultravioleta o bien resulta en una función que no es monótona, que toma valores negativos y cuya derivada con respecto a la constante de acoplamiento en el punto fijo ultravioleta es infinita.

Como ya hemos mencionado, otra cantidad que coincide con $F$ en el punto fijo ultravioleta es la parte finita ---cambiada de signo--- de la entropía de entrelazamiento entre un círculo y su complemento, que puede ser obtenida como límite de las entropías de Rényi $S_q$ \cite{Casini:2010kt,Hung:2011nu}:
\begin{align}
S_{\mathrm{ent}} = \lim_{q\rightarrow1} S_q := \lim_{q\rightarrow1}\frac{q\Gamma_{S^3}-\Gamma_{qS^3}}{1-q}\,,
\end{align}
donde $\Gamma_{qS^3}$ es la acción efectiva sobre el cubrimiento de orden $q$ de la esfera.\footnote{ \,Esta acción efectiva se calcula para valores enteros de $q$ y luego se extiende analíticamente a la recta real \cite{Calabrese:2004eu,Holzhey:1994we}.} Analizando esta cantidad para teorías libres masivas encontramos un comportamiento similar al de la acción efectiva. Consideramos entonces una variante de esta cantidad: la entropía de entrelazamiento renormalizada \cite{Casini:2012ei,Liu:2012eea}
\begin{align}
S_{\mathrm{ren}} = am \frac{dS_{\mathrm{ent}}}{d(am)} - S_{\mathrm{ent}}\,,
\end{align}
y vimos que en el caso de Dirac satisface la condición \ref{item:reqcr2}, pero que en el caso escalar no es una función monótona de la masa, y toma además valores negativos, como se vio también en \cite{Ben-Ami:2015zsa}.

Finalmente, consideramos la modificación de la acción efectiva
\begin{align}
\tilde{F} = \Gamma -\frac{n}{3}g\frac{d\Gamma}{dg}\,,
\end{align}
donde $n$ es el orden del operador de fluctuaciones cuánticas de la teoría y $g$ la constante de acoplamiento relevante, y encontramos que en ambos casos es una función positiva y monótonamente decreciente a lo largo del flujo, pero su derivada con respecto a la constante de acoplamiento en el punto fijo ultravioleta no se anula. Invitamos al lector interesado en explorar los detalles de este problema a consultar el artículo \cite{Beneventano:2017eyu}.

\vfill\pagebreak

\subsection{La entropía topológica}
\label{sec:holonomia:c:topologica}

En el capítulo \ref{sec:termo}, cuando estudiamos las propiedades termodinámicas de una teoría escalar sin masa acoplada conformemente a la métrica de distintos espacios esféricos, encontramos que es posible definir a partir de la entropía de dicha teoría una cantidad que llamamos \textit{entropía topológica}, que es positiva en todos los casos, y que no depende de la temperatura ni del volumen del espacio. Dicha cantidad puede obtenerse directamente de las acciones efectivas de la teoría sobre la parte espacial del espacio esférico correspondiente y sobre la esfera. Ahora bien; la teoría sin masa es el punto fijo ultravioleta en el flujo de masa de la teoría masiva, cuyo punto fijo infrarrojo corresponde al límite de masa infinita, donde la acción efectiva ---y por lo tanto también la entropía topológica--- se anula en todos los espacios. De este modo, los resultados que obtuvimos en los capítulos anteriores implican que para dicho flujo de masa del campo escalar sobre los distintos espacios esféricos, la entropía topológica satisface la condición \ref{item:reqc1} ---que es, como ya hemos dicho, la condición más débil que debe exigirse a una cantidad para que pueda considerarse en algún sentido como una medida de los grados de libertad de la teoría---.

Inspirándonos en esta observación, analizaremos a continuación las propiedades de dicha cantidad a lo largo de todo el flujo de masa. Para eso usaremos los cálculos del capítulo \ref{sec:masivo} para el caso masivo. En las teorías intermedias la cantidad correspondiente dependerá no sólo de la topología del espacio sino también de la masa adimensionalizada $am$, y entonces cambiaremos su nombre por el de \emph{entropía de holonomía}:
\begin{align}
S_{\mathrm{hol},\esp{H}} := \lim_{\beta\rightarrow 0} \left[ S_{\esp{H}}\left(am,\beta/a\right) - \frac{1}{|H|} S_{S^3}\left(am,\beta/a\right) \right]\,.
\label{eq:hol}
\end{align}

Como ya hemos mencionado, esta cantidad puede obtenerse a partir de las acciones efectivas de la teoría en tres dimensiones sobre la esfera y el espacio esférico correspondiente como
\begin{align}
S_{\mathrm{hol},\esp{H}} = \frac{1}{|H|} \Gamma_{S^3} - \Gamma_{\esp{H}}\,.
\label{eq:holseff}
\end{align}

\vfill\pagebreak

\section{El campo escalar masivo}

Calculamos entonces la entropía de holonomía \eqref{eq:hol} a partir de las expresiones para la entropía del campo escalar masivo sobre los distintos espacios esféricos obtenidas en el capítulo \ref{sec:masivo} en el régimen en el que la temperatura es la escala más grande. Los cálculos son directos: escribimos los resultados a continuación.

Para los espacios lente de orden impar tenemos
\begin{align}\label{eq:sholodd}
S_{\mathrm{hol},\esp{Z_{2q+1}}} =-\frac{1}{\pi}\sum_{r=1}^{q}r\sum_{n=1}^{\infty}\frac{1}{n^2}\left(\frac{2\pi amn}{2q+1}+1\right)\sen\!\left(\frac{4\pi rn}{2q+1}\right)e^{-\frac{2\pi amn}{2q+1}}\,,
\end{align}
mientras que para los espacios lente de orden par resulta
\begin{align}\label{eq:sholeven}
S_{\mathrm{hol},\esp{Z_{2q}}} =& - \frac{1}{8\pi^2 q}\sum_{k=1}^{\infty} \frac{1}{k^3} \left[ 1 + 2\pi am k + \frac12 (2\pi amk)^2 \right] e^{-2\pi am k} \\ \nonumber
&+ \frac{1}{\pi^2 q}\sum_{k=1}^{\infty} \frac{1}{k^3} \left[ 1 + \pi ma k + \frac12 (\pi amk)^2 \right] e^{-\pi am k} \\ \nonumber
&- \frac{1}{\pi} \sum_{r=1}^{q-1} r \sum_{n=1}^{\infty} \frac{1}{n^2} \left(\frac{2\pi amn}{2q}+1\right) \sen\!\left(2\pi\frac{2r+1}{2q}n\right)e^{-\frac{2\pi amn}{2q}}\,.
\end{align}

En el caso de los espacios prisma, tenemos
\begin{align}\label{eq:sholprism}
S_{\mathrm{hol},\esp{D_p^*}} =& {} \frac12 S_{\mathrm{hol},\esp{Z_{2q}}} +\frac{1}{\pi}\sum_{\substack{k=1\\ k \mathrm{\,impar}}}^{\infty} \frac{(-1)^{\lfloor k/2\rfloor}}{k^2} \left(1+\frac{\pi am k}{2}\right) e^{-\frac{\pi amk}{2}}\,.
\end{align}
Finalmente, en los espacios poliédricos la entropía de holonomía se obtiene como combinación de la misma cantidad en ciertos espacios lente de orden par, de acuerdo a las expresiones \eqref{eq:degpolycyclic}. 

\begin{figure}[h]
	\centering
	\begin{minipage}{.4\textwidth}
		\centering
		\includegraphics[height=38mm]{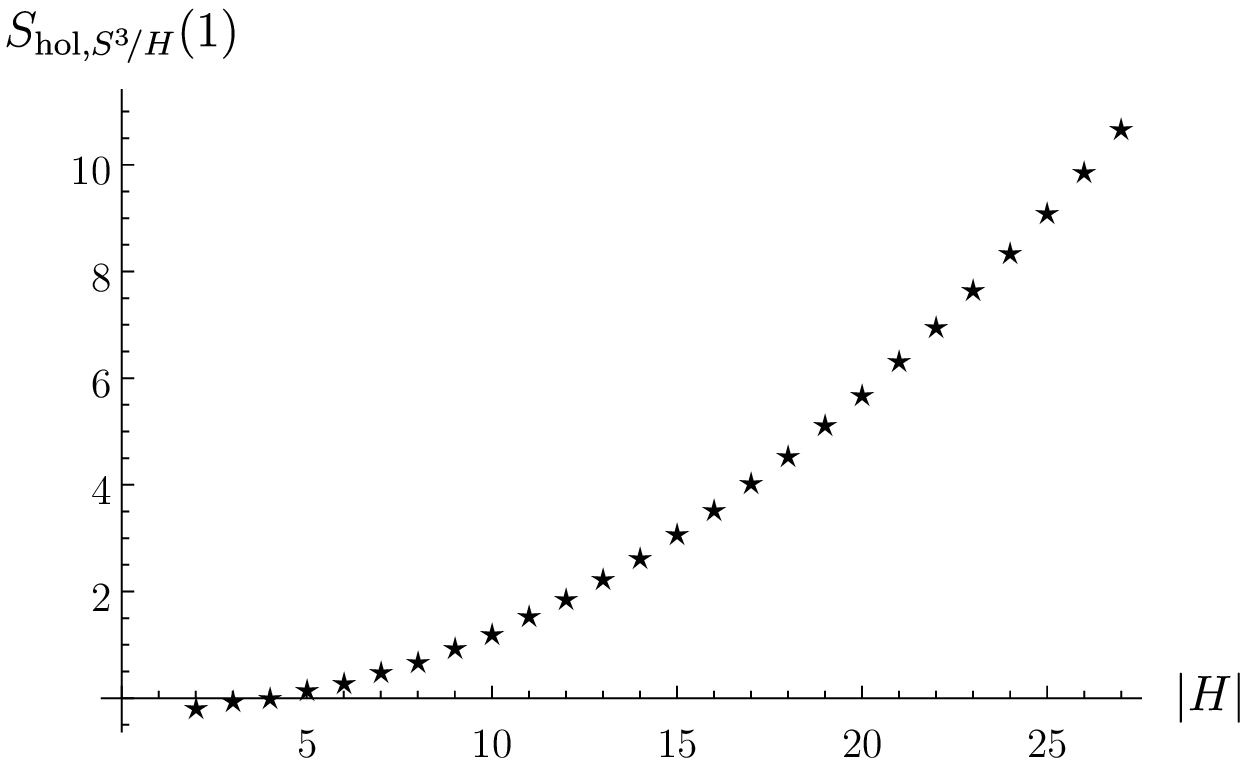}
	\end{minipage}
	\hspace{0.05\textwidth}
	\begin{minipage}{.4\textwidth}
		\centering
		\includegraphics[height=38mm]{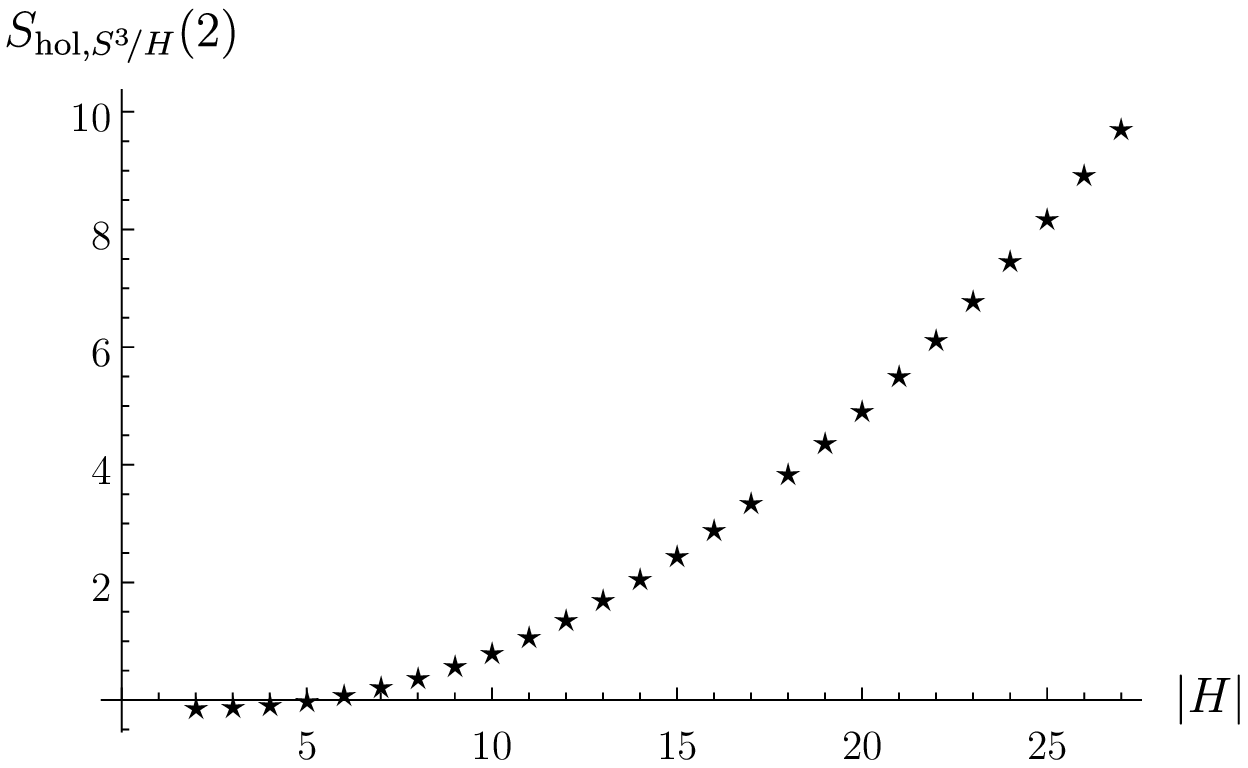}
	\end{minipage}\\
	\begin{minipage}{.4\textwidth}
		\centering
		\includegraphics[height=38mm]{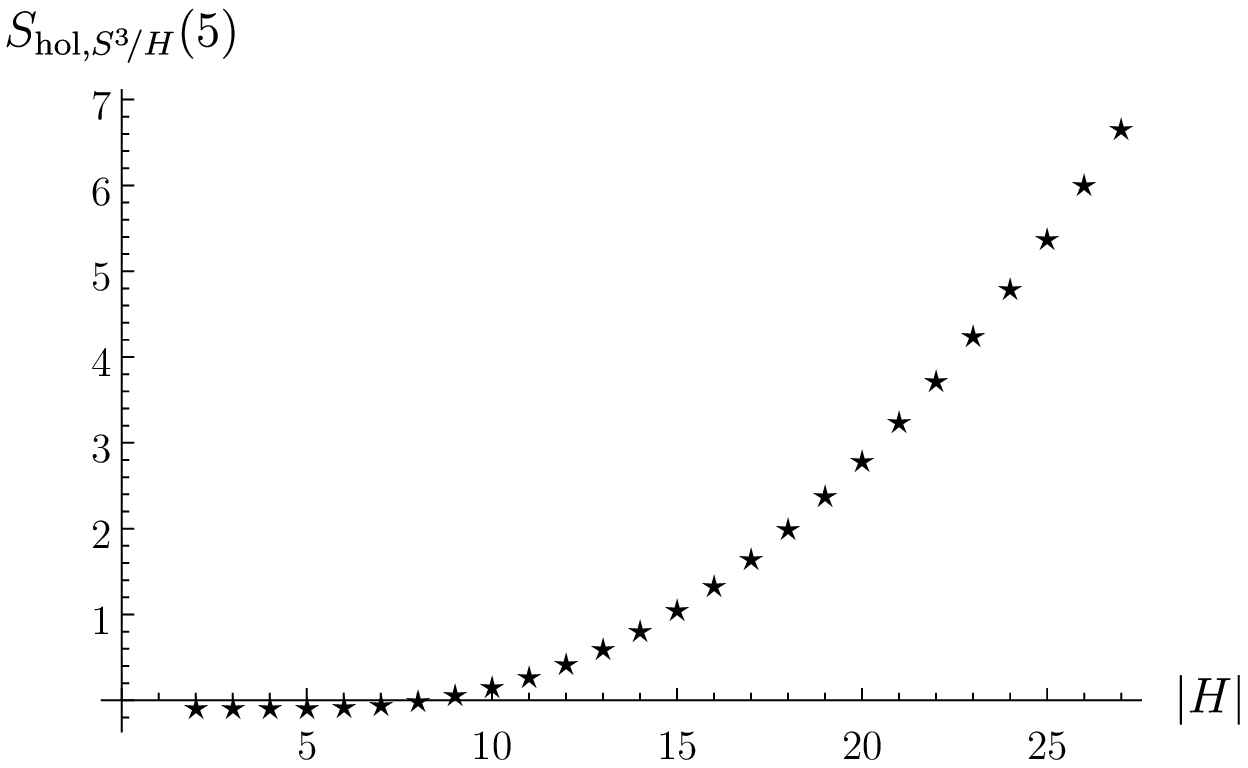}
	\end{minipage}
	\hspace{0.05\textwidth}
	\begin{minipage}{.4\textwidth}
		\centering
		\includegraphics[height=38mm]{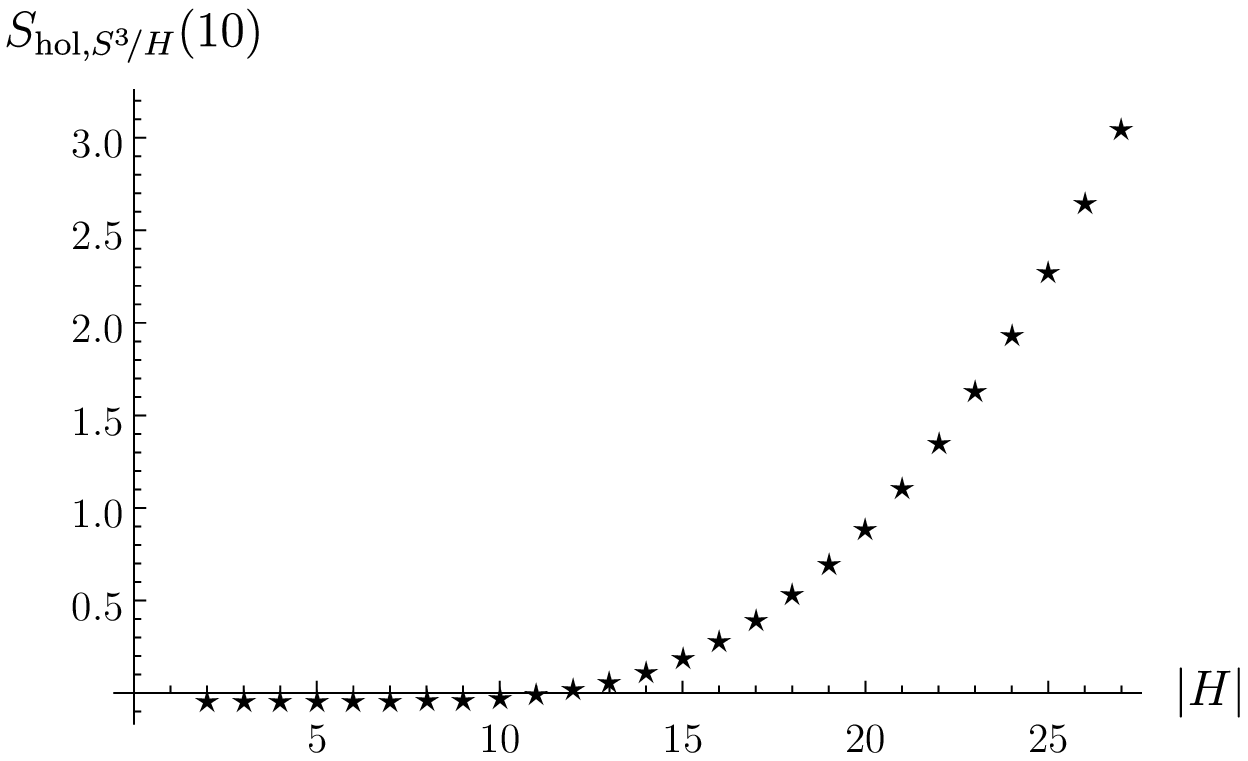}
	\end{minipage}
	\caption{\small Entropía de holonomía sobre espacios lente como función del orden del grupo para distintos valores de $ma$.}
	\label{fig:shollentemass}
\end{figure}

Notamos en primer lugar que, como ilustramos en la figura \ref{fig:shollentemass}, la dependencia del valor de la entropía de holonomía con el orden del grupo para el caso de los espacios lente es monótona independientemente del valor de la masa. En este sentido, esta cantidad es mejor que el término constante en la interpretación del orden del grupo como una especie de temperatura, como habíamos mencionado al final del capítulo \ref{sec:termo}.

\begin{figure}[h!]
	\centering
	\includegraphics[height=6cm]{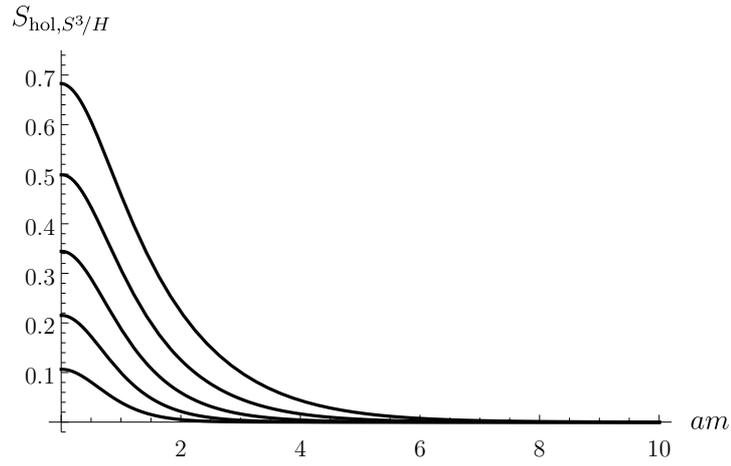}
	\caption{\small Entropía de holonomía como función de la masa para los primeros espacios lente. El corte con el eje vertical crece con el orden del grupo de isotropía de la variedad.}
	\label{fig:shollens}
\end{figure}

\begin{figure}[h!]
	\centering
	\begin{minipage}{.4\textwidth}
		\centering
		\includegraphics[height=40mm]{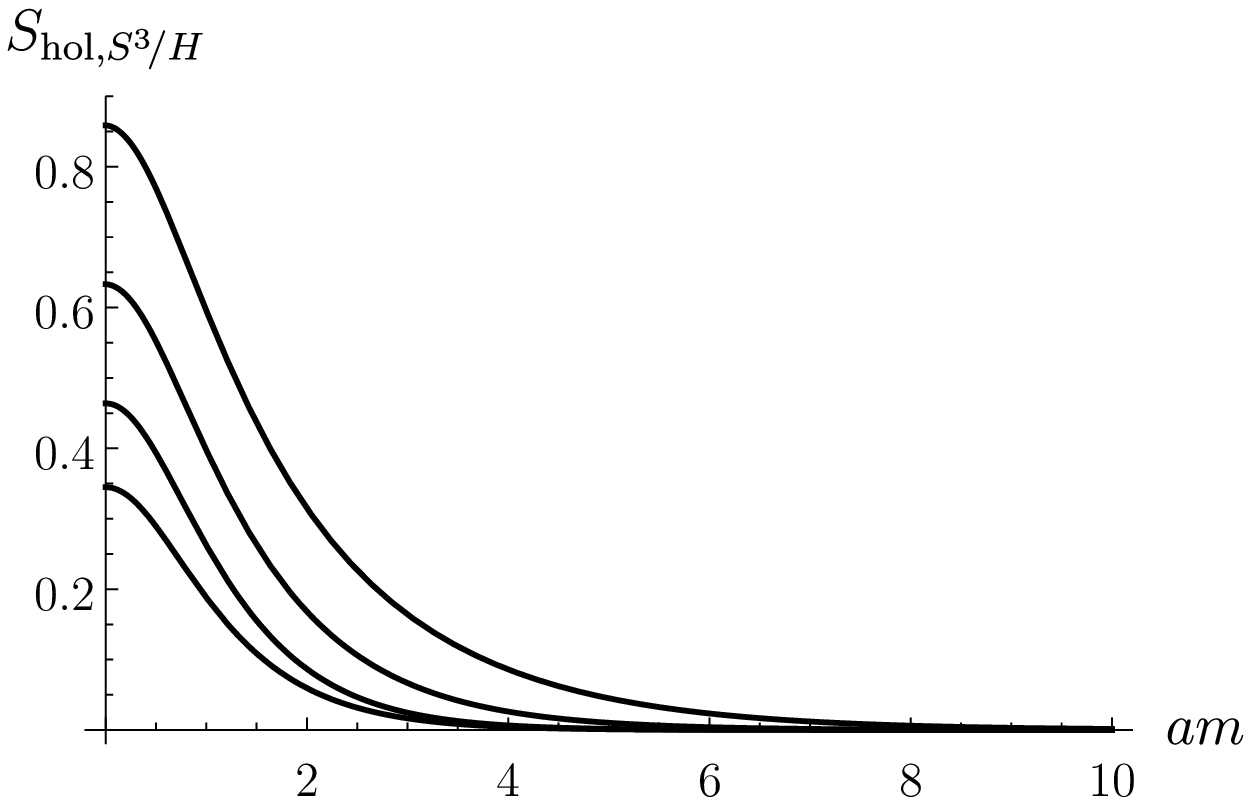}
	\end{minipage}
	\hspace{0.05\textwidth}
	\begin{minipage}{.4\textwidth}
		\centering
		\includegraphics[height=40mm]{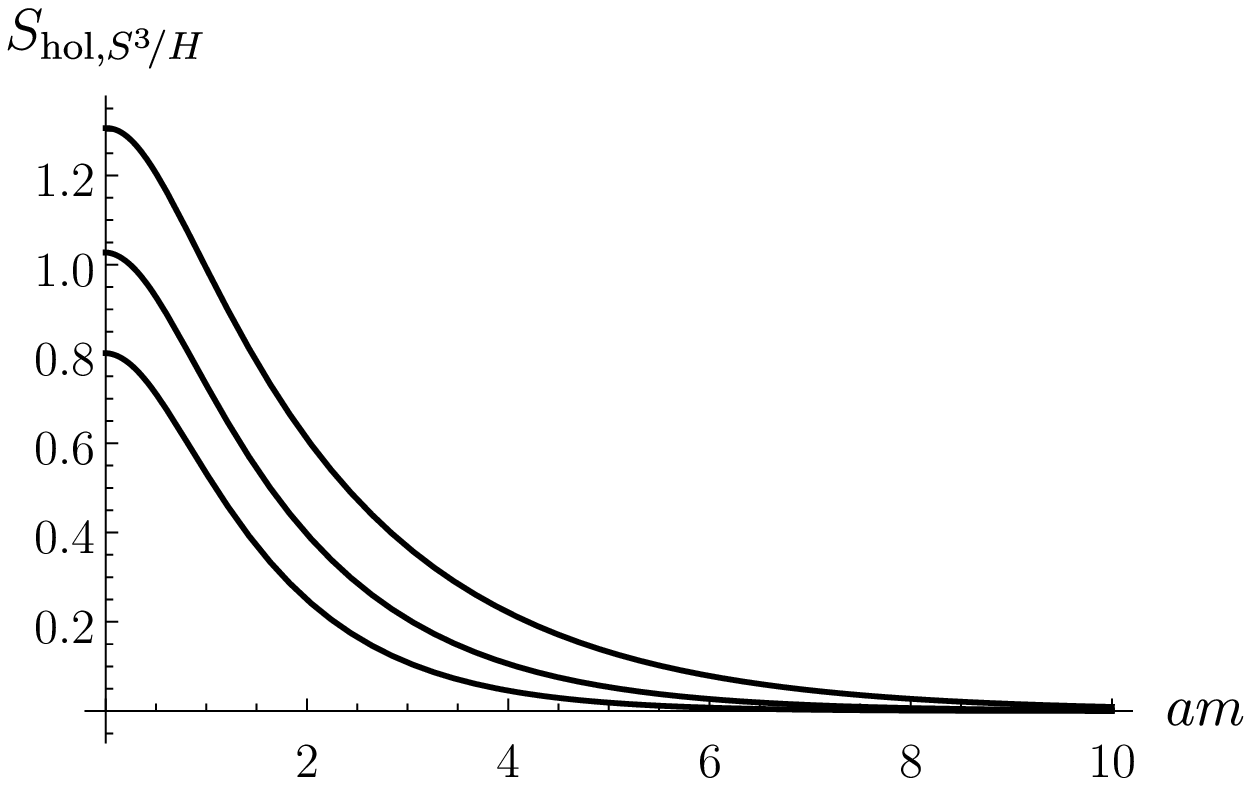}
	\end{minipage}
	\caption{\small Entropía de holonomía como función de la masa para los primeros espacios prisma (izquierda) y para los espacios poliédricos (derecha). En ambos casos, el corte con el eje vertical crece con el orden del grupo de isotropía de la variedad.}
	\label{fig:sholmass}
\end{figure}

\begin{figure}[h!]
	\centering
	\includegraphics[height=6cm]{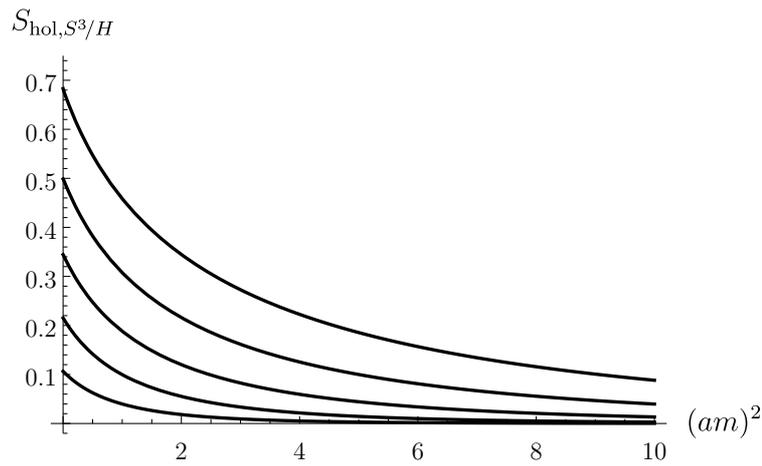}
	\caption{\small Entropía de holonomía como función del cuadrado de la masa para los primeros espacios lente. El corte con el eje vertical crece con el orden del grupo de isotropía de la variedad.}
	\label{fig:sholm2lens}
\end{figure}

Puede demostrarse que en todos los casos la entropía de holonomía es una función decreciente de la masa: no haremos aquí la demostración analítica, que no es complicada pero sí engorrosa, y que consideramos poco iluminadora; en cambio, mostramos este comportamiento con gráficos numéricos de las expresiones anteriores, en las figuras \ref{fig:shollens} y \ref{fig:sholmass}. De ellas podemos observar además que la entropía de holonomía se anula en todos los casos en el límite de masa infinita, lo cual está de acuerdo con el resultado que obtendríamos para dicho límite usando las extensiones analíticas a la región de parámetros en la que la masa es la escala más grande.
Con esto vemos que se verifica la condición \ref{item:reqcr2}, al menos parcialmente, puesto que la entropía de holonomía satisface la ecuación \eqref{eq:reqcr2}. Sin embargo, esta condición no se cumple totalmente: en efecto, un cálculo inmediato muestra que, si bien la derivada de la entropía de holonomía con respecto a la masa adimensionalizada se anula en $am=0$ para todos los espacios, su derivada con respecto a la constante de acoplamiento $(am)^2$ de la teoría es negativa. Ilustramos este comportamiento para los espacios lente en la figura \ref{fig:sholm2lens}. De manera algo rebuscada, podríamos definir a partir de la entropía de holonomía una nueva cantidad cuya derivada con respecto a la constante de acoplamiento se anule en ambos puntos fijos considerando por ejemplo la diferencia $2S_{\mathrm{hol}}(m') - S_{\mathrm{hol}}(m)$, donde $m'^2=m^2/2$.

Notamos que la entropía de holonomía es siempre positiva, lo cual es consecuencia de la subaditividad del término independiente de la temperatura en el desarrollo a altas temperaturas. Digamos aquí que si, en cambio, consideráramos por ejemplo la energía libre $F$ sobre estos espacios, no tendríamos esa propiedad \cite{Asorey:2014gsa}. Lo mismo sucede con la contribución topológica a la energía de vacío \cite{Asorey:2011cm,Dowker:2004nh}.

Habiendo mostrado que la entropía de holonomía posee un comportamiento compatible con el teorema $C$ para teorías escalares en tres dimensiones, tenemos dos caminos a seguir. Por un lado, podemos preguntarnos acerca de su validez como cantidad $C$ en tres dimensiones en un contexto más general; en particular, a falta de una prueba general, nos interesa examinar la propuesta en otras teorías. Por otro lado, podemos estudiar el mismo flujo de masa de la teoría escalar en dimensiones más altas. En ese caso podremos considerar el espacio proyectivo $S^d\!/Z_2$, que de todos los espacios esféricos es el único que puede ser definido en todas las dimensiones. En lo que sigue abordaremos las dos posibilidades.

\section{Generalización a dimensiones mayores}

Analizamos entonces la teoría escalar libre con masa sobre esferas en dimensión arbitraria. En primer lugar haremos un comentario acerca de la dimensión, que completaremos de alguna manera luego de calcular la entropía de holonomía para el campo escalar masivo acoplado conformemente a la métrica de la esfera tridimensional.

\subsection{Un comentario acerca de la dimensión}

Llegados a este punto, en el que estamos proponiendo a la entropía de holonomía como una cantidad $C$ en tres dimensiones, podemos objetar de la propuesta el hecho de que la teoría original que consideramos tuviera en realidad cuatro dimensiones ---siendo en nuestro caso la cuarta dimensión la que identificamos con la temperatura---. Para responder a la objeción, digamos que las cantidades que nos interesan se obtienen del límite de altas temperaturas de la función de partición, que resulta coincidir con la función de partición de la teoría en tres dimensiones. Sin embargo, aún luego de esa observación notamos que quedan rastros de la cuarta dimensión en el hecho de que el acoplamiento conforme que consideramos sea el correspondiente a la teoría en cuatro dimensiones. En lo que sigue calcularemos la entropía de holonomía para campos en dimensión arbitraria con el acoplamiento conforme correspondiente a esa dimensión. Consideramos por separado y en primer lugar el caso tridimensional.

\subsection[Acoplamiento conforme en tres dimensiones]{Campo escalar masivo con acoplamiento conforme en tres dimensiones}

\begin{figure}[h]
	\centering
	\includegraphics[height=6cm]{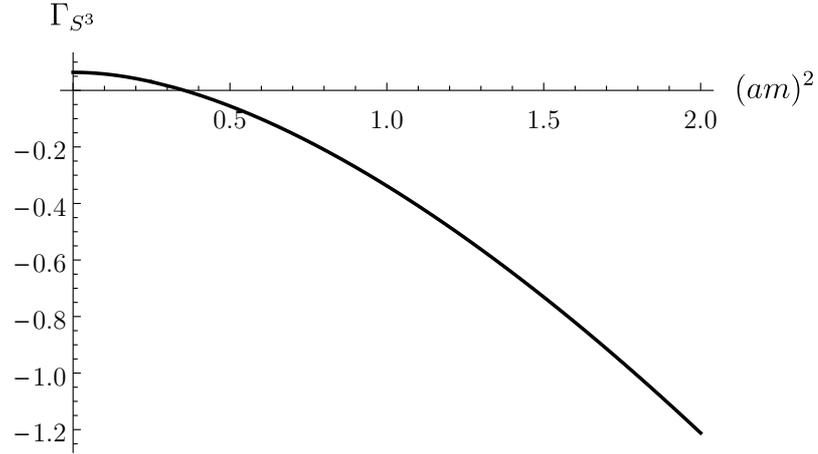}
	\caption{\small Acción efectiva para el campo conforme en tres dimensiones sobre la esfera como función del cuadrado de la masa adimensionalizada.}
	\label{fig:seff3mass}
\end{figure}

Consideremos un campo escalar con masa acoplado de manera conforme a la métrica de la esfera tridimensional; de la acción clásica leemos el operador de fluctuaciones cuánticas $\triangle + 3/4a^2+m^2$, cuyos autovalores difieren de los que ya habíamos considerado en una cantidad constante, manteniendo las mismas degeneraciones. La función zeta se escribe
\begin{align}\label{eq:zeta3mass}
\zeta_{S^3}(s) = (\mu a)^{2s}\sum_{k=1}^{\infty} k^2\left[k^2-\frac14+(am)^2\right]^{-s}\,.
\end{align}
Para obtener una extensión analítica de esta función zeta que nos permita tomar la derivada con respecto a $s$ en $s=0$, usamos un desarrollo del binomio de la potencia de la cantidad entre paréntesis para escribir 
\begin{align}
\zeta_{S^3}(s) = (\mu a)^{2s}\sum_{n=0}^{\infty} \frac{\Gamma(-s+1)}{n!\,\Gamma(-s+1-n)}\left[(am)^2-\frac14\right]^{n}\zeta_R(2s+2n-2)\,,
\end{align}
expresión que es válida siempre que $\left|(am)^2-1/4\right|<1$. Utilizando ahora la identidad $\Gamma(s)\Gamma(-s+1)=\pi/\sen\pi s$ podemos reescribir esta cantidad como
\begin{align}
\zeta_{S^3}(s) = \frac{(\mu a)^{2s}}{\Gamma(s)}\sum_{n=0}^{\infty} \frac{(-1)^n\,\Gamma(s+n)}{n!}\left[(am)^2-\frac14\right]^{n}\zeta_R(2s+2n-2)\,,
\end{align}
de donde vemos que ---como es bien conocido que sucede en dimensión impar--- la función zeta se anula en $s=0$ y la acción efectiva resulta
\begin{align}\label{eq:seff3mass}
\Gamma_{S^3}(am) = \frac{\zeta_R(3)}{4\pi^2}+\frac12 \sum_{n=0}^{\infty} \frac{(-1)^n}{n+1}\left[(am)^2-\frac14\right]^{n+1}\zeta_R(2n)\,.
\end{align}
El comportamiento de esta acción efectiva con la masa está ilustrado en la figura \ref{fig:seff3mass} que dejamos para el próximo apartado\footnote{ Para obtener dicho comportamiento hemos utilizado otra extensión analítica de la función zeta, que es válida para cualquier valor de la masa y además resulta más conveniente para la evaluación numérica. Obtendremos esa expresión en el apartado que sigue.}. Su valor cuando la masa se anula puede calcularse fácilmente de la expresión \eqref{eq:seff3mass}: evaluando para $am=0$ obtenemos
\begin{align}\label{eq:seff3mass0}
\Gamma_{S^3}(0) = \frac{\zeta_R(3)}{4\pi^2}-\frac18 \sum_{n=0}^{\infty} \frac{\zeta_R(2n)}{(n+1)2^{2n}}\,,
\end{align}
que podemos simplificar teniendo en cuenta que \cite[página 318]{srivastava}
\begin{align}
\sum_{n=1}^{\infty} \frac{\zeta_R(2n)}{(n+1)2^{2n}} = \frac12 +\log\left(\frac{B^{14}}{2}\right)\,,
\end{align}
con $\log B=\zeta_R(3)/4\pi^2$ \cite{srivastavaB}. Luego
\begin{align}
\Gamma_{S^3}(0) = -\frac{3\zeta_R(3)}{16\pi^2}-\frac{\log2}{8}\approx 0.0638\,,
\end{align}
que coincide con el valor reportado en \cite{Beneventano:2017eyu,Klebanov:2011gs}. Por otra parte, el valor de la derivada de la acción efectiva con respecto al cuadrado de la masa adimensionalizada puede obtenerse también mediante la expresión encontrada:
\begin{align}\label{eq:derseff3mass}
\frac{d\Gamma_{S^3}}{d(am)^2} = \frac12 \sum_{n=0}^{\infty} (-1)^n\left[(am)^2-\frac14\right]^{n}\zeta_R(2n)\,,
\end{align}
y usando la expresión \cite[página 271]{srivastava}
\begin{align}
\sum_{n=1}^{\infty} \frac{\zeta_R(2n)}{2^{2n-1}} = 1\,
\end{align}
puede verse que su valor a masa nula es cero.
por otra parte, en el espacio proyectivo $\esp{Z_2}$ la función zeta es
\begin{align}\label{eq:zetaproj3mass}
\zeta_{\esp{Z_2}}(s) = (\mu a)^{2s}\sum_{\substack{k=1\\ k\mathrm{ \,impar}}}^{\infty} k^2\left[k^2-\frac14+(am)^2\right]^{-s}\,,
\end{align}
que puede reescribirse como
\begin{align}\nonumber
\zeta_{\esp{Z_2}}(s) &= \zeta_{S^3}(s) - 4^{1-s}(\mu a)^{2s}\sum_{k=1}^{\infty} k^2\left[k^2-\frac{1}{16}+\frac{(am)^2}{4}\right]^{-s}\\
&= \zeta_{S^3}(s)\big\vert_{am} - 2^{2-2s}\zeta_{S^3}(s)\big\vert_{\tilde{m}} \,,
\end{align}
donde $4 \tilde{m}^2 = (am)^2+3/4$. Como la de la esfera, esta función zeta se anula en $s=0$ y tenemos para la acción efectiva
\begin{align}
\Gamma_{\esp{Z_2}}(am) = \Gamma_{S^3}(am) - 4\Gamma_{S^3}(\tilde{m})\,.
\end{align}

\vfill\pagebreak

\begin{figure}[h]
	\centering
	\includegraphics[height=6cm]{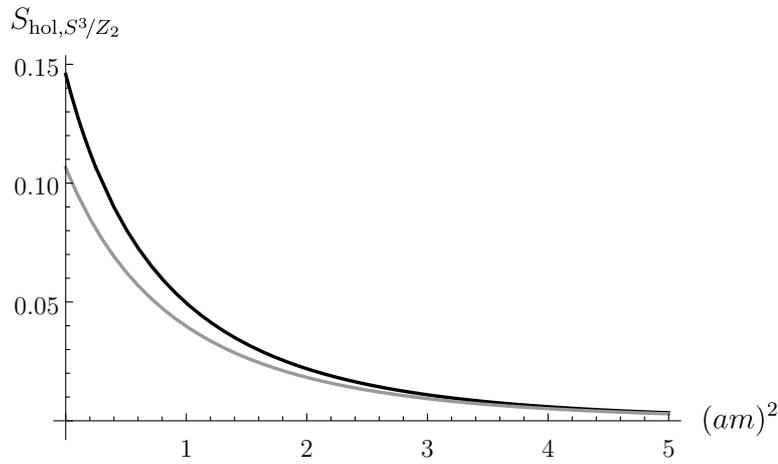}
	\caption{\small Entropía de holonomía en $\esp{Z_2}$ para el campo conforme en tres dimensiones como función del cuadrado de la masa adimensionalizada (curva negra). Se incluye, a modo de comparación, la misma cantidad para el campo con acoplamiento conforme en cuatro dimensiones (curva gris).}
	\label{fig:shol3mass}
\end{figure}

La entropía de holonomía tiene entonces la expresión
\begin{align}
S_{\mathrm{hol},\esp{Z_2}}(am) = 4\Gamma_{S^3} (\tilde{m}) -\frac12 \Gamma_{S^3} (am)
\end{align}
El comportamiento de la entropía de holonomía con el cuadrado de la masa puede verse en la figura \ref{fig:shol3mass}. El valor a masa cero de su derivada con respecto al cuadrado de la masa adimensionalizada no se anula; en efecto,
\begin{align}
\frac{d}{d(am)^2}  S_{\mathrm{hol},\esp{Z_2}}(am) = \frac{d}{d\tilde{m}^2} \Gamma_{S^3}(\tilde{m})\,,
\end{align}
de donde obtenemos,
\begin{align}
\left.\frac{d}{d(am)^2}S_{\mathrm{hol},\esp{Z_2}}(am)\right\vert_{am=0} = \left.\frac{d}{d\tilde{m}^2} \Gamma_{S^3}(\tilde{m})\right\vert_{\tilde{m}^2=3/16}\,.
\end{align}
Para calcular esta cantidad evaluamos una vez más la expresión \eqref{eq:derseff3mass}, de lo que obtenemos \cite[página 271]{srivastava}
\begin{align}
\left.\frac{d}{d\tilde{m}^2} \Gamma_{S^3}(\tilde{m})\right\vert_{\tilde{m}^2=3/16} = \frac12\sum_{n=0}^\infty\frac{\zeta_R(2n)}{4^{2n}}=-\frac{\pi}{16}\,.
\end{align}
La derivada de la entropía de holonomía resulta entonces
\begin{align}
\left.\frac{d}{d(am)^2}  S_{\mathrm{hol},\esp{Z_2}}(am) \right\vert_{am=0} = -\frac{\pi}{4}\,.
\end{align}

Vemos que, al igual que en el caso del campo escalar acoplado conformemente a la métrica tetradimensional del espacio $S^1\times S^3$, esta entropía de holonomía es una función monótonamente decreciente de la constante de acoplamiento, pero su derivada con respecto a la misma en el punto fijo ultravioleta no se anula. De esto concluimos que cualquiera de las dos posibilidades cumple parcialmente el requisito \ref{item:reqcr2} para una cantidad $C$.

Cualquiera de ellas podría entonces ser propuesta como la cantidad relevante en un teorema $C$, aunque, como ya hemos mencionado, nos inclinaremos por el acoplamiento conforme en la dimensión de interés. En la sección siguiente veremos que existe una conexión entre ambas cantidades.

\vfill\pagebreak

\subsubsection{Otra expresión para la entropía de holonomía}

Partiendo de las funciones zeta \eqref{eq:zeta3mass} y \eqref{eq:zetaproj3mass} es posible obtener una expresión diferente para la entropía de holonomía si en lugar de usar un desarrollo binomial separamos $k^2-M^2=(k+M)(k-M)$, donde definimos $M:=\sqrt{\tfrac14 -(am)^2}$, y calculamos por separado las funciones zeta correspondientes a cada factor. Como en tres dimensiones no hay anomalía multiplicativa del determinante \cite{Cognola:2014pha}, el logaritmo del determinante que da la acción efectiva puede calcularse como la suma de las dos funciones zeta correspondientes a cada uno de los factores,
\begin{align}
\zeta_{\pm}(s) = (\mu a)^{2s} \sum_{k=1}^\infty k^2 \left(k\pm M\right)^{-s}\,.
\end{align}
Completando el factor $k^2$ como $(k\pm M)^2\mp2M(k\pm M)+M^2$, podemos reescribir estas funciones zeta como
\begin{align}\label{eq:zetapm3}
\zeta_{\pm}(s) ={} & (\mu a)^{2s}\big[\zeta_H\!\left(s-2,1\pm M\right) \mp 2M \zeta_H\!\left(s-1,1\pm M\right) \\ \nonumber
& \qquad\qquad\qquad\qquad\qquad \qquad\qquad\qquad\qquad+ M^2 \zeta_H\!\left(s,1\pm M\right)\big]\,.
\end{align}
lo que deja para la acción efectiva sobre la esfera la expresión
\begin{align}\nonumber
\Gamma_{S^3}(m) = & -\frac12 \Big\{ \zeta'_H\!\left(-2,1+M\right) + \zeta'_H\!\left(-2,1-M\right) -2M \big[\zeta'_H\!\left(-1,1+M\right) \\ \nonumber & - \zeta'_H\!\left(-1,1-M\right)\big]  + M^2 \big[ \zeta'_H\!\left(0,1+M\right) + \zeta'_H\!\left(0,1-M\right) \big] \Big\}\,.
\end{align}
De aquí el valor a masa cero se obtiene fácilmente a partir de la representación de la función zeta de Hurwitz como suma de senos y cosenos \cite[página 1037]{Gradshteyn}, resultando el mismo valor que ya obtuvimos.

Esta expresión puede evaluarse para cualquier valor de la masa ---si bien puede implicar valores imaginarios de $M$, estos son tratables en el segundo argumento de la zeta de Hurwitz---, como muestra una simple evaluación numérica. En acuerdo con el valor nulo de la anomalía multiplicativa, puede verse que esta evaluación coincide además, en el rango de masas correspondiente, con la de la expresión \eqref{eq:seff3mass}. Para finalizar la sección mostramos los gráficos de la acción efectiva y la entropía de holonomía en función del cuadrado de la masa, a los que nos hemos referido en un apartado anterior.

\vfill\pagebreak

\subsection{Otro comentario acerca de la dimensión}

Como último dato en la discusión acerca de la dimensión, mostramos a continuación que existe una relación, en principio inesperada, entre la  entropía topológica ---que, recordemos, no es otra cosa que el valor de la entropía de holonomía en el límite no masivo--- para el campo escalar sobre espacios lente con acoplamiento conforme en cuatro dimensiones y la energía libre $F$ de la teoría tridimensional sobre la esfera. Podemos resumir dicha relación en la igualdad
\begin{align}
\partial_p S^{(4)}_{\mathrm{top},\esp{Z_p}}\Big|_{p=1} =2 F^{(3)}_{S^3}\,.
\label{eq:rel34}
\end{align}
Las implicaciones de esta identidad no son evidentes pero, si la entropía de holonomía fuera una buena cantidad $C$, su estudio podría echar luz al problema de la relación entre cantidades $C$ en  dimensiones diferentes \cite{Giombi:2014xxa}.

Para probar la igualdad anterior, comenzamos por dar sentido a su lado izquierdo notando que es posible extender los resultados obtenidos para la acción efectiva sobre espacios lente $\esp{Z_p}$ a valores no enteros de $p$ por medio de una representación de las degeneraciones en términos de una función generatriz \cite{Dowker:2004nh}; el resultado es \cite{Dowker:2013ia}
\begin{align}\label{eq:stop4cont}
S_{\mathrm{top},\esp{Z_p}}^{(4)} ={} &
\frac 1{4} \int_0^\infty  dx\  \Re \ \frac{p \senh z + \cosh z \senh pz}{z \senh^2 z \senh^2\frac{pz}{2}} \\ \nonumber
& \qquad
- \frac 1{4 p} \int_0^\infty  dx\ \Re \ \frac{\senh z\, ( 1+ \cosh z)}{z \senh^2 z \senh^2\frac{z}{2}}\,,
\end{align}
donde $z=x+i\Delta$, con $0<\Delta<2\pi/p$ \cite{Dowker:2012rp}. De esta expresión podemos tomar la derivada con respecto a $p$, con lo que tenemos para el lado izquierdo de la ecuación \eqref{eq:rel34}
\begin{align}
\left.\partial_p S^{(4)}_{\mathrm{top},\esp{Z_p}}\right\vert_{p=1} =& {} \frac{1}{4} \int_0^\infty \! dx\,  \Re \frac{\senh z}{z \senh^2 z \senh^2\frac {z}{2}} + \frac 1{4} \int_0^\infty  \!dx\, \Re \frac{\senh z\, ( 1+ \cosh z)}{z \senh^2 z \senh^2\frac{z}{2}}  \nonumber \\ \nonumber
&+ \frac 1{4} \int_0^\infty  dx\ \Re \frac{\cosh^2 z- \coth\frac{z}{2}\senh z\, (1+ \cosh z)}{\senh^2 z \senh^2\frac {z}{2}} \\
=:& {} I_1 + I_2 + I_3\,.
\end{align}
Veremos ahora cómo calcular las integrales $I_i$, $i=1,2,3$. Notemos en primer lugar que la tercera integral en la expresión anterior es nula. En efecto, haciendo uso de la identidad
\begin{align}\nonumber
\cosh^2z-\coth\frac{z}{2}\senh z \,(1+\cosh z) = -1-2\cosh z,
\end{align}
es posible escribir
\begin{align}\nonumber
I_3 = -\frac 1{4} \int_0^\infty  dx\ \Re \frac{1+ 2\cosh z}{\senh^2 z \senh^2\frac {z}{2}}\,.
\end{align}
Puede verse que el valor de esta integral no depende de $\Delta$, siempre que $0<\Delta<\pi$. Tomando $\Delta= \pi/2$ tenemos
\begin{align}\nonumber
I_3 = -\frac 1{4} \int_0^\infty  dx\ \Re \frac{1+ 2\cosh \left(x + i\frac{\pi}{2}\right)}{\senh^2 \left(x + i\frac{\pi}{2}\right) \senh^2\left(\frac{x}{2} + i\frac{\pi}{4}\right)}\,,
\end{align}
\vfill\pagebreak

\noindent que podemos reescribir usando 
\begin{align}\nonumber
\Re \frac{1+ 2\cosh \left(x + i\frac{\pi}{2}\right)}{\senh^2 \left(x + i\frac{\pi}{2}\right) \senh^2\left(\frac{x}{2} + i\frac{\pi}{4}\right)} = 2\left(\frac{3}{\cosh^4x} -\frac{2}{\cosh^2x}\right)\,,
\end{align}
y se anula, puesto que 
\begin{align}\nonumber
\int_0^\infty \frac{dx}{\cosh^2x} = 1\,,\,\,\,\, \int_0^\infty \frac{dx}{\cosh^4x} = \frac{2}{3}\,.
\end{align}
Luego,
\begin{align}\nonumber
\left.\partial_p S^{(4)}_{\mathrm{top},\esp{Z_p}}\right\vert_{p=1} = I_1+I_2 =  \frac 1{4} \int_0^\infty  dx\  \Re \frac{2+\cosh z}{z \senh z \senh^2\frac {z}{2}}\,.
\end{align}
Nuevamente, tomando $\Delta=\pi/2$ podemos reescribir la integral para obtener
\begin{align}\label{eq:dpstop3}
\left.\partial_p S^{(4)}_{\mathrm{top},\esp{Z_p}}\right\vert_{p=1} = \frac 1{4} \int_0^\infty  dx\  \frac{2\pi-6x\senh x -\pi\senh^2x}{(x^2+\frac{\pi^2}{4})\cosh^3x}\,.
\end{align}

Por otra parte, la energía libre sobre la esfera en el caso conforme en tres dimensiones puede escribirse como \cite{Dowker:2013ia}
\begin{align}\label{eq:f3cont}
F^{(3)} = -\frac 1{4} \int_0^\infty  dx\ \Re \frac{\cosh\frac{z}{2}\,(1+\cosh z)}{z\senh^2z \senh^2\frac{z}{2}}\,,
\end{align}
donde ahora $z=x+i \Delta$ con $0<\Delta<2\pi$. Si elegimos $\Delta = \pi$, esta integral puede ponerse en la forma
\begin{align}\label{eq:f3}
F^{(3)} = -\frac \pi{16} \int_0^\infty  dx\ \frac{1-\cosh 2x}{(x^2+\frac{\pi^2}{4})\cosh^3x}\,.
\end{align}
Podemos entonces combinar las integrales obtenidas para escribir la diferencia entre \eqref{eq:dpstop3} y el doble de \eqref{eq:f3} como
%
%
%
\begin{align}
\frac{1}{2} \int_0^\infty \!\! dx\ \frac{4-2\cosh^2\frac{\pi x}{2}}{(x^2+1)\cosh^3\frac{\pi x}{2}} -\frac{1}{2} \int_0^\infty \!\! dx\ \frac{3 x \senh\frac{\pi x}{2}}{(x^2+1)\cosh^3\frac{\pi x}{2}} = \tilde{I}_1 + \tilde{I}_2\,.
\end{align}
Las integrales $\tilde{I}_1$ and $\tilde{I}_2$ pueden evaluarse fácilmente por medio de las transformadas de Laplace
\begin{align}\nonumber
\frac{1}{1+x^2} = \int_0^\infty dt\ e^{-t} \cos tx\,,\,\,\,\,\,\, \frac{x}{1+x^2} = \int_0^\infty dt\ e^{-t} \sen tx\,,
\end{align}
con lo que obtenemos finalmente
\begin{align}\nonumber
\tilde{I}_1 = \int_0^\infty dt\ e^{-t} \int_0^\infty dx\ \frac{2-\cosh^2\frac{\pi x}{2}}{\cosh^3\frac{\pi x}{2}} \cos tx  = \frac{3 \zeta_R(3)}{2\pi^2}\,,
\end{align}
y
\begin{align}\nonumber
\tilde{I}_2 = -\frac{3}{2}\int_0^\infty dt\ e^{-t} \int_0^\infty dx\ \frac{\senh\frac{\pi x}{2}}{\cosh^3\frac{\pi x}{2}} \sen tx  = -\frac{3 \zeta_R(3)}{2\pi^2}\,,
\end{align}
lo que completa la prueba de la identidad \eqref{eq:rel34}.

\subsection{Campo escalar masivo conforme en dimensión arbitraria}

Consideramos ahora el caso del campo escalar masivo conforme en $d$ dimensiones sobre la esfera $S^{d}$ y el espacio proyectivo $S^{d}\!/Z_2$. De la acción clásica para tal teoría obtenemos el operador de fluctuaciones cuánticas
\begin{align}
\triangle^{\!(d)}+\frac{d(d-2)}{4a^2}+m^2\,,
\end{align}
con $\triangle^{\!(d)}$ el laplaciano sobre la esfera $d$-dimensional. Teniendo en cuenta que el espectro de $\triangle^{\!(d)}$ puede ser obtenido a partir de los autovalores y degeneraciones del Casimir cuadrático de $SO(d+1)$ sobre las representaciones irreducibles del grupo, podemos ver que los autovalores para la teoría sobre ambos espacios son \cite{Camporesi:1990wm}
\begin{align}
a^2 \lambda_n = \left(n+\frac{d-1}{2}\right)^2 -\frac14 +(am)^2\qquad n\in\mathbb{N}\,,
\end{align}
siendo sus respectivas degeneraciones
\begin{align}\label{eq:degdsph}
d_n = \frac{(2n+d-1)(n+d-2)!}{n!(d-1)!}\,,
\end{align}
donde en el caso de $S^{d}\!/Z_2$ tenemos que restringir el índice $n$ a sus valores pares. A continuación obtendremos las acciones efectivas y la entropía de holonomía por separado para el caso de dimensión par e impar.

\vfill\pagebreak

\subsection{Dimensiones impares}

Consideramos en primer lugar el caso en que la dimensión de la esfera es impar, $d=2k+1$. Renombrando $k+n\rightarrow n$, los autovalores son $\lambda_n=n^2-M^2$, con $n=k,k+1,\ldots$, y las degeneraciones pueden reescribirse como
\begin{align}\label{eq:degdimp}
d_n = \frac{2n (n+k-1)!}{(2k)!(n-k)!}\,.
\end{align}
Notando que los factores en la expresión expandida de $(n+k-1)!/(n-k)!$ pueden reordenarse como
\begin{align}
\frac{(n+k-1)!}{(n-k)!} = (n^2-(k-1)^2)(n^2-(k-2)^2)\cdots(n^2-1) n
\,,
\end{align}
que es un polinomio de orden impar en $n$, vemos que podemos escribir esta cantidad en la forma
\begin{align}\label{eq:coefimpar}
\frac{(n+k-1)!}{(n-k)!} = \sum_{l=1}^{k} c^{(2k+1)}_l\, n^{2l-1} \,,
\end{align}
con algunos coeficientes enteros $c^{(2k+1)}_l$.

La función zeta sobre la esfera $S^{2k+1}$ se escribe entonces
\begin{align}\label{eq:zetadimpar}
\zeta_{S^{2k+1}}(s) = \frac{2(\mu a)^{2s}}{(2k)!} \sum_{n=1}^\infty\sum_{l=1}^k c^{(2k+1)}_l n^{2l} \left[n^2 + (am)^2-1/4 \right]^{-s}\,,
\end{align}
donde la suma sobre $n$ puede ser extendida de modo de comenzar en $n=1$, puesto que las degeneraciones \eqref{eq:degdimp} se anulan para $n=1,2,\ldots,k-1$. Al igual que en el caso tridimensional, tenemos aquí dos posibilidades: podemos utilizar el desarrollo binomial de la potencia entre corchetes ---habiendo extraído previamente el factor $n^2$ en los autovalores--- y obtener para la acción efectiva una expresión en términos de una serie de potencias de $(am)^2-1/4$, que no será entonces válida fuera del rango $-3/4<(am)^2<5/4$, o podemos aprovechar el hecho de que en dimensión impar no hay anomalía del determinante \cite{Cognola:2014pha} y calcular la acción efectiva utilizando la suma de las dos funciones zeta que se obtienen de factorizar los autovalores como diferencia de cuadrados. Eligiendo esta última opción, escribimos 
\begin{align}
\Gamma_{S^{2k+1}} = -\frac12 \left.\frac{d}{ds} \left[\zeta_{+}(s)+\zeta_{-}(s)\right]\right\vert_{s=0}\,,
\end{align}
con 
\begin{align}
\zeta_{\pm}(s) = \frac{2(\mu a)^{2s}}{(2k)!} \sum_{n=1}^\infty\sum_{l=1}^k c^{(2k+1)}_l n^{2l} (n\pm M)^{-s}\,.
\end{align}
Si usamos el hecho de que
\begin{align}
n^{2l}=\sum_{p=0}^{2l}\frac{(2l)!}{p!(2l-p)!} (n\pm M)^{2l-p}(\mp M)^p
\end{align}
podemos poner estas funciones zeta en la forma
\begin{align}\nonumber
\zeta_{\pm}(s) = \frac{2(\mu a)^{2s}}{(2k)!} \sum_{l=1}^k c^{(2k+1)}_l \sum_{p=0}^{2l} \frac{(2l)!}{p!(2l-p)!}(\mp M)^p \zeta_H\!\left(s-2l+p,1\pm M\right)\,,
\end{align}
que para $k=1$ se reducen a las expresiones \eqref{eq:zetapm3} obtenidas en la sección anterior para el caso de la esfera tridimensional.
La acción efectiva resulta
\begin{align}
\Gamma_{S^{2k+1}}(m) = &{} -\frac{1}{(2k)!}\sum_{l=1}^k c^{(2k+1)}_l \sum_{p=0}^{2l} \frac{(2l)!}{p!(2l-p)!}M^p \\ \nonumber
& \qquad\qquad\times \left[ \zeta'_H(p-2l,1-M) + (-1)^p \zeta'_H(p-2l,1+M) \right]\,.
\end{align}
%

En el caso del espacio proyectivo $S^{2k+1}\!/Z_2$, los términos de la serie en \eqref{eq:zetadimpar} que permanecen son los de $n$ (im)par si $k$ es (im)par. Escribimos entonces
\begin{align}
\zeta_{S^{2k+1}\!/Z_2}(s) = \frac{1+(-1)^{k+1}}{2}\zeta_{S^{2k+1}}(s) + (-1)^k \Delta\zeta(s)\,,
\end{align}
donde $\Delta\zeta(s)$ tiene la misma forma que \eqref{eq:zetadimpar}, con la diferencia de que la suma sobre el índice $n$ se hace sólo sobre sus valores pares. Escribiendo como antes $\Delta\zeta(s)=\left[\zeta_{+}(s)+\zeta_{-}(s)\right]$, tenemos
\begin{align}\nonumber
\zeta_{\pm}(s) = \frac{2^{1-2s}(\mu a)^{2s}}{(2k)!} \sum_{l=1}^k c^{(2k+1)}_l 2^{2l} \sum_{p=0}^{2l} \frac{(2l)!}{p!(2l-p)!} \frac{(\mp M)^p}{2^p} \zeta_H\!\left(s-2l+p,1\pm \tfrac{M}{2}\right).
\end{align}
Luego de juntar ambos resultados, la entropía de holonomía se escribe 
\begin{align}
S_{\mathrm{hol},S^{2k+1}\!/Z_2} = &{} \frac{(-1)^{k+1}}{(2k)!}\sum_{l=1}^k c^{(2k+1)}_l \sum_{p=0}^{2l} \frac{(2l)!}{p!(2l-p)!}M^p \\ \nonumber
& \quad\times \Big\{\frac12\left[ \zeta'_H(p-2l,1-M) + (-1)^p \zeta'_H(p-2l,1+M) \right]\\\nonumber
& \quad\phantom{\times\Big\{} -2^{2l-p}\left[ \zeta'_H\!\left(p-2l,1-\tfrac{M}{2}\right) + (-1)^p \zeta'_H\!\left(p-2l,1+\tfrac{M}{2}\right) \right]\Big\}\,.
\end{align}

\begin{figure}[h!]
	\centering
	\includegraphics[height=6cm]{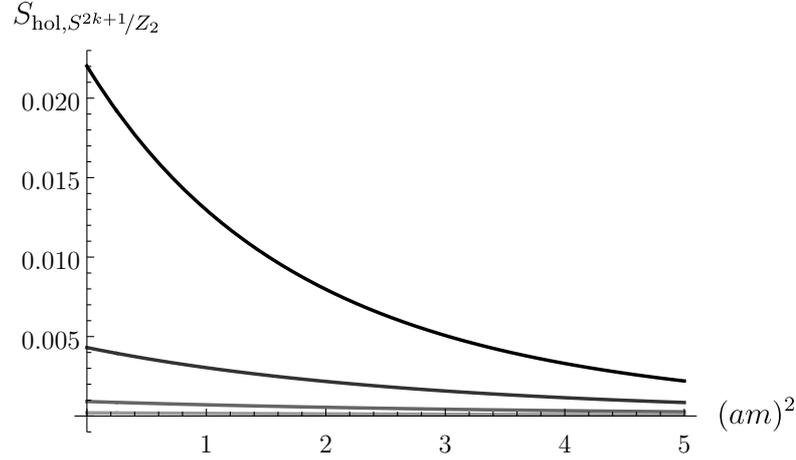}
	\caption{\small Entropía de holonomía para el campo conforme sobre la esfera $(2k+1)$-dimensional como función del cuadrado de la masa adimensionalizada para los primeros valores de $k$. El corte con el eje vertical es menor para los valores más altos de $k$.}
	\label{fig:sholdmass}
\end{figure}

En la figura \ref{fig:sholdmass} puede verse el comportamiento de esta entropía de holonomía con el cuadrado de la masa adimensionalizada para las dimensiones impares más bajas. Observamos que dicho comportamiento es similar al que obtuvimos en el caso tridimensional, lo cual está de acuerdo con nuestra hipótesis. En particular, al igual que en aquel caso, la derivada con respecto a la constante de acoplamiento en el punto fijo ultravioleta no se anula.

\subsection{Dimensiones pares}

En el caso en que la dimensión es par, $d=2k$, reescribimos las degeneraciones \eqref{eq:degdsph} como
\begin{align}
d_n = \frac{2}{(2k-1)!} \left(n+k-\frac12\right) \frac{(n+2k-2)!}{n!}\,.
\end{align}
Definiendo $\tilde{n}= n+k-\frac12$, vemos que
\begin{align}
\frac{(n+2k-2)!}{n!} = \frac{(\tilde{n}+k-\frac32)!}{(\tilde{n}-k+\frac12)!} = \prod_{l=0}^{k-2} \left[ \tilde{n}^2-\left(l+\frac12\right)^2 \right]\,,
\end{align}
que, siendo un polinomio de orden par en la variable $\tilde{n}$, puede escribirse como
\begin{align}\label{eq:coefpar}
\frac{(n+2k-2)!}{n!} = \sum_{l=0}^{k-1} c^{(2k)}_l\, \tilde{n}^{2l} \,,
\end{align}
donde $c^{(2k)}_l$ son números racionales.

Puede verse que en este caso la anomalía multiplicativa del determinante no se anula \cite{Cognola:2014pha}. Por simplicidad, en lugar de tenerla en cuenta en la descomposición, consideraremos el procedimiento que involucra un desarrollo binomial, que, aunque no válido en todo el rango de masas, nos permitirá al menos verificar si el comportamiento para valores pequeños de la masa es el esperado. Extrayendo convenientemente un factor en los autovalores, la función zeta sobre $S^{2k}$ se escribe
\begin{align}
\zeta_{S^{2k}}(s) = \frac{2\,(\mu a)^{2s}}{(2k-1)!} \sum_{l=0}^{k-1} c^{(2k)}_l \sum_{n=k}^\infty \left( n-\frac12 \right)^{\!\!1+2l-2s} \left[1+\frac{(am)^2-1/4}{(n-\frac12)^2}\right]^{-s}\,,
\end{align}
de donde, luego del desarrollo binomial de la potencia entre corchetes, y notando que por la forma de las degeneraciones  la suma sobre $n$ se puede extender de modo de comenzar en $n=2$, obtenemos para $|am|<1/2$
\begin{align}\label{eq:zetasdeven}
\zeta_{S^{2k}}(s) ={} &  \frac{2\,(\mu a)^{2s}}{(2k-1)!} \sum_{l=0}^{k-1} c^{(2k)}_l \sum_{p=0}^\infty \,\frac{\Gamma(-s+1)}{p!\,\Gamma(-s+1-p)} \left[(am)^2-\frac14\right]^p \\ \nonumber
& \qquad\qquad\qquad\qquad\qquad\qquad\qquad\qquad\times \zeta_H\!\left(2s+2p-2l-1,\tfrac32\right)\,.
\end{align}
Puede verse que esta función zeta no se anula en $s=0$, como es usual en el caso de problemas en dimensión par. En efecto, si usamos la identidad $\Gamma(1-x)\Gamma(x)=\pi/\sen\pi x$ podemos ver que para cada valor de $l$ los únicos términos de la suma sobre $p$ que no se anulan en el límite $s\rightarrow0$ son $p=0$ y $p=l+1$. El límite resulta
\begin{align}\nonumber
\zeta_{S^{2k}}(0) = \frac{2}{(2k-1)!} \sum_{l=0}^{k-1} c^{(2k)}_l \left\{ \zeta_H\!\left(-2l-1,\tfrac32\right) + \frac{1}{2(l+1)}\left[\frac14-(am)^2\right]^{l+1}\right\}\,.
\end{align}

Para el caso del espacio proyectivo con $k$ par, teniendo en cuenta la reducción en el espectro con respecto al de la esfera, podemos obtener para la función zeta la expresión
\begin{align}\nonumber
\zeta_{S^{2k}\!/Z_2}(s) = \frac{4(\mu a)^{2s}}{(2k-1)!} \sum_{l=0}^{k-1} c^{(2k)}_l\,2^{-2s+2l} \sum_{n=0}^\infty \!\left( n-\frac14 \right)^{\!\!1+2l-2s} \left[1+ \frac{(am)^2-1/4}{4(n-\frac14)^2}\right]^{-s},
\end{align}
que, para $|am|<1/4$, puede ser reescrita como
\begin{align}\label{eq:zeta2deven}
\zeta_{S^{2k}\!/Z_2}(s) ={} & \frac{(\mu a)^{2s}}{(2k-1)!} \sum_{l=0}^{k-1} c^{(2k)}_l \sum_{p=0}^\infty \,\frac{\Gamma(s+p)}{p!\,\Gamma(s)}\, 2^{2+2l-2s-2p} \left[\frac14-(am)^2\right]^p \\ \nonumber
&\qquad\qquad\qquad\qquad\qquad\qquad\qquad \times \zeta_H\!\left(2s+2p-2l-1,\tfrac34\right)\,.
\end{align}

\noindent El valor de esta función zeta en $s=0$ se puede calcular fácilmente y resulta
\begin{align}\nonumber
\zeta_{S^{2k}\!/Z_2}(0) \!= \!\frac{1}{(2k-1)!}\! \sum_{l=0}^{k-1} \!c^{(2k)}_l \!\left\{ 2^{2+2l}\zeta_H\!\left(-2l-1,\tfrac34\right) \!+\! \frac{1}{2(l+1)}\!\left[\frac14-(am)^2\right]^{l+1}\right\}.
\end{align}

En el caso en que $k$ es impar, la expresión correspondiente para la función zeta puede obtenerse como la diferencia entre la función zeta sobre la esfera y esta última expresión.

Es posible mostrar que, si bien las funciones zeta \eqref{eq:zetasdeven} y \eqref{eq:zeta2deven} no se anulan en $s=0$, su ``diferencia holonómica''
\begin{align}
\zeta^{\mathrm{hol}}(s):= \zeta_{S^{2k}\!/Z_2}(s)-\frac12 \zeta_{S^{2k}}(s)
\end{align}
sí lo hace, de modo que la acción efectiva no depende del regulador.

La entropía de holonomía resulta finalmente
\begin{align}\nonumber
S_{\mathrm{hol},S^{2k}\!/Z_2} ={} & \frac{(-1)^{k+1}}{(2k-1)!} \sum_{l=0}^{k-1} c^{(2k)}_l \Bigg\{ \frac{\left[\frac14-(am)^2\right]^{l+1}}{2(l+1)} \left[ \log{2} + \psi(3/4) - \psi(3/2) \right] \\ \nonumber 
& - 2^{2+2l} \left[ -\log{2}\, \zeta_H\!\left(-2l-1,\tfrac34\right) + \zeta'_H\!\left(-2l-1,\tfrac34\right) \right] \\ \nonumber
& - \sum_{\substack{p=1\\ p\neq l+1}}^{\infty} \frac{\left[\frac14-(am)^2\right]^p}{2p} \big[ 2^{2+2l-2p}\zeta_H\!\left(2p-2l-1,\tfrac34\right) \\[-4mm] \nonumber
& \qquad\qquad\qquad\qquad - \zeta_H\!\left(2p-2l-1,\tfrac32\right) \big]  + \zeta'_H\!\left(-2l-1,\tfrac32\right) \!\Bigg\}\,.
\end{align}

\begin{figure}[h!]
	\centering
	\includegraphics[height=6cm]{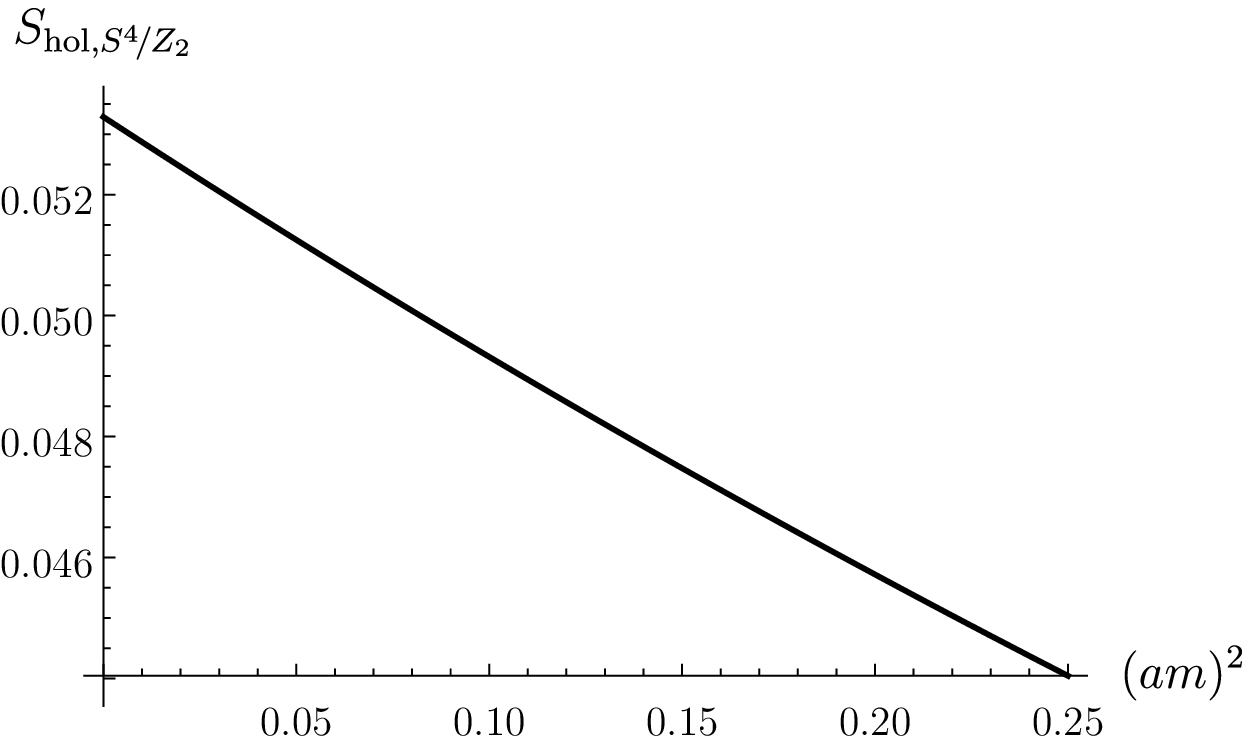}
	\caption{\small Entropía de holonomía para el campo conforme sobre la esfera $4$-dimensional como función del cuadrado de la masa adimensionalizada.}
	\label{fig:sholdmass4}
\end{figure}

\begin{figure}[h!]
	\centering
	\includegraphics[height=6cm]{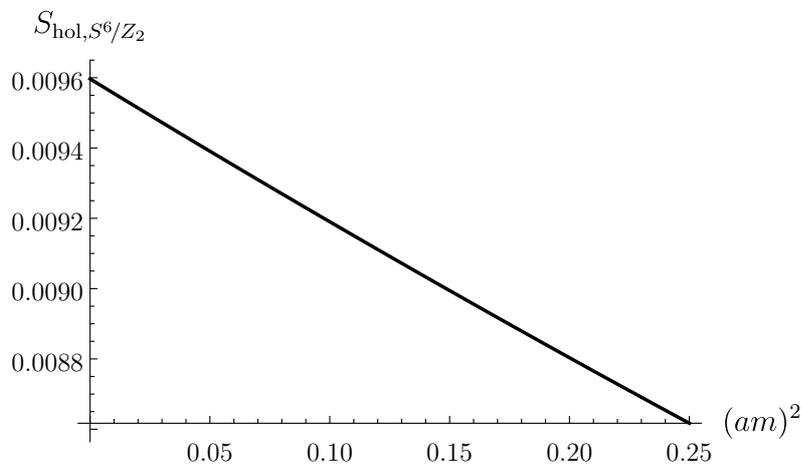}
	\caption{\small Entropía de holonomía para el campo conforme sobre la esfera $6$-dimensional como función del cuadrado de la masa adimensionalizada.}
	\label{fig:sholdmass6}
\end{figure}

En las figuras \ref{fig:sholdmass4} y \ref{fig:sholdmass6} ilustramos el comportamiento de esta entropía de holonomía con el cuadrado de la masa adimensionalizada para las esferas en 4 y 6 dimensiones respectivamente. De las curvas se desprende que esta cantidad se comporta como una cantidad $C$ para el rango de masas en el que los cálculos son válidos.

Vemos entonces que el testeo de la entropía de holonomía como cantidad $C$ para flujos de masa de campos escalares libres en dimensión arbitraria ha dado una respuesta parcialmente satisfactoria: en todos los casos y rangos analizados esta cantidad es monótonamente decreciente como función de la constante de acoplamiento, pero su derivada con respecto a esta variable en el punto fijo ultravioleta no se anula, cumpliendo entonces parcialmente, del mismo modo como sucedía para el caso tridimensional, el requisito \ref{item:reqcr2} para una cantidad $C$. Por otra parte, observamos que el orden de magnitud de los valores de la entropía de holonomía decrece sustancialmente con la dimensión del espacio, lo que hace que para dimensiones mayores los valores tengan un significado menos claro.

\vfill\pagebreak
 
\section{El campo de Dirac sobre espacios proyectivos}

Consideramos ahora un segundo testeo de la entropía de holonomía como cantidad $C$: el flujo de masa de un campo de Dirac libre sobre el espacio proyectivo $S^d\!/{Z_2}$. La acción clásica euclídea para dicha teoría se escribe 
\begin{align}
S[\psi] = \frac{1}{2}\int d^dx\sqrt{g} \, \bar{\psi}\left(\partial\!\!\!\slash+m\right)\psi\,,
\end{align}
donde la integral se realiza sobre la esfera $S^d$ o el espacio proyectivo $S^d\!/Z_2$ según sea el caso, y donde $g$ es el determinante de la métrica de la esfera. Una vez más, calcularemos las acciones efectivas a partir del determinante funcional del operador de fluctuaciones cuánticas de la teoría.

El espectro del operador en cuestión sobre la esfera $S^d$ se compone de los autovalores y degeneraciones \cite{Trautman:1995fr}
\begin{align}\label{eq:diracspectrum}
\lambda_n^{\pm} = \pm \frac{i}{a}\left(n+\frac{d}{2}\right)+m\,, \qquad d_n\equiv d_n^{\pm} = 2^{\lfloor\frac{d}{2}\rfloor} \binom{d+n-1}{n}\,,
\end{align}
donde $n$ es un número entero no negativo.

En el caso de los espacios proyectivos hay que hacer la siguiente salvedad: los espacios de dimensión par $S^{2k}\!/Z_2$ no son variedades de spin, puesto que no son orientables \cite{bar2003dirac}. No obstante, es posible dotarlas de una estructura pin \cite{Trautman:1995fr}. Las variedades $S^{2k+1}\!/Z_2$, en cambio, son orientables, pero puede verse que son spin sólo si $k$ es impar. En lo que sigue haremos un abuso de lenguaje y nos referiremos genéricamente a estas estructuras como \emph{de spin}. Cuando la dimensión $d$ es par o $d=3\mod 4$, la construcción de Clifford sobre $S^d\!/Z_2$ lleva a dos posibilidades para la estructura de spin que, a su vez, dan dos elecciones del operador de Dirac, que en el caso no masivo tienen espectros opuestos entre sí: llamaremos $\mathcal{D}^{(+)}$ y $\mathcal{D}^{(-)}$ a estos operadores. Los autovalores de $\mathcal{D}^{(+)}$ son \cite{bar1996dirac,Camporesi:1995fb,Trautman:1995fr}
\begin{align}\label{eq:spectrumdiracproj}
\lambda_n^{+} = \frac{i}{a}\left(n+\frac{d}{2}\right) +m\,,\quad n\ \mathrm{par}\,, \qquad \lambda_n^{-} = -\frac{i}{a}\left(n+\frac{d}{2}\right)+m\,,\quad n\ \mathrm{impar}\,,
\end{align} 
donde $n$ es nuevamente un entero no negativo. Las degeneraciones son las mismas que las de los autovalores de la esfera correspondientes.

Como los autovalores no son reales positivos tenemos, como mencionamos en el capítulo \ref{sec:mate}, una ambigüedad en la definición de la función zeta de los operadores $\mathcal{D}^{(\pm)}$. A lo largo de los cálculos, cuando fijemos un valor para la fase de alguna potencia compleja estaremos eligiendo el corte que hace que el campo se desacople en el límite $m\rightarrow\infty$ \cite{Deser:1997gp}.

Debido a la dependencia de los autovalores con la paridad del índice entero que los caracteriza, será útil dividir el cálculo en dos partes, dependiendo de la paridad de la dimensión del espacio. Sin embargo, haremos primero el cálculo en el caso tridimensional, puesto que éste nos permitirá tener algo más de control sobre las expresiones finales, con lo que intentaremos dar sentido a la definición de la acción efectiva sobre los espacios proyectivos.

\subsection{El caso tridimensional}

Veremos a continuación que la acción efectiva sobre el espacio proyectivo $\esp{Z_2}$ con una de las estructuras de spin ---digamos $\mathcal{D}^{(+)}$--- tiene una parte imaginaria no nula, que no se anula en su diferencia ``holonómica'' ---con la mitad de la acción efectiva sobre la esfera tridimensional---. En efecto, consideremos la cantidad
\begin{align}
\zeta_{\mathrm{hol}}^{(+)}(s):=\zeta_{\esp{Z_2}}(s)-\frac12\zeta_{S^3}(s)\,.
\end{align}

Utilizando la expresión para los autovalores y degeneraciones del operador de Dirac sobre ambos espacios que, aunque estemos considerando el caso particular de $d=3$, llamaremos como en \eqref{eq:diracspectrum}, podemos escribir
\begin{align}\nonumber
\zeta_{\mathrm{hol}}^{(+)}(s)=\frac{\mu^s}2\left\{\sum_{\substack{n=0\\ n \mathrm{\,par}}}^\infty d_n \left[ (\lambda_n^+)^{-s} - (\lambda_n^-)^{-s} \right] - \sum_{\substack{n=0\\ n \mathrm{\,impar}}}^\infty d_n \left[ (\lambda_n^+)^{-s} - (\lambda_n^-)^{-s} \right]\right\}\,,
\end{align}
que, fijando la fase de la potencia compleja en los autovalores mediante la elección $i^s=\exp(i\pi s/2)$ y reescribiendo las degeneraciones como polinomios de $\lambda_n^+$ o $\lambda_n^-$ según sea el caso, puede ponerse en la forma
\vfill\pagebreak

\begin{align}\label{eq:zetahol}
\zeta_{\mathrm{hol}}^{(+)}(s) ={} (\mu a)^s 2^{1-s}\Big\{ e^{-i\frac{\pi}{2}s}\delta\zeta_H\!\left(s,am\right) - e^{i\frac{\pi}{2}s}\delta\zeta_H\!\left(s,-am\right)\Big\}\,,
\end{align}
donde hemos denotado para abreviar
\begin{align}
\delta\zeta_H\!\left(s,am\right) :={}&  \tilde{\delta}\zeta_H\!\left(s-2,am\right) + iam\,\tilde{\delta}\zeta_H\!\left(s-1,am\right) \\ \nonumber
& \qquad\qquad\qquad\qquad\qquad\qquad -\frac{(am)^2+1/4}{4}\, \tilde{\delta}\zeta_H\!\left(s,am\right)\,,
\end{align}
con $\tilde{\delta}\zeta_H\!\left(s,am\right) := \zeta_H\!\left(s,\tfrac34-\tfrac{iam}{2}\right) - \zeta_H\!\left(s,\tfrac54-\tfrac{iam}{2}\right)$.

Escribiendo las funciones zeta de Hurwitz en enteros negativos en términos de polinomios de Bernoulli \cite[página 1037]{Gradshteyn} y haciendo las combinaciones correspondientes, puede verse que $\delta\zeta_H\!\left(0,\pm am\right)=1/16$, por lo que la función zeta en \eqref{eq:zetahol} se anula en $s=0$. La correspondiente diferencia de acciones efectivas tiene un término que depende de estas contribuciones ---que es el que proviene de derivar las exponenciales---, que no depende de la masa, y términos con derivadas de funciones zeta de Hurwitz con respecto a su primer argumento evaluadas en enteros negativos:
\begin{align}
\Gamma_{\mathrm{hol}}^{(+)} (am) = -i\frac{\pi}{8} + 2\,\delta\zeta_H'(0,am) - 2\,\delta\zeta_H'(0,-am)\,.
\end{align}

Si tomamos por ejemplo $am=0$, vemos que los términos en $\delta\zeta_H'(0,am)$ y $\delta\zeta_H'(0,-am)$ se cancelan entre sí, por lo que la acción efectiva resulta
\begin{align}
\Gamma_{\mathrm{hol}}^{(+)} (0) = -i\frac{\pi}{8}\,.
\end{align}
Esto es, la diferencia ``holonómica'' de acciones efectivas es imaginaria en el punto fijo ultravioleta, y puede verse que lo mismo es cierto para valores arbitrarios de la masa. Esto es consecuencia de la presencia de una asimetría espectral no nula. Es posible mostrar que lo mismo sucede para la otra estructura de spin, donde se obtiene
\begin{align}
\Gamma_{\mathrm{hol}}^{(-)} (am) = -i\frac{\pi}{8} - 2\,\delta\zeta_H'(0,am) + 2\,\delta\zeta_H'\left(0,-am\right)\,.
\end{align}
Esto implica que las diferencias ``holonómicas'' de acciones efectivas con una y otra estructura de spin son conjugadas entre sí.
Luego, una forma de tener para la teoría sobre el espacio proyectivo una acción efectiva real, de modo de recuperar la unitariedad de la teoría, es tomar el promedio sobre las posibles estructuras de spin ---como es usual en Teoría de Cuerdas \cite{Seiberg:1986by}---
\begin{align}
Z^{S^3\!/Z_2} := \frac12 \left(Z^{(+)} + Z^{(-)}\right)\,,
\end{align}
lo que corresponde a separar la integral funcional que define la función de partición sobre $S^3\!/Z_2$ en la suma de dos integrales funcionales diferentes ---una sobre cada sector---, que hemos llamado $Z^{(+)}$ y $Z^{(-)}$, con una normalización apropiada para la medida.
Con esta prescripción, la acción efectiva es
\begin{align}\label{eq:seffdiracrp3}
\Gamma^{S^3\!/Z_2} \equiv -\log Z^{S^3\!/Z_2} = -\log\left(\frac{e^{-\Gamma^{(+)}}+e^{-\Gamma^{(-)}}}{2}\right)
\end{align}
y, teniendo en cuenta que las acciones efectivas resultan opuestas una de la otra y su parte real coincide con la mitad de la acción efectiva sobre la esfera, tenemos para la entropía de holonomía
\begin{align}\label{eq:shol3dirac}
S_{\mathrm{hol},\esp{Z_2}} = \log\cos\Im\left(\Gamma^{S^3\!/Z_2,(+)}\right)\,.
\end{align}
En lo que sigue, calcularemos las entropías de holonomía para teorías de Dirac masivas sobre espacios proyectivos en dimensión arbitraria, tomando como definición de la acción efectiva sobre dichos espacios la expresión \eqref{eq:seffdiracrp3}
\begin{align}\label{eq:seffprojdirac}
\Gamma^{S^d\!/Z_2} := -\log\left(\frac{e^{-\Gamma^{(+)}}+e^{-\Gamma^{(-)}}}{2}\right)\,.
\end{align}

\subsection{Entropía de holonomía en dimensión par}

Comenzamos analizando el caso en el que la dimensión del espacio es par, $d=2k+2$. Reescribimos el espectro \eqref{eq:diracspectrum} haciendo el cambio  $n\rightarrow n-k-1$. Los autovalores se escriben entonces
\begin{align}
\lambda_n^{\pm} = \pm \frac{i}{a}\,n+m\,,\qquad n=k+1,k+2,\ldots\,,
\end{align}
y las correspondientes degeneraciones son
\begin{align}\label{eq:degevendirac}
d_n = \frac{2^{k+1}}{(2k+1)!} \frac{(n+k)!}{(n-k-1)!}\,.
\end{align}
Usaremos un procedimiento similar al del caso escalar, para lo cual escribimos
\begin{align}\nonumber
\frac{(n+k)!}{(n-k-1)!} & = (n+k)(n+k-1)(n+k-2)\ldots(n-k+1)(n-k) \\ \label{eq:diracfraceven}
& = n(n^2-1)(n^2-2^2)\ldots(n^2-k^2)
\end{align}
Esta expresión es un polinomio impar en la variable $n$, con lo que podemos escribirla como en aquel caso en términos de coeficientes enteros $c^{(2k+2)}_l$:
\begin{align}
d_n = \frac{2^{k+1}}{(2k+1)!}\sum_{l=0}^{k}c^{(2k+2)}_l n^{1+2l}\,.
\end{align}
Notando que la forma de las degeneraciones \eqref{eq:degevendirac} permite extender el rango del índice $n$ a todos los valores enteros no negativos, podemos escribir la función zeta para esta teoría como
\begin{align}
\zeta_{S^{2k+2}}(s) = (\mu a)^s\frac{2^{k+1}}{(2k+1)!}\sum_{l=0}^{k}c^{(2k+2)}_l \Bigg\{ & \sum_{n=1}^{\infty}n^{1+2l}\left(in+am\right)^{-s} +\\ \nonumber
& + (-1)^{-s} \sum_{n=0}^{\infty}n^{1+2l}\left(in-am\right)^{-s} \Bigg\}\,.
\end{align}
Tomemos ahora la primera suma entre llaves: usando un desarrollo binomial podemos escribir
\begin{align}
\sum_{n=1}^{\infty}n^{1+2l}  \left(in+am\right)^{-s} & = i^{-s}\sum_{n=1}^{\infty}\left(n-iam+iam\right)^{1+2l}\left(n-iam\right)^{-s}\\ \nonumber
& = i^{-s}\sum_{p=0}^{2l+1} \binom{2l+1}{p}(iam)^{2l+1-p}\zeta_H(s-p,1-iam)\,.
\end{align}
Esta expresión, junto con una similar para la otra suma sobre $n$, nos permite obtener
\begin{align}
\zeta_{S^{2k+2}}(s) ={} & {}  (\mu a)^s \frac{2^{k+1}}{(2k+1)!}\sum_{l=0}^{k} c^{(2k+2)}_l  \sum_{p=0}^{2l+1}  \binom{2l+1}{p}(iam)^{2l+1-p} \\ \nonumber
& \times \Big\{ i^{-s} \zeta_H(s-p,1-iam) + (-i)^{-s} (-1)^{1-s-p}\zeta_H(s-p,iam) \Big\}\,.
\end{align}
Puede verse que esta función zeta no se anula en $s=0$, por lo que la correspondiente acción efectiva tendrá un término proporcional a $\log(\mu a)$. Eligiendo otra vez la fase desacoplante, tenemos 
\vfill\pagebreak

\begin{align}
\Gamma^{S^{2k+2}} ={} & {} \frac{2^{k+1}}{(2k+1)!}\sum_{l=0}^{k} c^{(2k+2)}_l  \sum_{p=0}^{2l+1}  \binom{2l+1}{p}(iam)^{2l+1-p} \\ \nonumber 
&\times \Big\{ \log(\mu a) \left[ \zeta_H(-p,1-iam)+(-1)^{1+p}\zeta_H(-p,iam)\right] \\ \nonumber
& +\zeta'_H(-p,1-iam)+(-1)^{1+p}\zeta'_H(-p,iam) \\ \nonumber
& -i\pi\left[ \zeta_H(-p,1-iam)-(-1)^{1+p}\zeta_H(-p,iam)\right] \Big\}\,,
\end{align}
que, haciendo uso de la propiedad $\zeta_H(-p,1-x)=(-1)^{1+p}\zeta_H(-p,x)$, podemos finalmente reescribir como
\begin{align}\label{eq:seffevendirac}
\Gamma^{S^{2k+2}} = & {} \frac{2^{k+1}}{(2k+1)!}\sum_{l=0}^{k} c^{(2k+2)}_l  \sum_{p=0}^{2l+1}  \binom{2l+1}{p}(iam)^{2l+1-p} \\ \nonumber
&\times \Big\{2(-1)^{1+p}\zeta_H(-p,iam)\log(\mu a) +\zeta'_H(-p,1-iam)\\ \nonumber
& \qquad +(-1)^{1+p}\zeta'_H(-p,iam) \Big\}\,.
\end{align}
Observamos que, como sucedía en el caso del campo escalar masivo, la acción efectiva depende del regulador $\mu$, lo cual es un problema, puesto que su valor es ambiguo. Sin embargo, veremos a continuación que en la combinación ``holonómica'' que nos interesa el término problemático se cancelará con el correspondiente en el espacio proyectivo. El coeficiente de este término en la acción efectiva es proporcional a la anomalía de traza del tensor de energía impulso de la teoría; al final de esta sección reportaremos los valores de dicha cantidad para esferas en dimensión arbitraria, para compararlos con los presentes en la literatura para las dimensiones más bajas.

En el caso de los espacios proyectivos $S^{2k+2}\!/Z_2$ en dimensión par con las distintas estructuras de spin, donde el espectro está reducido a los autovalores \eqref{eq:spectrumdiracproj}, un cálculo similar al que hicimos para la esfera nos permite obtener
\begin{align}\label{eq:seffprojevendirac} 
\Gamma^{S^{2k+2}\!/Z_2,(\pm)} = & {} \frac{2^{k+2}}{(2k+1)!}\sum_{l=0}^{k} c^{(2k+2)}_l \sum_{p=0}^{2l+1}  \binom{2l+1}{p}2^p(iam)^{2l+1-p} \\ \nonumber 
&\times \Big\{ \log\frac{\mu a}{2} \left[\zeta_H\!\left(-p,\tfrac14\mp\tfrac14-\tfrac{iam}{2}\right) -(-1)^{p}\zeta_H\!\left(-p,\tfrac14\pm \tfrac14+\tfrac{iam}{2}\right)\right]  \\ \nonumber 
&  +\zeta'_H\!\left(-p,\tfrac14\mp \tfrac14-\tfrac{iam}{2}\right)-(-1)^{p}\zeta'_H\!\left(-p,\tfrac14\pm \tfrac14+\tfrac{iam}{2}\right) \\  \nonumber
&  - \frac{i\pi}{2} \left[\zeta_H\!\left(-p,\tfrac14\mp \tfrac14-\tfrac{iam}{2}\right) + (-1)^{p}\zeta_H\!\left(-p,\tfrac14\pm \tfrac14+\tfrac{iam}{2}\right)\right] \Big\}\,.
\end{align}

\vfill\pagebreak

Las dos acciones efectivas $\Gamma^{S^{2k+2}\!/Z_2,(\pm)}$ para una dimensión dada resultan ser complejas conjugadas una de la otra, como se desprende del hecho de que sus invariantes eta son opuestos \cite{bar2000dependence}. Además, su parte real es la mitad de la parte real de la acción efectiva \eqref{eq:seffevendirac} sobre la esfera. Con esto y la definición \eqref{eq:seffprojdirac}, la entropía de holonomía resulta
\begin{align}\label{eq:sholevendirac}
S_{\mathrm{hol},S^{2k+2}\!/Z_2} = \log\cos\Im\left(\Gamma^{S^{2k+2}\!/Z_2,(+)}\right)\,.
\end{align}

\begin{figure}[h!]
	\centering
	\includegraphics[height=6cm]{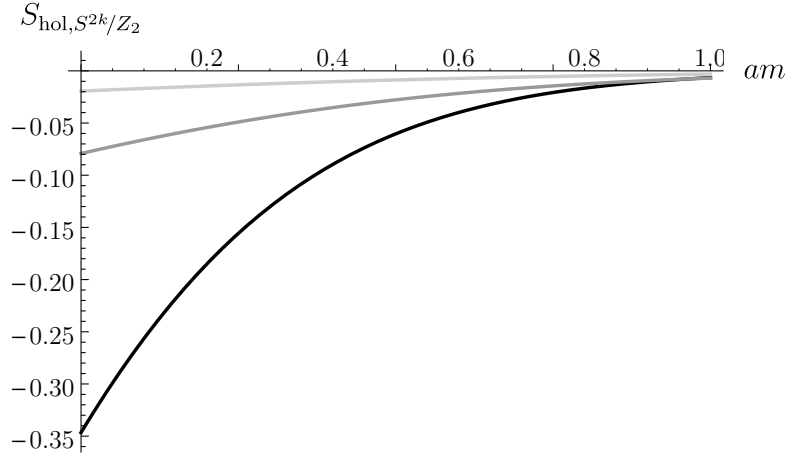}
	\caption{Entropía de holonomía como función de la masa adimensionalizada para espacios proyectivos en dimensión 2 (curva negra), 4 (curva gris oscura) y 6 (curva gris clara).} \label{figure:diracevend}
\end{figure}

En la figura \ref{figure:diracevend} puede verse el comportamiento de esta entropía de holonomía como función de la constante de acoplamiento ---que para esta teoría es la masa--- adimensionalizada, para las primeras dimensiones pares. Vemos que su valor es siempre negativo, lo cual derriba la posibilidad de considerarla como una cantidad $C$, en contra de lo que sucedía con el campo escalar. Sin embargo, podemos observar que su dependencia con la masa es monótona, lo que resulta en sí mismo no trivial.

\subsection[La anomalía de traza para fermiones en dimensión par]{Digresión: la anomalía de traza para fermiones en dimensión par}

Apartándonos de la línea principal del capítulo, aprovechamos los cálculos realizados sobre esferas en dimensión par para hacer algunos comentarios con respecto al coeficiente del término $\log(\mu a)$ en la acción efectiva \eqref{eq:seffevendirac} ---que, como es bien conocido, coincide a menos de un factor con la anomalía de traza del tensor de energía impulso de la teoría---,  

\begin{align}\label{eq:anom}
\mathcal{A}^{(2k+2)}(m) = & {} \frac{2^{k+2}}{(2k+1)!}\sum_{l=0}^{k}c_l^{(2k+2)}  \sum_{p=0}^{2l+1}  \binom{2l+1}{p}(iam)^{2l+1-p} \\\nonumber 
& \qquad\qquad\qquad\qquad\qquad\qquad\times(-1)^{1+p}\zeta_H(-p,iam)\,.
\end{align}
Notemos en primer lugar que $\mathcal{A}$ es un número real. En efecto, escribiendo la función zeta de Hurwitz en términos de polinomios de Bernoulli \cite[página 1037]{Gradshteyn} y usando la definición de éstos en términos de los números de Bernoulli $B_q$ \cite[página 1041]{Gradshteyn} obtenemos
\begin{align}\label{eq:terms}
(iam)^{2l+1-p}\zeta_H(-p,iam) = -\frac{1}{p+1}\sum_{q=0}^{p+1}\binom{p+1}{q} (iam)^{2l+2-q} B_q \,.
\end{align}
De aquí vemos inmediatamente que los términos con $q$ par son números reales. Con los términos correspondientes a valores impares de $q$ tendremos que tener alguna precaución: como los números de Bernoulli con índice impar son nulos a excepción de $B_1=-1/2$, podemos escribir
\begin{align}
-\frac{1}{p+1}\sum_{\substack{q=0\\q\mathrm{\, impar}}}^{p+1}\binom{p+1}{q} (iam)^{2l+2-q} B_q  = -\frac{1}{p+1}\binom{p+1}{1}  (iam)^{2l+1} B_1\,,
\end{align}
que una vez reemplazado en la ecuación \eqref{eq:anom} conduce a 
\begin{align}
\frac{2^{k+1}}{(2k+1)!}\sum_{l=0}^{k} c_l^{(2k+2)}  \sum_{p=0}^{2l+1}  \binom{2l+1}{p}(-1)^{1+p} (iam)^{2l+1}
\end{align}
o, lo que es lo mismo,
\begin{align}
-\frac{2^{k+1}}{(2k+1)!}\sum_{l=0}^{k}c_l^{(2k+2)}  (iam)^{2l+1}  (1-1)^{2l+1}\,,
\end{align}
lo que completa la prueba de que $\mathcal{A}$ es real.
En el caso no masivo este coeficiente asume la expresión más simple
\begin{align}
\mathcal{A}^{(2k+2)} = \frac{2^{k+2}}{(2k+1)!}\sum_{l=0}^{k} c_l^{(2k+2)}\zeta_R(-2l-1)\,,
\end{align}
o, en términos de números de Bernoulli, 
\begin{align}\label{eq:traceanom}
\mathcal{A}^{(2k+2)} = -\frac{2^{k+2}}{(2k+1)!}\sum_{l=0}^{k}c_l^{(2k+2)}\frac{B_{2l+2}}{2l+2}\,.
\end{align}

En la Tabla \ref{tabla:traceanom} pueden verse los valores de $\mathcal{A}$ para las dimensiones pares más bajas. El valor para $d=12$ está en desacuerdo con el reportado en \cite{Copeland:1985ua}, pero coincide con el calculado en \cite{Cappelli:2000fe}. Para el resto de las dimensiones, los valores obtenidos coinciden con los de ambas referencias.

\begin{table}[h]
	\centering
	\def\arraystretch{1.5}
	\begin{tabular}{l|ccccccccccc}
		$d$ & 2 & 4 & 6 & 8 & 10 & 12 & 14 \\
		\hline
		$\mathcal{A}^{(d)}$ & $-\frac{1}{3}$ & $\frac{11}{90}$ & $-\frac{191}{3780}$ & $\frac{2497}{113400}$ & $-\frac{14797}{1496880}$ & $\frac{92427157}{20432412000}$ & $-\frac{36740617}{17513496000}$ 
	\end{tabular}\\[2mm]
	\caption{Anomalía de traza para las dimensiones pares más bajas.}
	\label{tabla:traceanom}
\end{table}

\subsection{Entropía de holonomía en dimensión impar}

Volviendo a la entropía de holonomía, consideramos ahora el caso de una teoría de Dirac sobre un espacio proyectivo en dimensión impar $d=2k+1$. Luego del desplazamiento $n\rightarrow n-k$ en el espectro \eqref{eq:diracspectrum}, los autovalores se escriben
\begin{align}
\lambda_n^{\pm} = \pm \frac{i}{a}\left( n+\frac12\right)+m\,,\qquad n=k,k+1,k+2,\ldots\,,
\end{align}
con las correspondientes degeneraciones
\begin{align}\label{eq:degodddirac}
d_n = \frac{2^k}{(2k)!} \frac{(n+k)!}{(n-k)!}\,.
\end{align}

Una vez más, escribimos la cantidad
\begin{align}\nonumber
\frac{(n+k)!}{(n-k)!} & = (n+k)(n+k-1)(n+k-2)\ldots(n-k+1) \\ \label{eq:diracfracodd}
& = \left(n+\tfrac12+k-\tfrac12\right)\left(n+\tfrac12+k-\tfrac32\right)\ldots\left(n+\tfrac12-k+\tfrac12\right)\\ \nonumber
& = \left[\left(n+\tfrac12\right)^2-\left(k-\tfrac12\right)^2\right]\left[\left(n+\tfrac12\right)^2-\left(k-\tfrac32\right)^2\right]\ldots\left[\left(n+\tfrac12\right)^2-\tfrac14\right]
\end{align}
en términos de coeficientes racionales $c_l^{(2k+1)}$ como
\begin{align}
d_n = \frac{2^k}{(2k)!} \sum_{l=0}^{k}c_l^{(2k+1)}\left(n+\tfrac12\right)^{2l}\,.
\end{align}

\vfill\pagebreak

La función zeta correspondiente puede escribirse como
\begin{align}
\zeta_{S^{2k+1}}(s) = (\mu a)^s\frac{2^k}{(2k)!}\sum_{l=0}^{k}c_l^{(2k+1)} \Bigg\{ & i^{-s} \sum_{n=0}^{\infty}\left(n+\tfrac12\right)^{2l} \left(n+\tfrac12-iam\right)^{-s} \\ \nonumber
& + (-i)^{-s} \sum_{n=0}^{\infty}\left(n+\tfrac12\right)^{2l}\left(n+\tfrac12+iam\right)^{-s} \Bigg\}\,.
\end{align}
De manera análoga al caso par, usamos un desarrollo binomial para obtener
\begin{align}
\zeta_{S^{2k+1}}(s) ={} & (\mu a)^s\frac{2^k}{(2k)!}\sum_{l=0}^{k}c_l^{(2k+1)}  \sum_{p=0}^{2l}  \binom{2l}{p} (iam)^{2l-p} \\ \nonumber
& \times \Big\{  i^{-s} \zeta_H\!\left(s-p,\tfrac12-iam\right) + (-i)^{-s}(-1)^p\zeta_H\!\left(s-p,\tfrac12+iam\right) \Big\}\,.
\end{align}

Puede verse que esta función zeta se anula en $s=0$, como es de esperar en dimensión impar. Tomando una vez más el corte de la función logaritmo de modo que $i^s=\exp(i\pi s/2)$, podemos escribir
\begin{align}
\Gamma^{S^{2k+1}} ={} & \frac{2^k}{(2k)!}\sum_{l=0}^{k} c_l^{(2k+1)}  \sum_{p=0}^{2l}  \binom{2l}{p} (iam)^{2l-p}\\ \nonumber
&\times\Big\{ \zeta'_H\!\left(-p,\tfrac12-iam\right) + (-1)^{p}\zeta'_H\!\left(-p,\tfrac12+iam\right)\\\nonumber
&\qquad\qquad\qquad\qquad\qquad\qquad\qquad-i \pi \zeta_H\!\left(-p,\tfrac12-iam\right) \!\Big\}\,.
\end{align}

\begin{figure}[h!]
	\centering
	\includegraphics[height=6cm]{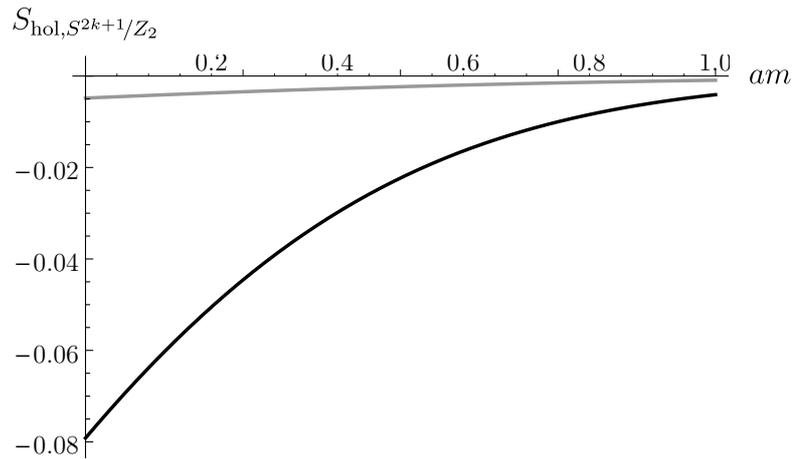}
	\caption{Entropía de holonomía como función de la masa adimensionalizada para espacios proyectivos en dimensión 3 (curva negra) y 7 (curva gris).} \label{fig:oddddirac}
\end{figure}

Sobre el espacio proyectivo $S^{2k+1}\!/Z_2$, un cálculo similar al del caso par permite obtener para las acciones efectivas correspondientes a las dos estructuras de spin las expresiones
\begin{align}\label{eq:seffdiracodd}
\Gamma^{S^{2k+1}\!/Z_2,(\pm)} ={} & \frac{2^k}{(2k)!}\sum_{l=0}^{k} c_l^{(2k+1)}  \sum_{p=0}^{2l}  \binom{2l}{p}2^p(iam)^{2l-p} \\ \nonumber
&\times \Big\{ \zeta'_H\!\left(-p,\tfrac12-\tfrac{iam}{2}\mp\tfrac14\right) + (-1)^{p} \zeta'_H\!\left(-p,\tfrac12+\tfrac{iam}{2}\pm\tfrac14\right) \\ \nonumber 
& -i\pi \zeta_H\!\left(-p,\tfrac12-\tfrac{iam}{2}\mp\tfrac14\right)\! \Big\}\,.
\end{align}
Al igual que antes, usamos el hecho de que estas acciones efectivas son complejas conjugadas entre sí y su parte real coincide con la mitad del valor de la acción efectiva sobre la correspondiente esfera para obtener la entropía de holonomía de la misma expresión que en el caso par, como en la ecuación \eqref{eq:sholevendirac}. En la figura \ref{fig:oddddirac} puede verse su comportamiento con la masa adimensionalizada para las dimensiones más bajas, que es similar al del caso par. Como en aquel caso, la dependencia con $am$ es monótona y ---como consecuencia de la elección que hicimos para la fase--- el campo se desacopla en el límite $am\rightarrow\infty$, pero la función es siempre negativa. Queda abierta la pregunta acerca de la posibilidad de definir la acción efectiva sobre el espacio proyectivo de otra forma, de modo de tener, por un lado, una acción efectiva real, pero además una entropía de holonomía positiva.

\begin{figure}[h!]
	\centering
	\includegraphics[height=6cm]{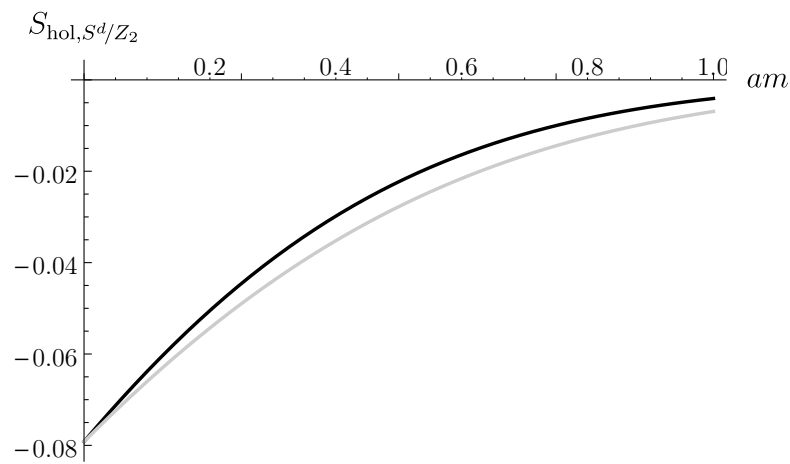}
	\caption{Entropía de holonomía como función de la masa adimensionalizada para espacios proyectivos en dimensión 3 (curva negra) y 4 (curva gris).} \label{fig:comparisondirac}
\end{figure}

Finalmente, puede verse que el valor de la entropía de holonomía para la teoría sin masa en dimensión $2k+1$ coincide con el valor de la misma cantidad en dimensión $2k+2$, aunque son diferentes para valores no nulos de la masa. 
Ilustramos este comportamiento con un ejemplo en la figura \ref{fig:comparisondirac}.

 %
%
\chapter{Conclusiones}
\label{sec:conclusiones}

\begin{minipage}{.885\textwidth}%
	\begin{flushright}
		\begin{minipage}{.66\textwidth}%
			\begin{flushleft}
				\emph{Mas depois?...\\
					Ah, meus poetas, meus poemas --- e depois?\\
					O pior é sempre o depois...}
			\end{flushleft}
			\begin{flushright}
				--- \textrm{Álvaro de Campos (Fernando Pessoa)} \\
				\emph{Tantos poemas contemporâneos!}
			\end{flushright}
		\end{minipage}%
	\end{flushright}
\end{minipage}%

\bigskip

La obtención de magnitudes cuánticas al orden de un loop en Teoría Cuántica de Campos está relacionada con el cálculo de determinantes funcionales. En esta tesis hemos utilizado para obtener esos determinantes la regularización analítica que se define a partir de la función zeta de los operadores correspondientes. Como aplicación principal de dicha regularización, calculamos las correcciones a un loop a las acciones efectivas de teorías escalares libres sobre los llamados \emph{espacios esféricos}, que se obtienen a partir de la esfera tridimensional mediante el cociente con ciertos subgrupos finitos de su grupo de isometrías.

En primer lugar, consideramos la teoría a temperatura finita $1/\beta$ de un campo escalar sin masa acoplado conformemente a la métrica en cada uno de los espacios esféricos, para la que obtuvimos dos expresiones analíticas diferentes de la acción efectiva, ambas válidas en todo el rango de temperaturas; cada una de estas expresiones es conveniente para tomar uno de los límites de temperatura \cite{Asorey:2012vp}. En ambos límites la dependencia con la temperatura está dada por potencias de la variable adimensional $\beta/a$ y correcciones adicionales suprimidas exponencialmente. En el límite de altas temperaturas la contribución dominante diverge como $\beta^{-3}$, de acuerdo con la ley de Stefan-Boltzmann, y tiene un coeficiente proporcional al volumen del espacio. La contribución subdominante es un término que en el caso de la entropía no depende de la temperatura o el volumen del espacio, siendo diferente para los distintos espacios esféricos. A partir de este término, que coincide con la acción efectiva de la teoría sin temperatura sobre la parte espacial de la variedad, definimos una cantidad que de algún modo mide la topología del espacio, a la que llamamos \emph{entropía topológica}. El resto del desarrollo se anula exponencialmente con $\beta/a$. En el límite de bajas temperaturas, por otra parte, la contribución dominante está dada por la energía de vacío, y el resto de los términos se anula exponencialmente cuando $\beta/a\rightarrow\infty$. Puede verse que allí la entropía se anula para todas las variedades consideradas, en acuerdo con la primera ley de la Termodinámica. En el caso de la esfera reproducimos la relación de dualidad entre los valores de la energía a altas y bajas temperaturas \cite{Dowker2002405}; en el resto de los espacios esféricos esta dualidad falla debido a la dependencia de la energía de vacío con la topología.

En el caso de los espacios lente ---cocientes de la esfera con grupos cíclicos $Z_p$--- encontramos que la entropía topológica crece con el orden $p$ del grupo, resultado que puede extenderse a valores continuos de $p$ como en \eqref{eq:stop4cont}. Esto sugiere la interpretación de $p$ como una especie de temperatura, aunque en nuestro análisis esta similitud no va más allá de la observación de que la entropía topológica es una función creciente de $p$, como en la segunda ley de la Termodinámica. Un problema que podría estar relacionado con esta idea es la dualidad formulada en \cite{Shaghoulian:2016gol}, que consiste en un intercambio entre $p$ y la temperatura en el límite en el que ambas cantidades tienden a infinito. Encontramos el mismo comportamiento con el orden del grupo para la familia de espacios que denominamos prisma ---en los que el cociente de la esfera es con subgrupos de $SU(2)$ que en $SO(3)$ corresponden a las simetrías de polígonos en el plano---.

Por otra parte, la entropía topológica es subaditiva en cada uno de los espacios esféricos, a la manera de una entropía de entrelazamiento. Teniendo en cuenta que una entropía de entrelazamiento fue utilizada para el estudio de un teorema $C$ en tres dimensiones \cite{Casini:2012ei}, como una aplicación de los cálculos sobre espacios esféricos nos preguntamos acerca de las propiedades de la entropía topológica a lo largo del flujo del grupo de renormalización. Analizamos entonces el caso más simple del flujo de masa de la teoría escalar libre sobre los distintos espacios esféricos, para la que calculamos las correspondientes acciones efectivas a temperatura finita \cite{Asorey:2014gsa}. Comenzamos analizando el caso de la esfera, en el que encontramos que la presencia de tres escalas independientes ---masa, radio del espacio y temperatura--- introduce una dependencia de la acción efectiva con el regulador $\mu$, que eliminamos por medio de una prescripción de renormalización. Luego de ese procedimiento obtenemos acciones efectivas en los diferentes regímenes de valores relativos de las escalas. A partir de las entropías de cada espacio esférico y de la esfera en el límite en el que la temperatura tiende a infinito más rápido que cualquier otro parámetro definimos, de manera similar a la que dio lugar a la entropía topológica en el caso no masivo, una diferencia que llamamos \emph{entropía de holonomía}, que depende de la masa y el radio de la esfera que cubre el espacio, y cuyo valor a masa cero es la entropía topológica del espacio en cuestión. Para la teoría escalar libre la constante de acoplamiento es el cuadrado de la masa del campo, y el flujo del grupo de renormalización se implementa variando esa cantidad. En todos los casos estudiados la entropía de holonomía es siempre positiva y decrece monótonamente con el flujo, anulándose en el punto fijo infrarrojo. Esto que se corresponde con el comportamiento de una cantidad $C$, al menos en una formulación débil \cite{Barnes:2004jj}: una cantidad $C$ como la de Zamolodchikov \cite{Zamolodchikov:1986gt} satisface además la condición de estabilidad en los puntos fijos del flujo, y la entropía de holonomía tiene en cambio una derivada con respecto a la constante de acoplamiento que es nula en el punto fijo infrarrojo ---correspondiente a la teoría desacoplada en el límite $m\rightarrow\infty$--- pero que es negativa en el punto fijo ultravioleta ---que corresponde a la teoría sin masa---. 

Con ese resultado, nos preguntamos si la entropía de holonomía define una cantidad $C$ más allá del ejemplo del flujo de masa del campo escalar libre en tres dimensiones. Dado que en ese caso los términos constantes en el desarrollo a altas temperaturas de las acciones efectivas pueden obtenerse a partir del determinante de la teoría a temperatura cero sobre la parte espacial de la variedad, estudiamos la versión tridimensional de la entropía de holonomía considerando la teoría con acoplamiento conforme en tres dimensiones a la variedad espacial, para la que encontramos un comportamiento análogo al de la anterior. Generalizamos luego estos cálculos a la esfera y el espacio proyectivo real en dimensión arbitraria, encontrando en todos los casos un comportamiento similar. 

Como primer ejemplo diferente de la teoría escalar consideramos el flujo de masa del campo de Dirac libre sobre esferas y espacios proyectivos en dimensión arbitraria. En el caso de los espacios proyectivos, la presencia de dos estructuras de (s)pin introduce una ambigüedad en la elección de los espinores. Debido a la presencia de una asimetría espectral no nula, las acciones efectivas sobre ambas estructuras resultan complejas, por lo que una teoría definida con sólo una de ellas no es unitaria. Por otra parte, dichas acciones efectivas son conjugadas entre sí y su parte real es la mitad de la acción efectiva sobre la esfera. Esto permite recuperar la unitariedad definiendo la función de partición sobre el espacio proyectivo como el promedio de las funciones de partición sobre las distintas estructuras de (s)pin. Con esa prescripción, los cálculos muestran que, a diferencia del caso escalar, la entropía de holonomía es negativa y creciente a lo largo del flujo, por lo que no es compatible con una cantidad $C$. Sin embargo, su comportamiento es monótono y presenta algunas similitudes con su contraparte escalar: su límite en el punto fijo infrarrojo $m\rightarrow\infty$ es nulo ---lo cual es consecuencia de la elección de la fase en el determinante de Dirac--- y su derivada con respecto a la constante de acoplamiento ---que en este caso es simplemente la masa--- en el punto fijo ultravioleta tiene el mismo signo que a lo largo del flujo. Dicho en otras palabras, puede verse que la entropía de holonomía cambiada de signo se comporta con la constante de acoplamiento para la teoría fermiónica sobre espacios proyectivos reales de la misma forma que la entropía de holonomía para la teoría escalar.

Los cálculos para las teorías escalar y de Dirac sobre la esfera tridimensional nos permitieron investigar \cite{Beneventano:2017eyu} el comportamiento frente al flujo de masa de algunas cantidades relacionadas con la acción efectiva que en el punto fijo ultravioleta de dicho flujo coinciden con una cantidad $C$ ya conocida \cite{Klebanov:2011gs}. De ese análisis encontramos que, en el mejor de los casos, la función interpolante se comporta de la misma forma que la entropía de holonomía.

Con estos resultados se presentan varias direcciones de acción. Por un lado, se impone un análisis más detallado de la entropía de holonomía en el caso fermiónico que justifique, de ser posible, el cambio de signo con respecto al caso bosónico. Podría buscarse, por ejemplo, una definición de la acción efectiva sobre espacios proyectivos que, siendo real, dé lugar a una entropía de holonomía con las propiedades requeridas. Otra posibilidad es la búsqueda de una relación entre ésta y alguna propiedad ya estudiada en el marco de los teoremas $C$, como la que obtuvimos en la ecuación \eqref{eq:rel34} para el caso escalar. Por otra parte, el estudio de teorías en interacción podría arrojar alguna información adicional acerca de las propiedades de la entropía de holonomía; el modelo $O(N)$ en tres dimensiones es de particular interés por haber constituido el primer ejemplo de la no monotonicidad a lo largo del flujo del grupo de renormalización del coeficiente de la potencia de la temperatura en una expresión de la entropía térmica \cite{Sachdev:1993pr}. 

Desde un punto de vista más general, si la entropía de holonomía fuera una buena medida del número de grados de libertad podríamos intentar reformular su definición para poder considerar teorías sobre otros espacios. Para ello, una posibilidad sería explorar la estructura de la acción efectiva, aislando los términos que correspondan a las contribuciones a la integral funcional de los caminos cerrados no contraíbles sobre la variedad, para poder identificarlos con contribuciones topológicas en el caso conforme \cite{Asorey:2011cm}.

Finalmente, queremos destacar la simplicidad con que la regularización zeta permite calcular cantidades físicas en teorías sobre espacios con topología no trivial. A los ejemplos considerados podrían también agregarse espacios que no tengan curvatura positiva: en el caso tridimensional han sido estudiados, entre otros \cite{Lima:2006rr,Sutter:2006dj}, el toro \cite{Kleinert:1996sc} y la botella de Klein \cite{DeWitt:1979dd,Dowker:1978vy}, y en espacios hiperbólicos la función de partición está relacionada con la entropía de entrelazamiento en esferas \cite{Casini:2011kv,Myers:2010tj}.

\cleardoublepage

{%
\setstretch{1.1}
\renewcommand{\bibfont}{\normalfont\small}
\setlength{\biblabelsep}{4pt}
\setlength{\bibitemsep}{0.5\baselineskip plus 0.5\baselineskip}
\printbibliography
\nocite{*}
}


\vfill

\pagebreak

%
\pagestyle{empty}
\hfill
\vfill
\pdfbookmark[0]{Epílogo}{Epílogo}



\null
\vfill

\begin{spacing}{1}
\noindent{\small The composition of the text of this thesis was done in \LaTeX, using the \TeX studio editor, in a style adapted from \texttt{cleanthesis}. Figures were drawn in 
\textit{Mathematica} 11 using the package MaTeX. 
}
\end{spacing}

\clearpage

\end{document}